\documentclass[11pt]{article}

\usepackage{pstricks}
\usepackage{pst-plot}
\usepackage{pst-node}
\usepackage{amsmath}
\usepackage{pst-grad}
\usepackage{amssymb}
\usepackage{xspace}
\usepackage{pst-math}
\usepackage{pst-xkey}
\usepackage{pst-coil}
\usepackage{pst-3dplot}
\usepackage[T1]{fontenc}
\usepackage{ae}
\usepackage[ansinew]{inputenc}

\newrgbcolor{darkred}{0.8 0 0}
\newrgbcolor{darkerred}{0.7 0 0}
\newrgbcolor{darkestred}{0.5 0 0}
\newrgbcolor{darkgreen}{0 0.5 0.5}
\newrgbcolor{darkergreen}{0 0.6 0}
\newrgbcolor{darkestgreen}{0 0.5 0}
\newrgbcolor{darkblue}{0 0 0.6}
\newrgbcolor{darkestblue}{0 0 0.5}
\newrgbcolor{redgreen}{0.8 0.8 0}
\newrgbcolor{bluegreen}{0 0.4 0.7}
\newrgbcolor{bluewhite}{0 0.9 1}
\newrgbcolor{lightgreen}{0.6 1 0.6}
\newrgbcolor{lightred}{1 0.8 0.8}
\newrgbcolor{lightyred}{1 0.4 0.4}
\newrgbcolor{lightyblue}{0.5 0.5 1}
\newrgbcolor{lightblue}{0.8 1 1}
\newrgbcolor{lilas}{0.5 0.3 0.7}
\newrgbcolor{lilaswhite}{1 0.8 1}
\newrgbcolor{redy}{1 0.6 0.6}
\newrgbcolor{bluish}{0.6 0.6 1}
\newrgbcolor{sand}{0.9 0.9 0.7}
\newrgbcolor{brown}{0.4 0.4 0}
\newrgbcolor{blue1}{0 0.35 0.65}
\newrgbcolor{blue2}{0 0.4 0.6}
\newrgbcolor{blue3}{0 0.45 0.55}
\newrgbcolor{blue4}{0 0.5 0.5}
\newrgbcolor{blue5}{0 0.55 0.45}
\newrgbcolor{blue6}{0 0.6 0.4}
\newrgbcolor{blue7}{0 0.65 0.35}

\newcommand{\R }{\mathbb R}
\newcommand{\N }{\mathbb N}
\newcommand{\Z }{\mathbb Z}
\newcommand{\Q }{\mathbb Q}
\newcommand{\C }{\mathbb C}
\newcommand{\sen }{\ \! {\rm sen}\ \! }
\newcommand{\tg }{\ \! {\rm tg}\ \! }
\newcommand{\cosec }{\ \! {\rm cosec}\ \! }
\newcommand{\cotg }{\ \! {\rm cotg}\ \! }
\newcommand{\dx }{\ \! dx}
\newcommand{\e }{\ \! e}
\newcommand{\senh }{\ \! {\rm senh}\ \! }
\newcommand{\tgh }{\ \! {\rm tgh}\ \! }
\newcommand{\cotgh }{\ \! {\rm cotgh}\ \! }
\newcommand{\sech }{\ \! {\rm sech}\ \! }
\newcommand{\cosech }{\ \! {\rm cosech}\ \! }
\newcommand{\arcsen }{\ \! {\rm arcsen}\ \! }
\newcommand{\arctg }{\ \! {\rm arctg}\ \! }
\newcommand{\arccotg }{\ \! {\rm arccotg}\ \! }
\newcommand{\arcsenh }{\ \! {\rm arcsenh}\ \! }
\newcommand{\arccosh }{\ \! {\rm arccosh}\ \! }
\newcommand{\arctgh }{\ \! {\rm arctgh}\ \! }
\newcommand{\arccosech }{\ \! {\rm arccosech}\ \! }
\newcommand{\arcsech }{\ \! {\rm arcsech}\ \! }
\newcommand{\arcsec }{\ \! {\rm arcsec}\ \! }
\newcommand{\arccosec }{\ \! {\rm arccosec}\ \! }
\renewcommand{\arctgh }{\ \! {\rm arctgh}\ \! }
\newcommand{\arccotgh }{\ \! {\rm arccotgh}\ \! }
\newcommand{\nega }{\neg \ }
\newcommand{\eq }{\Leftrightarrow }

\begin{document}

\title{Correlation of financial markets in times of crisis}

\author{Leonidas Sandoval Junior\thanks{E-mail: leonidassj@insper.org.br (corresponding author)} \\ Italo De Paula Franca \\ \\ Insper, Instituto de Ensino e Pesquisa}

\maketitle

\begin{abstract}
Using the eigenvalues and eigenvectors of correlations matrices of some of the main financial market indices in the world, we show that high volatility of markets is directly linked with strong correlations between them. This means that markets tend to behave as one during great crashes. In order to do so, we investigate financial market crises that occurred in the years 1987 (Black Monday), 1998 (Russian crisis), 2001 (Burst of the dot-com bubble and September 11), and 2008 (Subprime Mortgage Crisis), which mark some of the largest downturns of financial markets in the last three decades. \end{abstract}

\noindent PACS numbers: 89.65.Gh, 89.75.-k.

\section{Introduction}

The study of why many world financial markets crash simultaneously is of central importance, particularly after the recent worldwide downturn of the major markets in 2007 and 2008. Economists have been studying the reasons why markets crash, and why there is propagation of volatility from one market to another, since a long time. After the crash of 1987, many studies have been published on transmision of volatility (contagion) between markets using econometric models \cite{cont1}-\cite{cont15}, on how the correlation between world markets change with time \cite{time1}-\cite{time4}, and how correlation tends to increase in times of high volatility \cite{vol1}-\cite{vol15}. This issue is of particular importance if one wishes to build portfolios of international assets which can withstand times of crisis \cite{port1}-\cite{port13}. Many models were proposed by both economists and physicists in order to explain the correlation of international financial markets \cite{mod1}-\cite{mod19}, which is considered a complex system with many relations which are difficult to identify and quantify.

One tool that was first developed in nuclear physics for studying complex systems with unknown correlation structure is random matrix theory \cite{rmt1}-\cite{rmt4}, which confronts the results obtained for the eigenvalues of the correlation matrix of a real system with those of the correlation matrix obtained from a pure random matrix. This approach was successfully applied to a large number of financial markets \cite{rmt5}-\cite{rmt34}, and also to the relation between world markets \cite{world1}-\cite{world2}. This approach was also used in the construction of hierarchical structures between different assets of financial markets \cite{h1}-\cite{h40}.

Our work uses the tools of random matrix theory to analyze the correlation of world financial markets in times of crisis. In order to do so, we use data from some of the largest worldwide crashes since 1980, namely the 1987 Black Monday, the 1998 Russian Crisis, the Burst of the dot-com bubble of 2001, the shock after September, 11, 2001, and the USA subprime mortgage crisis of 2008. We start by defining a global financial crisis (Section 2) based on evidence of some financial markets chosen from diverse parts of the world. Then, we discuss some of the main theoretical results on Random Matrix Theory (Section 3). Then, in Section 4, we make a quick discussion on how we collected the data and how it was treated.

In the next three sections we study the correlation matrices between the log-returns of a number of financial market indices chosen so as to represent many geographical parts of the world and a diversity of economies (sections 5 to 8). In each section, we calculate the eigenvalues of the correlation matrix of the chosen indices and then study the eigenvector that corresponds to its largest eigenvalue, which is usually related with a {\sl market mode}, which is a co-movement of all indices. We then calculate correlation matrices in running windows and compare the average correlation between markets with the volatility and average volatility of the market mode obtained previously, showing that times of large volatility are strongly linked with strong correlations between world financial indices.

Section 9 looks more closely at the probability distribution of the correlation coefficients in different intervals of time and tests the hypothesis that it becomes closer to a normal during periods of crises.

Since the study of world stock exchanges involve dealing with different operating times, in Section 10, we compare the results obtained in the previous sections with results obtained by using the log-returns of Western markets with the log-returns of the next day in Asian markets. We also compare the results obtained in the main text of the article with those obtained by using Spearman's rank correlation instead of Pearson's correlation.

We finish by discussing recent methods for studying random matrices obtained from non-Gaussian distributions, shuch as t-Student distributions, which represent more closely the probability density distributions obtained from financial data \cite{ext1}-\cite{ext4}, in assossiation with other measures of co-movement of financial indices that are more appropriate for systems with very strong correlations, like it happens in times of financial crises \cite{ext5} \cite{ext6}.

Since world stock market indices (countries) are easier to relate with than equities in a stock market (companies), one of the aims of this article is to be a pedagogical introduction to most of the techniques that are used when Random Matrix Theory is applied to financial data. Hence we also supply an ample bibliography on the subject.

\section{Defining a global financial crisis}

Before studying periods of financial crises, we must make it clear what we consider to be a global crash of the financial markets. In order to adopt a more precise definition, we considered the time series of 15 financial markets representing different regions of the world from the beginning of 1985 until the end of 2010. Looking at the closing indices of every day in which there was negotiation, we considered the log-returns, given by
\begin{equation}
S_t=\ln (P_t)-\ln (P_{t-1})\approx \frac{P_t-P_{t-1}}{P_t}\ ,
\end{equation}
what makes it easier to compare the variations of the many indices. After that, the 10 most negative variations were chosen.

In order to illustrate the procedure, we consider the Dow Jones index of the New York Stock Exchange (NYSE). Figure 1 shows the log-density distribution for this index with data from 01/02/1985 to 12/31/2008. The log-density, defined as \begin{equation} \text{log-density}=\ln (1+\text{density})\ ,\end{equation}
is used instead of simple density in order to better visualize the most extreme points.

\begin{pspicture}(-10,-0.5)(3,4.6)
\psset{xunit=30,yunit=2.5} \psline{->}(-0.3,0)(0.2,0) \psline{->}(-0.3,0)(-0.3,1.6) \rput(0.23,0){interval} \rput(-0.262,1.6){log-density}
\scriptsize \psline(-0.3,-0.04)(-0.3,0.04) \rput(-0.3,-0.12){$-0.3$} \psline(-0.2,-0.04)(-0.2,0.04) \rput(-0.2,-0.12){$-0.2$} \psline(-0.1,-0.04)(-0.1,0.04) \rput(-0.1,-0.12){$-0.1$} \psline(0,-0.04)(0,0.04) \rput(0,-0.12){$0$} \psline(0.1,-0.04)(0.1,0.04) \rput(0.1,-0.12){$0.1$} \psline(-0.303,0)(-0.297,0) \rput(-0.315,0){0} \psline(-0.303,0.4)(-0.297,0.4) \rput(-0.315,0.4){$0.4$} \psline(-0.303,0.8)(-0.297,0.8) \rput(-0.315,0.8){$0.8$} \psline(-0.303,1.2)(-0.297,1.2) \rput(-0.315,1.2){$1.2$}
\pspolygon*[linecolor=lightred](-0.26,0)(-0.26,0.1477)(-0.256,0.1477)(-0.256,0)
\pspolygon[linecolor=red](-0.26,0)(-0.26,0.1477)(-0.256,0.1477)(-0.256,0)
\pspolygon*[linecolor=lightred](-0.084,0)(-0.084,0.2955)(-0.08,0.2955)(-0.08,0)
\pspolygon[linecolor=red](-0.084,0)(-0.084,0.2955)(-0.08,0.2955)(-0.08,0)
\pspolygon*[linecolor=lightred](-0.08,0)(-0.08,0.1477)(-0.076,0.1477)(-0.076,0)
\pspolygon[linecolor=red](-0.08,0)(-0.08,0.1477)(-0.076,0.1477)(-0.076,0)
\pspolygon*[linecolor=lightred](-0.076,0)(-0.076,0.2955)(-0.072,0.2955)(-0.072,0)
\pspolygon[linecolor=red](-0.076,0)(-0.076,0.2955)(-0.072,0.2955)(-0.072,0)
\pspolygon*[linecolor=lightred](-0.072,0)(-0.072,0.2342)(-0.068,0.2342)(-0.068,0)
\pspolygon[linecolor=red](-0.072,0)(-0.072,0.2342)(-0.068,0.2342)(-0.068,0)
\pspolygon*[linecolor=lightred](-0.068,0)(-0.068,0.1477)(-0.064,0.1477)(-0.064,0)
\pspolygon[linecolor=red](-0.068,0)(-0.068,0.1477)(-0.064,0.1477)(-0.064,0)
\pspolygon*[linecolor=lightred](-0.06,0)(-0.06,0.2955)(-0.056,0.2955)(-0.056,0)
\pspolygon[linecolor=red](-0.06,0)(-0.06,0.2955)(-0.056,0.2955)(-0.056,0)
\pspolygon*[linecolor=lightred](-0.056,0)(-0.056,0.2342)(-0.052,0.2342)(-0.052,0)
\pspolygon[linecolor=red](-0.056,0)(-0.056,0.2342)(-0.052,0.2342)(-0.052,0)
\pspolygon*[linecolor=lightred](-0.052,0)(-0.052,0.3430)(-0.048,0.3430)(-0.048,0)
\pspolygon[linecolor=red](-0.052,0)(-0.052,0.3430)(-0.048,0.3430)(-0.048,0)
\pspolygon*[linecolor=lightred](-0.048,0)(-0.048,0.3819)(-0.044,0.3819)(-0.044,0)
\pspolygon[linecolor=red](-0.048,0)(-0.048,0.3819)(-0.044,0.3819)(-0.044,0)
\pspolygon*[linecolor=lightred](-0.044,0)(-0.044,0.4148)(-0.04,0.4148)(-0.04,0)
\pspolygon[linecolor=red](-0.044,0)(-0.044,0.4148)(-0.04,0.4148)(-0.04,0)
\pspolygon*[linecolor=lightred](-0.04,0)(-0.04,0.5296)(-0.036,0.5296)(-0.036,0)
\pspolygon[linecolor=red](-0.04,0)(-0.04,0.5296)(-0.036,0.5296)(-0.036,0)
\pspolygon*[linecolor=lightred](-0.036,0)(-0.036,0.5111)(-0.032,0.5111)(-0.032,0)
\pspolygon[linecolor=red](-0.036,0)(-0.036,0.5111)(-0.032,0.5111)(-0.032,0)
\pspolygon*[linecolor=lightred](-0.032,0)(-0.032,0.6944)(-0.028,0.6944)(-0.028,0)
\pspolygon[linecolor=red](-0.032,0)(-0.032,0.6944)(-0.028,0.6944)(-0.028,0)
\pspolygon*[linecolor=lightred](-0.028,0)(-0.028,0.8017)(-0.024,0.8017)(-0.024,0)
\pspolygon[linecolor=red](-0.028,0)(-0.028,0.8017)(-0.024,0.8017)(-0.024,0)
\pspolygon*[linecolor=lightred](-0.024,0)(-0.024,0.9340)(-0.02,0.9340)(-0.02,0)
\pspolygon[linecolor=red](-0.024,0)(-0.024,0.9340)(-0.02,0.9340)(-0.02,0)
\pspolygon*[linecolor=lightred](-0.02,0)(-0.02,1.0375)(-0.016,1.0375)(-0.016,0)
\pspolygon[linecolor=red](-0.02,0)(-0.02,1.0375)(-0.016,1.0375)(-0.016,0)
\pspolygon*[linecolor=lightred](-0.016,0)(-0.016,1.1477)(-0.012,1.1477)(-0.012,0)
\pspolygon[linecolor=red](-0.016,0)(-0.016,1.1477)(-0.012,1.1477)(-0.012,0)
\pspolygon*[linecolor=lightred](-0.012,0)(-0.012,1.2770)(-0.008,1.2770)(-0.008,0)
\pspolygon[linecolor=red](-0.012,0)(-0.012,1.2770)(-0.008,1.2770)(-0.008,0)
\pspolygon*[linecolor=lightred](-0.008,0)(-0.008,1.4029)(-0.004,1.4029)(-0.004,0)
\pspolygon[linecolor=red](-0.008,0)(-0.008,1.4029)(-0.004,1.4029)(-0.004,0)
\pspolygon*[linecolor=lightred](-0.004,0)(-0.004,1.5105)(0,1.5105)(0,0)
\pspolygon[linecolor=red](-0.004,0)(-0.004,1.5105)(0,1.5105)(0,0)
\pspolygon*[linecolor=lightred](0,0)(0,1.5224)(0.004,1.5224)(0.004,0)
\pspolygon[linecolor=red](0,0)(0,1.5224)(0.004,1.5224)(0.004,0)
\pspolygon*[linecolor=lightred](0.004,0)(0.004,1.4369)(0.008,1.4369)(0.008,0)
\pspolygon[linecolor=red](0.004,0)(0.004,1.4369)(0.008,1.4369)(0.008,0)
\pspolygon*[linecolor=lightred](0.008,0)(0.008,1.3271)(0.012,1.3271)(0.012,0)
\pspolygon[linecolor=red](0.008,0)(0.008,1.3271)(0.012,1.3271)(0.012,0)
\pspolygon*[linecolor=lightred](0.012,0)(0.012,1.1844)(0.016,1.1844)(0.016,0)
\pspolygon[linecolor=red](0.012,0)(0.012,1.1844)(0.016,1.1844)(0.016,0)
\pspolygon*[linecolor=lightred](0.016,0)(0.016,1.0637)(0.02,1.0637)(0.02,0)
\pspolygon[linecolor=red](0.016,0)(0.016,1.0637)(0.02,1.0637)(0.02,0)
\pspolygon*[linecolor=lightred](0.02,0)(0.02,0.9145)(0.024,0.9145)(0.024,0)
\pspolygon[linecolor=red](0.02,0)(0.02,0.9145)(0.024,0.9145)(0.024,0)
\pspolygon*[linecolor=lightred](0.024,0)(0.024,0.7638)(0.028,0.7638)(0.028,0)
\pspolygon[linecolor=red](0.024,0)(0.024,0.7638)(0.028,0.7638)(0.028,0)
\pspolygon*[linecolor=lightred](0.028,0)(0.028,0.7025)(0.032,0.7025)(0.032,0)
\pspolygon[linecolor=red](0.028,0)(0.028,0.7025)(0.032,0.7025)(0.032,0)
\pspolygon*[linecolor=lightred](0.032,0)(0.032,0.5906)(0.036,0.5906)(0.036,0)
\pspolygon[linecolor=red](0.032,0)(0.032,0.5906)(0.036,0.5906)(0.036,0)
\pspolygon*[linecolor=lightred](0.036,0)(0.036,0.3819)(0.04,0.3819)(0.04,0)
\pspolygon[linecolor=red](0.036,0)(0.036,0.3819)(0.04,0.3819)(0.04,0)
\pspolygon*[linecolor=lightred](0.04,0)(0.04,0.3819)(0.044,0.3819)(0.044,0)
\pspolygon[linecolor=red](0.04,0)(0.04,0.3819)(0.044,0.3819)(0.044,0)
\pspolygon*[linecolor=lightred](0.044,0)(0.044,0.4432)(0.048,0.4432)(0.048,0)
\pspolygon[linecolor=red](0.044,0)(0.044,0.4432)(0.048,0.4432)(0.048,0)
\pspolygon*[linecolor=lightred](0.048,0)(0.048,0.3819)(0.052,0.3819)(0.052,0)
\pspolygon[linecolor=red](0.048,0)(0.048,0.3819)(0.052,0.3819)(0.052,0)
\pspolygon*[linecolor=lightred](0.052,0)(0.052,0.1477)(0.056,0.1477)(0.056,0)
\pspolygon[linecolor=red](0.052,0)(0.052,0.1477)(0.056,0.1477)(0.056,0)
\pspolygon*[linecolor=lightred](0.056,0)(0.056,0.1477)(0.06,0.1477)(0.06,0)
\pspolygon[linecolor=red](0.056,0)(0.056,0.1477)(0.06,0.1477)(0.06,0)
\pspolygon*[linecolor=lightred](0.06,0)(0.06,0.2342)(0.064,0.2342)(0.064,0)
\pspolygon[linecolor=red](0.06,0)(0.06,0.2342)(0.064,0.2342)(0.064,0)
\pspolygon*[linecolor=lightred](0.064,0)(0.064,0.1477)(0.068,0.1477)(0.068,0)
\pspolygon[linecolor=red](0.064,0)(0.064,0.1477)(0.068,0.1477)(0.068,0)
\pspolygon*[linecolor=lightred](0.096,0)(0.096,0.1477)(0.1,0.1477)(0.1,0)
\pspolygon[linecolor=red](0.096,0)(0.096,0.1477)(0.1,0.1477)(0.1,0)
\pspolygon*[linecolor=lightred](0.1,0)(0.1,0.1477)(0.104,0.1477)(0.104,0)
\pspolygon[linecolor=red](0.1,0)(0.1,0.1477)(0.104,0.1477)(0.104,0)
\pspolygon*[linecolor=lightred](0.104,0)(0.104,0.1477)(0.108,0.1477)(0.108,0)
\pspolygon[linecolor=red](0.104,0)(0.104,0.1477)(0.108,0.1477)(0.108,0)
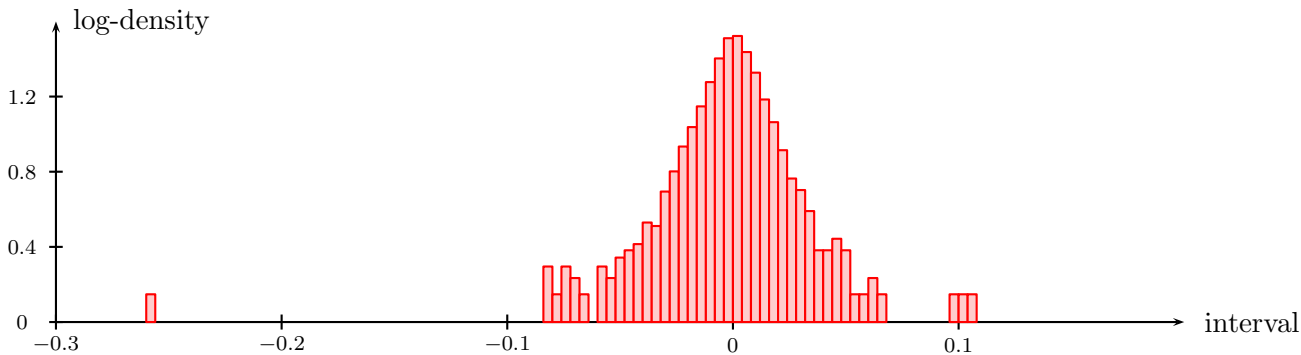
\rput(-0.03,-0.4){\small Figure 1: log-density distribution of the Dow Jones index of the NYSE, from 01/02/1985 to 12/31/2008.}
\end{pspicture}

\vskip 0.9 cm

The ten most negative values of the log-returns are bellow $-0.07$. These events occured in the following occasions: 10/19/1987 (22.61\%), 10/26/1987 (8.04\%), 01/08/1988 (6.85\%), 10/13/1989 (6.90\%), 10/27/1997 (7.18\%), 09/17/2001 (7.13\%), 09/29/2008 (6.98\%), 10/09/2008 (7.33\%), 10/15/2008 (7.87\%), and 12/01/2008 (7.70\%). These dates include the 1987 Black Monday, part of the Asian Crisis of 1997, the 1998 Russian Crisis, the aftermath of September 11, 2001, and the Subprime Mortgage Crisis of 2008.

The same technique was used for the Nasdaq (USA), S\&P/TSX Composite from Canada (Can), Ibovespa (Brazil), FTSE 100 (UK), DAX (Germany), ISEQ (Ireland), AEX (Netherlands), SENSEX 30 (India), Colombo All-Share (Sri Lanka), Nikkei (Japan), Hang Seng (Hong Kong), TAIEX (Taiwan), Kospi (South Korea), Kuala Lumpur Composite (Malaysia), and Jakarta Composite (Indonesia). The next table displays the years of major drops (between the beginning of 1985, and the end of 2010) and the quantity of markets which presented those falls. When a market drops substantially more than once in the same year, these are counted more than once as well, in order to gauge the depth of the shocks. This helps identity the times where there were major crashes around the world.

\vskip -0.3 cm

\[ \begin{array}{|c|ccccccccccccc|} \hline \text{Year} & 1985 & 1986 & 1987 & 1988 & 1989 & 1990 & 1991 & 1992 & 1993 & 1994 & 1995 & 1996 & 1997 \\ \hline \text{Occurences} & 3 & 0 & 29 & 2 & 9 & 13 & 2 & 4 & 0 & 0 & 0 & 1 & 10 \\ \hline \hline \text{Year} & 1998 & 1999 & 2000 & 2001 & 2002 & 2003 & 2004 & 2005 & 2006 & 2007 & 2008 & 2009 & 2010 \\ \hline \text{Occurences} & 11 & 1 & 4 & 8 & 1 & 3 & 4 & 3 & 0 & 0 & 50 & 2 & 0 \\ \hline \end{array} \]
\hskip 0.9 cm Table 1: number of occurences per year of major drops in fifteen diverse stock markets in the world.

\vskip 0.3 cm

It is possible to pinpoint two major crises in 1987 and 2008, and minor crises in 1989, 1990, 1997, 1998, and in 2001. The crisis of 1987 corresponds to the so called Black Monday, the one in 1989 is the USA saving and loan crisis, 1990 are the Japanese asset price bubble and the Scandinavian banking crisis, 1992 is the so-called Black Wednesday, 1997 is the Asian financial crisis, 1998 is the Russian crisis, 2000 and 2001 correspond to the Burst of the dot-com bubble, and 2008 corresponds to the Subprime Mortgage Crisis in the USA.

We shall apply a theory called {\sl Random Matrix Theory} in order to analyze four of these crises. The next section gives a pragmatic introduction to this theory.

\section{Random matrix theory}

Random matrix theory had its origins in 1953, in the work of the German physicist Eugene Wigner \cite{rmt1} \cite{rmt2}. He was studying the energy levels of complex atomic nuclei, such as uranium, and had no means of calculating the distance between those levels. He then assumed that those distances were random, and arranged the random number in a matrix which expressed the connections between the many energy levels. Surprisingly, he could then be able to make sensible predictions about how the energy levels related to one another.

This method also found connections with the study of the Riemann zeta function, which is of primordial importance to the study of prime numbers, used for coding and decoding information, for example. The theory was later developed, with many and surprising results arising. Today, Random Matrix Theory is applied to quantum physics, nanotechnology, quantum gravity, the study of the structure of crystals, and may have applications in ecology, linguistics, and many other fields where a large amount of apparently unrelated information may be understood as being somehow connected (for a recent book on the subject, see \cite{rmt4}). The theory was also applied to finance in a series of works dealing with the correlation matrices of stock prices, and also to risk management in portfolios \cite{rmt5}-\cite{rmt34} (for recent reviews on the subject, see \cite{rmt27} and \cite{rmt28}).

In this section, we shall focus on the results that are most important to the present work, which is studying the correlations between world financial markets in times of crisis. The first result of the theory that we shall mention is that, given an $L\times N$ matrix with random numbers built on a Gaussian distribution with average zero and standard deviation $\sigma $, then, in the limit $L\to \infty $ and $N\to \infty$ such that $Q=L/N$ remains finite and greater than 1, the eigenvalues $\lambda$ of such a matrix will have the following probability density function, called a Mar\v{c}enku-Pastur distribution \cite{rmt3}:
\begin{equation}
\label{dist}
\rho(\lambda )=\frac{Q}{2\pi \sigma ^2}\frac{\sqrt{(\lambda_+-\lambda )(\lambda -\lambda_-)}}{\lambda }\ ,
\end{equation}
where
\begin{equation}
\lambda_-=\sigma ^2\left( 1+\frac{1}{Q}-2\sqrt{\frac{1}{Q}}\right) \ \ ,\ \ \lambda_+=\sigma ^2\left( 1+\frac{1}{Q}+2\sqrt{\frac{1}{Q}}\right) \ ,
\end{equation}
and $\lambda $ is restricted to the interval $\left[ \lambda_-,\lambda_+\right] $.

Since the distribution (\ref{dist}) is only valid for the limit $L\to \infty $ and $N\to \infty $, finite distributions will present differences from this behavior. In figure 2, we compare the theoretical distribution for $Q=10$ and $\sigma =1$ to distributions of the eigenvalues of three correlation matrices generated from finite $L\times N$ matrices such that $Q=L/M=10$, and the elements of the matrices are random numbers with mean zero and standard deviation 1.

\begin{pspicture}(-1,0)(3.5,3.8)
\psset{xunit=1,yunit=2}
\pspolygon*[linecolor=lightblue](0.4,0)(0.4,0.5)(0.6,0.5)(0.6,0)
\pspolygon[linecolor=blue](0.4,0)(0.4,0.5)(0.6,0.5)(0.6,0)
\pspolygon*[linecolor=lightblue](0.6,0)(0.6,1)(0.8,1)(0.8,0)
\pspolygon[linecolor=blue](0.6,0)(0.6,1)(0.8,1)(0.8,0)
\pspolygon*[linecolor=lightblue](0.8,0)(0.8,1.5)(1,1.5)(1,0)
\pspolygon[linecolor=blue](0.8,0)(0.8,1.5)(1,1.5)(1,0)
\pspolygon*[linecolor=lightblue](1,0)(1,0.5)(1.2,0.5)(1.2,0)
\pspolygon[linecolor=blue](1,0)(1,0.5)(1.2,0.5)(1.2,0)
\pspolygon*[linecolor=lightblue](1.2,0)(1.2,1)(1.4,1)(1.4,0)
\pspolygon[linecolor=blue](1.2,0)(1.2,1)(1.4,1)(1.4,0)
\pspolygon*[linecolor=lightblue](1.4,0)(1.4,0.5)(1.6,0.5)(1.6,0)
\pspolygon[linecolor=blue](1.4,0)(1.4,0.5)(1.6,0.5)(1.6,0)
\psline{->}(0,0)(3,0) \psline{->}(0,0)(0,1.5) \rput(3.3,0){$\lambda $} \rput(0.45,1.65){$\rho (\lambda )$} \scriptsize \psline(1,-0.05)(1,0.05) \rput(1,-0.15){1} \psline(2,-0.05)(2,0.05) \rput(2,-0.15){2} \psline(-0.1,0.5)(0.1,0.5) \rput(-0.5,0.5){$0,5$} \psline(-0.1,1)(0.1,1) \rput(-0.3,1){1}
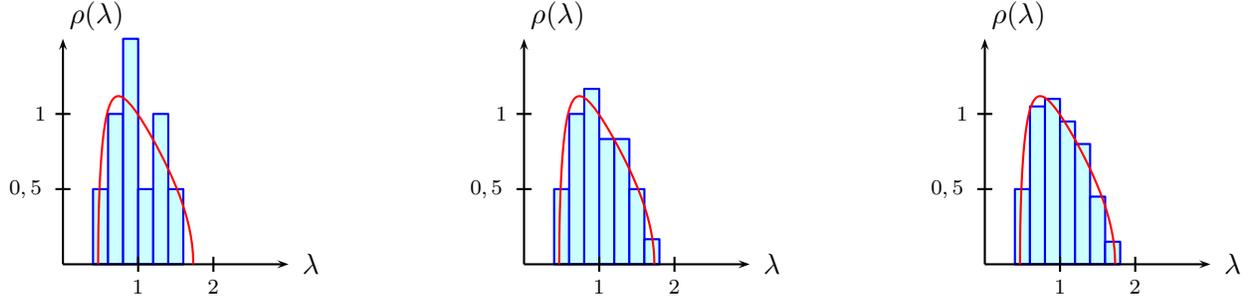
\psplot[linecolor=red,plotpoints=500]{0.467544}{1.732456}{1.732456 x sub x 0.467544 sub mul 0.5 exp x -1 exp mul 1.591549 mul} \small \rput(7.9,-0.5){Figure 2: theoretical and sample finite distributions for a random matrix with $N=10$ and $L=100$,} \rput(4,-0.7){$N=30$ and $L=300$, and $N=100$ and $L=1000$.}
\end{pspicture}
\begin{pspicture}(-2.5,0)(3.5,3.8)
\psset{xunit=1,yunit=2}
\pspolygon*[linecolor=lightblue](0.4,0)(0.4,0.5)(0.6,0.5)(0.6,0)
\pspolygon[linecolor=blue](0.4,0)(0.4,0.5)(0.6,0.5)(0.6,0)
\pspolygon*[linecolor=lightblue](0.6,0)(0.6,1)(0.8,1)(0.8,0)
\pspolygon[linecolor=blue](0.6,0)(0.6,1)(0.8,1)(0.8,0)
\pspolygon*[linecolor=lightblue](0.8,0)(0.8,1.16667)(1,1.16667)(1,0)
\pspolygon[linecolor=blue](0.8,0)(0.8,1.16667)(1,1.16667)(1,0)
\pspolygon*[linecolor=lightblue](1,0)(1,0.8333)(1.2,0.8333)(1.2,0)
\pspolygon[linecolor=blue](1,0)(1,0.8333)(1.2,0.8333)(1.2,0)
\pspolygon*[linecolor=lightblue](1.2,0)(1.2,0.8333)(1.4,0.8333)(1.4,0)
\pspolygon[linecolor=blue](1.2,0)(1.2,0.8333)(1.4,0.8333)(1.4,0)
\pspolygon*[linecolor=lightblue](1.4,0)(1.4,0.5)(1.6,0.5)(1.6,0)
\pspolygon[linecolor=blue](1.4,0)(1.4,0.5)(1.6,0.5)(1.6,0)
\pspolygon*[linecolor=lightblue](1.6,0)(1.6,0.16667)(1.8,0.16667)(1.8,0)
\pspolygon[linecolor=blue](1.6,0)(1.6,0.16667)(1.8,0.16667)(1.8,0)
\psline{->}(0,0)(3,0) \psline{->}(0,0)(0,1.5) \rput(3.3,0){$\lambda $} \rput(0.45,1.65){$\rho (\lambda )$} \scriptsize \psline(1,-0.05)(1,0.05) \rput(1,-0.15){1} \psline(2,-0.05)(2,0.05) \rput(2,-0.15){2} \psline(-0.1,0.5)(0.1,0.5) \rput(-0.5,0.5){$0,5$} \psline(-0.1,1)(0.1,1) \rput(-0.3,1){1}
\psplot[linecolor=red,plotpoints=500]{0.467544}{1.732456}{1.732456 x sub x 0.467544 sub mul 0.5 exp x -1 exp mul 1.591549 mul}
\end{pspicture}
\begin{pspicture}(-2.5,0)(3.5,3.8)
\psset{xunit=1,yunit=2}
\pspolygon*[linecolor=lightblue](0.4,0)(0.4,0.5)(0.6,0.5)(0.6,0)
\pspolygon[linecolor=blue](0.4,0)(0.4,0.5)(0.6,0.5)(0.6,0)
\pspolygon*[linecolor=lightblue](0.6,0)(0.6,1.05)(0.8,1.05)(0.8,0)
\pspolygon[linecolor=blue](0.6,0)(0.6,1.05)(0.8,1.05)(0.8,0)
\pspolygon*[linecolor=lightblue](0.8,0)(0.8,1.1)(1,1.1)(1,0)
\pspolygon[linecolor=blue](0.8,0)(0.8,1.1)(1,1.1)(1,0)
\pspolygon*[linecolor=lightblue](1,0)(1,0.95)(1.2,0.95)(1.2,0)
\pspolygon[linecolor=blue](1,0)(1,0.95)(1.2,0.95)(1.2,0)
\pspolygon*[linecolor=lightblue](1.2,0)(1.2,0.8)(1.4,0.8)(1.4,0)
\pspolygon[linecolor=blue](1.2,0)(1.2,0.8)(1.4,0.8)(1.4,0)
\pspolygon*[linecolor=lightblue](1.4,0)(1.4,0.45)(1.6,0.45)(1.6,0)
\pspolygon[linecolor=blue](1.4,0)(1.4,0.45)(1.6,0.45)(1.6,0)
\pspolygon*[linecolor=lightblue](1.6,0)(1.6,0.15)(1.8,0.15)(1.8,0)
\pspolygon[linecolor=blue](1.6,0)(1.6,0.15)(1.8,0.15)(1.8,0)
\psline{->}(0,0)(3,0) \psline{->}(0,0)(0,1.5) \rput(3.3,0){$\lambda $} \rput(0.45,1.65){$\rho (\lambda )$} \scriptsize \psline(1,-0.05)(1,0.05) \rput(1,-0.15){1} \psline(2,-0.05)(2,0.05) \rput(2,-0.15){2} \psline(-0.1,0.5)(0.1,0.5) \rput(-0.5,0.5){$0,5$} \psline(-0.1,1)(0.1,1) \rput(-0.3,1){1}
\psplot[linecolor=red,plotpoints=500]{0.467544}{1.732456}{1.732456 x sub x 0.467544 sub mul 0.5 exp x -1 exp mul 1.591549 mul}
\end{pspicture}

\vskip 1.8 cm

Consequently, real data will deviate from the theoretical probability distribution. Nevertheless, the theoretical result may serve as a parameter to the results obtained experimentally.

Another source of deviations is the fact that financial time series are better described by non-Gaussian distributions, such as t Student or Tsallis distribution. This can be seen from figure 1: a Gaussian distribution would be represented by a parabola, what is clearly not the case. Recent studies \cite{ext1}-\cite{ext6} developed part of the theoretical framework in which finite series and series with fat tales, as is the case of financial time series of returns, can be studied.

Since Random Matrix Theory is based on random matrices with a single standard deviation $\sigma $, we must compensate the data obtained from the many indices so that all series have average zero and the same standard deviation, which we chose to be equal to one. This can be done using the formula
\begin{equation}
\label{padronized}
X_t=\frac{S_t-\left< S\right> }{\sigma }\ ,
\end{equation}
where $\left< S\right> $ is the average of the time series used, and $\sigma $ is its standard deviation.

\section{Data}

We shall work with one stock market index of each country (with the expception of the USA, with two indices). The indices were chosen among the ones that are considered benchmarks in each stock market. The data were collected from Bloomberg, from 1980, or the first available date, until the end of 2010. For 1987, we collected 23 indices (4 from the Americas, 9 from Europe, and 10 from Asia); for 1998, we have 63 indices (12 from America, 24 from Europe, 19 from Asia, 1 from Oceania, and 7 from Africa); from 2001, 79 indices (13 from America, 29 from Europe, 26 from Asia, 2 from Oceania, and 9 from Africa); for 2008, we have 92 indices (14 from America, 35 from Europe, 30 from Asia, 2 from Oceania, and 11 from Africa). The number of indices collected grew in time both due to the adoption of the indices by their respective stock markets, teh availability of data, and by the emergence of new countries.

This work is motivated by the will to understand how each index affects the others, so as to later attempt to build a model of how crises propagate in a network of indices. Thus, we do not consider here the indices normalized to a single currency, which would be useful, as an example, for building portfolios for investors. That is because we want the numbers to be the ones to which agents operating in their own stock markets react when they take decisions. That is also the reason we are not using indices that are standardized in terms of methodology, like the ones calculated by Morgan and Stanley Capital International (MSCI), which are used mainly by researchers and international investors, but that are not the ones usually published by the press, or seen on the news broadcasts. One of the authors (LSJr) has some research underway using MCSI indices.

When analyzing the data, we had to be careful with the differences in public holidays or weekends among countries. Particular care had to be taken with Israel, Palestine, Jordan, Saudi Arabia, Kwait, Bahrein, Qatar, the United Arab Emirates, Ohman, Bangladesh, and Egypt, for which weekends do not occur on Saturdays and Sundays, but on Fridays and Saturdays, for example. For these countries, we shifted data so as not to lose information, making missing data due to weekends coincide with the other indices. Our general rule was that, when more than 30\% of the markets didn't open on a certain day, we removed that day from our data, and when that number was bellow 30\% , we kept the existing indices and repeated the last computed index for each of the remaining ones. We did not make linear extrapolations of missing indices, for we could then lose the effects of drops, like the one that occured after September 11, when the stock exchanges of the USA remained closed for some days.

Another pressing problem was that markets do not operate at the same time zones, so we had to decide wether to consider the data concerning American countries at the same day as Asian-Pacific countries, for example, or to shift the data for Asian countries so as to compare indices from the USA with the next day index from Japan, for example. There is even some evidence \cite{vol8} that the correlations of Asian with the USA indices increase when one considers the correlation of the USA indices with the next day indices of the Asian markets.

We decided to consider all indices taken at the same date. This was motivated by some comparisons between the correlations among indices: we compared the individual correlations among most of the indices and checked if the correlation increased or decreased shifting the relevant data. The result was inconclusive, for there was an almost equal number of correlations which increased taking Western indices one day before their Eastern counterparts as there were other correlations that remained higher taking the same day indices (some correlations were higher when we took the Western indices one day after the Eastern ones). In order to gauge the effects of such a shift, we recalculated all our results by shifting the indices from Asia and Oceania in one day, without many changes in the outcomes. Those results are better explained in Section 10.

One option, frequently adopted, to avoid the problem of different operating hours between international markets is to consider weekly data instead of daily data. We didn't adopt that approach so as not to miss major changes in markets, which tend to occur during a small interval of days. Instead, we preffered to compare results with and without shifting part of the data by one day.

In the next four sections, we shall consider the years 1987, 1998, 2001, and 2008 in the light of Random Matrix Theory. The indices we used, together with their countries of affiliation, the symbols used for them, and their codes in Bloomberg, are placed in table 2, in the Appendix.

\section{1987, the Black Monday}

In 1987, the financial world lived a time of panic, much like the one of the great crash of 1920. In a matter of 3 days, most of the stock markets in the world lost about 30\% of their value and trillions of dollars evaporated, leaving a trace of destruction that affected what is reffered to as real economy for many years. The day the first and major collapse occured, a Monday, was later called the {\sl Black Monday}.

In order to analyze that crisis, we consider now the correlation matrix of 23 indices of stock exchanges around the world: the S\&P 500 from the New York Stock Exchange (S\&P), and NASDAQ (Nasd), both from the USA, S\&P TSX from Canada (Cana), Ibovespa from Brazil (Braz), FTSE 100 from the United Kingdom (UK), ISEQ from Ireland (Irel), DAX from Germany (Germ), or West Germany in 1987, ATX from Austria (Autr), AEX from the Netherlands (Neth), OMX from Sweden (Swed), OMX Helsinki from Finland (Finl), IBEX 35 from Spain (Spai), ASE General Index from Greece (Gree), SENSEX from India (Indi), Colombo All Share from Sri Lanka (SrLa), DSE General Index from Bangladesh (Bang), Nikkei 25 from Japan (Japa), Hang Seng from Hong Kong (HoKo), TAIEX from Taiwan (Taiw), Kospi from South Korea (SoKo), Kuala Lumpur Composite from Malaysia (Mala), JCI from Indonesia (Indo), and the PSEi from the Philipines (Phil). So, we have three indices from North America, one from South America, nine from Europe, and ten from Asia, with a total of 23 indices. These offer a good variety of indices worldwide. In subsequent years, we shall increase this number, mainly due to the appearance of new indices and countries, and to the access to data about them.

We shall use 1987 as an example for the other years, and because of that we will be showing more details in the calculations for that year. Calculating the correlation matrix for the indices that are being considered, one obtains a $23\times 23$ matrix. The average of the values of this correlation matrix is a good measure of the overall correlation between the many indices (the average is taken over the elements of the correlation matrix for which $i<j$). For the present correlation matrix, the average is given by $<C>=0.16$, with standard deviation $\sigma =0.04$. Since the correlation matrix is $23\times 23$, and the number of days considered in calculating it is 256, we then have $Q=L/M=256/23\approx 11.130$, and the upper and lower bounds
\begin{equation} \lambda_-=0.490\ \ ,\ \ \lambda_+=1.689 \end{equation}
for the eigenvalues that constitute the bulk of the eigenvalue distribution due to noise.

A frequency distribution of the 23 eigenvalues of the correlation matrix is shown in figure 3, with the theoretical distribution of an infinite random matrix for $Q=11.130$ with mean zero and standard deviation one superimposed on it. In figure 4, the eigenvalues are plotted in order of magnitude. The shaded area indicates the region predicted by theory were the data related with a purely random behavior of the normalized log-returns.

\begin{pspicture}(-0.5,0)(3.5,6.1)
\psset{xunit=1.07,yunit=4}
\pspolygon*[linecolor=lightblue](0,0)(0,0.652)(0.2,0.652)(0.2,0)
\pspolygon*[linecolor=lightblue](0.2,0)(0.2,0.870)(0.4,0.870)(0.4,0)
\pspolygon*[linecolor=lightblue](0.4,0)(0.4,0.435)(0.6,0.435)(0.6,0)
\pspolygon*[linecolor=lightblue](0.6,0)(0.6,0.435)(0.8,0.435)(0.8,0)
\pspolygon*[linecolor=lightblue](0.8,0)(0.8,1.087)(1,1.087)(1,0)
\pspolygon*[linecolor=lightblue](1,0)(1,0.652)(1.2,0.652)(1.2,0)
\pspolygon*[linecolor=lightblue](1.2,0)(1.2,0.217)(1.4,0.217)(1.4,0)
\pspolygon*[linecolor=lightblue](1.4,0)(1.4,0.217)(1.6,0.217)(1.6,0)
\pspolygon*[linecolor=lightblue](1.8,0)(1.8,0.217)(2,0.217)(2,0)
\pspolygon*[linecolor=lightblue](6.4,0)(6.4,0.217)(6.6,0.217)(6.6,0)
\pspolygon[linecolor=blue](0,0)(0,0.652)(0.2,0.652)(0.2,0)
\pspolygon[linecolor=blue](0.2,0)(0.2,0.870)(0.4,0.870)(0.4,0)
\pspolygon[linecolor=blue](0.4,0)(0.4,0.435)(0.6,0.435)(0.6,0)
\pspolygon[linecolor=blue](0.6,0)(0.6,0.435)(0.8,0.435)(0.8,0)
\pspolygon[linecolor=blue](0.8,0)(0.8,1.087)(1,1.087)(1,0)
\pspolygon[linecolor=blue](1,0)(1,0.652)(1.2,0.652)(1.2,0)
\pspolygon[linecolor=blue](1.2,0)(1.2,0.217)(1.4,0.217)(1.4,0)
\pspolygon[linecolor=blue](1.4,0)(1.4,0.217)(1.6,0.217)(1.6,0)
\pspolygon[linecolor=blue](1.8,0)(1.8,0.217)(2,0.217)(2,0)
\pspolygon[linecolor=blue](6.4,0)(6.4,0.217)(6.6,0.217)(6.6,0)
\psline{->}(0,0)(7,0) \psline{->}(0,0)(0,1.4) \rput(7.3,0){$\lambda $} \rput(0.5,1.4){$\rho (\lambda )$} \scriptsize \psline(1,-0.025)(1,0.025) \rput(1,-0.075){1} \psline(2,-0.025)(2,0.025) \rput(2,-0.075){2} \psline(3,-0.025)(3,0.025) \rput(3,-0.075){3}  \psline(4,-0.02)(4,0.02) \rput(4,-0.075){4} \psline(5,-0.02)(5,0.02) \rput(5,-0.075){5} \psline(6,-0.025)(6,0.025) \rput(6,-0.075){6} \psline(-0.1,0.2)(0.1,0.2) \rput(-0.4,0.2){$0.2$} \psline(-0.1,0.4)(0.1,0.4) \rput(-0.4,0.4){$0.4$} \psline(-0.1,0.6)(0.1,0.6) \rput(-0.4,0.6){$0.6$} \psline(-0.1,0.8)(0.1,0.8) \rput(-0.4,0.8){$0.8$} \psline(-0.1,1)(0.1,1) \rput(-0.4,1){$1$} \psline(-0.1,1.2)(0.1,1.2) \rput(-0.4,1.2){$1.2$}
\psplot[linecolor=red,plotpoints=500]{0.490}{1.689}{1.689 x sub x 0.490 sub mul 0.5 exp x -1 exp mul 1.771 mul} \small \rput(3.5,-0.2){Figure 3: frequency distribution of the eigenvalues of the} \rput(3.4,-0.3){correlation matrix for 1987. The theoretical distribution}  \rput(2.5,-0.4){for a random matrix is superimposed on it.}
\end{pspicture}
\begin{pspicture}(-5.5,0)(3.5,6.1)
\psset{xunit=1.07,yunit=2.2}
\pspolygon*[linecolor=lightgray](0.490,0)(0.490,0.8)(1.689,0.8)(1.689,0)
\psline{->}(0,0)(7,0)  \rput(7.3,0){$\lambda $} \scriptsize \psline(1,-0.05)(1,0.05) \rput(1,-0.15){1} \psline(2,-0.05)(2,0.05) \rput(2,-0.15){2} \psline(3,-0.05)(3,0.05) \rput(3,-0.15){3}  \psline(4,-0.05)(4,0.05) \rput(4,-0.15){4} \psline(5,-0.05)(5,0.05) \rput(5,-0.15){5} \psline(6,-0.05)(6,0.05) \rput(6,-0.15){6}
\psline[linewidth=1pt](0.090,0)(0.090,0.5) \psline[linewidth=1pt](0.119,0)(0.119,0.5) \psline[linewidth=1pt](0.156,0)(0.156,0.5) \psline[linewidth=1pt](0.224,0)(0.224,0.5) \psline[linewidth=1pt](0.278,0)(0.278,0.5)\psline[linewidth=1pt](0.308,0)(0.308,0.5) \psline[linewidth=1pt](0.384,0)(0.384,0.5)\psline[linewidth=1pt](0.525,0)(0.525,0.5) \psline[linewidth=1pt](0.556,0)(0.556,0.5)\psline[linewidth=1pt](0.671,0)(0.671,0.5) \psline[linewidth=1pt](0.735,0)(0.735,0.5)\psline[linewidth=1pt](0.803,0)(0.803,0.5) \psline[linewidth=1pt](0.832,0)(0.832,0.5) \psline[linewidth=1pt](0.869,0)(0.869,0.5)
\psline[linewidth=1pt](0.959,0)(0.959,0.5) \psline[linewidth=1pt](0.997,0)(0.997,0.5)
\psline[linewidth=1pt](1.034,0)(1.034,0.5) \psline[linewidth=1pt](1.104,0)(1.104,0.5)
\psline[linewidth=1pt](1.117,0)(1.117,0.5) \psline[linewidth=1pt](1.292,0)(1.292,0.5)
\psline[linewidth=1pt](1.502,0)(1.502,0.5) \psline[linewidth=1pt](1.946,0)(1.946,0.5)
\psline[linewidth=1pt](6.500,0)(6.500,0.5)
\small \rput(3.45,-0.4){Figure 4: eigenvalues in order of magnitude. The} \rput(3.7,-0.6){shaded area corresponds to the eigenvalues predicted} \rput(1.45,-0.8){for a random matrix.}
\end{pspicture}

\vskip 2.3 cm

Only 60\% of the eigenvalues fall inside the region predicted by Random Matrix Theory. Note that the highest eigenvalue stands out from all the others, being more than three times bigger than the uppermost limit $\lambda_+$ of the theoretical distribution. This is in agreement with many other results, obtained for a great number of financial institutions to which this same formalism has been already applied \cite{rmt5}-\cite{rmt34}. It is believed that this eigenvalue corresponds to the action of a single market, which influences all the other members of the correlation matrix. Usually, for the correlation matrix of individual stocks in a single market, this eigenvalue is much larger, some times 25 times larger, than the largest eigenvalue predicted for the correlation matrix of a random time series, although the size of the sample directly influenciates that as well. In our case, it responds for about 28\% of the collective behavior of the time series being considered, which is the ratio of the largest eigenvalue and the sum of the eigenvalues of the correlation matrix.

Figure 5 shows the contributions of the many indices which we are considering in our study in some of the eingenvectors of the correlation matrix. The blue bars represent positive values and the red bars represent negative ones.

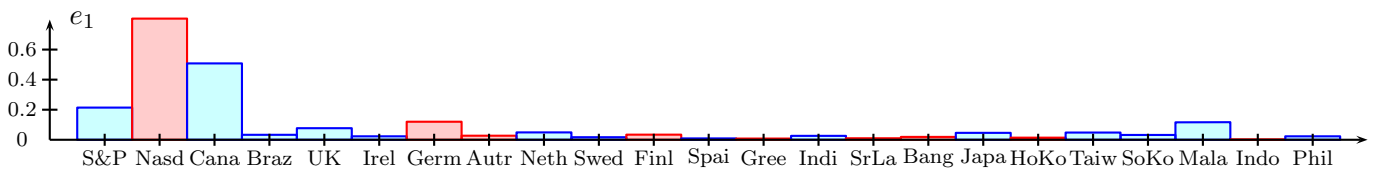
\begin{pspicture}(-0.2,0)(3.5,2)
\psset{xunit=0.73,yunit=2}
\pspolygon*[linecolor=lightblue](0.5,0)(0.5,0.214)(1.5,0.214)(1.5,0)
\pspolygon*[linecolor=lightred](1.5,0)(1.5,0.806)(2.5,0.806)(2.5,0)
\pspolygon*[linecolor=lightblue](2.5,0)(2.5,0.508)(3.5,0.508)(3.5,0)
\pspolygon*[linecolor=lightblue](3.5,0)(3.5,0.033)(4.5,0.033)(4.5,0)
\pspolygon*[linecolor=lightblue](4.5,0)(4.5,0.077)(5.5,0.077)(5.5,0)
\pspolygon*[linecolor=lightblue](5.5,0)(5.5,0.023)(6.5,0.023)(6.5,0)
\pspolygon*[linecolor=lightred](6.5,0)(6.5,0.120)(7.5,0.120)(7.5,0)
\pspolygon*[linecolor=lightred](7.5,0)(7.5,0.027)(8.5,0.027)(8.5,0)
\pspolygon*[linecolor=lightblue](8.5,0)(8.5,0.049)(9.5,0.049)(9.5,0)
\pspolygon*[linecolor=lightblue](9.5,0)(9.5,0.017)(10.5,0.017)(10.5,0)
\pspolygon*[linecolor=lightred](10.5,0)(10.5,0.034)(11.5,0.034)(11.5,0)
\pspolygon*[linecolor=lightblue](11.5,0)(11.5,0.009)(12.5,0.009)(12.5,0)
\pspolygon*[linecolor=lightred](12.5,0)(12.5,0.008)(13.5,0.008)(13.5,0)
\pspolygon*[linecolor=lightblue](13.5,0)(13.5,0.026)(14.5,0.026)(14.5,0)
\pspolygon*[linecolor=lightred](14.5,0)(14.5,0.010)(15.5,0.010)(15.5,0)
\pspolygon*[linecolor=lightred](15.5,0)(15.5,0.019)(16.5,0.019)(16.5,0)
\pspolygon*[linecolor=lightblue](16.5,0)(16.5,0.046)(17.5,0.046)(17.5,0)
\pspolygon*[linecolor=lightred](17.5,0)(17.5,0.014)(18.5,0.014)(18.5,0)
\pspolygon*[linecolor=lightblue](18.5,0)(18.5,0.048)(19.5,0.048)(19.5,0)
\pspolygon*[linecolor=lightblue](19.5,0)(19.5,0.032)(20.5,0.032)(20.5,0)
\pspolygon*[linecolor=lightblue](20.5,0)(20.5,0.116)(21.5,0.117)(21.5,0)
\pspolygon*[linecolor=lightred](21.5,0)(21.5,0.001)(22.5,0.001)(22.5,0)
\pspolygon*[linecolor=lightblue](22.5,0)(22.5,0.023)(23.5,0.023)(23.5,0)
\pspolygon[linecolor=blue](0.5,0)(0.5,0.214)(1.5,0.214)(1.5,0)
\pspolygon[linecolor=red](1.5,0)(1.5,0.806)(2.5,0.806)(2.5,0)
\pspolygon[linecolor=blue](2.5,0)(2.5,0.508)(3.5,0.508)(3.5,0)
\pspolygon[linecolor=blue](3.5,0)(3.5,0.033)(4.5,0.033)(4.5,0)
\pspolygon[linecolor=blue](4.5,0)(4.5,0.077)(5.5,0.077)(5.5,0)
\pspolygon[linecolor=blue](5.5,0)(5.5,0.023)(6.5,0.023)(6.5,0)
\pspolygon[linecolor=red](6.5,0)(6.5,0.120)(7.5,0.120)(7.5,0)
\pspolygon[linecolor=red](7.5,0)(7.5,0.027)(8.5,0.027)(8.5,0)
\pspolygon[linecolor=blue](8.5,0)(8.5,0.049)(9.5,0.049)(9.5,0)
\pspolygon[linecolor=blue](9.5,0)(9.5,0.017)(10.5,0.017)(10.5,0)
\pspolygon[linecolor=red](10.5,0)(10.5,0.034)(11.5,0.034)(11.5,0)
\pspolygon[linecolor=blue](11.5,0)(11.5,0.009)(12.5,0.009)(12.5,0)
\pspolygon[linecolor=red](12.5,0)(12.5,0.008)(13.5,0.008)(13.5,0)
\pspolygon[linecolor=blue](13.5,0)(13.5,0.026)(14.5,0.026)(14.5,0)
\pspolygon[linecolor=red](14.5,0)(14.5,0.010)(15.5,0.010)(15.5,0)
\pspolygon[linecolor=red](15.5,0)(15.5,0.019)(16.5,0.019)(16.5,0)
\pspolygon[linecolor=blue](16.5,0)(16.5,0.046)(17.5,0.046)(17.5,0)
\pspolygon[linecolor=red](17.5,0)(17.5,0.014)(18.5,0.014)(18.5,0)
\pspolygon[linecolor=blue](18.5,0)(18.5,0.048)(19.5,0.048)(19.5,0)
\pspolygon[linecolor=blue](19.5,0)(19.5,0.032)(20.5,0.032)(20.5,0)
\pspolygon[linecolor=blue](20.5,0)(20.5,0.116)(21.5,0.117)(21.5,0)
\pspolygon[linecolor=red](21.5,0)(21.5,0.001)(22.5,0.001)(22.5,0)
\pspolygon[linecolor=blue](22.5,0)(22.5,0.023)(23.5,0.023)(23.5,0)
\psline{->}(0,0)(24,0) \psline{->}(0,0)(0,0.8) \rput(0.6,0.8){$e_{1}$} \scriptsize \psline(1,-0.04)(1,0.04) \rput(1,-0.12){S\&P} \psline(2,-0.04)(2,0.04) \rput(2,-0.12){Nasd} \psline(3,-0.04)(3,0.04) \rput(3,-0.12){Cana} \psline(4,-0.04)(4,0.04) \rput(4,-0.12){Braz} \psline(5,-0.04)(5,0.04) \rput(5,-0.12){UK} \psline(6,-0.04)(6,0.04) \rput(6,-0.12){Irel} \psline(7,-0.04)(7,0.04) \rput(7,-0.12){Germ} \psline(8,-0.04)(8,0.04) \rput(8,-0.12){Autr} \psline(9,-0.04)(9,0.04) \rput(9,-0.12){Neth} \psline(10,-0.04)(10,0.04) \rput(10,-0.12){Swed} \psline(11,-0.04)(11,0.04) \rput(11,-0.12){Finl} \psline(12,-0.04)(12,0.04) \rput(12,-0.12){Spai} \psline(13,-0.04)(13,0.04) \rput(13,-0.12){Gree} \psline(14,-0.04)(14,0.04) \rput(14,-0.12){Indi} \psline(15,-0.04)(15,0.04) \rput(15,-0.12){SrLa} \psline(16,-0.04)(16,0.04) \rput(16,-0.12){Bang} \psline(17,-0.04)(17,0.04) \rput(17,-0.12){Japa} \psline(18,-0.04)(18,0.04) \rput(18,-0.12){HoKo} \psline(19,-0.04)(19,0.04) \rput(19,-0.12){Taiw} \psline(20,-0.04)(20,0.04) \rput(20,-0.12){SoKo} \psline(21,-0.04)(21,0.04) \rput(21,-0.12){Mala} \psline(22,-0.04)(22,0.04) \rput(22,-0.12){Indo} \psline(23,-0.04)(23,0.04) \rput(23,-0.12){Phil} \rput(-0.5,0){$0$} \psline(-0.14,0.2)(0.14,0.2) \rput(-0.5,0.2){$0.2$} \psline(-0.14,0.4)(0.14,0.4) \rput(-0.5,0.4){$0.4$} \psline(-0.14,0.6)(0.14,0.6) \rput(-0.5,0.6){$0.6$}
\end{pspicture}

\begin{pspicture}(-0.2,0)(3.5,2.4)
\psset{xunit=0.73,yunit=2}
\pspolygon*[linecolor=lightblue](0.5,0)(0.5,0.050)(1.5,0.050)(1.5,0)
\pspolygon*[linecolor=lightblue](1.5,0)(1.5,0.054)(2.5,0.054)(2.5,0)
\pspolygon*[linecolor=lightblue](2.5,0)(2.5,0.029)(3.5,0.029)(3.5,0)
\pspolygon*[linecolor=lightred](3.5,0)(3.5,0.137)(4.5,0.137)(4.5,0)
\pspolygon*[linecolor=lightred](4.5,0)(4.5,0.019)(5.5,0.019)(5.5,0)
\pspolygon*[linecolor=lightred](5.5,0)(5.5,0.008)(6.5,0.008)(6.5,0)
\pspolygon*[linecolor=lightblue](6.5,0)(6.5,0.115)(7.5,0.115)(7.5,0)
\pspolygon*[linecolor=lightblue](7.5,0)(7.5,0.099)(8.5,0.099)(8.5,0)
\pspolygon*[linecolor=lightred](8.5,0)(8.5,0.099)(9.5,0.099)(9.5,0)
\pspolygon*[linecolor=lightred](9.5,0)(9.5,0.107)(10.5,0.107)(10.5,0)
\pspolygon*[linecolor=lightblue](10.5,0)(10.5,0.126)(11.5,0.126)(11.5,0)
\pspolygon*[linecolor=lightblue](11.5,0)(11.5,0.122)(12.5,0.122)(12.5,0)
\pspolygon*[linecolor=lightred](12.5,0)(12.5,0.076)(13.5,0.076)(13.5,0)
\pspolygon*[linecolor=lightblue](13.5,0)(13.5,0.024)(14.5,0.024)(14.5,0)
\pspolygon*[linecolor=lightblue](14.5,0)(14.5,0.314)(15.5,0.314)(15.5,0)
\pspolygon*[linecolor=lightred](15.5,0)(15.5,0.508)(16.5,0.508)(16.5,0)
\pspolygon*[linecolor=lightblue](16.5,0)(16.5,0.104)(17.5,0.104)(17.5,0)
\pspolygon*[linecolor=lightred](17.5,0)(17.5,0.098)(18.5,0.098)(18.5,0)
\pspolygon*[linecolor=lightblue](18.5,0)(18.5,0.064)(19.5,0.064)(19.5,0)
\pspolygon*[linecolor=lightred](19.5,0)(19.5,0.450)(20.5,0.450)(20.5,0)
\pspolygon*[linecolor=lightblue](20.5,0)(20.5,0.265)(21.5,0.265)(21.5,0)
\pspolygon*[linecolor=lightred](21.5,0)(21.5,0.146)(22.5,0.146)(22.5,0)
\pspolygon*[linecolor=lightred](22.5,0)(22.5,0.467)(23.5,0.467)(23.5,0)
\pspolygon[linecolor=blue](0.5,0)(0.5,0.050)(1.5,0.050)(1.5,0)
\pspolygon[linecolor=blue](1.5,0)(1.5,0.054)(2.5,0.054)(2.5,0)
\pspolygon[linecolor=blue](2.5,0)(2.5,0.029)(3.5,0.029)(3.5,0)
\pspolygon[linecolor=red](3.5,0)(3.5,0.137)(4.5,0.137)(4.5,0)
\pspolygon[linecolor=red](4.5,0)(4.5,0.019)(5.5,0.019)(5.5,0)
\pspolygon[linecolor=red](5.5,0)(5.5,0.008)(6.5,0.008)(6.5,0)
\pspolygon[linecolor=blue](6.5,0)(6.5,0.115)(7.5,0.115)(7.5,0)
\pspolygon[linecolor=blue](7.5,0)(7.5,0.099)(8.5,0.099)(8.5,0)
\pspolygon[linecolor=red](8.5,0)(8.5,0.099)(9.5,0.099)(9.5,0)
\pspolygon[linecolor=red](9.5,0)(9.5,0.107)(10.5,0.107)(10.5,0)
\pspolygon[linecolor=blue](10.5,0)(10.5,0.126)(11.5,0.126)(11.5,0)
\pspolygon[linecolor=blue](11.5,0)(11.5,0.122)(12.5,0.122)(12.5,0)
\pspolygon[linecolor=red](12.5,0)(12.5,0.076)(13.5,0.076)(13.5,0)
\pspolygon[linecolor=blue](13.5,0)(13.5,0.024)(14.5,0.024)(14.5,0)
\pspolygon[linecolor=blue](14.5,0)(14.5,0.314)(15.5,0.314)(15.5,0)
\pspolygon[linecolor=red](15.5,0)(15.5,0.508)(16.5,0.508)(16.5,0)
\pspolygon[linecolor=blue](16.5,0)(16.5,0.104)(17.5,0.104)(17.5,0)
\pspolygon[linecolor=red](17.5,0)(17.5,0.098)(18.5,0.098)(18.5,0)
\pspolygon[linecolor=blue](18.5,0)(18.5,0.064)(19.5,0.064)(19.5,0)
\pspolygon[linecolor=red](19.5,0)(19.5,0.450)(20.5,0.450)(20.5,0)
\pspolygon[linecolor=blue](20.5,0)(20.5,0.265)(21.5,0.265)(21.5,0)
\pspolygon[linecolor=red](21.5,0)(21.5,0.146)(22.5,0.146)(22.5,0)
\pspolygon[linecolor=red](22.5,0)(22.5,0.467)(23.5,0.467)(23.5,0)
\psline{->}(0,0)(24,0) \psline{->}(0,0)(0,0.8) \rput(0.7,0.8){$e_{15}$} \scriptsize \psline(1,-0.04)(1,0.04) \rput(1,-0.12){S\&P} \psline(2,-0.04)(2,0.04) \rput(2,-0.12){Nasd} \psline(3,-0.04)(3,0.04) \rput(3,-0.12){Cana} \psline(4,-0.04)(4,0.04) \rput(4,-0.12){Braz} \psline(5,-0.04)(5,0.04) \rput(5,-0.12){UK} \psline(6,-0.04)(6,0.04) \rput(6,-0.12){Irel} \psline(7,-0.04)(7,0.04) \rput(7,-0.12){Germ} \psline(8,-0.04)(8,0.04) \rput(8,-0.12){Autr} \psline(9,-0.04)(9,0.04) \rput(9,-0.12){Neth} \psline(10,-0.04)(10,0.04) \rput(10,-0.12){Swed} \psline(11,-0.04)(11,0.04) \rput(11,-0.12){Finl} \psline(12,-0.04)(12,0.04) \rput(12,-0.12){Spai} \psline(13,-0.04)(13,0.04) \rput(13,-0.12){Gree} \psline(14,-0.04)(14,0.04) \rput(14,-0.12){Indi} \psline(15,-0.04)(15,0.04) \rput(15,-0.12){SrLa} \psline(16,-0.04)(16,0.04) \rput(16,-0.12){Bang} \psline(17,-0.04)(17,0.04) \rput(17,-0.12){Japa} \psline(18,-0.04)(18,0.04) \rput(18,-0.12){HoKo} \psline(19,-0.04)(19,0.04) \rput(19,-0.12){Taiw} \psline(20,-0.04)(20,0.04) \rput(20,-0.12){SoKo} \psline(21,-0.04)(21,0.04) \rput(21,-0.12){Mala} \psline(22,-0.04)(22,0.04) \rput(22,-0.12){Indo} \psline(23,-0.04)(23,0.04) \rput(23,-0.12){Phil} \rput(-0.5,0){$0$} \psline(-0.14,0.2)(0.14,0.2) \rput(-0.5,0.2){$0.2$} \psline(-0.14,0.4)(0.14,0.4) \rput(-0.5,0.4){$0.4$} \psline(-0.14,0.6)(0.14,0.6) \rput(-0.5,0.6){$0.6$}
\end{pspicture}

\begin{pspicture}(-0.2,0)(3.5,2.4)
\psset{xunit=0.73,yunit=2}
\pspolygon*[linecolor=lightblue](0.5,0)(0.5,0.227)(1.5,0.227)(1.5,0)
\pspolygon*[linecolor=lightblue](1.5,0)(1.5,0.317)(2.5,0.317)(2.5,0)
\pspolygon*[linecolor=lightblue](2.5,0)(2.5,0.315)(3.5,0.315)(3.5,0)
\pspolygon*[linecolor=lightblue](3.5,0)(3.5,0.140)(4.5,0.140)(4.5,0)
\pspolygon*[linecolor=lightblue](4.5,0)(4.5,0.093)(5.5,0.093)(5.5,0)
\pspolygon*[linecolor=lightblue](5.5,0)(5.5,0.251)(6.5,0.251)(6.5,0)
\pspolygon*[linecolor=lightblue](6.5,0)(6.5,0.310)(7.5,0.310)(7.5,0)
\pspolygon*[linecolor=lightblue](7.5,0)(7.5,0.259)(8.5,0.259)(8.5,0)
\pspolygon*[linecolor=lightblue](8.5,0)(8.5,0.280)(9.5,0.280)(9.5,0)
\pspolygon*[linecolor=lightblue](9.5,0)(9.5,0.130)(10.5,0.130)(10.5,0)
\pspolygon*[linecolor=lightblue](10.5,0)(10.5,0.240)(11.5,0.240)(11.5,0)
\pspolygon*[linecolor=lightblue](11.5,0)(11.5,0.193)(12.5,0.193)(12.5,0)
\pspolygon*[linecolor=lightblue](12.5,0)(12.5,0.062)(13.5,0.062)(13.5,0)
\pspolygon*[linecolor=lightblue](13.5,0)(13.5,0.085)(14.5,0.085)(14.5,0)
\pspolygon*[linecolor=lightblue](14.5,0)(14.5,0.235)(15.5,0.235)(15.5,0)
\pspolygon*[linecolor=lightblue](15.5,0)(15.5,0.197)(16.5,0.197)(16.5,0)
\pspolygon*[linecolor=lightblue](16.5,0)(16.5,0.019)(17.5,0.019)(17.5,0)
\pspolygon*[linecolor=lightblue](17.5,0)(17.5,0.225)(18.5,0.225)(18.5,0)
\pspolygon*[linecolor=lightblue](18.5,0)(18.5,0.236)(19.5,0.236)(19.5,0)
\pspolygon*[linecolor=lightred](19.5,0)(19.5,0.015)(20.5,0.015)(20.5,0)
\pspolygon*[linecolor=lightred](20.5,0)(20.5,0.017)(21.5,0.017)(21.5,0)
\pspolygon*[linecolor=lightblue](21.5,0)(21.5,0.022)(22.5,0.022)(22.5,0)
\pspolygon*[linecolor=lightblue](22.5,0)(22.5,0.303)(23.5,0.303)(23.5,0)
\pspolygon[linecolor=blue](0.5,0)(0.5,0.227)(1.5,0.227)(1.5,0)
\pspolygon[linecolor=blue](1.5,0)(1.5,0.317)(2.5,0.317)(2.5,0)
\pspolygon[linecolor=blue](2.5,0)(2.5,0.315)(3.5,0.315)(3.5,0)
\pspolygon[linecolor=blue](3.5,0)(3.5,0.140)(4.5,0.140)(4.5,0)
\pspolygon[linecolor=blue](4.5,0)(4.5,0.093)(5.5,0.093)(5.5,0)
\pspolygon[linecolor=blue](5.5,0)(5.5,0.251)(6.5,0.251)(6.5,0)
\pspolygon[linecolor=blue](6.5,0)(6.5,0.310)(7.5,0.310)(7.5,0)
\pspolygon[linecolor=blue](7.5,0)(7.5,0.259)(8.5,0.259)(8.5,0)
\pspolygon[linecolor=blue](8.5,0)(8.5,0.280)(9.5,0.280)(9.5,0)
\pspolygon[linecolor=blue](9.5,0)(9.5,0.130)(10.5,0.130)(10.5,0)
\pspolygon[linecolor=blue](10.5,0)(10.5,0.240)(11.5,0.240)(11.5,0)
\pspolygon[linecolor=blue](11.5,0)(11.5,0.193)(12.5,0.193)(12.5,0)
\pspolygon[linecolor=blue](12.5,0)(12.5,0.062)(13.5,0.062)(13.5,0)
\pspolygon[linecolor=blue](13.5,0)(13.5,0.085)(14.5,0.085)(14.5,0)
\pspolygon[linecolor=blue](14.5,0)(14.5,0.235)(15.5,0.235)(15.5,0)
\pspolygon[linecolor=blue](15.5,0)(15.5,0.197)(16.5,0.197)(16.5,0)
\pspolygon[linecolor=blue](16.5,0)(16.5,0.019)(17.5,0.019)(17.5,0)
\pspolygon[linecolor=blue](17.5,0)(17.5,0.225)(18.5,0.225)(18.5,0)
\pspolygon[linecolor=blue](18.5,0)(18.5,0.236)(19.5,0.236)(19.5,0)
\pspolygon[linecolor=red](19.5,0)(19.5,0.015)(20.5,0.015)(20.5,0)
\pspolygon[linecolor=red](20.5,0)(20.5,0.017)(21.5,0.017)(21.5,0)
\pspolygon[linecolor=blue](21.5,0)(21.5,0.022)(22.5,0.022)(22.5,0)
\pspolygon[linecolor=blue](22.5,0)(22.5,0.303)(23.5,0.303)(23.5,0)
\psline{->}(0,0)(24,0) \psline{->}(0,0)(0,0.8) \rput(0.7,0.8){$e_{23}$} \scriptsize \psline(1,-0.04)(1,0.04) \rput(1,-0.12){S\&P} \psline(2,-0.04)(2,0.04) \rput(2,-0.12){Nasd} \psline(3,-0.04)(3,0.04) \rput(3,-0.12){Cana} \psline(4,-0.04)(4,0.04) \rput(4,-0.12){Braz} \psline(5,-0.04)(5,0.04) \rput(5,-0.12){UK} \psline(6,-0.04)(6,0.04) \rput(6,-0.12){Irel} \psline(7,-0.04)(7,0.04) \rput(7,-0.12){Germ} \psline(8,-0.04)(8,0.04) \rput(8,-0.12){Autr} \psline(9,-0.04)(9,0.04) \rput(9,-0.12){Neth} \psline(10,-0.04)(10,0.04) \rput(10,-0.12){Swed} \psline(11,-0.04)(11,0.04) \rput(11,-0.12){Finl} \psline(12,-0.04)(12,0.04) \rput(12,-0.12){Spai} \psline(13,-0.04)(13,0.04) \rput(13,-0.12){Gree} \psline(14,-0.04)(14,0.04) \rput(14,-0.12){Indi} \psline(15,-0.04)(15,0.04) \rput(15,-0.12){SrLa} \psline(16,-0.04)(16,0.04) \rput(16,-0.12){Bang} \psline(17,-0.04)(17,0.04) \rput(17,-0.12){Japa} \psline(18,-0.04)(18,0.04) \rput(18,-0.12){HoKo} \psline(19,-0.04)(19,0.04) \rput(19,-0.12){Taiw} \psline(20,-0.04)(20,0.04) \rput(20,-0.12){SoKo} \psline(21,-0.04)(21,0.04) \rput(21,-0.12){Mala} \psline(22,-0.04)(22,0.04) \rput(22,-0.12){Indo} \psline(23,-0.04)(23,0.04) \rput(23,-0.12){Phil} \rput(-0.5,0){$0$} \psline(-0.14,0.2)(0.14,0.2) \rput(-0.5,0.2){$0.2$} \psline(-0.14,0.4)(0.14,0.4) \rput(-0.5,0.4){$0.4$} \psline(-0.14,0.6)(0.14,0.6) \rput(-0.5,0.6){$0.6$}
\end{pspicture}

\vskip 0.6 cm

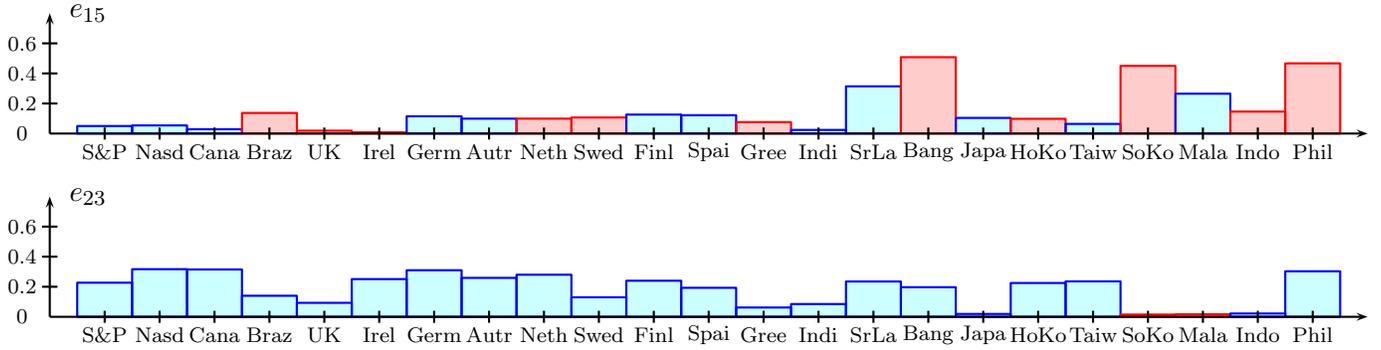
\small Figure 5: contributions of the stock market indices to eigenvectors $e_1$, $e_{15}$, and $e_{23}$. Blue bars indicate positive values,

and red bars correspond to negative values. \normalsize

\vskip 0.4 cm

One can see that the eigenvector corresponding to the largest eigenvalue is qualitatively different from the others. Nearly all markets (with the exception of Bangladesh and Indonesia) have positive representations. That is a compelling reason to believe that it represents a global market that is the result of the interactions of all local markets, or may also be the result of external news on the market as a whole. Figure 6 compares the time series of an index built using eigenvector 23 (in blue) with the world index calculated by the MSCI (Morgan Stanley Capital International), in red. Both indices are normalized so as to have mean two and standard deviation one.

\begin{pspicture}(-2,-2.7)(3.5,2.7)
\psset{xunit=2.5,yunit=1}
\psline{->}(0,-2)(5.4,-2) \psline{->}(0,-2.5)(0,2) \rput(5.6,-2){day} \rput(0.3,2){Indices}\scriptsize \psline(0.02,-2.1)(0.02,-1.9) \rput(0.02,-2.3){01/02} \psline(0.44,-2.1)(0.44,-1.9) \rput(0.44,-2.3){02/02} \psline(0.84,-2.1)(0.84,-1.9) \rput(0.84,-2.3){03/02} \psline(1.28,-2.1)(1.28,-1.9) \rput(1.28,-2.3){04/01} \psline(1.72,-2.1)(1.72,-1.9) \rput(1.72,-2.3){05/01} \psline(2.14,-2.1)(2.14,-1.9) \rput(2.14,-2.3){06/01} \psline(2.58,-2.1)(2.58,-1.9) \rput(2.58,-2.3){07/01} \psline(3.04,-2.1)(3.04,-1.9) \rput(3.04,-2.3){08/03} \psline(3.46,-2.1)(3.46,-1.9) \rput(3.46,-2.3){09/01} \psline(3.90,-2.1)(3.90,-1.9) \rput(3.90,-2.3){10/01} \psline(4.34,-2.1)(4.34,-1.9) \rput(4.34,-2.3){11/02} \psline(4.76,-2.1)(4.76,-1.9) \rput(4.76,-2.3){12/01} \psline(5.20,-2.1)(5.20,-1.9) \rput(5.20,-2.3){12/31} \psline(-0.043,-2)(0.043,-2) \rput(-0.17,-2){$0$} \psline(-0.043,-1)(0.043,-1) \rput(-0.17,-1){$1$} \psline(-0.043,0)(0.043,0) \rput(-0.17,0){$2$} \psline(-0.043,1)(0.043,1) \rput(-0.17,1){$3$}
\psline[linecolor=blue](0,-1.761) (0.02,-1.770) (0.04,-1.745) (0.06,-1.683) (0.08,-1.677) (0.1,-1.685) (0.12,-1.668) (0.14,-1.662) (0.16,-1.691) (0.18,-1.631) (0.2,-1.608) (0.22,-1.514) (0.24,-1.529) (0.26,-1.526) (0.28,-1.460) (0.3,-1.450) (0.32,-1.427) (0.34,-1.424) (0.36,-1.355) (0.38,-1.326) (0.4,-1.291) (0.42,-1.249) (0.44,-1.217) (0.46,-1.244) (0.48,-1.212) (0.5,-1.220) (0.52,-1.231) (0.54,-1.211) (0.56,-1.214) (0.58,-1.196) (0.6,-1.175) (0.62,-1.220) (0.64,-1.224) (0.66,-1.168) (0.68,-1.082) (0.7,-0.982) (0.72,-1.003) (0.74,-1.061) (0.76,-1.029) (0.78,-0.988) (0.8,-0.918) (0.82,-0.902) (0.84,-0.762) (0.86,-0.729) (0.88,-0.696) (0.9,-0.665) (0.92,-0.681) (0.94,-0.677) (0.96,-0.676) (0.98,-0.645) (1,-0.581) (1.02,-0.564) (1.04,-0.608) (1.06,-0.573) (1.08,-0.477) (1.1,-0.500) (1.12,-0.460) (1.14,-0.450) (1.16,-0.473) (1.18,-0.464) (1.2,-0.446) (1.22,-0.302) (1.24,-0.466) (1.26,-0.509) (1.28,-0.397) (1.3,-0.286) (1.32,-0.261) (1.34,-0.182) (1.36,-0.149) (1.38,-0.106) (1.4,-0.110) (1.42,-0.050) (1.44,-0.156) (1.46,-0.040) (1.48,-0.020) (1.5,0.102) (1.52,0.122) (1.54,0.120) (1.56,0.192) (1.58,0.164) (1.6,0.144) (1.62,-0.169) (1.64,-0.197) (1.66,-0.190) (1.68,-0.056) (1.7,-0.036) (1.72,-0.014) (1.74,0.189) (1.76,0.270) (1.78,0.403) (1.8,0.441) (1.82,0.414) (1.84,0.386) (1.86,0.479) (1.88,0.498) (1.9,0.376) (1.92,0.304) (1.94,0.081) (1.96,0.157) (1.98,0.267) (2,0.389) (2.02,0.395) (2.04,0.393) (2.06,0.377) (2.08,0.480) (2.1,0.546) (2.12,0.523) (2.14,0.593) (2.16,0.689) (2.18,0.683) (2.2,0.747) (2.22,0.726) (2.24,0.817) (2.26,0.868) (2.28,0.897) (2.3,0.878) (2.32,0.884) (2.34,0.956) (2.36,0.891) (2.38,0.786) (2.4,0.590) (2.42,0.638) (2.44,0.685) (2.46,0.725) (2.48,0.757) (2.5,0.628) (2.52,0.541) (2.54,0.556) (2.56,0.715) (2.58,0.705) (2.6,0.570) (2.62,0.593) (2.64,0.533) (2.66,0.615) (2.68,0.719) (2.7,0.751) (2.72,0.732) (2.74,0.771) (2.76,0.820) (2.78,0.837) (2.8,0.681) (2.82,0.594) (2.84,0.470) (2.86,0.533) (2.88,0.773) (2.9,0.807) (2.92,0.924) (2.94,0.982) (2.96,1.026) (2.98,1.045) (3,1.043) (3.02,0.965) (3.04,1.001) (3.06,1.073) (3.08,1.109) (3.1,1.216) (3.12,1.296) (3.14,1.379) (3.16,1.376) (3.18,1.354) (3.2,1.306) (3.22,1.249) (3.24,1.176) (3.26,1.223) (3.28,1.283) (3.3,1.321) (3.32,1.315) (3.34,1.402) (3.36,1.439) (3.38,1.446) (3.4,1.473) (3.42,1.502) (3.44,1.442) (3.46,1.353) (3.48,1.368) (3.5,1.162) (3.52,1.189) (3.54,1.096) (3.56,1.086) (3.58,1.129) (3.6,1.178) (3.62,1.157) (3.64,1.146) (3.66,1.111) (3.68,1.148) (3.7,1.163) (3.72,1.164) (3.74,1.198) (3.76,1.206) (3.78,1.277) (3.8,1.501) (3.82,1.541) (3.84,1.576) (3.86,1.531) (3.88,1.584) (3.9,1.643) (3.92,1.623) (3.94,1.573) (3.96,1.659) (3.98,1.646) (4,1.638) (4.02,1.659) (4.04,1.690) (4.06,1.571) (4.08,1.490) (4.1,0.850) (4.12,-0.481) (4.14,0.287) (4.16,0.267) (4.18,-0.102) (4.2,-0.838) (4.22,-0.653) (4.24,-0.835) (4.26,-1.028) (4.28,-0.699) (4.3,-0.518) (4.32,-0.599) (4.34,-0.786) (4.36,-0.936) (4.38,-0.874) (4.4,-1.045) (4.42,-1.344) (4.44,-1.491) (4.46,-1.217) (4.48,-0.943) (4.5,-0.856) (4.52,-0.984) (4.54,-0.887) (4.56,-0.945) (4.58,-0.968) (4.6,-0.951) (4.62,-0.857) (4.64,-0.749) (4.66,-0.728) (4.68,-0.731) (4.7,-0.993) (4.72,-0.958) (4.74,-0.928) (4.76,-1.013) (4.78,-1.127) (4.8,-1.122) (4.82,-0.954) (4.84,-0.949) (4.86,-0.847) (4.88,-0.893) (4.9,-0.875) (4.92,-0.874) (4.94,-0.840) (4.96,-0.818) (4.98,-0.812) (5,-0.712) (5.02,-0.777) (5.04,-0.746) (5.06,-0.757) (5.08,-1.087) (5.1,-1.112) (5.12,-1.094)
\psline[linecolor=red](0.02,-2.1089)(0.04,-2.0073)(0.06,-1.9222)(0.08,-1.8933)(0.1,-1.8809)(0.12,-1.8613)(0.14,-1.8168)
(0.16,-1.8138)(0.18,-1.6320)(0.2,-1.5349)(0.22,-1.4115)(0.24,-1.2348)(0.26,-1.3137)(0.28,-1.3439)(0.3,-1.1491)
(0.32,-1.2420)(0.34,-1.2569)(0.36,-1.1235)(0.38,-1.0227)(0.4,-1.0680)(0.42,-1.1159)(0.44,-1.0009)(0.46,-1.0079)
(0.48,-1.0041)(0.5,-1.0211)(0.52,-1.0752)(0.54,-1.0494)(0.56,-1.0569)(0.58,-1.0548)(0.6,-1.0499)(0.62,-1.0739)
(0.64,-1.0661)(0.66,-0.9394)(0.68,-0.9146)(0.7,-0.8723)(0.72,-0.8465)(0.74,-0.9518)(0.76,-0.8890)(0.78,-0.8354)
(0.8,-0.7961)(0.82,-0.7796)(0.84,-0.7206)(0.86,-0.6853)(0.88,-0.5622)(0.9,-0.5363)(0.92,-0.5479)(0.94,-0.6010)
(0.96,-0.5522)(0.98,-0.5606)(1,-0.5034)(1.02,-0.4991)(1.04,-0.5053)(1.06,-0.3755)(1.08,-0.3264)(1.1,-0.3571)
(1.12,-0.2601)(1.14,-0.1381)(1.16,-0.0680)(1.18,-0.0839)(1.2,-0.0543)(1.22,0.0527)(1.24,-0.0901)(1.26,-0.1125)
(1.28,-0.0580)(1.3,0.0007)(1.32,0.1637)(1.34,0.2443)(1.36,0.1947)(1.38,0.1858)(1.4,0.2017)(1.42,0.3933)(1.44,0.2298)
(1.46,0.3656)(1.48,0.4346)(1.5,0.5394)(1.52,0.5819)(1.54,0.7291)(1.56,0.6337)(1.58,0.6299)(1.6,0.6520)(1.62,0.4478)
(1.64,0.3901)(1.66,0.3866)(1.68,0.5574)(1.7,0.6744)(1.72,0.7706)(1.74,0.9131)(1.76,0.9142)(1.78,0.9255)(1.8,0.9972)
(1.82,0.9737)(1.84,0.9848)(1.86,0.9242)(1.88,0.9964)(1.9,0.9163)(1.92,0.7687)(1.94,0.6135)(1.96,0.4168)(1.98,0.4418)
(2,0.5302)(2.02,0.6137)(2.04,0.6035)(2.06,0.5361)(2.08,0.5348)(2.1,0.5566)(2.12,0.6345)(2.14,0.6585)(2.16,0.7266)
(2.18,0.7770)(2.2,0.7388)(2.22,0.8743)(2.24,0.9050)(2.26,1.0266)(2.28,1.0716)(2.3,1.0982)(2.32,1.0376)(2.34,1.0708)
(2.36,1.0848)(2.38,1.0284)(2.4,0.9126)(2.42,0.7741)(2.44,0.7735)(2.46,0.8069)(2.48,0.8301)(2.5,0.8231)(2.52,0.6754)
(2.54,0.5313)(2.56,0.5027)(2.58,0.6167)(2.6,0.5671)(2.62,0.4095)(2.64,0.4464)(2.66,0.3869)(2.68,0.4553)(2.7,0.5237)
(2.72,0.5097)(2.74,0.5326)(2.76,0.6135)(2.78,0.6698)(2.8,0.5911)(2.82,0.4022)(2.84,0.2950)(2.86,0.2004)(2.88,0.2958)
(2.9,0.5270)(2.92,0.5658)(2.94,0.6280)(2.96,0.7056)(2.98,0.7250)(3,0.7603)(3.02,0.7226)(3.04,0.5722)(3.06,0.6426)
(3.08,0.7355)(3.1,0.7326)(3.12,0.8813)(3.14,1.0050)(3.16,1.0756)(3.18,1.1117)(3.2,1.1608)(3.22,1.1263)(3.24,1.1263)
(3.26,1.1432)(3.28,1.2982)(3.3,1.4499)(3.32,1.5035)(3.34,1.5065)(3.36,1.4766)(3.38,1.5183)(3.4,1.4483)(3.42,1.4868)
(3.44,1.4189)(3.46,1.3903)(3.48,1.3200)(3.5,1.2772)(3.52,1.1459)(3.54,1.1160)(3.56,1.0716)(3.58,1.0411)(3.6,1.1489)
(3.62,1.0923)(3.64,1.0500)(3.66,1.0012)(3.68,1.0082)(3.7,1.0344)(3.72,0.9406)(3.74,1.0578)(3.76,1.0988)(3.78,1.0775)
(3.8,1.1311)(3.82,1.3747)(3.84,1.2661)(3.86,1.2362)(3.88,1.2316)(3.9,1.2979)(3.92,1.3149)(3.94,1.2270)(3.96,1.2117)
(3.98,1.2640)(4,1.2384)(4.02,1.2012)(4.04,1.2661)(4.06,1.2087)(4.08,1.0395)(4.1,0.7479)(4.12,-0.4916)(4.14,-1.4309)
(4.16,-0.5563)(4.18,-0.7866)(4.2,-1.0001)(4.22,-1.6484)(4.24,-1.4549)(4.26,-1.4886)(4.28,-1.4167)(4.3,-1.0036)
(4.32,-0.8061)(4.34,-0.9187)(4.36,-1.1186)(4.38,-1.0389)(4.4,-1.0418)(4.42,-1.2994)(4.44,-1.5115)(4.46,-1.5775)
(4.48,-1.3102)(4.5,-1.1900)(4.52,-1.1590)(4.54,-1.2383)(4.56,-1.1243)(4.58,-1.2229)(4.6,-1.1852)(4.62,-1.1677)
(4.64,-1.0623)(4.66,-0.9836)(4.68,-0.9669)(4.7,-1.0127)(4.72,-1.2870)(4.74,-1.3113)(4.76,-1.2736)(4.78,-1.4614)
(4.8,-1.5535)(4.82,-1.5075)(4.84,-1.3083)(4.86,-1.2396)(4.88,-1.1291)(4.9,-1.1046)(4.92,-0.9640)(4.94,-0.9429)
(4.96,-0.8551)(4.98,-0.8669)(5,-0.8133)(5.02,-0.7535)(5.04,-0.7759)(5.06,-0.7301)(5.08,-0.7198)(5.1,-0.8917)
(5.12,-0.9887)(5.14,-0.9049)(5.16,-0.8497)
\end{pspicture}

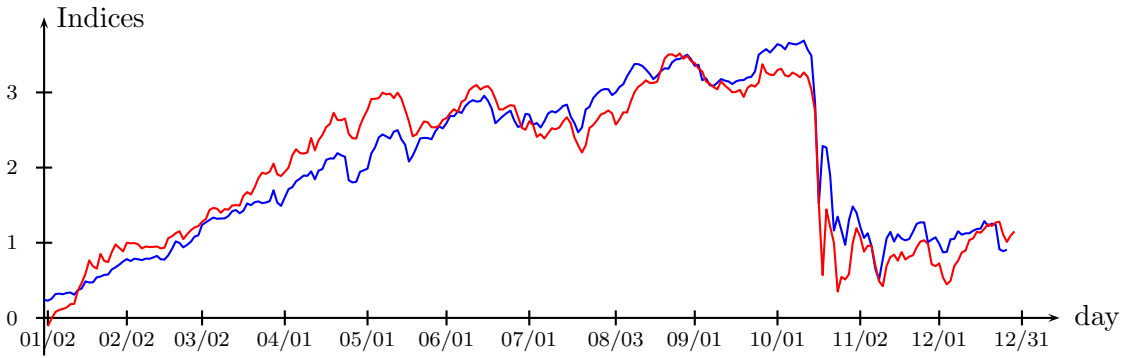
{\small \noindent Figure 6: time series of the market mode calculated using the eigenvector related with the largest eingenvalue of the correlation matrix (blue) plotted against the world index calculated by the MSCI (red). Both indices are normalized so as to have mean two and standard deviation one.}

\vskip 0.3 cm

In terms of portfolio theory, as stated by Markowitz' ideas \cite{Elton}, \cite{Bodie}, the eigenvector corresponding to the largest eigenvalue represents the riskier portfolio one may build, as most of the indices vary in the same way. In constrast, some of the smaller eigenvectors represent portfolios with less risk, as, for example, eigenvector $e_1$, which basically consists on ``buying'' S\&P 500 (USA) and S\&P TSX (Canada) and ``short-selling'' Nasdaq (USA), which are three very closely connected indices. Eigenvector $e_{15}$ corresponds to one of the eigenvectors that are within the region considered as noise, and should represent just a random combination of stock market indices.

More differences between eigenvector $e_{23}$ and the other eigenvectors can be seen if we build probability distribution of frequencies graphs for the twenty three eigenvalues. All distributions, except the one for eigenvector $e_{23}$, have average near zero and standard deviation around 0.21, while this is not the case for eigenvector $e_{23}$. The elements of eigenvector $e_{23}$ have mean $0.17$ and standard deviation $0.13$.

Some recent works discussed how finite sized data and log-return distributions that are not Gaussian could affect the probability distribution of the eigenvalues of an empirical correlation matrix. Some of the results imply that the usual Mar\v{e}nko-Pastur distribution acquires a fat tail in the direction of the largest eigenvalue.

A last analysis which shows the difference between the highest eigenvalues and the eigenvalues belonging to the range associated with noise may be done using the so called {\sl Inverse Participation Ratio} (IPR),
\begin{equation}
\label{ipr}
IPR_k=\sum_{i=1}^{N}\left( e_k^i\right) ^4\ ,
\end{equation}
where $e_k^i$ is the $i$-th element of eigenvector $e_k$, and $N$ is the total number of eigenvectors. Its inverse gives the average number of stocks which contribute significantly to a portfolio built with such eigenvector. The next two figures show the IPR for the 23 eigenvectors, in ascending order from the left to the right (figure 7), and its inverse, $PR_k=1/IPR_k$, for Participation Ratio (figure 8).

\begin{pspicture}(-2,0)(3.5,3.6)
\psset{xunit=0.5,yunit=5}
\psdots[linecolor=blue](1,0.490)(2,0.317)(3,0.383)(4,0.248)(5,0.130)(6,0.170)(7,0.164)(8,0.237)(9,0.221)(10,0.112)
(11,0.197)(12,0.120)(13,0.122)(14,0.160)(15,0.172)(16,0.371)(17,0.178)(18,0.150)(19,0.099)(20,0.223)(21,0.109)
(22,0.098)(23,0.084)
\psline(1,0.490)(2,0.317)(3,0.383)(4,0.248)(5,0.130)(6,0.170)(7,0.164)(8,0.237)(9,0.221)(10,0.112)
(11,0.197)(12,0.120)(13,0.122)(14,0.160)(15,0.172)(16,0.371)(17,0.178)(18,0.150)(19,0.099)(20,0.223)(21,0.109)
(22,0.098)(23,0.084)
\psline{->}(0,0)(24,0) \rput(24.6,0){$\lambda $} \psline{->}(0,0)(0,0.6) \rput(1.1,0.6){$IPR$} \scriptsize \psline(-0.2,0.1)(0.2,0.1) \rput(-0.8,0.1){$0.1$} \psline(-0.2,0.2)(0.2,0.2) \rput(-0.8,0.2){$0.2$} \psline(-0.2,0.3)(0.2,0.3) \rput(-0.8,0.3){$0.3$} \psline(-0.2,0.4)(0.2,0.4) \rput(-0.8,0.4){$0.4$} \psline(-0.2,0.5)(0.2,0.5) \rput(-0.8,0.5){$0.5$}
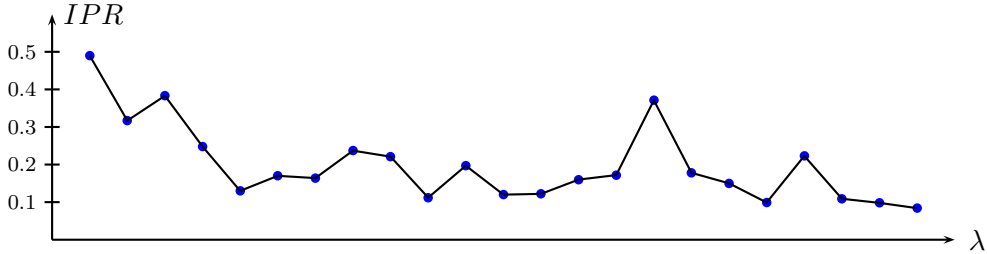
\small \rput(12.5,-0.13){Figure 7: inverse participation ratio of the eigenvectors of the correlation matrix.}
\end{pspicture}

\begin{pspicture}(-2,0)(3.5,6.3)
\psset{xunit=0.5,yunit=0.3}
\pspolygon*[linecolor=lightred](0,0)(0,2.040)(1,2.040)(1,0)
\pspolygon*[linecolor=lightred](1,0)(1,3.158)(2,3.158)(2,0)
\pspolygon*[linecolor=lightred](2,0)(2,2.609)(3,2.609)(3,0)
\pspolygon*[linecolor=lightred](3,0)(3,4.035)(4,4.035)(4,0)
\pspolygon*[linecolor=lightred](4,0)(4,7.720)(5,7.720)(5,0)
\pspolygon*[linecolor=lightred](5,0)(5,5.869)(6,5.869)(6,0)
\pspolygon*[linecolor=lightred](6,0)(6,6.100)(7,6.100)(7,0)
\pspolygon*[linecolor=lightred](7,0)(7,4.224)(8,4.224)(8,0)
\pspolygon*[linecolor=lightred](8,0)(8,4.535)(9,4.535)(9,0)
\pspolygon*[linecolor=lightred](9,0)(9,8.900)(10,8.900)(10,0)
\pspolygon*[linecolor=lightred](10,0)(10,5.076)(11,5.076)(11,0)
\pspolygon*[linecolor=lightred](11,0)(11,8.352)(12,8.352)(12,0)
\pspolygon*[linecolor=lightred](12,0)(12,8.230)(13,8.230)(13,0)
\pspolygon*[linecolor=lightred](13,0)(13,6.254)(14,6.254)(14,0)
\pspolygon*[linecolor=lightred](14,0)(14,5.814)(15,5.814)(15,0)
\pspolygon*[linecolor=lightred](15,0)(15,2.695)(16,2.695)(16,0)
\pspolygon*[linecolor=lightred](16,0)(16,5.634)(17,5.634)(17,0)
\pspolygon*[linecolor=lightred](17,0)(17,6.665)(18,6.665)(18,0)
\pspolygon*[linecolor=lightred](18,0)(18,10.135)(19,10.135)(19,0)
\pspolygon*[linecolor=lightred](19,0)(19,4.493)(20,4.493)(20,0)
\pspolygon*[linecolor=lightred](20,0)(20,9.188)(21,9.188)(21,0)
\pspolygon*[linecolor=lightred](21,0)(21,10.171)(22,10.171)(22,0)
\pspolygon*[linecolor=lightred](22,0)(22,11.951)(23,11.951)(23,0)
\pspolygon[linecolor=red](0,0)(0,2.040)(1,2.040)(1,0)
\pspolygon[linecolor=red](1,0)(1,3.158)(2,3.158)(2,0)
\pspolygon[linecolor=red](2,0)(2,2.609)(3,2.609)(3,0)
\pspolygon[linecolor=red](3,0)(3,4.035)(4,4.035)(4,0)
\pspolygon[linecolor=red](4,0)(4,7.720)(5,7.720)(5,0)
\pspolygon[linecolor=red](5,0)(5,5.869)(6,5.869)(6,0)
\pspolygon[linecolor=red](6,0)(6,6.100)(7,6.100)(7,0)
\pspolygon[linecolor=red](7,0)(7,4.224)(8,4.224)(8,0)
\pspolygon[linecolor=red](8,0)(8,4.535)(9,4.535)(9,0)
\pspolygon[linecolor=red](9,0)(9,8.900)(10,8.900)(10,0)
\pspolygon[linecolor=red](10,0)(10,5.076)(11,5.076)(11,0)
\pspolygon[linecolor=red](11,0)(11,8.352)(12,8.352)(12,0)
\pspolygon[linecolor=red](12,0)(12,8.230)(13,8.230)(13,0)
\pspolygon[linecolor=red](13,0)(13,6.254)(14,6.254)(14,0)
\pspolygon[linecolor=red](14,0)(14,5.814)(15,5.814)(15,0)
\pspolygon[linecolor=red](15,0)(15,2.695)(16,2.695)(16,0)
\pspolygon[linecolor=red](16,0)(16,5.634)(17,5.634)(17,0)
\pspolygon[linecolor=red](17,0)(17,6.665)(18,6.665)(18,0)
\pspolygon[linecolor=red](18,0)(18,10.135)(19,10.135)(19,0)
\pspolygon[linecolor=red](19,0)(19,4.493)(20,4.493)(20,0)
\pspolygon[linecolor=red](20,0)(20,9.188)(21,9.188)(21,0)
\pspolygon[linecolor=red](21,0)(21,10.171)(22,10.171)(22,0)
\pspolygon[linecolor=red](22,0)(22,11.951)(23,11.951)(23,0)
\psline[linestyle=dashed](0,6.254)(24,6.254) \rput(26,6.4){Mean PR.}
\psline{->}(0,0)(25,0) \rput(25.6,0){$\lambda $} \psline{->}(0,0)(0,16) \rput(1,16){$PR$} \scriptsize \psline(-0.2,2)(0.2,2) \rput(-0.8,2){$2$} \psline(-0.2,4)(0.2,4) \rput(-0.8,4){$4$} \psline(-0.2,6)(0.2,6) \rput(-0.8,6){$6$} \psline(-0.2,8)(0.2,8) \rput(-0.8,8){$8$} \psline(-0.2,10)(0.2,10) \rput(-0.8,10){$10$} \psline(-0.2,12)(0.2,12) \rput(-0.8,12){$12$} \psline(-0.2,14)(0.2,14) \rput(-0.8,14){$14$} \small \rput(12,-1.9){Figure 8: participation ratio of the eigenvectors of the correlation matrix.}
\end{pspicture}

\vskip 1.1 cm

Note that, for eigenvector $e_{23}$, the number of participating indices is larger than the average, which is about 6. Most of the eigenvectors corresponding to noise fall around that average number, but this is not true for the eigenvectors corresponding to the lowest eigenvalues, which have a very small number of participating indices.

One important result of this theoretical treatment is that the largest eigenvalue, associated with a {\sl market mode}, is like another matrix that is added to the true correlation matrix of the log-returns. In order to study the remaining eigenvalues, one must first clean the empirical correlation matrix from the market mode. The process is known as {\sl single index model}, and is widely used by theoreticians and practitioners of financial markets in order to remove the market mode of stocks negotiated in the same stock exchange \cite{Elton}. This is done in order to study clusters of stocks that move together as blocks in stock markets.

\vskip 0.2 cm

We now measure the average of the correlation matrices in a moving window of 30 days, changing one day at a time. The results are displayed in figure 9, where the average correlation is plotted together with the volatility of the market mode, which we consider here as the absolute value of $S_t$, where $S_t$ is a linear combination of all indices with the elements of eigenvector $e_{23}$ as the coefficients. The plot represents the average correlation of each window as a function of the last day of the window, so that events that occur after the date to which the average correlation is assigned do not influence its value. Volatility is in blue, and the average correlation is in red.

Figure 10 shows the same information, but now both the average correlation and the volatility of the market mode are normalized so as to have mean two and standard deviation one. This is done in order to best compare both values, and it will be more useful for comparisons made for the other crises.

Figure 11 shows the average correlation and the average volatility of the market mode, both calculated in a running window of 30 days, and normalized so as to have mean two and standard deviation one. In this picture, the rise of valatility seems to be preceded by a rise in the correlation between international stock market indices, although that is not a conclusion that may be taken, since we are using averages here over a large period of time.

\begin{pspicture}(-2,-0.6)(3.5,5.6)
\psset{xunit=2.5,yunit=10}
\psline{->}(0,0)(5.4,0) \psline{->}(0,0)(0,0.5) \rput(5.6,0){day} \rput(0.4,0.5){$<C>$,$\text{vol}$}\scriptsize \psline(0.02,-0.01)(0.02,0.01) \rput(0.02,-0.03){01/02} \psline(0.44,-0.01)(0.44,0.01) \rput(0.44,-0.03){02/02} \psline(0.84,-0.01)(0.84,0.01) \rput(0.84,-0.03){03/02} \psline(1.28,-0.01)(1.28,0.01) \rput(1.28,-0.03){04/01} \psline(1.7,-0.01)(1.7,0.01) \rput(1.7,-0.03){05/01} \psline(2.12,-0.01)(2.12,0.01) \rput(2.12,-0.03){06/01} \psline(2.56,-0.01)(2.56,0.01) \rput(2.56,-0.03){07/01} \psline(3.02,-0.01)(3.02,0.01) \rput(3.02,-0.03){08/03} \psline(3.44,-0.01)(3.44,0.01) \rput(3.44,-0.03){09/01} \psline(3.88,-0.01)(3.88,0.01) \rput(3.88,-0.03){10/01} \psline(4.32,-0.01)(4.32,0.01) \rput(4.32,-0.03){11/02} \psline(4.74,-0.01)(4.74,0.01) \rput(4.74,-0.03){12/01} \psline(5.16,-0.01)(5.16,0.01) \rput(5.16,-0.03){12/31} \psline(-0.043,0)(0.043,0) \rput(-0.17,0){$0$} \psline(-0.043,0.1)(0.043,0.1) \rput(-0.17,0.1){$0.1$} \psline(-0.043,0.2)(0.043,0.2) \rput(-0.17,0.2){$0.2$} \psline(-0.043,0.3)(0.043,0.3) \rput(-0.17,0.3){$0.3$} \psline(-0.043,0.4)(0.043,0.4) \rput(-0.17,0.4){$0.4$}
\psline[linecolor=blue](0.62,0.004) (0.64,0.019) (0.66,0.015) (0.68,0.018) (0.7,0.019) (0.72,0.018) (0.74,0.014) (0.76,0.010) (0.78,0.005) (0.8,0.013) (0.82,0.016) (0.84,0.016) (0.86,0.005) (0.88,0.004) (0.9,0.021) (0.92,0.012) (0.94,0.003) (0.96,0.007) (0.98,0.004) (1,0.004) (1.02,0.006) (1.04,0.012) (1.06,0.019) (1.08,0.002) (1.1,0.025) (1.12,0.008) (1.14,0.028) (1.16,0.018) (1.18,0.010) (1.2,0.006) (1.22,0.006) (1.24,0.019) (1.26,0.042) (1.28,0.011) (1.3,0.005) (1.32,0.022) (1.34,0.020) (1.36,0.037) (1.38,0.017) (1.4,0.003) (1.42,0.008) (1.44,0.014) (1.46,0.021) (1.48,0.022) (1.5,0.031) (1.52,0.031) (1.54,0.002) (1.56,0.012) (1.58,0.018) (1.6,0.003) (1.62,0.010) (1.64,0.047) (1.66,0.015) (1.68,0.003) (1.7,0.022) (1.72,0.000) (1.74,0.012) (1.76,0.018) (1.78,0.007) (1.8,0.017) (1.82,0.010) (1.84,0.011) (1.86,0.016) (1.88,0.010) (1.9,0.006) (1.92,0.028) (1.94,0.009) (1.96,0.049) (1.98,0.011) (2,0.007) (2.02,0.010) (2.04,0.014) (2.06,0.011) (2.08,0.009) (2.1,0.016) (2.12,0.012) (2.14,0.008) (2.16,0.015) (2.18,0.006) (2.2,0.003) (2.22,0.003) (2.24,0.005) (2.26,0.007) (2.28,0.021) (2.3,0.021) (2.32,0.014) (2.34,0.034) (2.36,0.020) (2.38,0.005) (2.4,0.011) (2.42,0.002) (2.44,0.011) (2.46,0.016) (2.48,0.015) (2.5,0.013) (2.52,0.007) (2.54,0.010) (2.56,0.002) (2.58,0.035) (2.6,0.027) (2.62,0.003) (2.64,0.022) (2.66,0.013) (2.68,0.010) (2.7,0.017) (2.72,0.023) (2.74,0.018) (2.76,0.026) (2.78,0.032) (2.8,0.030) (2.82,0.002) (2.84,0.007) (2.86,0.018) (2.88,0.019) (2.9,0.006) (2.92,0.004) (2.94,0.047) (2.96,0.023) (2.98,0.022) (3,0.015) (3.02,0.001) (3.04,0.004) (3.06,0.028) (3.08,0.009) (3.1,0.006) (3.12,0.016) (3.14,0.029) (3.16,0.001) (3.18,0.006) (3.2,0.000) (3.22,0.019) (3.24,0.030) (3.26,0.023) (3.28,0.012) (3.3,0.014) (3.32,0.000) (3.34,0.001) (3.36,0.014) (3.38,0.002) (3.4,0.008) (3.42,0.007) (3.44,0.013) (3.46,0.013) (3.48,0.009) (3.5,0.005) (3.52,0.018) (3.54,0.000) (3.56,0.008) (3.58,0.022) (3.6,0.021) (3.62,0.018) (3.64,0.024) (3.66,0.013) (3.68,0.012) (3.7,0.026) (3.72,0.003) (3.74,0.016) (3.76,0.024) (3.78,0.011) (3.8,0.023) (3.82,0.029) (3.84,0.012) (3.86,0.018) (3.88,0.028) (3.9,0.010) (3.92,0.014) (3.94,0.021) (3.96,0.005) (3.98,0.002) (4,0.008) (4.02,0.003) (4.04,0.012) (4.06,0.022) (4.08,0.044) (4.1,0.056) (4.12,0.363) (4.14,0.301) (4.16,0.213) (4.18,0.105) (4.2,0.060) (4.22,0.332) (4.24,0.011) (4.26,0.134) (4.28,0.016) (4.3,0.147) (4.32,0.022) (4.34,0.077) (4.36,0.109) (4.38,0.038) (4.4,0.011) (4.42,0.102) (4.44,0.122) (4.46,0.032) (4.48,0.155) (4.5,0.043) (4.52,0.045) (4.54,0.067) (4.56,0.001) (4.58,0.049) (4.6,0.041) (4.62,0.034) (4.64,0.045) (4.66,0.001) (4.68,0.022) (4.7,0.012) (4.72,0.116) (4.74,0.006) (4.76,0.008) (4.78,0.041) (4.8,0.064) (4.82,0.005) (4.84,0.060) (4.86,0.031) (4.88,0.001) (4.9,0.022) (4.92,0.031) (4.94,0.036) (4.96,0.044) (4.98,0.008) (5,0.021) (5.02,0.073) (5.04,0.024) (5.06,0.032) (5.08,0.021) (5.1,0.085) (5.12,0.018)
\psline[linecolor=red](0.62,0.008) (0.64,0.010) (0.66,0.011) (0.68,0.009) (0.7,0.008) (0.72,0.009) (0.74,0.010) (0.76,0.002) (0.78,0.003) (0.8,0.002) (0.82,0.006) (0.84,0.005) (0.86,0.005) (0.88,0.005) (0.9,0.005) (0.92,0.010) (0.94,0.008) (0.96,-0.001) (0.98,0.002) (1,0.001) (1.02,0.004) (1.04,0.010) (1.06,0.008) (1.08,0.008) (1.1,0.005) (1.12,0.003) (1.14,0.009) (1.16,0.009) (1.18,-0.003) (1.2,-0.002) (1.22,-0.004) (1.24,0.014) (1.26,0.014) (1.28,0.017) (1.3,0.018) (1.32,0.015) (1.34,0.021) (1.36,0.028) (1.38,0.028) (1.4,0.030) (1.42,0.030) (1.44,0.022) (1.46,0.023) (1.48,0.025) (1.5,0.026) (1.52,0.027) (1.54,0.025) (1.56,0.024) (1.58,0.025) (1.6,0.026) (1.62,0.043) (1.64,0.041) (1.66,0.038) (1.68,0.038) (1.7,0.037) (1.72,0.035) (1.74,0.033) (1.76,0.029) (1.78,0.032) (1.8,0.028) (1.82,0.030) (1.84,0.030) (1.86,0.031) (1.88,0.034) (1.9,0.039) (1.92,0.037) (1.94,0.055) (1.96,0.050) (1.98,0.048) (2,0.052) (2.02,0.052) (2.04,0.057) (2.06,0.060) (2.08,0.063) (2.1,0.061) (2.12,0.057) (2.14,0.056) (2.16,0.052) (2.18,0.053) (2.2,0.054) (2.22,0.052) (2.24,0.031) (2.26,0.037) (2.28,0.040) (2.3,0.040) (2.32,0.046) (2.34,0.046) (2.36,0.041) (2.38,0.038) (2.4,0.031) (2.42,0.034) (2.44,0.033) (2.46,0.036) (2.48,0.043) (2.5,0.043) (2.52,0.030) (2.54,0.031) (2.56,0.007) (2.58,0.013) (2.6,0.012) (2.62,0.016) (2.64,0.015) (2.66,0.014) (2.68,0.011) (2.7,0.014) (2.72,0.014) (2.74,0.007) (2.76,0.011) (2.78,0.014) (2.8,0.011) (2.82,0.021) (2.84,0.037) (2.86,0.044) (2.88,0.041) (2.9,0.041) (2.92,0.052) (2.94,0.050) (2.96,0.050) (2.98,0.049) (3,0.049) (3.02,0.049) (3.04,0.049) (3.06,0.046) (3.08,0.046) (3.1,0.048) (3.12,0.048) (3.14,0.046) (3.16,0.045) (3.18,0.042) (3.2,0.044) (3.22,0.061) (3.24,0.064) (3.26,0.064) (3.28,0.066) (3.3,0.066) (3.32,0.062) (3.34,0.062) (3.36,0.061) (3.38,0.058) (3.4,0.053) (3.42,0.054) (3.44,0.052) (3.46,0.048) (3.48,0.048) (3.5,0.046) (3.52,0.042) (3.54,0.026) (3.56,0.030) (3.58,0.027) (3.6,0.033) (3.62,0.037) (3.64,0.037) (3.66,0.035) (3.68,0.046) (3.7,0.046) (3.72,0.041) (3.74,0.037) (3.76,0.038) (3.78,0.045) (3.8,0.047) (3.82,0.047) (3.84,0.037) (3.86,0.038) (3.88,0.041) (3.9,0.042) (3.92,0.045) (3.94,0.047) (3.96,0.045) (3.98,0.043) (4,0.040) (4.02,0.038) (4.04,0.047) (4.06,0.065) (4.08,0.078) (4.1,0.260) (4.12,0.291) (4.14,0.263) (4.16,0.248) (4.18,0.249) (4.2,0.275) (4.22,0.266) (4.24,0.266) (4.26,0.251) (4.28,0.258) (4.3,0.255) (4.32,0.256) (4.34,0.254) (4.36,0.251) (4.38,0.250) (4.4,0.246) (4.42,0.250) (4.44,0.236) (4.46,0.225) (4.48,0.220) (4.5,0.215) (4.52,0.212) (4.54,0.212) (4.56,0.211) (4.58,0.211) (4.6,0.217) (4.62,0.198) (4.64,0.200) (4.66,0.197) (4.68,0.196) (4.7,0.193) (4.72,0.192) (4.74,0.174) (4.76,0.169) (4.78,0.181) (4.8,0.179) (4.82,0.125) (4.84,0.137) (4.86,0.131) (4.88,0.147) (4.9,0.120) (4.92,0.123) (4.94,0.126) (4.96,0.124) (4.98,0.133) (5,0.133) (5.02,0.123) (5.04,0.096) (5.06,0.104) (5.08,0.103) (5.1,0.103) (5.12,0.105)
\end{pspicture}

\vskip 0.1 cm

{\noindent \small Figure 9: volatility of the market mode (blue) and average correlation (red) based on the log-returns for 1987, both calculated in a moving window of 30 days.}

\vskip 0.3 cm

\begin{pspicture}(-2,-1)(3.5,5.6)
\psset{xunit=2.5,yunit=0.5}
\psline{->}(0,-2)(5.4,-2) \psline{->}(0,-2)(0,10) \rput(5.6,-2){day} \rput(0.5,10){$<C>_n$,$\text{vol}_n$}\scriptsize \psline(0.02,-2.2)(0.02,-1.8) \rput(0.02,-2.6){01/02} \psline(0.44,-2.2)(0.44,-1.8) \rput(0.44,-2.6){02/02} \psline(0.84,-2.2)(0.84,-1.8) \rput(0.84,-2.6){03/02} \psline(1.28,-2.2)(1.28,-1.8) \rput(1.28,-2.6){04/01} \psline(1.7,-2.2)(1.7,-1.8) \rput(1.7,-2.6){05/01} \psline(2.12,-2.2)(2.12,-1.8) \rput(2.12,-2.6){06/01} \psline(2.56,-2.2)(2.56,-1.8) \rput(2.56,-2.6){07/01} \psline(3.02,-2.2)(3.02,-1.8) \rput(3.02,-2.6){08/03} \psline(3.44,-2.2)(3.44,-1.8) \rput(3.44,-2.6){09/01} \psline(3.88,-2.2)(3.88,-1.8) \rput(3.88,-2.6){10/01} \psline(4.32,-2.2)(4.32,-1.8) \rput(4.32,-2.6){11/02} \psline(4.74,-2.2)(4.74,-1.8) \rput(4.74,-2.6){12/01} \psline(5.16,-2.2)(5.16,-1.8) \rput(5.16,-2.6){12/31} \psline(-0.043,0)(0.043,0) \rput(-0.17,0){$2$} \psline(-0.043,2)(0.043,2) \rput(-0.17,2){$4$} \psline(-0.043,4)(0.043,4) \rput(-0.17,4){$6$} \psline(-0.043,6)(0.043,6) \rput(-0.17,6){$8$} \psline(-0.043,8)(0.043,8) \rput(-0.17,8){$10$}
\psline[linecolor=blue](0.62,-0.508) (0.64,-0.165) (0.66,-0.260) (0.68,-0.182) (0.7,-0.173) (0.72,-0.194) (0.74,-0.275) (0.76,-0.359) (0.78,-0.478) (0.8,-0.308) (0.82,-0.224) (0.84,-0.228) (0.86,-0.486) (0.88,-0.502) (0.9,-0.121) (0.92,-0.320) (0.94,-0.522) (0.96,-0.433) (0.98,-0.503) (1,-0.510) (1.02,-0.452) (1.04,-0.332) (1.06,-0.167) (1.08,-0.545) (1.1,-0.025) (1.12,-0.400) (1.14,0.033) (1.16,-0.179) (1.18,-0.360) (1.2,-0.457) (1.22,-0.456) (1.24,-0.160) (1.26,0.343) (1.28,-0.348) (1.3,-0.475) (1.32,-0.100) (1.34,-0.137) (1.36,0.223) (1.38,-0.218) (1.4,-0.512) (1.42,-0.399) (1.44,-0.268) (1.46,-0.117) (1.48,-0.109) (1.5,0.091) (1.52,0.100) (1.54,-0.543) (1.56,-0.324) (1.58,-0.187) (1.6,-0.520) (1.62,-0.369) (1.64,0.462) (1.66,-0.246) (1.68,-0.511) (1.7,-0.103) (1.72,-0.579) (1.74,-0.314) (1.76,-0.184) (1.78,-0.428) (1.8,-0.222) (1.82,-0.369) (1.84,-0.349) (1.86,-0.241) (1.88,-0.356) (1.9,-0.460) (1.92,0.025) (1.94,-0.384) (1.96,0.490) (1.98,-0.341) (2,-0.436) (2.02,-0.372) (2.04,-0.282) (2.06,-0.353) (2.08,-0.380) (2.1,-0.224) (2.12,-0.332) (2.14,-0.406) (2.16,-0.263) (2.18,-0.452) (2.2,-0.517) (2.22,-0.514) (2.24,-0.482) (2.26,-0.423) (2.28,-0.127) (2.3,-0.112) (2.32,-0.279) (2.34,0.162) (2.36,-0.134) (2.38,-0.484) (2.4,-0.347) (2.42,-0.545) (2.44,-0.348) (2.46,-0.229) (2.48,-0.264) (2.5,-0.306) (2.52,-0.432) (2.54,-0.373) (2.56,-0.547) (2.58,0.178) (2.6,0.019) (2.62,-0.520) (2.64,-0.109) (2.66,-0.310) (2.68,-0.376) (2.7,-0.209) (2.72,-0.083) (2.74,-0.194) (2.76,-0.015) (2.78,0.121) (2.8,0.086) (2.82,-0.550) (2.84,-0.439) (2.86,-0.185) (2.88,-0.157) (2.9,-0.465) (2.92,-0.492) (2.94,0.442) (2.96,-0.080) (2.98,-0.109) (3,-0.266) (3.02,-0.555) (3.04,-0.498) (3.06,0.024) (3.08,-0.382) (3.1,-0.460) (3.12,-0.243) (3.14,0.058) (3.16,-0.562) (3.18,-0.457) (3.2,-0.578) (3.22,-0.161) (3.24,0.079) (3.26,-0.086) (3.28,-0.327) (3.3,-0.270) (3.32,-0.577) (3.34,-0.562) (3.36,-0.283) (3.38,-0.545) (3.4,-0.420) (3.42,-0.428) (3.44,-0.305) (3.46,-0.297) (3.48,-0.385) (3.5,-0.475) (3.52,-0.190) (3.54,-0.587) (3.56,-0.407) (3.58,-0.110) (3.6,-0.129) (3.62,-0.184) (3.64,-0.062) (3.66,-0.294) (3.68,-0.320) (3.7,-0.018) (3.72,-0.532) (3.74,-0.236) (3.76,-0.059) (3.78,-0.348) (3.8,-0.076) (3.82,0.049) (3.84,-0.330) (3.86,-0.181) (3.88,0.036) (3.9,-0.356) (3.92,-0.278) (3.94,-0.122) (3.96,-0.468) (3.98,-0.536) (4,-0.400) (4.02,-0.523) (4.04,-0.320) (4.06,-0.098) (4.08,0.381) (4.1,0.646) (4.12,7.445) (4.14,6.067) (4.16,4.128) (4.18,1.729) (4.2,0.746) (4.22,6.757) (4.24,-0.345) (4.26,2.387) (4.28,-0.224) (4.3,2.669) (4.32,-0.100) (4.34,1.107) (4.36,1.831) (4.38,0.243) (4.4,-0.349) (4.42,1.670) (4.44,2.119) (4.46,0.113) (4.48,2.840) (4.5,0.361) (4.52,0.402) (4.54,0.891) (4.56,-0.574) (4.58,0.501) (4.6,0.312) (4.62,0.154) (4.64,0.420) (4.66,-0.559) (4.68,-0.100) (4.7,-0.318) (4.72,1.987) (4.74,-0.444) (4.76,-0.402) (4.78,0.317) (4.8,0.833) (4.82,-0.486) (4.84,0.744) (4.86,0.088) (4.88,-0.564) (4.9,-0.111) (4.92,0.089) (4.94,0.210) (4.96,0.379) (4.98,-0.410) (5,-0.117) (5.02,1.023) (5.04,-0.057) (5.06,0.111) (5.08,-0.129) (5.1,1.296) (5.12,-0.184)
\psline[linecolor=red](0.62,-0.828) (0.64,-0.792) (0.66,-0.787) (0.68,-0.812) (0.7,-0.827) (0.72,-0.814) (0.74,-0.799) (0.76,-0.901) (0.78,-0.894) (0.8,-0.906) (0.82,-0.856) (0.84,-0.861) (0.86,-0.866) (0.88,-0.860) (0.9,-0.870) (0.92,-0.799) (0.94,-0.827) (0.96,-0.945) (0.98,-0.904) (1,-0.916) (1.02,-0.877) (1.04,-0.795) (1.06,-0.830) (1.08,-0.830) (1.1,-0.860) (1.12,-0.890) (1.14,-0.815) (1.16,-0.808) (1.18,-0.966) (1.2,-0.958) (1.22,-0.980) (1.24,-0.744) (1.26,-0.747) (1.28,-0.711) (1.3,-0.695) (1.32,-0.729) (1.34,-0.649) (1.36,-0.561) (1.38,-0.559) (1.4,-0.529) (1.42,-0.534) (1.44,-0.641) (1.46,-0.628) (1.48,-0.603) (1.5,-0.586) (1.52,-0.577) (1.54,-0.596) (1.56,-0.608) (1.58,-0.599) (1.6,-0.588) (1.62,-0.360) (1.64,-0.381) (1.66,-0.424) (1.68,-0.423) (1.7,-0.441) (1.72,-0.462) (1.74,-0.493) (1.76,-0.547) (1.78,-0.511) (1.8,-0.559) (1.82,-0.539) (1.84,-0.527) (1.86,-0.520) (1.88,-0.481) (1.9,-0.409) (1.92,-0.440) (1.94,-0.201) (1.96,-0.263) (1.98,-0.294) (2,-0.244) (2.02,-0.238) (2.04,-0.177) (2.06,-0.133) (2.08,-0.095) (2.1,-0.117) (2.12,-0.181) (2.14,-0.189) (2.16,-0.243) (2.18,-0.223) (2.2,-0.211) (2.22,-0.240) (2.24,-0.518) (2.26,-0.438) (2.28,-0.398) (2.3,-0.398) (2.32,-0.327) (2.34,-0.317) (2.36,-0.391) (2.38,-0.431) (2.4,-0.515) (2.42,-0.482) (2.44,-0.496) (2.46,-0.447) (2.48,-0.358) (2.5,-0.364) (2.52,-0.531) (2.54,-0.524) (2.56,-0.842) (2.58,-0.756) (2.6,-0.772) (2.62,-0.720) (2.64,-0.726) (2.66,-0.743) (2.68,-0.789) (2.7,-0.748) (2.72,-0.743) (2.74,-0.840) (2.76,-0.782) (2.78,-0.740) (2.8,-0.780) (2.82,-0.656) (2.84,-0.443) (2.86,-0.350) (2.88,-0.389) (2.9,-0.389) (2.92,-0.239) (2.94,-0.272) (2.96,-0.262) (2.98,-0.276) (3,-0.281) (3.02,-0.279) (3.04,-0.286) (3.06,-0.325) (3.08,-0.322) (3.1,-0.301) (3.12,-0.298) (3.14,-0.316) (3.16,-0.333) (3.18,-0.378) (3.2,-0.353) (3.22,-0.127) (3.24,-0.089) (3.26,-0.083) (3.28,-0.055) (3.3,-0.054) (3.32,-0.115) (3.34,-0.111) (3.36,-0.119) (3.38,-0.162) (3.4,-0.232) (3.42,-0.209) (3.44,-0.237) (3.46,-0.289) (3.48,-0.298) (3.5,-0.320) (3.52,-0.370) (3.54,-0.585) (3.56,-0.539) (3.58,-0.574) (3.6,-0.499) (3.62,-0.447) (3.64,-0.441) (3.66,-0.472) (3.68,-0.326) (3.7,-0.320) (3.72,-0.381) (3.74,-0.439) (3.76,-0.421) (3.78,-0.332) (3.8,-0.302) (3.82,-0.305) (3.84,-0.434) (3.86,-0.422) (3.88,-0.383) (3.9,-0.372) (3.92,-0.341) (3.94,-0.302) (3.96,-0.331) (3.98,-0.364) (4,-0.399) (4.02,-0.431) (4.04,-0.310) (4.06,-0.069) (4.08,0.097) (4.1,2.511) (4.12,2.921) (4.14,2.551) (4.16,2.352) (4.18,2.359) (4.2,2.709) (4.22,2.590) (4.24,2.589) (4.26,2.395) (4.28,2.483) (4.3,2.440) (4.32,2.453) (4.34,2.432) (4.36,2.392) (4.38,2.372) (4.4,2.323) (4.42,2.383) (4.44,2.196) (4.46,2.053) (4.48,1.976) (4.5,1.917) (4.52,1.876) (4.54,1.874) (4.56,1.860) (4.58,1.867) (4.6,1.942) (4.62,1.694) (4.64,1.714) (4.66,1.670) (4.68,1.670) (4.7,1.624) (4.72,1.612) (4.74,1.367) (4.76,1.306) (4.78,1.460) (4.8,1.443) (4.82,0.725) (4.84,0.889) (4.86,0.806) (4.88,1.012) (4.9,0.661) (4.92,0.702) (4.94,0.733) (4.96,0.717) (4.98,0.827) (5,0.831) (5.02,0.700) (5.04,0.345) (5.06,0.451) (5.08,0.439) (5.1,0.437) (5.12,0.453)
\end{pspicture}

\vskip 0.7 cm

{\noindent \small Figure 10: volatility of the market mode (blue) and average correlation (red) based on the log-returns for 1987, calculated in a moving window of 30 days and normalized so as to have mean two and standard deviation one.}

\vskip 0.3 cm

\begin{pspicture}(-2,-1)(3.5,2.6)
\psset{xunit=2.5,yunit=0.5}
\psline{->}(0,-2)(5.4,-2) \psline{->}(0,-2)(0,4) \rput(5.6,-2){day} \rput(0.7,4){$<C>_n$,$<\text{vol}>_n$}\scriptsize \psline(0.02,-2.2)(0.02,-1.8) \rput(0.02,-2.6){01/02} \psline(0.44,-2.2)(0.44,-1.8) \rput(0.44,-2.6){02/02} \psline(0.84,-2.2)(0.84,-1.8) \rput(0.84,-2.6){03/02} \psline(1.28,-2.2)(1.28,-1.8) \rput(1.28,-2.6){04/01} \psline(1.7,-2.2)(1.7,-1.8) \rput(1.7,-2.6){05/01} \psline(2.12,-2.2)(2.12,-1.8) \rput(2.12,-2.6){06/01} \psline(2.56,-2.2)(2.56,-1.8) \rput(2.56,-2.6){07/01} \psline(3.02,-2.2)(3.02,-1.8) \rput(3.02,-2.6){08/03} \psline(3.44,-2.2)(3.44,-1.8) \rput(3.44,-2.6){09/01} \psline(3.88,-2.2)(3.88,-1.8) \rput(3.88,-2.6){10/01} \psline(4.32,-2.2)(4.32,-1.8) \rput(4.32,-2.6){11/02} \psline(4.74,-2.2)(4.74,-1.8) \rput(4.74,-2.6){12/01} \psline(5.16,-2.2)(5.16,-1.8) \rput(5.16,-2.6){12/31} \psline(-0.043,0)(0.043,0) \rput(-0.17,0){$2$} \psline(-0.043,2)(0.043,2) \rput(-0.17,2){$4$}
\psline[linecolor=blue](0.62,-0.630) (0.64,-0.608) (0.66,-0.614) (0.68,-0.614) (0.7,-0.611) (0.72,-0.586) (0.74,-0.568) (0.76,-0.576) (0.78,-0.584) (0.8,-0.566) (0.82,-0.578) (0.84,-0.569) (0.86,-0.569) (0.88,-0.565) (0.9,-0.536) (0.92,-0.540) (0.94,-0.537) (0.96,-0.573) (0.98,-0.590) (1,-0.597) (1.02,-0.598) (1.04,-0.601) (1.06,-0.586) (1.08,-0.584) (1.1,-0.557) (1.12,-0.581) (1.14,-0.557) (1.16,-0.533) (1.18,-0.543) (1.2,-0.572) (1.22,-0.569) (1.24,-0.569) (1.26,-0.529) (1.28,-0.540) (1.3,-0.560) (1.32,-0.554) (1.34,-0.544) (1.36,-0.506) (1.38,-0.489) (1.4,-0.502) (1.42,-0.514) (1.44,-0.517) (1.46,-0.492) (1.48,-0.466) (1.5,-0.452) (1.52,-0.425) (1.54,-0.426) (1.56,-0.419) (1.58,-0.398) (1.6,-0.399) (1.62,-0.393) (1.64,-0.341) (1.66,-0.346) (1.68,-0.344) (1.7,-0.349) (1.72,-0.361) (1.74,-0.384) (1.76,-0.384) (1.78,-0.388) (1.8,-0.373) (1.82,-0.367) (1.84,-0.380) (1.86,-0.418) (1.88,-0.419) (1.9,-0.418) (1.92,-0.409) (1.94,-0.426) (1.96,-0.408) (1.98,-0.416) (2,-0.411) (2.02,-0.409) (2.04,-0.410) (2.06,-0.426) (2.08,-0.444) (2.1,-0.465) (2.12,-0.493) (2.14,-0.484) (2.16,-0.480) (2.18,-0.497) (2.2,-0.497) (2.22,-0.507) (2.24,-0.569) (2.26,-0.581) (2.28,-0.555) (2.3,-0.556) (2.32,-0.536) (2.34,-0.505) (2.36,-0.501) (2.38,-0.505) (2.4,-0.514) (2.42,-0.525) (2.44,-0.525) (2.46,-0.524) (2.48,-0.518) (2.5,-0.508) (2.52,-0.538) (2.54,-0.537) (2.56,-0.606) (2.58,-0.572) (2.6,-0.542) (2.62,-0.551) (2.64,-0.540) (2.66,-0.537) (2.68,-0.537) (2.7,-0.536) (2.72,-0.519) (2.74,-0.505) (2.76,-0.489) (2.78,-0.451) (2.8,-0.412) (2.82,-0.414) (2.84,-0.411) (2.86,-0.395) (2.88,-0.397) (2.9,-0.421) (2.92,-0.435) (2.94,-0.416) (2.96,-0.413) (2.98,-0.388) (3,-0.383) (3.02,-0.383) (3.04,-0.393) (3.06,-0.377) (3.08,-0.384) (3.1,-0.395) (3.12,-0.382) (3.14,-0.354) (3.16,-0.355) (3.18,-0.396) (3.2,-0.436) (3.22,-0.412) (3.24,-0.400) (3.26,-0.385) (3.28,-0.382) (3.3,-0.386) (3.32,-0.418) (3.34,-0.443) (3.36,-0.460) (3.38,-0.504) (3.4,-0.537) (3.42,-0.529) (3.44,-0.521) (3.46,-0.528) (3.48,-0.543) (3.5,-0.544) (3.52,-0.524) (3.54,-0.592) (3.56,-0.613) (3.58,-0.613) (3.6,-0.604) (3.62,-0.580) (3.64,-0.551) (3.66,-0.572) (3.68,-0.568) (3.7,-0.539) (3.72,-0.558) (3.74,-0.577) (3.76,-0.544) (3.78,-0.537) (3.8,-0.504) (3.82,-0.490) (3.84,-0.517) (3.86,-0.523) (3.88,-0.499) (3.9,-0.505) (3.92,-0.485) (3.94,-0.456) (3.96,-0.468) (3.98,-0.468) (4,-0.467) (4.02,-0.473) (4.04,-0.474) (4.06,-0.461) (4.08,-0.410) (4.1,-0.336) (4.12,0.167) (4.14,0.606) (4.16,0.904) (4.18,1.026) (4.2,1.083) (4.22,1.541) (4.24,1.522) (4.26,1.699) (4.28,1.705) (4.3,1.882) (4.32,1.911) (4.34,1.999) (4.36,2.124) (4.38,2.163) (4.4,2.145) (4.42,2.252) (4.44,2.413) (4.46,2.432) (4.48,2.617) (4.5,2.664) (4.52,2.709) (4.54,2.776) (4.56,2.769) (4.58,2.838) (4.6,2.884) (4.62,2.929) (4.64,2.978) (4.66,2.947) (4.68,2.916) (4.7,2.852) (4.72,2.493) (4.74,2.063) (4.76,1.765) (4.78,1.672) (4.8,1.678) (4.82,1.200) (4.84,1.272) (4.86,1.120) (4.88,1.098) (4.9,0.915) (4.92,0.927) (4.94,0.868) (4.96,0.772) (4.98,0.729) (5,0.745) (5.02,0.702) (5.04,0.559) (5.06,0.559) (5.08,0.363) (5.1,0.424) (5.12,0.386)
\psline[linecolor=red](0.62,-0.828) (0.64,-0.792) (0.66,-0.787) (0.68,-0.812) (0.7,-0.827) (0.72,-0.814) (0.74,-0.799) (0.76,-0.901) (0.78,-0.894) (0.8,-0.906) (0.82,-0.856) (0.84,-0.861) (0.86,-0.866) (0.88,-0.860) (0.9,-0.870) (0.92,-0.799) (0.94,-0.827) (0.96,-0.945) (0.98,-0.904) (1,-0.916) (1.02,-0.877) (1.04,-0.795) (1.06,-0.830) (1.08,-0.830) (1.1,-0.860) (1.12,-0.890) (1.14,-0.815) (1.16,-0.808) (1.18,-0.966) (1.2,-0.958) (1.22,-0.980) (1.24,-0.744) (1.26,-0.747) (1.28,-0.711) (1.3,-0.695) (1.32,-0.729) (1.34,-0.649) (1.36,-0.561) (1.38,-0.559) (1.4,-0.529) (1.42,-0.534) (1.44,-0.641) (1.46,-0.628) (1.48,-0.603) (1.5,-0.586) (1.52,-0.577) (1.54,-0.596) (1.56,-0.608) (1.58,-0.599) (1.6,-0.588) (1.62,-0.360) (1.64,-0.381) (1.66,-0.424) (1.68,-0.423) (1.7,-0.441) (1.72,-0.462) (1.74,-0.493) (1.76,-0.547) (1.78,-0.511) (1.8,-0.559) (1.82,-0.539) (1.84,-0.527) (1.86,-0.520) (1.88,-0.481) (1.9,-0.409) (1.92,-0.440) (1.94,-0.201) (1.96,-0.263) (1.98,-0.294) (2,-0.244) (2.02,-0.238) (2.04,-0.177) (2.06,-0.133) (2.08,-0.095) (2.1,-0.117) (2.12,-0.181) (2.14,-0.189) (2.16,-0.243) (2.18,-0.223) (2.2,-0.211) (2.22,-0.240) (2.24,-0.518) (2.26,-0.438) (2.28,-0.398) (2.3,-0.398) (2.32,-0.327) (2.34,-0.317) (2.36,-0.391) (2.38,-0.431) (2.4,-0.515) (2.42,-0.482) (2.44,-0.496) (2.46,-0.447) (2.48,-0.358) (2.5,-0.364) (2.52,-0.531) (2.54,-0.524) (2.56,-0.842) (2.58,-0.756) (2.6,-0.772) (2.62,-0.720) (2.64,-0.726) (2.66,-0.743) (2.68,-0.789) (2.7,-0.748) (2.72,-0.743) (2.74,-0.840) (2.76,-0.782) (2.78,-0.740) (2.8,-0.780) (2.82,-0.656) (2.84,-0.443) (2.86,-0.350) (2.88,-0.389) (2.9,-0.389) (2.92,-0.239) (2.94,-0.272) (2.96,-0.262) (2.98,-0.276) (3,-0.281) (3.02,-0.279) (3.04,-0.286) (3.06,-0.325) (3.08,-0.322) (3.1,-0.301) (3.12,-0.298) (3.14,-0.316) (3.16,-0.333) (3.18,-0.378) (3.2,-0.353) (3.22,-0.127) (3.24,-0.089) (3.26,-0.083) (3.28,-0.055) (3.3,-0.054) (3.32,-0.115) (3.34,-0.111) (3.36,-0.119) (3.38,-0.162) (3.4,-0.232) (3.42,-0.209) (3.44,-0.237) (3.46,-0.289) (3.48,-0.298) (3.5,-0.320) (3.52,-0.370) (3.54,-0.585) (3.56,-0.539) (3.58,-0.574) (3.6,-0.499) (3.62,-0.447) (3.64,-0.441) (3.66,-0.472) (3.68,-0.326) (3.7,-0.320) (3.72,-0.381) (3.74,-0.439) (3.76,-0.421) (3.78,-0.332) (3.8,-0.302) (3.82,-0.305) (3.84,-0.434) (3.86,-0.422) (3.88,-0.383) (3.9,-0.372) (3.92,-0.341) (3.94,-0.302) (3.96,-0.331) (3.98,-0.364) (4,-0.399) (4.02,-0.431) (4.04,-0.310) (4.06,-0.069) (4.08,0.097) (4.1,2.511) (4.12,2.921) (4.14,2.551) (4.16,2.352) (4.18,2.359) (4.2,2.709) (4.22,2.590) (4.24,2.589) (4.26,2.395) (4.28,2.483) (4.3,2.440) (4.32,2.453) (4.34,2.432) (4.36,2.392) (4.38,2.372) (4.4,2.323) (4.42,2.383) (4.44,2.196) (4.46,2.053) (4.48,1.976) (4.5,1.917) (4.52,1.876) (4.54,1.874) (4.56,1.860) (4.58,1.867) (4.6,1.942) (4.62,1.694) (4.64,1.714) (4.66,1.670) (4.68,1.670) (4.7,1.624) (4.72,1.612) (4.74,1.367) (4.76,1.306) (4.78,1.460) (4.8,1.443) (4.82,0.725) (4.84,0.889) (4.86,0.806) (4.88,1.012) (4.9,0.661) (4.92,0.702) (4.94,0.733) (4.96,0.717) (4.98,0.827) (5,0.831) (5.02,0.700) (5.04,0.345) (5.06,0.451) (5.08,0.439) (5.1,0.437) (5.12,0.453)
\end{pspicture}

\vskip 0.7 cm

{\noindent \small Figure 11: average volatility of the market mode (blue) and average correlation (red) based on the log-returns for 1987, both calculated in a moving window of 30 days and normalized so as to have mean two and standard deviation one.}

\vskip 0.3 cm

It is quite clear that there is a strong correspondence between global market volatility and the correlation of the market indices. The correlation between the two variables along this period is 0.62. One can also note that markets are much more correlated after the period of crisis, and this behavior tends to endure for some time after the crash \cite{out6}, although one must take into account that the averaging procedure for the average correlation makes the curve smoother and thus decreasing less steeply. Figure 12 shows the evolution of the covariance between volatility and $<C>$ in time, calculated in a moving window of 30 days, starting from 02/12/1987 (the first day we assign an average correlation). A clear peak can be seen on the days of greatest volatility (we plot the covariance at the end of the time interval considered for each calculation). Although the covariance is influenced by the value of the volatility, so we expect to have large covariance when volatility is high, it has shown to be more efficient in determining periods of crisis than the correlation, that being the reason we are using it.

\begin{pspicture}(-2,-0.3)(3.5,4.7)
\psset{xunit=2.5,yunit=0.4}
\psline{->}(0,0)(5.4,0) \psline{->}(0,-0.5)(0,10) \rput(5.6,0){day} \rput(0.3,10){covar.}\scriptsize \psline(0.02,-0.25)(0.02,0.25) \rput(0.02,-1){01/02} \psline(0.44,-0.25)(0.44,0.25) \rput(0.44,-1){02/02} \psline(0.84,-0.25)(0.84,0.25) \rput(0.84,-1){03/02} \psline(1.28,-0.25)(1.28,0.25) \rput(1.28,-1){04/01} \psline(1.7,-0.25)(1.7,0.25) \rput(1.7,-1){05/01} \psline(2.12,-0.25)(2.12,0.25) \rput(2.12,-1){06/01} \psline(2.56,-0.25)(2.56,0.25) \rput(2.56,-1){07/01} \psline(3.02,-0.25)(3.02,0.25) \rput(3.02,-1){08/03} \psline(3.44,-0.25)(3.44,0.25) \rput(3.44,-1){09/01} \psline(3.88,-0.25)(3.88,0.25) \rput(3.88,-1){10/01} \psline(4.32,-0.25)(4.32,0.25) \rput(4.32,-1){11/02} \psline(4.74,-0.25)(4.74,0.25) \rput(4.74,-1){12/01} \psline(5.16,-0.25)(5.16,0.25) \rput(5.16,-1){12/31} \psline(-0.043,2)(0.043,2) \rput(-0.2,2){$0.002$} \psline(-0.043,4)(0.043,4) \rput(-0.2,4){$0.004$} \psline(-0.043,6)(0.043,6) \rput(-0.2,6){$0.006$} \psline(-0.043,8)(0.043,8) \rput(-0.2,8){$0.008$}
\psline[linecolor=red](1.22,0.011) (1.24,0.014) (1.26,0.015) (1.28,0.023) (1.3,0.021) (1.32,0.018) (1.34,0.020) (1.36,0.024) (1.38,0.040) (1.4,0.040) (1.42,0.033) (1.44,0.030) (1.46,0.031) (1.48,0.032) (1.5,0.033) (1.52,0.042) (1.54,0.049) (1.56,0.041) (1.58,0.036) (1.6,0.032) (1.62,0.021) (1.64,0.011) (1.66,0.034) (1.68,0.034) (1.7,0.019) (1.72,0.026) (1.74,0.013) (1.76,0.017) (1.78,0.018) (1.8,0.011) (1.82,0.001) (1.84,-0.011) (1.86,-0.010) (1.88,0.003) (1.9,0.000) (1.92,-0.008) (1.94,-0.002) (1.96,-0.005) (1.98,0.018) (2,0.016) (2.02,0.009) (2.04,0.005) (2.06,0.003) (2.08,0.001) (2.1,-0.001) (2.12,0.007) (2.14,0.013) (2.16,0.004) (2.18,0.003) (2.2,0.003) (2.22,-0.008) (2.24,-0.010) (2.26,-0.003) (2.28,-0.002) (2.3,-0.005) (2.32,-0.003) (2.34,-0.007) (2.36,-0.007) (2.38,-0.005) (2.4,-0.006) (2.42,-0.003) (2.44,0.000) (2.46,0.000) (2.48,0.000) (2.5,-0.001) (2.52,-0.003) (2.54,0.005) (2.56,0.008) (2.58,0.014) (2.6,-0.009) (2.62,-0.023) (2.64,-0.013) (2.66,-0.021) (2.68,-0.019) (2.7,-0.014) (2.72,-0.020) (2.74,-0.025) (2.76,-0.024) (2.78,-0.033) (2.8,-0.035) (2.82,-0.032) (2.84,-0.019) (2.86,-0.021) (2.88,-0.016) (2.9,-0.017) (2.92,-0.024) (2.94,-0.033) (2.96,-0.021) (2.98,-0.018) (3,-0.009) (3.02,-0.010) (3.04,-0.018) (3.06,-0.025) (3.08,-0.019) (3.1,-0.022) (3.12,-0.027) (3.14,-0.027) (3.16,-0.020) (3.18,-0.040) (3.2,-0.032) (3.22,-0.028) (3.24,-0.033) (3.26,-0.016) (3.28,-0.013) (3.3,-0.024) (3.32,-0.025) (3.34,-0.027) (3.36,-0.032) (3.38,-0.020) (3.4,-0.003) (3.42,0.018) (3.44,0.006) (3.46,0.003) (3.48,0.004) (3.5,0.007) (3.52,0.006) (3.54,0.004) (3.56,0.018) (3.58,0.020) (3.6,0.012) (3.62,0.007) (3.64,0.004) (3.66,-0.001) (3.68,0.000) (3.7,0.000) (3.72,-0.001) (3.74,0.001) (3.76,0.001) (3.78,-0.004) (3.8,-0.005) (3.82,-0.007) (3.84,-0.009) (3.86,-0.018) (3.88,-0.025) (3.9,-0.025) (3.92,-0.026) (3.94,-0.018) (3.96,-0.009) (3.98,-0.009) (4,-0.003) (4.02,0.000) (4.04,0.005) (4.06,0.005) (4.08,0.011) (4.1,0.047) (4.12,0.327) (4.14,3.014) (4.16,4.765) (4.18,5.737) (4.2,6.024) (4.22,6.045) (4.24,7.580) (4.26,7.232) (4.28,7.462) (4.3,7.116) (4.32,7.375) (4.34,6.987) (4.36,6.837) (4.38,6.802) (4.4,6.460) (4.42,6.085) (4.44,5.982) (4.46,5.785) (4.48,5.420) (4.5,5.252) (4.52,4.849) (4.54,4.463) (4.56,4.116) (4.58,3.657) (4.6,3.170) (4.62,2.687) (4.64,2.189) (4.66,1.748) (4.68,1.463) (4.7,1.300) (4.72,1.442) (4.74,0.875) (4.76,0.771) (4.78,0.794) (4.8,0.799) (4.82,0.803) (4.84,0.542) (4.86,0.597) (4.88,0.543) (4.9,0.700) (4.92,0.600) (4.94,0.685) (4.96,0.645) (4.98,0.511) (5,0.581) (5.02,0.681) (5.04,0.478) (5.06,0.344) (5.08,0.369) (5.1,0.181) (5.12,0.068) (5.14,0.077) (5.16,0.014)
\end{pspicture}

\vskip 0.3 cm

\hskip 2 cm {\small Figure 12: covariance between volatility and average correlation as a function of time.}

\vskip 0.4 cm

\section{1998, Russian Crisis}

The Asian Financial Crisis, which ocurred in 1997, made the demand for raw materials fall worldwide, affecting Russia in particular, which is one of the major world exporters of commodities. With the war in Chechnya, and the transition to a capitalist economy, Russia showed signs of decline in its economy. By May, 1998, the fears concerning the Russian economy brought most of the world's financial markets down, since many countries had a good amount of money invested in that country.

In order to analyze that crisis, we added to the previous indices the following: IPC from Mexico (Mexi), BCP Corp Costa Rica from Costa Rica (CoRi), Bermuda SX Index (Bermuda), Jamaica SX Market Index from Jamaica (Jama), MERVAL from Argentina (Arge), IPSA from Chile (Chil), IBVC from Venezuela (Vene), IGBVL from Peru (Peru), CAC 40 from France (Fran), SMI from Switzerland (Swit), FTSE MIB from Italy (Ital), BEL 20 from Belgium (Belg), OMX Copenhagen 20 from Denmark (Denm), OBX from Norway (Norw), OMX Iceland All-Share Index from Iceland (Icel), PSI 20 from Portugal (Port), PX from the Czech Republic (CzRe), PX from Slovakia (Slok), Budapest SX Index from Hungary (Hung), WIG from Poland (Pola), BET 10 from Romania (Roma), OMXT from Estonia (Esto), PFTS from Ukraine (Ukra), MICEX from Russia (Russ), ISE National 100 from Turkey (Turk), TA 25 from Israel (Isra), BLOM from Lebanon (Leba), TASI from Saudi Arabia (SaAr), MSM 30 from Ohman (Ohma), Karachi 100 from Pakistan (Paki), SSE Composite from China (Chin), SET from Thailand (Thai), S\&P/ASX 200 from Australia, CFG 25 from Morocco (Moro), EGX 30 from Egypt (Egyp), Ghana All Share from Ghana (Ghan), NSE ASI from Nigeria (Nige), NSE 20 from Kenya (Keny), FTSE/JSE Africa All Shares from South Africa (SoAf), and SEMDEX from Mauritius (Maur). So, now we have a total of 63 indices, 5 from North America (if we include Bermuda), 2 from Central America and the Caribbean, 5 from South America, 24 from Europe, 2 from Eurasia, 17 from Asia, 1 from Oceania, and 7 from Africa, where we are considering Russia and Turkey as part of Eurasia, for both countries are located in both continents. This offers a good degree of diversification, and includes Russia, which was of paramount importance in that particular crisis.

Using the modified log-returns (\ref{padronized}) based on the closing indices from 01/02/1998 to 12/30/1998, we built a $63\times 63$ correlation matrix between those. This matrix has average correlation $<C>=0.17$, standard deviation $\sigma =0.04$, and is based on $L=257$ days for the $M=63$ indices, which gives $Q=L/M=257/63\approx 4.079$.

The upper and lower bounds of the eigenvalues of the Mar\v{e}nko-Pastur distribution (\ref{dist}) are
\begin{equation}
\lambda_-=0.255 \ \ \text{and}\ \ \lambda_+=2.235 \ .
\end{equation}

The frequency distribution of the eigenvalues is displayed bellow (figure 13), plotted against the theoretical Mar\v{e}nko-Pastur distribution were it an infinite random matrix with mean zero and standard deviation 1. Figure 14 shows the eigenvalues in order of magnitude, with the area corresponding to noise shaded.

Note that the largest eigenvalue is completely out of scale. We also have several eigenvalues that are bellow the minimum theoretical eingenvalue and two other eigenvalues above the maximum theoretical eigenvalue.

The next picture (figure 15) shows eigenvector $e_{63}$, which corresponds to a combination of all indices in a market movement that explains about 36\% of the collective movement of all indices.

\begin{pspicture}(-0.5,0)(3.5,5)
\psset{xunit=1.07,yunit=3.8}
\pspolygon*[linecolor=lightblue](0,0)(0,0.952)(0.2,0.952)(0.2,0)
\pspolygon*[linecolor=lightblue](0.2,0)(0.2,0.873)(0.4,0.873)(0.4,0)
\pspolygon*[linecolor=lightblue](0.4,0)(0.4,0.714)(0.6,0.714)(0.6,0)
\pspolygon*[linecolor=lightblue](0.6,0)(0.6,0.556)(0.8,0.556)(0.8,0)
\pspolygon*[linecolor=lightblue](0.8,0)(0.8,0.476)(1,0.476)(1,0)
\pspolygon*[linecolor=lightblue](1,0)(1,0.476)(1.2,0.476)(1.2,0)
\pspolygon*[linecolor=lightblue](1.2,0)(1.2,0.238)(1.4,0.238)(1.4,0)
\pspolygon*[linecolor=lightblue](1.4,0)(1.4,0.238)(1.6,0.238)(1.6,0)
\pspolygon*[linecolor=lightblue](1.6,0)(1.6,0.179)(1.8,0.179)(1.8,0)
\pspolygon*[linecolor=lightblue](1.8,0)(1.8,0.079)(2,0.079)(2,0)
\pspolygon*[linecolor=lightblue](3,0)(3,0.079)(3.2,0.079)(3.2,0)
\pspolygon*[linecolor=lightblue](3.4,0)(3.4,0.079)(3.6,0.079)(3.6,0)
\pspolygon*[linecolor=lightblue](5.8,0)(5.8,0.079)(6,0.079)(6,0)
\pspolygon[linecolor=blue](0,0)(0,0.952)(0.2,0.952)(0.2,0)
\pspolygon[linecolor=blue](0.2,0)(0.2,0.873)(0.4,0.873)(0.4,0)
\pspolygon[linecolor=blue](0.4,0)(0.4,0.714)(0.6,0.714)(0.6,0)
\pspolygon[linecolor=blue](0.6,0)(0.6,0.556)(0.8,0.556)(0.8,0)
\pspolygon[linecolor=blue](0.8,0)(0.8,0.476)(1,0.476)(1,0)
\pspolygon[linecolor=blue](1,0)(1,0.476)(1.2,0.476)(1.2,0)
\pspolygon[linecolor=blue](1.2,0)(1.2,0.238)(1.4,0.238)(1.4,0)
\pspolygon[linecolor=blue](1.4,0)(1.4,0.238)(1.6,0.238)(1.6,0)
\pspolygon[linecolor=blue](1.6,0)(1.6,0.179)(1.8,0.179)(1.8,0)
\pspolygon[linecolor=blue](1.8,0)(1.8,0.079)(2,0.079)(2,0)
\pspolygon[linecolor=blue](3,0)(3,0.079)(3.2,0.079)(3.2,0)
\pspolygon[linecolor=blue](3.4,0)(3.4,0.079)(3.6,0.079)(3.6,0)
\pspolygon[linecolor=blue](5.8,0)(5.8,0.079)(6,0.079)(6,0)
\psline{->}(0,0)(7,0) \psline[linecolor=white,linewidth=2pt](4.4,0)(4.6,0) \psline(4.3,-0.05)(4.5,0.05) \psline(4.5,-0.05)(4.7,0.05) \psline{->}(0,0)(0,1.2) \rput(7.3,0){$\lambda $} \rput(0.5,1.2){$\rho (\lambda )$} \scriptsize \psline(1,-0.025)(1,0.025) \rput(1,-0.075){1} \psline(2,-0.025)(2,0.025) \rput(2,-0.075){2} \psline(3,-0.025)(3,0.025) \rput(3,-0.075){3}  \psline(4,-0.025)(4,0.025) \rput(4,-0.075){4} \psline(5,-0.025)(5,0.025) \rput(5,-0.075){16} \psline(6,-0.025)(6,0.025) \rput(6,-0.075){17} \psline(-0.1,0.2)(0.1,0.2) \rput(-0.4,0.2){$0.2$} \psline(-0.1,0.4)(0.1,0.4) \rput(-0.4,0.4){$0.4$} \psline(-0.1,0.6)(0.1,0.6) \rput(-0.4,0.6){$0.6$} \psline(-0.1,0.8)(0.1,0.8) \rput(-0.4,0.8){$0.8$} \psline(-0.1,1)(0.1,1) \rput(-0.4,1){$1$}
\psplot[linecolor=red,plotpoints=500]{0.225}{2.235}{2.235 x sub x 0.225 sub mul 0.5 exp x -1 exp mul 0.649 mul} \small \rput(3.5,-0.2){Figure 13: frequency distribution of the eigenvalues of the} \rput(3.4,-0.3){correlation matrix for 1998. The theoretical distribution}  \rput(1,-0.4){is superimposed on it.}
\end{pspicture}
\begin{pspicture}(-5.5,0)(3.5,2.1)
\psset{xunit=1.07,yunit=2.2}
\pspolygon*[linecolor=lightgray](0.225,0)(0.225,0.8)(2.235,0.8)(2.235,0)
\psline{->}(0,0)(7,0)  \psline[linecolor=white,linewidth=2pt](4.4,0)(4.6,0) \psline(4.3,-0.1)(4.5,0.1) \psline(4.5,-0.1)(4.7,0.1) \rput(7.3,0){$\lambda $} \scriptsize \psline(1,-0.05)(1,0.05) \rput(1,-0.15){1} \psline(2,-0.05)(2,0.05) \rput(2,-0.15){2} \psline(3,-0.05)(3,0.05) \rput(3,-0.15){3}  \psline(4,-0.05)(4,0.05) \rput(4,-0.15){4} \psline(5,-0.05)(5,0.05) \rput(5,-0.15){16} \psline(6,-0.05)(6,0.05) \rput(6,-0.15){17}
\psline[linewidth=1pt](0.060,0)(0.060,0.5) \psline[linewidth=1pt](0.091,0)(0.091,0.5) \psline[linewidth=1pt](0.097,0)(0.097,0.5) \psline[linewidth=1pt](0.107,0)(0.107,0.5) \psline[linewidth=1pt](0.115,0)(0.115,0.5)\psline[linewidth=1pt](0.121,0)(0.121,0.5) \psline[linewidth=1pt](0.127,0)(0.127,0.5)\psline[linewidth=1pt](0.139,0)(0.139,0.5) \psline[linewidth=1pt](0.151,0)(0.151,0.5)\psline[linewidth=1pt](0.174,0)(0.174,0.5) \psline[linewidth=1pt](0.183,0)(0.183,0.5)\psline[linewidth=1pt](0.194,0)(0.194,0.5) \psline[linewidth=1pt](0.203,0)(0.203,0.5) \psline[linewidth=1pt](0.226,0)(0.226,0.5)
\psline[linewidth=1pt](0.244,0)(0.244,0.5) \psline[linewidth=1pt](0.253,0)(0.253,0.5) \psline[linewidth=1pt](0.273,0)(0.273,0.5) \psline[linewidth=1pt](0.287,0)(0.287,0.5) \psline[linewidth=1pt](0.301,0)(0.301,0.5) \psline[linewidth=1pt](0.339,0)(0.339,0.5) \psline[linewidth=1pt](0.346,0)(0.346,0.5) \psline[linewidth=1pt](0.363,0)(0.363,0.5)
\psline[linewidth=1pt](0.374,0)(0.374,0.5) \psline[linewidth=1pt](0.411,0)(0.411,0.5)
\psline[linewidth=1pt](0.417,0)(0.417,0.5) \psline[linewidth=1pt](0.435,0)(0.435,0.5)
\psline[linewidth=1pt](0.464,0)(0.464,0.5) \psline[linewidth=1pt](0.485,0)(0.485,0.5)
\psline[linewidth=1pt](0.525,0)(0.525,0.5) \psline[linewidth=1pt](0.546,0)(0.546,0.5)
\psline[linewidth=1pt](0.563,0)(0.563,0.5) \psline[linewidth=1pt](0.594,0)(0.594,0.5) \psline[linewidth=1pt](0.662,0)(0.662,0.5) \psline[linewidth=1pt](0.671,0)(0.671,0.5) \psline[linewidth=1pt](0.693,0)(0.693,0.5) \psline[linewidth=1pt](0.719,0)(0.719,0.5)
\psline[linewidth=1pt](0.743,0)(0.743,0.5) \psline[linewidth=1pt](0.760,0)(0.760,0.5) \psline[linewidth=1pt](0.789,0)(0.789,0.5) \psline[linewidth=1pt](0.822,0)(0.822,0.5) \psline[linewidth=1pt](0.842,0)(0.842,0.5) \psline[linewidth=1pt](0.900,0)(0.900,0.5)
\psline[linewidth=1pt](0.914,0)(0.914,0.5) \psline[linewidth=1pt](0.945,0)(0.945,0.5)
\psline[linewidth=1pt](0.956,0)(0.956,0.5) \psline[linewidth=1pt](1.009,0)(1.009,0.5) \psline[linewidth=1pt](1.036,0)(1.036,0.5) \psline[linewidth=1pt](1.055,0)(1.055,0.5) \psline[linewidth=1pt](1.112,0)(1.112,0.5) \psline[linewidth=1pt](1.142,0)(1.142,0.5) \psline[linewidth=1pt](1.192,0)(1.192,0.5) \psline[linewidth=1pt](1.207,0)(1.207,0.5) \psline[linewidth=1pt](1.279,0)(1.279,0.5) \psline[linewidth=1pt](1.366,0)(1.366,0.5)
\psline[linewidth=1pt](1.551,0)(1.551,0.5) \psline[linewidth=1pt](1.612,0)(1.612,0.5) \psline[linewidth=1pt](1.627,0)(1.627,0.5) \psline[linewidth=1pt](1.823,0)(1.823,0.5)
\psline[linewidth=1pt](3.150,0)(3.150,0.5) \psline[linewidth=1pt](3.424,0)(3.424,0.5) \psline[linewidth=1pt](5.897,0)(5.897,0.5)
\small \rput(3.45,-0.36){Figure 14: eigenvalues in order of magnitude. The} \rput(3.7,-0.56){shaded area corresponds to the eigenvalues predicted} \rput(1.45,-0.76){for a random matrix.}
\end{pspicture}

\vskip 2.1 cm

\begin{pspicture}(-0.1,0)(3.5,2.5)
\psset{xunit=0.27,yunit=5}
\pspolygon*[linecolor=lightblue](0.5,0)(0.5,0.137)(1.5,0.137)(1.5,0)
\pspolygon*[linecolor=lightblue](1.5,0)(1.5,0.137)(2.5,0.137)(2.5,0)
\pspolygon*[linecolor=lightblue](2.5,0)(2.5,0.164)(3.5,0.164)(3.5,0)
\pspolygon*[linecolor=lightblue](3.5,0)(3.5,0.128)(4.5,0.128)(4.5,0)
\pspolygon*[linecolor=lightred](4.5,0)(4.5,0.024)(5.5,0.024)(5.5,0)
\pspolygon*[linecolor=lightblue](5.5,0)(5.5,0.021)(6.5,0.021)(6.5,0)
\pspolygon*[linecolor=lightred](6.5,0)(6.5,0.004)(7.5,0.004)(7.5,0)
\pspolygon*[linecolor=lightblue](7.5,0)(7.5,0.112)(8.5,0.112)(8.5,0)
\pspolygon*[linecolor=lightblue](8.5,0)(8.5,0.136)(9.5,0.136)(9.5,0)
\pspolygon*[linecolor=lightblue](9.5,0)(9.5,0.131)(10.5,0.131)(10.5,0)
\pspolygon*[linecolor=lightblue](10.5,0)(10.5,0.106)(11.5,0.106)(11.5,0)
\pspolygon*[linecolor=lightblue](11.5,0)(11.5,0.131)(12.5,0.131)(12.5,0)
\pspolygon*[linecolor=lightblue](12.5,0)(12.5,0.204)(13.5,0.204)(13.5,0)
\pspolygon*[linecolor=lightblue](13.5,0)(13.5,0.181)(14.5,0.181)(14.5,0)
\pspolygon*[linecolor=lightblue](14.5,0)(14.5,0.205)(15.5,0.205)(15.5,0)
\pspolygon*[linecolor=lightblue](15.5,0)(15.5,0.206)(16.5,0.206)(16.5,0)
\pspolygon*[linecolor=lightblue](16.5,0)(16.5,0.109)(17.5,0.109)(17.5,0)
\pspolygon*[linecolor=lightblue](17.5,0)(17.5,0.182)(18.5,0.182)(18.5,0)
\pspolygon*[linecolor=lightblue](18.5,0)(18.5,0.191)(19.5,0.191)(19.5,0)
\pspolygon*[linecolor=lightblue](19.5,0)(19.5,0.201)(20.5,0.201)(20.5,0)
\pspolygon*[linecolor=lightblue](20.5,0)(20.5,0.207)(21.5,0.207)(21.5,0)
\pspolygon*[linecolor=lightblue](21.5,0)(21.5,0.192)(22.5,0.192)(22.5,0)
\pspolygon*[linecolor=lightblue](22.5,0)(22.5,0.184)(23.5,0.184)(23.5,0)
\pspolygon*[linecolor=lightblue](23.5,0)(23.5,0.207)(24.5,0.207)(24.5,0)
\pspolygon*[linecolor=lightblue](24.5,0)(24.5,0.190)(25.5,0.190)(25.5,0)
\pspolygon*[linecolor=lightblue](25.5,0)(25.5,0.017)(26.5,0.017)(26.5,0)
\pspolygon*[linecolor=lightblue](26.5,0)(26.5,0.196)(27.5,0.196)(27.5,0)
\pspolygon*[linecolor=lightblue](27.5,0)(27.5,0.183)(28.5,0.183)(28.5,0)
\pspolygon*[linecolor=lightblue](28.5,0)(28.5,0.129)(29.5,0.129)(29.5,0)
\pspolygon*[linecolor=lightblue](29.5,0)(29.5,0.173)(30.5,0.173)(30.5,0)
\pspolygon*[linecolor=lightblue](30.5,0)(30.5,0.008)(31.5,0.008)(31.5,0)
\pspolygon*[linecolor=lightblue](31.5,0)(31.5,0.165)(32.5,0.165)(32.5,0)
\pspolygon*[linecolor=lightblue](32.5,0)(32.5,0.131)(33.5,0.131)(33.5,0)
\pspolygon*[linecolor=lightblue](33.5,0)(33.5,0.013)(34.5,0.013)(34.5,0)
\pspolygon*[linecolor=lightblue](34.5,0)(34.5,0.046)(35.5,0.046)(35.5,0)
\pspolygon*[linecolor=lightblue](35.5,0)(35.5,0.029)(36.5,0.029)(36.5,0)
\pspolygon*[linecolor=lightblue](36.5,0)(36.5,0.089)(37.5,0.089)(37.5,0)
\pspolygon*[linecolor=lightblue](37.5,0)(37.5,0.124)(38.5,0.124)(38.5,0)
\pspolygon*[linecolor=lightblue](38.5,0)(38.5,0.140)(39.5,0.140)(39.5,0)
\pspolygon*[linecolor=lightblue](39.5,0)(39.5,0.020)(40.5,0.020)(40.5,0)
\pspolygon*[linecolor=lightblue](40.5,0)(40.5,0.029)(41.5,0.029)(41.5,0)
\pspolygon*[linecolor=lightred](41.5,0)(41.5,0.020)(42.5,0.020)(42.5,0)
\pspolygon*[linecolor=lightblue](42.5,0)(42.5,0.025)(43.5,0.025)(43.5,0)
\pspolygon*[linecolor=lightblue](43.5,0)(43.5,0.067)(44.5,0.067)(44.5,0)
\pspolygon*[linecolor=lightblue](44.5,0)(44.5,0.030)(45.5,0.030)(45.5,0)
\pspolygon*[linecolor=lightred](45.5,0)(45.5,0.025)(46.5,0.025)(46.5,0)
\pspolygon*[linecolor=lightblue](46.5,0)(46.5,0.110)(47.5,0.110)(47.5,0)
\pspolygon*[linecolor=lightblue](47.5,0)(47.5,0.136)(48.5,0.136)(48.5,0)
\pspolygon*[linecolor=lightblue](48.5,0)(48.5,0.004)(49.5,0.004)(49.5,0)
\pspolygon*[linecolor=lightblue](49.5,0)(49.5,0.072)(50.5,0.072)(50.5,0)
\pspolygon*[linecolor=lightblue](50.5,0)(50.5,0.049)(51.5,0.049)(51.5,0)
\pspolygon*[linecolor=lightblue](51.5,0)(51.5,0.086)(52.5,0.086)(52.5,0)
\pspolygon*[linecolor=lightblue](52.5,0)(52.5,0.074)(53.5,0.074)(53.5,0)
\pspolygon*[linecolor=lightblue](53.5,0)(53.5,0.076)(54.5,0.076)(54.5,0)
\pspolygon*[linecolor=lightblue](54.5,0)(54.5,0.086)(55.5,0.086)(55.5,0)
\pspolygon*[linecolor=lightblue](55.5,0)(55.5,0.130)(56.5,0.130)(56.5,0)
\pspolygon*[linecolor=lightred](56.5,0)(56.5,0.008)(57.5,0.008)(57.5,0)
\pspolygon*[linecolor=lightblue](57.5,0)(57.5,0.002)(58.5,0.002)(58.5,0)
\pspolygon*[linecolor=lightblue](58.5,0)(58.5,0.029)(59.5,0.029)(59.5,0)
\pspolygon*[linecolor=lightblue](59.5,0)(59.5,0.003)(60.5,0.003)(60.5,0)
\pspolygon*[linecolor=lightred](60.5,0)(60.5,0.025)(61.5,0.025)(61.5,0)
\pspolygon*[linecolor=lightblue](61.5,0)(61.5,0.189)(62.5,0.189)(62.5,0)
\pspolygon*[linecolor=lightblue](62.5,0)(62.5,0.025)(63.5,0.025)(63.5,0)
\pspolygon[linecolor=blue](0.5,0)(0.5,0.137)(1.5,0.137)(1.5,0)
\pspolygon[linecolor=blue](1.5,0)(1.5,0.137)(2.5,0.137)(2.5,0)
\pspolygon[linecolor=blue](2.5,0)(2.5,0.164)(3.5,0.164)(3.5,0)
\pspolygon[linecolor=blue](3.5,0)(3.5,0.128)(4.5,0.128)(4.5,0)
\pspolygon[linecolor=red](4.5,0)(4.5,0.024)(5.5,0.024)(5.5,0)
\pspolygon[linecolor=blue](5.5,0)(5.5,0.021)(6.5,0.021)(6.5,0)
\pspolygon[linecolor=red](6.5,0)(6.5,0.004)(7.5,0.004)(7.5,0)
\pspolygon[linecolor=blue](7.5,0)(7.5,0.112)(8.5,0.112)(8.5,0)
\pspolygon[linecolor=blue](8.5,0)(8.5,0.136)(9.5,0.136)(9.5,0)
\pspolygon[linecolor=blue](9.5,0)(9.5,0.131)(10.5,0.131)(10.5,0)
\pspolygon[linecolor=blue](10.5,0)(10.5,0.106)(11.5,0.106)(11.5,0)
\pspolygon[linecolor=blue](11.5,0)(11.5,0.131)(12.5,0.131)(12.5,0)
\pspolygon[linecolor=blue](12.5,0)(12.5,0.204)(13.5,0.204)(13.5,0)
\pspolygon[linecolor=blue](13.5,0)(13.5,0.181)(14.5,0.181)(14.5,0)
\pspolygon[linecolor=blue](14.5,0)(14.5,0.205)(15.5,0.205)(15.5,0)
\pspolygon[linecolor=blue](15.5,0)(15.5,0.206)(16.5,0.206)(16.5,0)
\pspolygon[linecolor=blue](16.5,0)(16.5,0.109)(17.5,0.109)(17.5,0)
\pspolygon[linecolor=blue](17.5,0)(17.5,0.182)(18.5,0.182)(18.5,0)
\pspolygon[linecolor=blue](18.5,0)(18.5,0.191)(19.5,0.191)(19.5,0)
\pspolygon[linecolor=blue](19.5,0)(19.5,0.201)(20.5,0.201)(20.5,0)
\pspolygon[linecolor=blue](20.5,0)(20.5,0.207)(21.5,0.207)(21.5,0)
\pspolygon[linecolor=blue](21.5,0)(21.5,0.192)(22.5,0.192)(22.5,0)
\pspolygon[linecolor=blue](22.5,0)(22.5,0.184)(23.5,0.184)(23.5,0)
\pspolygon[linecolor=blue](23.5,0)(23.5,0.207)(24.5,0.207)(24.5,0)
\pspolygon[linecolor=blue](24.5,0)(24.5,0.190)(25.5,0.190)(25.5,0)
\pspolygon[linecolor=blue](25.5,0)(25.5,0.017)(26.5,0.017)(26.5,0)
\pspolygon[linecolor=blue](26.5,0)(26.5,0.196)(27.5,0.196)(27.5,0)
\pspolygon[linecolor=blue](27.5,0)(27.5,0.183)(28.5,0.183)(28.5,0)
\pspolygon[linecolor=blue](28.5,0)(28.5,0.129)(29.5,0.129)(29.5,0)
\pspolygon[linecolor=blue](29.5,0)(29.5,0.173)(30.5,0.173)(30.5,0)
\pspolygon[linecolor=blue](30.5,0)(30.5,0.008)(31.5,0.008)(31.5,0)
\pspolygon[linecolor=blue](31.5,0)(31.5,0.165)(32.5,0.165)(32.5,0)
\pspolygon[linecolor=blue](32.5,0)(32.5,0.131)(33.5,0.131)(33.5,0)
\pspolygon[linecolor=blue](33.5,0)(33.5,0.013)(34.5,0.013)(34.5,0)
\pspolygon[linecolor=blue](34.5,0)(34.5,0.046)(35.5,0.046)(35.5,0)
\pspolygon[linecolor=blue](35.5,0)(35.5,0.029)(36.5,0.029)(36.5,0)
\pspolygon[linecolor=blue](36.5,0)(36.5,0.089)(37.5,0.089)(37.5,0)
\pspolygon[linecolor=blue](37.5,0)(37.5,0.124)(38.5,0.124)(38.5,0)
\pspolygon[linecolor=blue](38.5,0)(38.5,0.140)(39.5,0.140)(39.5,0)
\pspolygon[linecolor=blue](39.5,0)(39.5,0.020)(40.5,0.020)(40.5,0)
\pspolygon[linecolor=blue](40.5,0)(40.5,0.029)(41.5,0.029)(41.5,0)
\pspolygon[linecolor=red](41.5,0)(41.5,0.020)(42.5,0.020)(42.5,0)
\pspolygon[linecolor=blue](42.5,0)(42.5,0.025)(43.5,0.025)(43.5,0)
\pspolygon[linecolor=blue](43.5,0)(43.5,0.067)(44.5,0.067)(44.5,0)
\pspolygon[linecolor=blue](44.5,0)(44.5,0.030)(45.5,0.030)(45.5,0)
\pspolygon[linecolor=red](45.5,0)(45.5,0.025)(46.5,0.025)(46.5,0)
\pspolygon[linecolor=blue](46.5,0)(46.5,0.110)(47.5,0.110)(47.5,0)
\pspolygon[linecolor=blue](47.5,0)(47.5,0.136)(48.5,0.136)(48.5,0)
\pspolygon[linecolor=blue](48.5,0)(48.5,0.004)(49.5,0.004)(49.5,0)
\pspolygon[linecolor=blue](49.5,0)(49.5,0.072)(50.5,0.072)(50.5,0)
\pspolygon[linecolor=blue](50.5,0)(50.5,0.049)(51.5,0.049)(51.5,0)
\pspolygon[linecolor=blue](51.5,0)(51.5,0.086)(52.5,0.086)(52.5,0)
\pspolygon[linecolor=blue](52.5,0)(52.5,0.074)(53.5,0.074)(53.5,0)
\pspolygon[linecolor=blue](53.5,0)(53.5,0.076)(54.5,0.076)(54.5,0)
\pspolygon[linecolor=blue](54.5,0)(54.5,0.086)(55.5,0.086)(55.5,0)
\pspolygon[linecolor=blue](55.5,0)(55.5,0.130)(56.5,0.130)(56.5,0)
\pspolygon[linecolor=red](56.5,0)(56.5,0.008)(57.5,0.008)(57.5,0)
\pspolygon[linecolor=blue](57.5,0)(57.5,0.002)(58.5,0.002)(58.5,0)
\pspolygon[linecolor=blue](58.5,0)(58.5,0.029)(59.5,0.029)(59.5,0)
\pspolygon[linecolor=blue](59.5,0)(59.5,0.003)(60.5,0.003)(60.5,0)
\pspolygon[linecolor=red](60.5,0)(60.5,0.025)(61.5,0.025)(61.5,0)
\pspolygon[linecolor=blue](61.5,0)(61.5,0.189)(62.5,0.189)(62.5,0)
\pspolygon[linecolor=blue](62.5,0)(62.5,0.025)(63.5,0.025)(63.5,0)
\psline{->}(0,0)(65,0) \psline{->}(0,0)(0,0.4) \rput(1.3,0.4){$e_{63}$} \scriptsize \psline(1,-0.02)(1,0.02) \rput(1,-0.06){S\&P} \psline(5,-0.02)(5,0.02) \rput(5,-0.06){CoRi} \psline(10,-0.02)(10,0.02) \rput(10,-0.06){Chil} \psline(15,-0.02)(15,0.02) \rput(15,-0.06){Fran} \psline(20,-0.02)(20,0.02) \rput(20,-0.06){Belg} \psline(25,-0.02)(25,0.02) \rput(25,-0.06){Norw} \psline(30,-0.02)(30,0.02) \rput(30,-0.06){CzRe} \psline(35,-0.02)(35,0.02) \rput(35,-0.06){Esto} \psline(40,-0.02)(40,0.02) \rput(40,-0.06){Leba} \psline(45,-0.02)(45,0.02) \rput(45,-0.06){SrLa} \psline(50,-0.02)(50,0.02) \rput(50,-0.06){Taiw} \psline(55,-0.02)(55,0.02) \rput(55,-0.06){Phil} \psline(60,-0.02)(60,0.02) \rput(60,-0.06){Nige} \psline(63,-0.02)(63,0.02) \rput(63,-0.06){Maur} \scriptsize \psline(-0.28,0.1)(0.28,0.1) \rput(-1.2,0.1){$0.1$} \psline(-0.28,0.2)(0.28,0.2) \rput(-1.2,0.2){$0.2$} \psline(-0.28,0.3)(0.28,0.3) \rput(-1.2,0.3){$0.3$}
\end{pspicture}

\vskip 0.6 cm

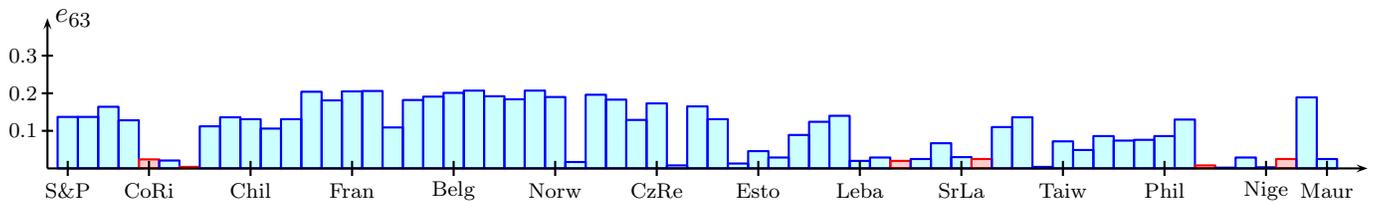
{\noindent \small Figure 15: contributions of the stock market indices to eigenvector $e_{48}$, corresponding to the largest eigenvalue of the correlation matrix. Blue bars indicate positive values, and red bars correspond to negative values. The indices are aligned in the following way: {\bf S\&P}, Nasd, Cana, Mexi, {\bf CoRi}, Berm, Jama, Bra, Arg, {\bf Chil}, Ven, Peru, UK, Irel, {\bf Fran}, Germ, Swit, Autr, Ital, {\bf Belg}, Neth, Swed, Denm, Finl, {\bf Norw}, Icel, Spai, Port, Gree, {\bf CzRe}, Slok, Hung, Pola, Roma, {\bf Esto}, Ukra, Russ, Turk, Isra, {\bf Leba}, SaAr, Ohma, Paki, Indi, {\bf SrLa}, Bang, Japa, HoKo, Chin, {\bf Taiw}, SoKo, Thai, Mala, Indo, {\bf Phil}, Aust, Moro, Egyp, Ghan, {\bf Nige}, Keny, SoAf, {\bf Maur}.}

\vskip 0.4 cm

Note that most indices have similar participations, with the USA and European indices appearing with the largest components for the eigenvector. The smallest participations, some of them with very small negative values, are the ones from Costa Rica, Bermuda, and Jamaica (Central America and the Caribbean), Iceland and Slovakia (Europe), all the Arab countries and most of the Southern Asia ones, China, and the African countries, with the exception of South Africa.

Figure 16 shows the market volatility, together with the average correlation between the indices for 1998, using a running window of 70 days, and representing the average correlation of each window as a correlation of the last day of that window. The window has been enlarged due to the increase in the number of indices so as to avoid too much statistical noise. The volatility of the market mode is in blue, and the average correlation is in red. Both are normalized so as to have mean two and standard deviation one. This is done in order to better compare both measures.

Note that the average correlation is high throughout the period, and it increases beginning in August, 1998, which is the start of the Russian crisis. The volatility of the market mode also grows higher during the same period, although it presents some peaks prior to that time. As the market was unstable due to the Asian crisis of the previous year, that can be explained as well, although there was a drop in correlation between the world stock markets around April, 1998.

Figure 17 shows the average volatility of the market mode (blue) and the average correlation, both normalized so as to have mean two and standard deviation one, and both calculated in a moving window of 70 days.

One can see that volatility and average volatility are correlated with the average correlation between the indices during the times of crisis. This dos not seem to be the case at the beginning of the year, when there was no crisis.

\vskip 0.4 cm

\begin{pspicture}(-1.5,-0.8)(3.5,4.2)
\psset{xunit=2.7,yunit=0.55}
\psline{->}(0,0)(5.6,0) \psline{->}(0,-1)(0,7) \rput(5.8,0){day} \rput(0.5,7){$<C>_n$,$\text{vol.}_n$}\scriptsize \psline(0.02,-0.18)(0.02,0.18) \rput(0.02,-0.54){01/02} \psline(0.44,-0.18)(0.44,0.18) \rput(0.44,-0.54){02/02} \psline(0.84,-0.18)(0.84,0.18) \rput(0.84,-0.54){03/02} \psline(1.28,-0.18)(1.28,0.18) \rput(1.28,-0.54){04/01} \psline(1.7,-0.18)(1.7,0.18) \rput(1.7,-0.54){05/04} \psline(2.1,-0.18)(2.1,0.18) \rput(2.1,-0.54){06/01} \psline(2.54,-0.18)(2.54,0.18) \rput(2.54,-0.54){07/01} \psline(3,-0.18)(3,0.18) \rput(3,-0.54){08/03} \psline(3.42,-0.18)(3.42,0.18) \rput(3.42,-0.54){09/01} \psline(3.86,-0.18)(3.86,0.18) \rput(3.86,-0.54){10/01} \psline(4.3,-0.18)(4.3,0.18) \rput(4.3,-0.54){11/02} \psline(4.72,-0.18)(4.72,0.18) \rput(4.72,-0.54){12/01} \psline(5.14,-0.18)(5.14,0.18) \rput(5.13,-0.54){12/31} \psline(-0.037,1)(0.037,1) \rput(-0.111,1){$1$} \psline(-0.037,2)(0.037,2) \rput(-0.111,2){$2$} \psline(-0.037,3)(0.037,3) \rput(-0.111,3){$3$} \psline(-0.037,4)(0.037,4) \rput(-0.111,4){$4$} \psline(-0.037,5)(0.037,5) \rput(-0.111,5){$5$} \psline(-0.037,6)(0.037,6) \rput(-0.111,6){$6$}
\psline[linecolor=blue](1.42,1.158) (1.44,0.944) (1.46,1.748) (1.48,1.360) (1.5,1.396) (1.52,1.065) (1.54,1.602) (1.56,1.110) (1.58,0.968) (1.6,1.656) (1.62,1.585) (1.64,3.702) (1.66,1.665) (1.68,0.978) (1.7,2.314) (1.72,2.371) (1.74,1.635) (1.76,1.462) (1.78,1.885) (1.8,1.299) (1.82,1.412) (1.84,1.736) (1.86,1.409) (1.88,1.138) (1.9,1.025) (1.92,2.437) (1.94,1.456) (1.96,1.508) (1.98,1.261) (2,0.973) (2.02,0.896) (2.04,1.687) (2.06,3.503) (2.08,1.015) (2.1,0.914) (2.12,2.472) (2.14,1.166) (2.16,1.581) (2.18,0.982) (2.2,1.924) (2.22,1.478) (2.24,1.315) (2.26,2.562) (2.28,2.139) (2.3,2.238) (2.32,3.361) (2.34,1.041) (2.36,3.445) (2.38,0.995) (2.4,1.357) (2.42,1.645) (2.44,1.760) (2.46,1.404) (2.48,1.464) (2.5,1.291) (2.52,1.312) (2.54,0.910) (2.56,2.193) (2.58,1.737) (2.6,1.488) (2.62,1.036) (2.64,1.205) (2.66,1.274) (2.68,1.073) (2.7,1.463) (2.72,1.368) (2.74,2.315) (2.76,1.728) (2.78,1.315) (2.8,1.944) (2.82,1.113) (2.84,1.362) (2.86,2.271) (2.88,1.901) (2.9,1.197) (2.92,2.118) (2.94,1.391) (2.96,1.316) (2.98,1.795) (3,1.419) (3.02,2.563) (3.04,1.680) (3.06,2.725) (3.08,1.934) (3.1,1.094) (3.12,2.559) (3.14,4.890) (3.16,1.807) (3.18,2.582) (3.2,2.184) (3.22,1.805) (3.24,2.466) (3.26,1.666) (3.28,2.239) (3.3,4.499) (3.32,1.468) (3.34,1.974) (3.36,3.480) (3.38,6.147) (3.4,3.126) (3.42,2.733) (3.44,2.342) (3.46,3.558) (3.48,3.155) (3.5,1.263) (3.52,3.886) (3.54,2.640) (3.56,2.213) (3.58,5.300) (3.6,2.058) (3.62,3.248) (3.64,1.092) (3.66,1.510) (3.68,4.572) (3.7,1.115) (3.72,4.213) (3.74,2.966) (3.76,4.169) (3.78,1.574) (3.8,2.474) (3.82,2.433) (3.84,1.332) (3.86,3.331) (3.88,5.805) (3.9,2.397) (3.92,2.999) (3.94,2.819) (3.96,1.490) (3.98,3.647) (4,3.135) (4.02,4.993) (4.04,1.614) (4.06,1.806) (4.08,3.001) (4.1,3.232) (4.12,0.968) (4.14,3.565) (4.16,1.313) (4.18,1.334) (4.2,0.875) (4.22,1.718) (4.24,2.510) (4.26,2.352) (4.28,1.364) (4.3,3.056) (4.32,3.298) (4.34,1.177) (4.36,2.733) (4.38,1.706) (4.4,1.211) (4.42,1.663) (4.44,2.196) (4.46,1.609) (4.48,1.806) (4.5,1.147) (4.52,2.678) (4.54,0.968) (4.56,1.205) (4.58,2.157) (4.6,2.694) (4.62,3.010) (4.64,1.481) (4.66,1.229) (4.68,2.183) (4.7,1.372) (4.72,2.820) (4.74,3.312) (4.76,1.021) (4.78,1.550) (4.8,1.459) (4.82,1.967) (4.84,1.012) (4.86,1.305) (4.88,1.401) (4.9,2.192) (4.92,2.360) (4.94,1.360) (4.96,1.678) (4.98,1.083) (5,1.482) (5.02,2.922) (5.04,0.963) (5.06,1.957) (5.08,1.151) (5.1,1.524) (5.12,0.897) (5.14,0.864)
\psline[linecolor=red](1.42,1.908) (1.44,1.902) (1.46,1.933) (1.48,1.853) (1.5,1.794) (1.52,1.449) (1.54,0.883) (1.56,-0.180) (1.58,-0.215) (1.6,-0.239) (1.62,0.595) (1.64,0.439) (1.66,0.224) (1.68,0.316) (1.7,0.284) (1.72,0.174) (1.74,0.251) (1.76,0.392) (1.78,0.419) (1.8,0.351) (1.82,0.439) (1.84,0.450) (1.86,-0.115) (1.88,-0.031) (1.9,0.160) (1.92,0.090) (1.94,0.065) (1.96,0.063) (1.98,0.062) (2,0.034) (2.02,0.071) (2.04,0.910) (2.06,0.879) (2.08,0.846) (2.1,1.238) (2.12,1.204) (2.14,1.088) (2.16,1.013) (2.18,1.006) (2.2,0.960) (2.22,0.894) (2.24,1.244) (2.26,1.269) (2.28,1.395) (2.3,1.810) (2.32,1.652) (2.34,1.858) (2.36,1.691) (2.38,1.644) (2.4,1.561) (2.42,1.600) (2.44,1.592) (2.46,1.518) (2.48,1.477) (2.5,1.495) (2.52,1.457) (2.54,1.408) (2.56,1.476) (2.58,1.462) (2.6,1.441) (2.62,1.441) (2.64,1.398) (2.66,1.384) (2.68,1.388) (2.7,1.364) (2.72,1.423) (2.74,1.447) (2.76,1.464) (2.78,1.577) (2.8,1.583) (2.82,1.590) (2.84,1.584) (2.86,1.594) (2.88,1.603) (2.9,1.664) (2.92,1.669) (2.94,1.665) (2.96,1.769) (2.98,1.740) (3,1.893) (3.02,1.869) (3.04,1.591) (3.06,1.657) (3.08,1.616) (3.1,1.719) (3.12,2.403) (3.14,2.387) (3.16,2.427) (3.18,2.401) (3.2,2.430) (3.22,2.456) (3.24,2.486) (3.26,2.501) (3.28,2.903) (3.3,2.858) (3.32,2.785) (3.34,2.948) (3.36,3.606) (3.38,3.687) (3.4,3.564) (3.42,3.454) (3.44,3.545) (3.46,3.272) (3.48,3.156) (3.5,3.423) (3.52,3.228) (3.54,3.247) (3.56,3.488) (3.58,3.292) (3.6,3.298) (3.62,3.093) (3.64,3.130) (3.66,3.048) (3.68,2.936) (3.7,2.907) (3.72,2.841) (3.74,3.124) (3.76,2.867) (3.78,2.885) (3.8,2.911) (3.82,2.931) (3.84,2.992) (3.86,3.220) (3.88,3.161) (3.9,3.177) (3.92,3.049) (3.94,2.943) (3.96,2.832) (3.98,2.809) (4,2.821) (4.02,2.785) (4.04,2.749) (4.06,2.812) (4.08,2.942) (4.1,2.934) (4.12,3.031) (4.14,2.986) (4.16,2.913) (4.18,2.844) (4.2,2.807) (4.22,2.831) (4.24,2.848) (4.26,2.854) (4.28,2.994) (4.3,3.053) (4.32,3.053) (4.34,3.124) (4.36,3.081) (4.38,3.018) (4.4,3.013) (4.42,2.963) (4.44,2.981) (4.46,2.983) (4.48,2.984) (4.5,3.082) (4.52,2.996) (4.54,2.801) (4.56,2.815) (4.58,2.879) (4.6,2.969) (4.62,2.891) (4.64,2.914) (4.66,2.887) (4.68,2.890) (4.7,2.671) (4.72,2.693) (4.74,2.701) (4.76,2.614) (4.78,2.172) (4.8,2.038) (4.82,1.997) (4.84,1.970) (4.86,1.887) (4.88,1.866) (4.9,1.885) (4.92,1.686) (4.94,1.783) (4.96,1.793) (4.98,1.502) (5,1.617) (5.02,1.615) (5.04,1.817) (5.06,1.842) (5.08,1.706) (5.1,1.728) (5.12,1.590) (5.14,1.604)
\end{pspicture}

{\noindent \small Figure 16: volatility of the market mode (blue) and average correlation (red) based on the log-returns for 1998, calculated in a moving window of 70 days and normalized so as to have mean two and standard deviation one.}

\vskip 0.7 cm

\begin{pspicture}(-1.5,-0.8)(3.5,2)
\psset{xunit=2.7,yunit=0.55}
\psline{->}(0,0)(5.6,0) \psline{->}(0,-1)(0,4) \rput(5.8,0){day} \rput(0.5,4){$<C>_n$,$\text{vol.}_n$}\scriptsize \psline(0.02,-0.18)(0.02,0.18) \rput(0.02,-0.54){01/02} \psline(0.44,-0.18)(0.44,0.18) \rput(0.44,-0.54){02/02} \psline(0.84,-0.18)(0.84,0.18) \rput(0.84,-0.54){03/02} \psline(1.28,-0.18)(1.28,0.18) \rput(1.28,-0.54){04/01} \psline(1.7,-0.18)(1.7,0.18) \rput(1.7,-0.54){05/04} \psline(2.1,-0.18)(2.1,0.18) \rput(2.1,-0.54){06/01} \psline(2.54,-0.18)(2.54,0.18) \rput(2.54,-0.54){07/01} \psline(3,-0.18)(3,0.18) \rput(3,-0.54){08/03} \psline(3.42,-0.18)(3.42,0.18) \rput(3.42,-0.54){09/01} \psline(3.86,-0.18)(3.86,0.18) \rput(3.86,-0.54){10/01} \psline(4.3,-0.18)(4.3,0.18) \rput(4.3,-0.54){11/02} \psline(4.72,-0.18)(4.72,0.18) \rput(4.72,-0.54){12/01} \psline(5.14,-0.18)(5.14,0.18) \rput(5.13,-0.54){12/31} \psline(-0.037,1)(0.037,1) \rput(-0.111,1){$1$} \psline(-0.037,2)(0.037,2) \rput(-0.111,2){$2$} \psline(-0.037,3)(0.037,3) \rput(-0.111,3){$3$}
\psline[linecolor=blue](1.42,1.184) (1.44,1.156) (1.46,1.148) (1.48,1.142) (1.5,1.135) (1.52,1.087) (1.54,1.029) (1.56,0.925) (1.58,0.858) (1.6,0.840) (1.62,0.848) (1.64,0.896) (1.66,0.863) (1.68,0.852) (1.7,0.890) (1.72,0.900) (1.74,0.919) (1.76,0.937) (1.78,0.942) (1.8,0.924) (1.82,0.929) (1.84,0.931) (1.86,0.863) (1.88,0.860) (1.9,0.838) (1.92,0.867) (1.94,0.868) (1.96,0.877) (1.98,0.889) (2,0.882) (2.02,0.853) (2.04,0.881) (2.06,0.971) (2.08,0.939) (2.1,0.930) (2.12,0.982) (2.14,0.983) (2.16,0.981) (2.18,0.975) (2.2,0.968) (2.22,0.963) (2.24,0.959) (2.26,0.972) (2.28,1.015) (2.3,1.039) (2.32,1.083) (2.34,1.036) (2.36,1.108) (2.38,1.094) (2.4,1.085) (2.42,1.110) (2.44,1.113) (2.46,1.119) (2.48,1.124) (2.5,1.129) (2.52,1.121) (2.54,1.087) (2.56,1.117) (2.58,1.134) (2.6,1.131) (2.62,1.131) (2.64,1.125) (2.66,1.126) (2.68,1.132) (2.7,1.137) (2.72,1.141) (2.74,1.177) (2.76,1.179) (2.78,1.174) (2.8,1.197) (2.82,1.195) (2.84,1.210) (2.86,1.228) (2.88,1.247) (2.9,1.240) (2.92,1.276) (2.94,1.269) (2.96,1.276) (2.98,1.304) (3,1.296) (3.02,1.330) (3.04,1.260) (3.06,1.297) (3.08,1.330) (3.1,1.288) (3.12,1.294) (3.14,1.407) (3.16,1.418) (3.18,1.443) (3.2,1.473) (3.22,1.487) (3.24,1.512) (3.26,1.521) (3.28,1.559) (3.3,1.679) (3.32,1.645) (3.34,1.663) (3.36,1.731) (3.38,1.900) (3.4,1.974) (3.42,2.037) (3.44,2.060) (3.46,2.062) (3.48,2.136) (3.5,2.148) (3.52,2.197) (3.54,2.247) (3.56,2.269) (3.58,2.418) (3.6,2.423) (3.62,2.484) (3.64,2.476) (3.66,2.440) (3.68,2.524) (3.7,2.485) (3.72,2.515) (3.74,2.581) (3.76,2.606) (3.78,2.626) (3.8,2.665) (3.82,2.692) (3.84,2.677) (3.86,2.744) (3.88,2.893) (3.9,2.932) (3.92,2.990) (3.94,3.056) (3.96,3.031) (3.98,3.097) (4,3.154) (4.02,3.291) (4.04,3.305) (4.06,3.323) (4.08,3.390) (4.1,3.451) (4.12,3.437) (4.14,3.480) (4.16,3.466) (4.18,3.467) (4.2,3.430) (4.22,3.451) (4.24,3.490) (4.26,3.493) (4.28,3.474) (4.3,3.539) (4.32,3.579) (4.34,3.572) (4.36,3.621) (4.38,3.618) (4.4,3.611) (4.42,3.580) (4.44,3.597) (4.46,3.559) (4.48,3.554) (4.5,3.556) (4.52,3.560) (4.54,3.425) (4.56,3.404) (4.58,3.390) (4.6,3.407) (4.62,3.449) (4.64,3.415) (4.66,3.400) (4.68,3.398) (4.7,3.290) (4.72,3.337) (4.74,3.383) (4.76,3.298) (4.78,3.139) (4.8,3.082) (4.82,3.055) (4.84,3.009) (4.86,2.932) (4.88,2.871) (4.9,2.903) (4.92,2.850) (4.94,2.806) (4.96,2.788) (4.98,2.642) (5,2.622) (5.02,2.611) (5.04,2.607) (5.06,2.622) (5.08,2.504) (5.1,2.518) (5.12,2.404) (5.14,2.331)
\psline[linecolor=red](1.42,1.908) (1.44,1.902) (1.46,1.933) (1.48,1.853) (1.5,1.794) (1.52,1.449) (1.54,0.883) (1.56,-0.180) (1.58,-0.215) (1.6,-0.239) (1.62,0.595) (1.64,0.439) (1.66,0.224) (1.68,0.316) (1.7,0.284) (1.72,0.174) (1.74,0.251) (1.76,0.392) (1.78,0.419) (1.8,0.351) (1.82,0.439) (1.84,0.450) (1.86,-0.115) (1.88,-0.031) (1.9,0.160) (1.92,0.090) (1.94,0.065) (1.96,0.063) (1.98,0.062) (2,0.034) (2.02,0.071) (2.04,0.910) (2.06,0.879) (2.08,0.846) (2.1,1.238) (2.12,1.204) (2.14,1.088) (2.16,1.013) (2.18,1.006) (2.2,0.960) (2.22,0.894) (2.24,1.244) (2.26,1.269) (2.28,1.395) (2.3,1.810) (2.32,1.652) (2.34,1.858) (2.36,1.691) (2.38,1.644) (2.4,1.561) (2.42,1.600) (2.44,1.592) (2.46,1.518) (2.48,1.477) (2.5,1.495) (2.52,1.457) (2.54,1.408) (2.56,1.476) (2.58,1.462) (2.6,1.441) (2.62,1.441) (2.64,1.398) (2.66,1.384) (2.68,1.388) (2.7,1.364) (2.72,1.423) (2.74,1.447) (2.76,1.464) (2.78,1.577) (2.8,1.583) (2.82,1.590) (2.84,1.584) (2.86,1.594) (2.88,1.603) (2.9,1.664) (2.92,1.669) (2.94,1.665) (2.96,1.769) (2.98,1.740) (3,1.893) (3.02,1.869) (3.04,1.591) (3.06,1.657) (3.08,1.616) (3.1,1.719) (3.12,2.403) (3.14,2.387) (3.16,2.427) (3.18,2.401) (3.2,2.430) (3.22,2.456) (3.24,2.486) (3.26,2.501) (3.28,2.903) (3.3,2.858) (3.32,2.785) (3.34,2.948) (3.36,3.606) (3.38,3.687) (3.4,3.564) (3.42,3.454) (3.44,3.545) (3.46,3.272) (3.48,3.156) (3.5,3.423) (3.52,3.228) (3.54,3.247) (3.56,3.488) (3.58,3.292) (3.6,3.298) (3.62,3.093) (3.64,3.130) (3.66,3.048) (3.68,2.936) (3.7,2.907) (3.72,2.841) (3.74,3.124) (3.76,2.867) (3.78,2.885) (3.8,2.911) (3.82,2.931) (3.84,2.992) (3.86,3.220) (3.88,3.161) (3.9,3.177) (3.92,3.049) (3.94,2.943) (3.96,2.832) (3.98,2.809) (4,2.821) (4.02,2.785) (4.04,2.749) (4.06,2.812) (4.08,2.942) (4.1,2.934) (4.12,3.031) (4.14,2.986) (4.16,2.913) (4.18,2.844) (4.2,2.807) (4.22,2.831) (4.24,2.848) (4.26,2.854) (4.28,2.994) (4.3,3.053) (4.32,3.053) (4.34,3.124) (4.36,3.081) (4.38,3.018) (4.4,3.013) (4.42,2.963) (4.44,2.981) (4.46,2.983) (4.48,2.984) (4.5,3.082) (4.52,2.996) (4.54,2.801) (4.56,2.815) (4.58,2.879) (4.6,2.969) (4.62,2.891) (4.64,2.914) (4.66,2.887) (4.68,2.890) (4.7,2.671) (4.72,2.693) (4.74,2.701) (4.76,2.614) (4.78,2.172) (4.8,2.038) (4.82,1.997) (4.84,1.970) (4.86,1.887) (4.88,1.866) (4.9,1.885) (4.92,1.686) (4.94,1.783) (4.96,1.793) (4.98,1.502) (5,1.617) (5.02,1.615) (5.04,1.817) (5.06,1.842) (5.08,1.706) (5.1,1.728) (5.12,1.590) (5.14,1.604)
\end{pspicture}

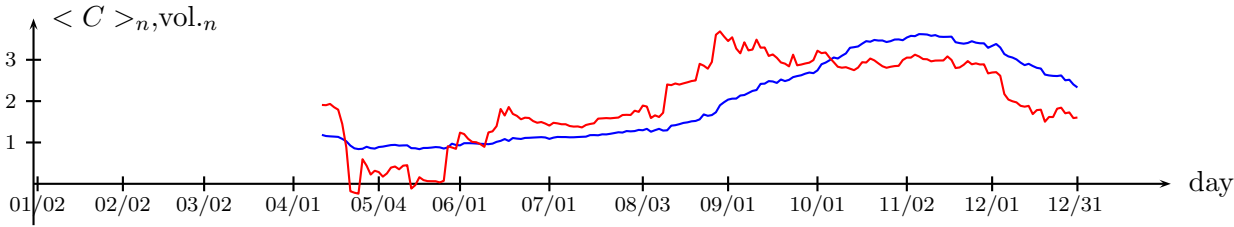
{\noindent \small Figure 17: average volatility of the market mode (blue) and average correlation (red) based on the log-returns for 1998, both calculated in a moving window of 70 days and normalized so as to have mean two and standard deviation one.}

\vskip 0.2 cm

The covariance between the volatility (not the average volatility) and the average covariance, in red, calculated in a moving window of 30 days, is plotted in figure 18. One can verify that the covariance between them increases during the Russian crisis.

\begin{pspicture}(-1.5,-0.6)(3.5,3.4)
\psset{xunit=2.7,yunit=3.5}
\psline{->}(0,0)(5.6,0) \psline{->}(0,-0.2)(0,0.8) \rput(5.8,0){day} \rput(0.3,0.8){covar.}\scriptsize \psline(0.02,-0.028)(0.02,0.028) \rput(0.02,-0.086){01/02} \psline(0.44,-0.028)(0.44,0.028) \rput(0.44,-0.086){02/02} \psline(0.84,-0.028)(0.84,0.028) \rput(0.84,-0.086){03/02} \psline(1.28,-0.028)(1.28,0.028) \rput(1.28,-0.086){04/01} \psline(1.7,-0.028)(1.7,0.028) \rput(1.7,-0.086){05/04} \psline(2.1,-0.028)(2.1,0.028) \rput(2.1,-0.086){06/01} \psline(2.54,-0.028)(2.54,0.028) \rput(2.54,-0.086){07/01} \psline(3,-0.028)(3,0.028) \rput(3,-0.086){08/03} \psline(3.42,-0.028)(3.42,0.028) \rput(3.42,-0.086){09/01} \psline(3.86,-0.028)(3.86,0.028) \rput(3.86,-0.086){10/01} \psline(4.3,-0.028)(4.3,0.028) \rput(4.3,-0.086){11/02} \psline(4.72,-0.028)(4.72,0.028) \rput(4.72,-0.086){12/01} \psline(5.14,-0.028)(5.14,0.028) \rput(5.14,-0.086){12/31} \psline(-0.043,0.2)(0.043,0.2) \rput(-0.24,0.2){$0.0002$} \psline(-0.043,0.4)(0.043,0.4) \rput(-0.24,0.4){$0.0004$} \psline(-0.043,0.6)(0.043,0.6) \rput(-0.24,0.6){$0.0006$}
\psline[linecolor=red](2.02,-0.065) (2.04,-0.024) (2.06,0.023) (2.08,0.056) (2.1,0.061) (2.12,0.045) (2.14,0.113) (2.16,0.096) (2.18,0.080) (2.2,0.040) (2.22,0.047) (2.24,0.044) (2.26,0.038) (2.28,0.077) (2.3,0.092) (2.32,0.139) (2.34,0.244) (2.36,0.208) (2.38,0.288) (2.4,0.262) (2.42,0.240) (2.44,0.233) (2.46,0.237) (2.48,0.213) (2.5,0.179) (2.52,0.139) (2.54,0.170) (2.56,0.149) (2.58,0.148) (2.6,0.123) (2.62,0.076) (2.64,0.020) (2.66,0.019) (2.68,0.065) (2.7,0.048) (2.72,0.042) (2.74,0.051) (2.76,0.044) (2.78,0.043) (2.8,0.026) (2.82,0.034) (2.84,0.026) (2.86,0.020) (2.88,0.034) (2.9,0.039) (2.92,0.027) (2.94,0.020) (2.96,0.027) (2.98,0.008) (3,0.015) (3.02,0.013) (3.04,0.029) (3.06,0.029) (3.08,0.035) (3.1,0.036) (3.12,0.030) (3.14,0.067) (3.16,0.183) (3.18,0.187) (3.2,0.216) (3.22,0.225) (3.24,0.212) (3.26,0.221) (3.28,0.200) (3.3,0.196) (3.32,0.302) (3.34,0.260) (3.36,0.264) (3.38,0.364) (3.4,0.651) (3.42,0.700) (3.44,0.683) (3.46,0.645) (3.48,0.691) (3.5,0.693) (3.52,0.584) (3.54,0.614) (3.56,0.562) (3.58,0.490) (3.6,0.524) (3.62,0.449) (3.64,0.448) (3.66,0.357) (3.68,0.348) (3.7,0.293) (3.72,0.191) (3.74,0.171) (3.76,0.240) (3.78,0.195) (3.8,0.198) (3.82,0.180) (3.84,0.150) (3.86,0.148) (3.88,0.112) (3.9,0.103) (3.92,0.127) (3.94,0.097) (3.96,0.087) (3.98,0.101) (4,-0.001) (4.02,-0.012) (4.04,-0.041) (4.06,-0.007) (4.08,-0.002) (4.1,-0.005) (4.12,0.026) (4.14,0.013) (4.16,0.014) (4.18,0.035) (4.2,0.003) (4.22,0.027) (4.24,0.028) (4.26,0.043) (4.28,0.051) (4.3,0.045) (4.32,0.047) (4.34,0.059) (4.36,0.042) (4.38,0.050) (4.4,0.044) (4.42,0.040) (4.44,0.040) (4.46,0.042) (4.48,0.028) (4.5,-0.011) (4.52,-0.021) (4.54,-0.024) (4.56,-0.016) (4.58,-0.014) (4.6,-0.005) (4.62,0.002) (4.64,0.023) (4.66,0.020) (4.68,0.021) (4.7,0.021) (4.72,0.029) (4.74,0.023) (4.76,0.002) (4.78,0.016) (4.8,0.028) (4.82,0.045) (4.84,0.045) (4.86,0.084) (4.88,0.109) (4.9,0.137) (4.92,0.103) (4.94,0.048) (4.96,0.084) (4.98,0.066) (5,0.105) (5.02,0.132) (5.04,0.089) (5.06,0.102) (5.08,0.104) (5.1,0.122) (5.12,0.155) (5.14,0.147)
\end{pspicture}

\vskip 0.2 cm

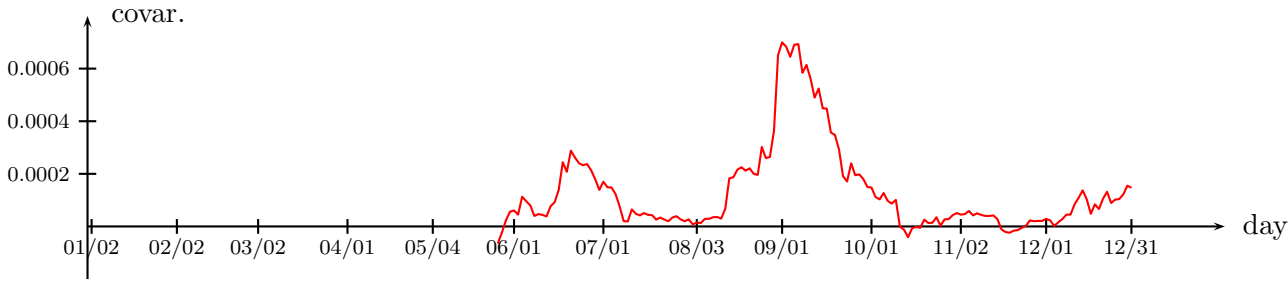
\hskip 2 cm {\small Figure 18: covariance between volatility (blue) and average correlation (red) as a function of time.}

\vskip 0.4 cm

\section{2001, Burst of the dot-com bubble and September 11}

On September, 11, 2001, the world was shocked, as the biggest terrorist atack in human history was perpetrated against the USA. The death toll was close to 3,000, when two hijacked airplanes were thrown into the Twin Towers of the World Trade Center, in New York, one hit the Pentagon, in Virginia, and another fell in Pennsylvania. Financial markets all over the world plummeted, in an uncertainty crisis that lasted a few days.

On that same year, closer to March, there was the end of a financial bubble centered on internet-based companies, the so-called burst of the dot-com companies. That crash affected most markets in the world and is believed to be a result of an escalation of speculation with companies whose true values were much bellow the prices their stocks were being negotiated with.

Here we analyze these two crises, one (September 11) which is a good exemple of a crisis which is caused by a completely exogenous cause, and the other (burst of the dot-com bubble) which is the result of high speculation on stock prices. For 2001, we use 53 indices, adding the KSE 100 from Pakistan (Paki), the Tunindex from Tunisia (Tuni), the SOFIX from Bulgaria (Bulg), the KASE from Kazakhstan (Kaza), and the NZSX 50 from New Zealand (NZ) to the ones already used for 1998.

Using the modified log-returns (\ref{padronized}) based on the closing indices from 01/02/2001 to 12/31/2001, we built a $79\times 79$ correlation matrix between those. This matrix has an average correlation $<C>=0.11$, standard deviation $\sigma =0.03$, and is based on $L=260$ days for the $M=79$ indices, which gives $Q=L/M=260/79\approx 3.29$.

The upper and lower bounds of the eigenvalues of the Mar\v{e}nko-Pastur distribution (\ref{dist}) are
\begin{equation}
\lambda_-=0.295 \ \ \text{and}\ \ \lambda_+=2.122 \ .
\end{equation}

The frequency distribution of the eigenvalues is displayed bellow (figure 19), plotted against the theoretical Mar\v{e}nko-Pastur distribution were it an infinite random matrix with mean zero and standard deviation 1. Figure 20 shows the eigenvalues in order of magnitude. The region associated with noise is shaded.

\begin{pspicture}(-0.5,0)(3.5,5.5)
\psset{xunit=1.07,yunit=4}
\pspolygon*[linecolor=lightblue](0,0)(0,0.696)(0.2,0.696)(0.2,0)
\pspolygon*[linecolor=lightblue](0.2,0)(0.2,0.949)(0.4,0.949)(0.4,0)
\pspolygon*[linecolor=lightblue](0.4,0)(0.4,0.696)(0.6,0.696)(0.6,0)
\pspolygon*[linecolor=lightblue](0.6,0)(0.6,0.506)(0.8,0.506)(0.8,0)
\pspolygon*[linecolor=lightblue](0.8,0)(0.8,0.570)(1,0.570)(1,0)
\pspolygon*[linecolor=lightblue](1,0)(1,0.443)(1.2,0.443)(1.2,0)
\pspolygon*[linecolor=lightblue](1.2,0)(1.2,0.380)(1.4,0.380)(1.4,0)
\pspolygon*[linecolor=lightblue](1.4,0)(1.4,0.190)(1.6,0.190)(1.6,0)
\pspolygon*[linecolor=lightblue](1.6,0)(1.6,0.190)(1.8,0.190)(1.8,0)
\pspolygon*[linecolor=lightblue](1.8,0)(1.8,0.063)(2.0,0.063)(2.0,0)
\pspolygon*[linecolor=lightblue](2.0,0)(2.0,0.127)(2.2,0.127)(2.2,0)
\pspolygon*[linecolor=lightblue](2.4,0)(2.4,0.063)(2.6,0.063)(2.6,0)
\pspolygon*[linecolor=lightblue](4.6,0)(4.6,0.063)(4.8,0.063)(4.8,0)
\pspolygon*[linecolor=lightblue](6.2,0)(6.2,0.063)(6.4,0.063)(6.4,0)
\pspolygon[linecolor=blue](0,0)(0,0.696)(0.2,0.696)(0.2,0)
\pspolygon[linecolor=blue](0.2,0)(0.2,0.949)(0.4,0.949)(0.4,0)
\pspolygon[linecolor=blue](0.4,0)(0.4,0.696)(0.6,0.696)(0.6,0)
\pspolygon[linecolor=blue](0.6,0)(0.6,0.506)(0.8,0.506)(0.8,0)
\pspolygon[linecolor=blue](0.8,0)(0.8,0.570)(1,0.570)(1,0)
\pspolygon[linecolor=blue](1,0)(1,0.443)(1.2,0.443)(1.2,0)
\pspolygon[linecolor=blue](1.2,0)(1.2,0.380)(1.4,0.380)(1.4,0)
\pspolygon[linecolor=blue](1.4,0)(1.4,0.190)(1.6,0.190)(1.6,0)
\pspolygon[linecolor=blue](1.6,0)(1.6,0.190)(1.8,0.190)(1.8,0)
\pspolygon[linecolor=blue](1.8,0)(1.8,0.063)(2.0,0.063)(2.0,0)
\pspolygon[linecolor=blue](2.0,0)(2.0,0.127)(2.2,0.127)(2.2,0)
\pspolygon[linecolor=blue](2.4,0)(2.4,0.063)(2.6,0.063)(2.6,0)
\pspolygon[linecolor=blue](4.6,0)(4.6,0.063)(4.8,0.063)(4.8,0)
\pspolygon[linecolor=blue](6.2,0)(6.2,0.063)(6.4,0.063)(6.4,0)
\psline{->}(0,0)(7,0) \psline[linecolor=white,linewidth=2pt](5.4,0)(5.6,0) \psline(5.3,-0.05)(5.5,0.05) \psline(5.5,-0.05)(5.7,0.05) \psline{->}(0,0)(0,1.2) \rput(7.3,0){$\lambda $} \rput(0.5,1.2){$\rho (\lambda )$} \scriptsize \psline(1,-0.025)(1,0.025) \rput(1,-0.075){1} \psline(2,-0.025)(2,0.025) \rput(2,-0.075){2} \psline(3,-0.025)(3,0.025) \rput(3,-0.075){3}  \psline(4,-0.025)(4,0.025) \rput(4,-0.075){4} \psline(5,-0.025)(5,0.025) \rput(5,-0.075){5} \psline(6,-0.025)(6,0.025) \rput(6,-0.075){15} \psline(-0.1,0.2)(0.1,0.2) \rput(-0.4,0.2){$0.2$} \psline(-0.1,0.4)(0.1,0.4) \rput(-0.4,0.4){$0.4$} \psline(-0.1,0.6)(0.1,0.6) \rput(-0.4,0.6){$0.6$} \psline(-0.1,0.8)(0.1,0.8) \rput(-0.4,0.8){$0.8$} \psline(-0.1,1)(0.1,1) \rput(-0.4,1){$1.0$}
\psplot[linecolor=red,plotpoints=500]{0.301}{2.107}{2.107 x sub x 0.301 sub mul 0.5 exp x -1 exp mul 0.781 mul} \small \rput(3.5,-0.2){Figure 19: frequency distribution of the eigenvalues of the} \rput(3.4,-0.3){correlation matrix for 2001. The theoretical distribution}  \rput(1,-0.4){is superimposed on it.}
\end{pspicture}
\begin{pspicture}(-5.5,0)(3.5,2.8)
\psset{xunit=1.07,yunit=2.2}
\pspolygon*[linecolor=lightgray](0.301,0)(0.301,0.8)(2.107,0.8)(2.107,0)
\psline{->}(0,0)(7,0)  \psline[linecolor=white,linewidth=2pt](5.4,0)(5.6,0) \psline(5.3,-0.1)(5.5,0.1) \psline(5.5,-0.1)(5.7,0.1) \rput(7.3,0){$\lambda $} \scriptsize \psline(1,-0.05)(1,0.05) \rput(1,-0.15){1} \psline(2,-0.05)(2,0.05) \rput(2,-0.15){2} \psline(3,-0.05)(3,0.05) \rput(3,-0.15){3}  \psline(4,-0.05)(4,0.05) \rput(4,-0.15){4} \psline(5,-0.05)(5,0.05) \rput(5,-0.15){5} \psline(6,-0.05)(6,0.05) \rput(6,-0.15){15}
\psline[linewidth=1pt](0.046,0)(0.046,0.5) \psline[linewidth=1pt](0.064,0)(0.064,0.5) \psline[linewidth=1pt](0.068,0)(0.068,0.5) \psline[linewidth=1pt](0.093,0)(0.093,0.5) \psline[linewidth=1pt](0.110,0)(0.110,0.5) \psline[linewidth=1pt](0.112,0)(0.112,0.5) \psline[linewidth=1pt](0.123,0)(0.123,0.5) \psline[linewidth=1pt](0.153,0)(0.153,0.5) \psline[linewidth=1pt](0.168,0)(0.168,0.5) \psline[linewidth=1pt](0.173,0)(0.173,0.5) \psline[linewidth=1pt](0.189,0)(0.189,0.5) \psline[linewidth=1pt](0.204,0)(0.204,0.5) \psline[linewidth=1pt](0.210,0)(0.210,0.5) \psline[linewidth=1pt](0.224,0)(0.224,0.5) \psline[linewidth=1pt](0.231,0)(0.231,0.5) \psline[linewidth=1pt](0.269,0)(0.269,0.5) \psline[linewidth=1pt](0.271,0)(0.271,0.5) \psline[linewidth=1pt](0.287,0)(0.287,0.5) \psline[linewidth=1pt](0.296,0)(0.296,0.5) \psline[linewidth=1pt](0.319,0)(0.319,0.5) \psline[linewidth=1pt](0.335,0)(0.335,0.5) \psline[linewidth=1pt](0.349,0)(0.349,0.5) \psline[linewidth=1pt](0.360,0)(0.360,0.5) \psline[linewidth=1pt](0.365,0)(0.365,0.5) \psline[linewidth=1pt](0.384,0)(0.384,0.5) \psline[linewidth=1pt](0.394,0)(0.394,0.5) \psline[linewidth=1pt](0.408,0)(0.408,0.5) \psline[linewidth=1pt](0.427,0)(0.427,0.5) \psline[linewidth=1pt](0.447,0)(0.447,0.5) \psline[linewidth=1pt](0.494,0)(0.494,0.5) \psline[linewidth=1pt](0.516,0)(0.516,0.5) \psline[linewidth=1pt](0.524,0)(0.524,0.5) \psline[linewidth=1pt](0.540,0)(0.540,0.5) \psline[linewidth=1pt](0.551,0)(0.551,0.5) \psline[linewidth=1pt](0.561,0)(0.561,0.5) \psline[linewidth=1pt](0.588,0)(0.588,0.5) \psline[linewidth=1pt](0.590,0)(0.590,0.5) \psline[linewidth=1pt](0.623,0)(0.623,0.5) \psline[linewidth=1pt](0.646,0)(0.646,0.5) \psline[linewidth=1pt](0.674,0)(0.674,0.5) \psline[linewidth=1pt](0.692,0)(0.692,0.5) \psline[linewidth=1pt](0.702,0)(0.702,0.5) \psline[linewidth=1pt](0.721,0)(0.721,0.5) \psline[linewidth=1pt](0.761,0)(0.761,0.5) \psline[linewidth=1pt](0.771,0)(0.771,0.5) \psline[linewidth=1pt](0.823,0)(0.823,0.5) \psline[linewidth=1pt](0.831,0)(0.831,0.5) \psline[linewidth=1pt](0.861,0)(0.861,0.5) \psline[linewidth=1pt](0.882,0)(0.882,0.5) \psline[linewidth=1pt](0.919,0)(0.919,0.5) \psline[linewidth=1pt](0.944,0)(0.944,0.5) \psline[linewidth=1pt](0.950,0)(0.950,0.5) \psline[linewidth=1pt](0.976,0)(0.976,0.5) \psline[linewidth=1pt](0.991,0)(0.991,0.5)
\psline[linewidth=1pt](1.010,0)(1.010,0.5) \psline[linewidth=1pt](1.063,0)(1.063,0.5) \psline[linewidth=1pt](1.089,0)(1.089,0.5) \psline[linewidth=1pt](1.105,0)(1.105,0.5)
\psline[linewidth=1pt](1.010,0)(1.010,0.5) \psline[linewidth=1pt](1.063,0)(1.063,0.5) \psline[linewidth=1pt](1.089,0)(1.089,0.5) \psline[linewidth=1pt](1.105,0)(1.105,0.5)
\psline[linewidth=1pt](1.142,0)(1.142,0.5) \psline[linewidth=1pt](1.173,0)(1.173,0.5) \psline[linewidth=1pt](1.190,0)(1.190,0.5) \psline[linewidth=1pt](1.210,0)(1.210,0.5)
\psline[linewidth=1pt](1.249,0)(1.249,0.5) \psline[linewidth=1pt](1.287,0)(1.287,0.5) \psline[linewidth=1pt](1.332,0)(1.332,0.5) \psline[linewidth=1pt](1.352,0)(1.352,0.5)
\psline[linewidth=1pt](1.388,0)(1.388,0.5) \psline[linewidth=1pt](1.415,0)(1.415,0.5)
\psline[linewidth=1pt](1.456,0)(1.456,0.5) \psline[linewidth=1pt](1.545,0)(1.545,0.5)
\psline[linewidth=1pt](1.614,0)(1.614,0.5) \psline[linewidth=1pt](1.710,0)(1.710,0.5)
\psline[linewidth=1pt](1.763,0)(1.763,0.5) \psline[linewidth=1pt](1.931,0)(1.931,0.5)
\psline[linewidth=1pt](2.072,0)(2.072,0.5) \psline[linewidth=1pt](2.126,0)(2.126,0.5)
\psline[linewidth=1pt](2.509,0)(2.509,0.5) \psline[linewidth=1pt](4.774,0)(4.774,0.5)
\psline[linewidth=1pt](6.284,0)(6.284,0.5)
\small \rput(3.45,-0.4){Figure 20: eigenvalues in order of magnitude. The} \rput(3.7,-0.6){shaded area corresponds to the eigenvalues predicted} \rput(1.45,-0.8){for a random matrix.}
\end{pspicture}

\vskip 2.3 cm

The largest eigenvalue is once more completely out of scale. We also have several eigenvalues that are bellow the minimum theoretical eingenvalue. The next picture (figure 21) shows eigenvector $e_{79}$, which corresponds to a combination of all indices in a market movement that explains about 19\% of the collective movement of all indices.

\begin{pspicture}(-0.1,0)(3.5,2.5)
\psset{xunit=0.215,yunit=5}
\pspolygon*[linecolor=lightblue](0.5,0)(0.5,0.144)(1.5,0.144)(1.5,0)
\pspolygon*[linecolor=lightblue](1.5,0)(1.5,0.117)(2.5,0.117)(2.5,0)
\pspolygon*[linecolor=lightblue](2.5,0)(2.5,0.153)(3.5,0.153)(3.5,0)
\pspolygon*[linecolor=lightblue](3.5,0)(3.5,0.146)(4.5,0.146)(4.5,0)
\pspolygon*[linecolor=lightblue](4.5,0)(4.5,0.006)(5.5,0.006)(5.5,0)
\pspolygon*[linecolor=lightred](5.5,0)(5.5,0.009)(6.5,0.009)(6.5,0)
\pspolygon*[linecolor=lightblue](6.5,0)(6.5,0.026)(7.5,0.026)(7.5,0)
\pspolygon*[linecolor=lightblue](7.5,0)(7.5,0.004)(8.5,0.004)(8.5,0)
\pspolygon*[linecolor=lightblue](8.5,0)(8.5,0.107)(9.5,0.107)(9.5,0)
\pspolygon*[linecolor=lightblue](9.5,0)(9.5,0.071)(10.5,0.071)(10.5,0)
\pspolygon*[linecolor=lightblue](10.5,0)(10.5,0.134)(11.5,0.134)(11.5,0)
\pspolygon*[linecolor=lightblue](11.5,0)(11.5,0.042)(12.5,0.042)(12.5,0)
\pspolygon*[linecolor=lightblue](12.5,0)(12.5,0.120)(13.5,0.120)(13.5,0)
\pspolygon*[linecolor=lightblue](13.5,0)(13.5,0.204)(14.5,0.204)(14.5,0)
\pspolygon*[linecolor=lightblue](14.5,0)(14.5,0.170)(15.5,0.170)(15.5,0)
\pspolygon*[linecolor=lightblue](15.5,0)(15.5,0.223)(16.5,0.223)(16.5,0)
\pspolygon*[linecolor=lightblue](16.5,0)(16.5,0.211)(17.5,0.211)(17.5,0)
\pspolygon*[linecolor=lightblue](17.5,0)(17.5,0.206)(18.5,0.206)(18.5,0)
\pspolygon*[linecolor=lightblue](18.5,0)(18.5,0.116)(19.5,0.116)(19.5,0)
\pspolygon*[linecolor=lightblue](19.5,0)(19.5,0.218)(20.5,0.218)(20.5,0)
\pspolygon*[linecolor=lightred](20.5,0)(20.5,0.006)(21.5,0.006)(21.5,0)
\pspolygon*[linecolor=lightblue](21.5,0)(21.5,0.189)(22.5,0.189)(22.5,0)
\pspolygon*[linecolor=lightblue](22.5,0)(22.5,0.222)(23.5,0.222)(23.5,0)
\pspolygon*[linecolor=lightblue](23.5,0)(23.5,0.146)(24.5,0.146)(24.5,0)
\pspolygon*[linecolor=lightblue](24.5,0)(24.5,0.200)(25.5,0.200)(25.5,0)
\pspolygon*[linecolor=lightblue](25.5,0)(25.5,0.182)(26.5,0.182)(26.5,0)
\pspolygon*[linecolor=lightblue](26.5,0)(26.5,0.178)(27.5,0.178)(27.5,0)
\pspolygon*[linecolor=lightblue](27.5,0)(27.5,0.186)(28.5,0.186)(28.5,0)
\pspolygon*[linecolor=lightblue](28.5,0)(28.5,0.051)(29.5,0.051)(29.5,0)
\pspolygon*[linecolor=lightblue](29.5,0)(29.5,0.208)(30.5,0.208)(30.5,0)
\pspolygon*[linecolor=lightblue](30.5,0)(30.5,0.177)(31.5,0.177)(31.5,0)
\pspolygon*[linecolor=lightblue](31.5,0)(31.5,0.124)(32.5,0.124)(32.5,0)
\pspolygon*[linecolor=lightblue](32.5,0)(32.5,0.141)(33.5,0.141)(33.5,0)
\pspolygon*[linecolor=lightblue](33.5,0)(33.5,0.032)(34.5,0.032)(34.5,0)
\pspolygon*[linecolor=lightblue](34.5,0)(34.5,0.173)(35.5,0.173)(35.5,0)
\pspolygon*[linecolor=lightblue](35.5,0)(35.5,0.147)(36.5,0.147)(36.5,0)
\pspolygon*[linecolor=lightblue](36.5,0)(36.5,0.031)(37.5,0.031)(37.5,0)
\pspolygon*[linecolor=lightblue](37.5,0)(37.5,0.021)(38.5,0.021)(38.5,0)
\pspolygon*[linecolor=lightblue](38.5,0)(38.5,0.108)(39.5,0.108)(39.5,0)
\pspolygon*[linecolor=lightred](39.5,0)(39.5,0.004)(40.5,0.004)(40.5,0)
\pspolygon*[linecolor=lightblue](40.5,0)(40.5,0.043)(41.5,0.043)(41.5,0)
\pspolygon*[linecolor=lightred](41.5,0)(41.5,0.014)(42.5,0.014)(42.5,0)
\pspolygon*[linecolor=lightblue](42.5,0)(42.5,0.107)(43.5,0.107)(43.5,0)
\pspolygon*[linecolor=lightblue](43.5,0)(43.5,0.008)(44.5,0.008)(44.5,0)
\pspolygon*[linecolor=lightblue](44.5,0)(44.5,0.095)(45.5,0.095)(45.5,0)
\pspolygon*[linecolor=lightblue](45.5,0)(45.5,0.154)(46.5,0.154)(46.5,0)
\pspolygon*[linecolor=lightred](46.5,0)(46.5,0.006)(47.5,0.006)(47.5,0)
\pspolygon*[linecolor=lightblue](47.5,0)(47.5,0.030)(48.5,0.030)(48.5,0)
\pspolygon*[linecolor=lightblue](48.5,0)(48.5,0.032)(49.5,0.032)(49.5,0)
\pspolygon*[linecolor=lightblue](49.5,0)(49.5,0.095)(50.5,0.095)(50.5,0)
\pspolygon*[linecolor=lightblue](50.5,0)(50.5,0.017)(51.5,0.017)(51.5,0)
\pspolygon*[linecolor=lightred](51.5,0)(51.5,0.016)(52.5,0.016)(52.5,0)
\pspolygon*[linecolor=lightblue](52.5,0)(52.5,0.026)(53.5,0.026)(53.5,0)
\pspolygon*[linecolor=lightblue](53.5,0)(53.5,0.086)(54.5,0.086)(54.5,0)
\pspolygon*[linecolor=lightblue](54.5,0)(54.5,0.028)(55.5,0.028)(55.5,0)
\pspolygon*[linecolor=lightblue](55.5,0)(55.5,0.010)(56.5,0.010)(56.5,0)
\pspolygon*[linecolor=lightblue](56.5,0)(56.5,0.085)(57.5,0.085)(57.5,0)
\pspolygon*[linecolor=lightblue](57.5,0)(57.5,0.137)(58.5,0.137)(58.5,0)
\pspolygon*[linecolor=lightblue](58.5,0)(58.5,0.003)(59.5,0.003)(59.5,0)
\pspolygon*[linecolor=lightred](59.5,0)(59.5,0.011)(60.5,0.011)(60.5,0)
\pspolygon*[linecolor=lightblue](60.5,0)(60.5,0.078)(61.5,0.078)(61.5,0)
\pspolygon*[linecolor=lightblue](61.5,0)(61.5,0.118)(62.5,0.118)(62.5,0)
\pspolygon*[linecolor=lightblue](62.5,0)(62.5,0.081)(63.5,0.081)(63.5,0)
\pspolygon*[linecolor=lightred](63.5,0)(63.5,0.005)(64.5,0.005)(64.5,0)
\pspolygon*[linecolor=lightblue](64.5,0)(64.5,0.041)(65.5,0.041)(65.5,0)
\pspolygon*[linecolor=lightblue](65.5,0)(65.5,0.131)(66.5,0.131)(66.5,0)
\pspolygon*[linecolor=lightblue](66.5,0)(66.5,0.039)(67.5,0.039)(67.5,0)
\pspolygon*[linecolor=lightblue](67.5,0)(67.5,0.009)(68.5,0.009)(68.5,0)
\pspolygon*[linecolor=lightblue](68.5,0)(68.5,0.093)(69.5,0.093)(69.5,0)
\pspolygon*[linecolor=lightblue](69.5,0)(69.5,0.083)(70.5,0.083)(70.5,0)
\pspolygon*[linecolor=lightred](70.5,0)(70.5,0.004)(71.5,0.004)(71.5,0)
\pspolygon*[linecolor=lightred](71.5,0)(71.5,0.022)(72.5,0.022)(72.5,0)
\pspolygon*[linecolor=lightblue](72.5,0)(72.5,0.012)(73.5,0.012)(73.5,0)
\pspolygon*[linecolor=lightred](73.5,0)(73.5,0.004)(74.5,0.004)(74.5,0)
\pspolygon*[linecolor=lightblue](74.5,0)(74.5,0.011)(75.5,0.011)(75.5,0)
\pspolygon*[linecolor=lightblue](75.5,0)(75.5,0.028)(76.5,0.028)(76.5,0)
\pspolygon*[linecolor=lightred](76.5,0)(76.5,0.033)(77.5,0.033)(77.5,0)
\pspolygon*[linecolor=lightblue](77.5,0)(77.5,0.146)(78.5,0.146)(78.5,0)
\pspolygon*[linecolor=lightblue](78.5,0)(78.5,0.007)(79.5,0.007)(79.5,0)
\pspolygon[linecolor=blue](0.5,0)(0.5,0.144)(1.5,0.144)(1.5,0)
\pspolygon[linecolor=blue](1.5,0)(1.5,0.117)(2.5,0.117)(2.5,0)
\pspolygon[linecolor=blue](2.5,0)(2.5,0.153)(3.5,0.153)(3.5,0)
\pspolygon[linecolor=blue](3.5,0)(3.5,0.146)(4.5,0.146)(4.5,0)
\pspolygon[linecolor=blue](4.5,0)(4.5,0.006)(5.5,0.006)(5.5,0)
\pspolygon[linecolor=red](5.5,0)(5.5,0.009)(6.5,0.009)(6.5,0)
\pspolygon[linecolor=blue](6.5,0)(6.5,0.026)(7.5,0.026)(7.5,0)
\pspolygon[linecolor=blue](7.5,0)(7.5,0.004)(8.5,0.004)(8.5,0)
\pspolygon[linecolor=blue](8.5,0)(8.5,0.107)(9.5,0.107)(9.5,0)
\pspolygon[linecolor=blue](9.5,0)(9.5,0.071)(10.5,0.071)(10.5,0)
\pspolygon[linecolor=blue](10.5,0)(10.5,0.134)(11.5,0.134)(11.5,0)
\pspolygon[linecolor=blue](11.5,0)(11.5,0.042)(12.5,0.042)(12.5,0)
\pspolygon[linecolor=blue](12.5,0)(12.5,0.120)(13.5,0.120)(13.5,0)
\pspolygon[linecolor=blue](13.5,0)(13.5,0.204)(14.5,0.204)(14.5,0)
\pspolygon[linecolor=blue](14.5,0)(14.5,0.170)(15.5,0.170)(15.5,0)
\pspolygon[linecolor=blue](15.5,0)(15.5,0.223)(16.5,0.223)(16.5,0)
\pspolygon[linecolor=blue](16.5,0)(16.5,0.211)(17.5,0.211)(17.5,0)
\pspolygon[linecolor=blue](17.5,0)(17.5,0.206)(18.5,0.206)(18.5,0)
\pspolygon[linecolor=blue](18.5,0)(18.5,0.116)(19.5,0.116)(19.5,0)
\pspolygon[linecolor=blue](19.5,0)(19.5,0.218)(20.5,0.218)(20.5,0)
\pspolygon[linecolor=red](20.5,0)(20.5,0.006)(21.5,0.006)(21.5,0)
\pspolygon[linecolor=blue](21.5,0)(21.5,0.189)(22.5,0.189)(22.5,0)
\pspolygon[linecolor=blue](22.5,0)(22.5,0.222)(23.5,0.222)(23.5,0)
\pspolygon[linecolor=blue](23.5,0)(23.5,0.146)(24.5,0.146)(24.5,0)
\pspolygon[linecolor=blue](24.5,0)(24.5,0.200)(25.5,0.200)(25.5,0)
\pspolygon[linecolor=blue](25.5,0)(25.5,0.182)(26.5,0.182)(26.5,0)
\pspolygon[linecolor=blue](26.5,0)(26.5,0.178)(27.5,0.178)(27.5,0)
\pspolygon[linecolor=blue](27.5,0)(27.5,0.186)(28.5,0.186)(28.5,0)
\pspolygon[linecolor=blue](28.5,0)(28.5,0.051)(29.5,0.051)(29.5,0)
\pspolygon[linecolor=blue](29.5,0)(29.5,0.208)(30.5,0.208)(30.5,0)
\pspolygon[linecolor=blue](30.5,0)(30.5,0.177)(31.5,0.177)(31.5,0)
\pspolygon[linecolor=blue](31.5,0)(31.5,0.124)(32.5,0.124)(32.5,0)
\pspolygon[linecolor=blue](32.5,0)(32.5,0.141)(33.5,0.141)(33.5,0)
\pspolygon[linecolor=blue](33.5,0)(33.5,0.032)(34.5,0.032)(34.5,0)
\pspolygon[linecolor=blue](34.5,0)(34.5,0.173)(35.5,0.173)(35.5,0)
\pspolygon[linecolor=blue](35.5,0)(35.5,0.147)(36.5,0.147)(36.5,0)
\pspolygon[linecolor=blue](36.5,0)(36.5,0.031)(37.5,0.031)(37.5,0)
\pspolygon[linecolor=blue](37.5,0)(37.5,0.021)(38.5,0.021)(38.5,0)
\pspolygon[linecolor=blue](38.5,0)(38.5,0.108)(39.5,0.108)(39.5,0)
\pspolygon[linecolor=red](39.5,0)(39.5,0.004)(40.5,0.004)(40.5,0)
\pspolygon[linecolor=blue](40.5,0)(40.5,0.043)(41.5,0.043)(41.5,0)
\pspolygon[linecolor=red](41.5,0)(41.5,0.014)(42.5,0.014)(42.5,0)
\pspolygon[linecolor=blue](42.5,0)(42.5,0.107)(43.5,0.107)(43.5,0)
\pspolygon[linecolor=blue](43.5,0)(43.5,0.008)(44.5,0.008)(44.5,0)
\pspolygon[linecolor=blue](44.5,0)(44.5,0.095)(45.5,0.095)(45.5,0)
\pspolygon[linecolor=blue](45.5,0)(45.5,0.154)(46.5,0.154)(46.5,0)
\pspolygon[linecolor=red](46.5,0)(46.5,0.006)(47.5,0.006)(47.5,0)
\pspolygon[linecolor=blue](47.5,0)(47.5,0.030)(48.5,0.030)(48.5,0)
\pspolygon[linecolor=blue](48.5,0)(48.5,0.032)(49.5,0.032)(49.5,0)
\pspolygon[linecolor=blue](49.5,0)(49.5,0.095)(50.5,0.095)(50.5,0)
\pspolygon[linecolor=blue](50.5,0)(50.5,0.017)(51.5,0.017)(51.5,0)
\pspolygon[linecolor=red](51.5,0)(51.5,0.016)(52.5,0.016)(52.5,0)
\pspolygon[linecolor=blue](52.5,0)(52.5,0.026)(53.5,0.026)(53.5,0)
\pspolygon[linecolor=blue](53.5,0)(53.5,0.086)(54.5,0.086)(54.5,0)
\pspolygon[linecolor=blue](54.5,0)(54.5,0.028)(55.5,0.028)(55.5,0)
\pspolygon[linecolor=blue](55.5,0)(55.5,0.010)(56.5,0.010)(56.5,0)
\pspolygon[linecolor=blue](56.5,0)(56.5,0.085)(57.5,0.085)(57.5,0)
\pspolygon[linecolor=blue](57.5,0)(57.5,0.137)(58.5,0.137)(58.5,0)
\pspolygon[linecolor=blue](58.5,0)(58.5,0.003)(59.5,0.003)(59.5,0)
\pspolygon[linecolor=red](59.5,0)(59.5,0.011)(60.5,0.011)(60.5,0)
\pspolygon[linecolor=blue](60.5,0)(60.5,0.078)(61.5,0.078)(61.5,0)
\pspolygon[linecolor=blue](61.5,0)(61.5,0.118)(62.5,0.118)(62.5,0)
\pspolygon[linecolor=blue](62.5,0)(62.5,0.081)(63.5,0.081)(63.5,0)
\pspolygon[linecolor=red](63.5,0)(63.5,0.005)(64.5,0.005)(64.5,0)
\pspolygon[linecolor=blue](64.5,0)(64.5,0.041)(65.5,0.041)(65.5,0)
\pspolygon[linecolor=blue](65.5,0)(65.5,0.131)(66.5,0.131)(66.5,0)
\pspolygon[linecolor=blue](66.5,0)(66.5,0.039)(67.5,0.039)(67.5,0)
\pspolygon[linecolor=blue](67.5,0)(67.5,0.009)(68.5,0.009)(68.5,0)
\pspolygon[linecolor=blue](68.5,0)(68.5,0.093)(69.5,0.093)(69.5,0)
\pspolygon[linecolor=blue](69.5,0)(69.5,0.083)(70.5,0.083)(70.5,0)
\pspolygon[linecolor=red](70.5,0)(70.5,0.004)(71.5,0.004)(71.5,0)
\pspolygon[linecolor=red](71.5,0)(71.5,0.022)(72.5,0.022)(72.5,0)
\pspolygon[linecolor=blue](72.5,0)(72.5,0.012)(73.5,0.012)(73.5,0)
\pspolygon[linecolor=red](73.5,0)(73.5,0.004)(74.5,0.004)(74.5,0)
\pspolygon[linecolor=blue](74.5,0)(74.5,0.011)(75.5,0.011)(75.5,0)
\pspolygon[linecolor=blue](75.5,0)(75.5,0.028)(76.5,0.028)(76.5,0)
\pspolygon[linecolor=red](76.5,0)(76.5,0.033)(77.5,0.033)(77.5,0)
\pspolygon[linecolor=blue](77.5,0)(77.5,0.146)(78.5,0.146)(78.5,0)
\pspolygon[linecolor=blue](78.5,0)(78.5,0.007)(79.5,0.007)(79.5,0)
\psline{->}(0,0)(81,0) \psline{->}(0,0)(0,0.4) \rput(1.7,0.4){$e_{79}$} \scriptsize \psline(1,-0.02)(1,0.02) \rput(1,-0.06){S\&P} \psline(5,-0.02)(5,0.02) \rput(5,-0.06){Pana} \psline(10,-0.02)(10,0.02) \rput(10,-0.06){Arge} \psline(15,-0.02)(15,0.02) \rput(15,-0.06){UK} \psline(20,-0.02)(20,0.02) \rput(20,-0.06){Autr} \psline(25,-0.02)(25,0.02) \rput(25,-0.06){Luxe} \psline(30,-0.02)(30,0.02) \rput(30,-0.06){Icel} \psline(35,-0.02)(35,0.02) \rput(35,-0.06){Slok} \psline(40,-0.02)(40,0.02) \rput(40,-0.06){Esto} \psline(45,-0.02)(45,0.02) \rput(45,-0.06){Kaza} \psline(50,-0.02)(50,0.02) \rput(49,-0.06){Jord} \psline(55,-0.02)(55,0.02) \rput(55,-0.06){Indi} \psline(60,-0.02)(60,0.02) \rput(60,-0.06){Chin} \psline(65,-0.02)(65,0.02) \rput(65,-0.06){Viet} \psline(70,-0.02)(70,0.02) \rput(70,-0.06){Aust} \psline(75,-0.02)(75,0.02) \rput(75,-0.06){Ghan} \psline(79,-0.02)(79,0.02) \rput(79,-0.06){Maur} \scriptsize \psline(-0.28,0.1)(0.28,0.1) \rput(-1.38,0.1){$0.1$} \psline(-0.28,0.2)(0.28,0.2) \rput(-1.38,0.2){$0.2$} \psline(-0.28,0.3)(0.28,0.3) \rput(-1.38,0.3){$0.3$}
\end{pspicture}

\vskip 0.6 cm

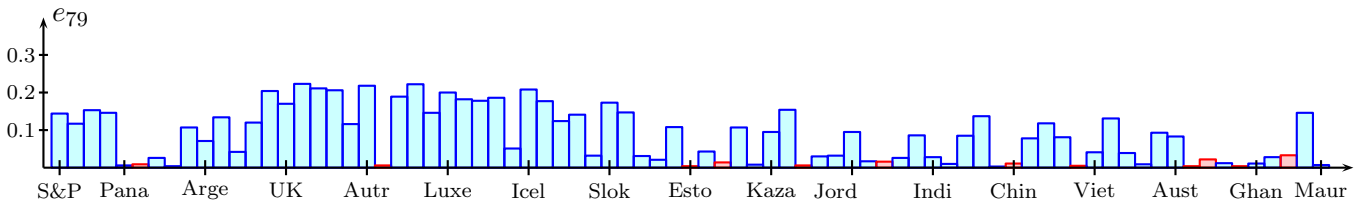
{\noindent \small Figure 21: contributions of the stock market indices to eigenvector $e_{79}$, corresponding to the largest eigenvalue of the correlation matrix. Blue bars indicate positive values, and red bars correspond to negative values. The indices are aligned in the following way: {\bf S\&P}, Nasd, Cana, Mexi, {\bf Pana}, CoRi, Berm, Jama, Braz, {\bf Arge}, Chil, Vene, Peru, {\bf UK}, Irel, Fran, Germ, Swit, {\bf Autr}, Ita, Malt, Belg, Neth, {\bf Luxe}, Swed, Denm, Finl, Norw, {\bf Icel}, Spai, Port, Gree, CzRe, {\bf Slok}, Hung, Pola, Roma, Bulg, {\bf Esto}, Latv, Lith, Ukra, Russ, {\bf Kaza}, Turk, Isra, Pale, Leba, Jord, SaAr, Qata, Ohma, Paki, {\bf Indi}, SrLa, Bang, Japa, HoKo, {\bf Chin}, Mong, Taiw, SoKo, Thai, {\bf Viet}, Mala, Sing, Indo, Phil, {\bf Aust}, NeZe, Moro, Tuni, Egyp, {\bf Ghan}, Nige, Keny, Bots, SoAf, {\bf Maur}.}

\vskip 0.4 cm

Many of the indices have a very small participation, which amounts to no participation, due to possible error bars, and many others have almost no participation. The indices that have participation smaller than $0,05$ are the ones from Central America, Bermuda, Venezuela, Malta, Slovakia, Romania, Bulgaria, Latvia, Lithuania, Ukraine, Kazakhstan, all the Arab countries, with the exception of Saudi Arabia, Sri Lanka, Bangladesh, China, Mongolia, Vietnam, Malaysia, Indonesia, Philipines, and all the African countries, with the exception of South Africa. Given the size of those markets, this is within what was expected. The major contributions come from the North American countries, the major South American ones, most of Western and Central Europe, the Czech Republic, Hungary, Poland, Estonia, Russia, Israel, Hong Kong, South Korea, Singapore, and South Africa.

Figure 22 shows the average correlation, calculated in a running window of 80 days, and the volatility of the market mode, both normalized so as to have mean 2 and standard deviation 1, since the correlation between both measures becomes more transparent in this framework. The normalized volatility is in blue, and the normalized average correlation is in red.

\begin{pspicture}(-2,0)(3.5,4.6)
\psset{xunit=2.5,yunit=0.5}
\psline{->}(0,0)(5.4,0) \psline{->}(0,0)(0,8) \rput(5.6,0){day} \rput(0.5,8){$<C>_n$,$\text{vol.}_n$}\scriptsize \psline(0.02,-0.2)(0.02,0.2) \rput(0.02,-0.6){01/02} \psline(0.46,-0.2)(0.46,0.2) \rput(0.46,-0.6){02/01} \psline(0.86,-0.2)(0.86,0.2) \rput(0.86,-0.6){03/01} \psline(1.3,-0.2)(1.3,0.2) \rput(1.3,-0.6){04/02} \psline(1.72,-0.2)(1.72,0.2) \rput(1.72,-0.6){05/01} \psline(2.18,-0.2)(2.18,0.2) \rput(2.18,-0.6){06/01} \psline(2.6,-0.2)(2.6,0.2) \rput(2.6,-0.6){07/02} \psline(3.04,-0.2)(3.04,0.2) \rput(3.04,-0.6){08/01} \psline(3.5,-0.2)(3.5,0.2) \rput(3.5,-0.6){09/03} \psline(3.9,-0.2)(3.9,0.2) \rput(3.9,-0.6){10/01} \psline(4.36,-0.2)(4.36,0.2) \rput(4.36,-0.6){11/01} \psline(4.8,-0.2)(4.8,0.2) \rput(4.8,-0.6){12/03} \psline(5.2,-0.2)(5.2,0.2) \rput(5.2,-0.6){12/31} \psline(-0.043,2)(0.043,2) \rput(-0.17,2){$2$} \psline(-0.043,4)(0.043,4) \rput(-0.17,4){$4$} \psline(-0.043,6)(0.043,6) \rput(-0.17,6){$6$}
\psline[linecolor=blue](1.62,2.450) (1.64,1.317) (1.66,0.950) (1.68,1.857) (1.7,2.454) (1.72,2.199) (1.74,0.907) (1.76,1.099) (1.78,2.205) (1.8,1.060) (1.82,1.308) (1.84,1.015) (1.86,1.629) (1.88,2.498) (1.9,1.180) (1.92,2.192) (1.94,1.236) (1.96,1.052) (1.98,2.872) (2,1.794) (2.02,1.870) (2.04,1.809) (2.06,1.542) (2.08,1.012) (2.1,1.510) (2.12,0.856) (2.14,1.565) (2.16,2.528) (2.18,1.202) (2.2,1.106) (2.22,1.522) (2.24,1.901) (2.26,1.054) (2.28,0.960) (2.3,1.029) (2.32,2.130) (2.34,3.038) (2.36,1.518) (2.38,2.470) (2.4,1.845) (2.42,2.333) (2.44,1.029) (2.46,1.587) (2.48,1.171) (2.5,1.208) (2.52,0.867) (2.54,2.292) (2.56,1.020) (2.58,1.736) (2.6,2.501) (2.62,1.529) (2.64,1.874) (2.66,1.616) (2.68,1.696) (2.7,3.280) (2.72,1.322) (2.74,1.694) (2.76,2.728) (2.78,2.231) (2.8,0.960) (2.82,1.904) (2.84,1.323) (2.86,1.987) (2.88,2.852) (2.9,1.815) (2.92,1.080) (2.94,2.060) (2.96,1.978) (2.98,1.660) (3,2.331) (3.02,1.474) (3.04,1.660) (3.06,1.495) (3.08,1.398) (3.1,1.466) (3.12,1.303) (3.14,1.427) (3.16,1.956) (3.18,2.638) (3.2,1.321) (3.22,1.350) (3.24,2.071) (3.26,1.473) (3.28,2.044) (3.3,2.614) (3.32,1.363) (3.34,1.310) (3.36,1.041) (3.38,1.137) (3.4,2.370) (3.42,1.342) (3.44,1.973) (3.46,1.411) (3.48,2.684) (3.5,1.120) (3.52,2.257) (3.54,1.376) (3.56,2.162) (3.58,2.860) (3.6,3.147) (3.62,2.550) (3.64,6.484) (3.66,2.769) (3.68,1.343) (3.7,6.140) (3.72,1.981) (3.74,1.309) (3.76,1.550) (3.78,5.728) (3.8,5.512) (3.82,5.930) (3.84,2.174) (3.86,2.035) (3.88,1.785) (3.9,4.154) (3.92,2.401) (3.94,1.612) (3.96,1.346) (3.98,4.167) (4,1.342) (4.02,1.721) (4.04,1.576) (4.06,3.556) (4.08,4.183) (4.1,1.630) (4.12,2.538) (4.14,2.691) (4.16,2.626) (4.18,2.706) (4.2,1.628) (4.22,2.711) (4.24,3.221) (4.26,1.807) (4.28,2.154) (4.3,2.789) (4.32,2.787) (4.34,3.663) (4.36,2.327) (4.38,1.430) (4.4,1.214) (4.42,3.388) (4.44,1.247) (4.46,1.378) (4.48,2.858) (4.5,1.515) (4.52,2.214) (4.54,3.962) (4.56,2.334) (4.58,1.862) (4.6,1.388) (4.62,2.475) (4.64,2.326) (4.66,1.195) (4.68,1.922) (4.7,1.305) (4.72,1.751) (4.74,1.808) (4.76,2.850) (4.78,1.043) (4.8,1.873) (4.82,1.598) (4.84,2.383) (4.86,4.515) (4.88,1.929) (4.9,0.888) (4.92,2.507) (4.94,0.923) (4.96,1.224) (4.98,2.988) (5,1.412) (5.02,2.177) (5.04,1.299) (5.06,1.133) (5.08,0.914) (5.1,1.904) (5.12,1.133) (5.14,0.961) (5.16,0.957) (5.18,2.723) (5.2,1.808)
\psline[linecolor=red](1.62,1.897) (1.64,1.944) (1.66,2.091) (1.68,1.844) (1.7,1.896) (1.72,1.900) (1.74,1.906) (1.76,1.931) (1.78,1.917) (1.8,1.841) (1.82,1.862) (1.84,1.880) (1.86,1.877) (1.88,1.873) (1.9,1.911) (1.92,1.969) (1.94,1.944) (1.96,2.067) (1.98,2.070) (2,2.048) (2.02,2.051) (2.04,2.049) (2.06,2.025) (2.08,2.031) (2.1,2.071) (2.12,2.042) (2.14,2.060) (2.16,2.056) (2.18,2.060) (2.2,2.114) (2.22,2.074) (2.24,2.082) (2.26,2.047) (2.28,2.020) (2.3,2.029) (2.32,2.037) (2.34,2.116) (2.36,2.063) (2.38,2.061) (2.4,2.080) (2.42,2.057) (2.44,2.061) (2.46,2.087) (2.48,2.126) (2.5,2.141) (2.52,2.190) (2.54,2.137) (2.56,2.110) (2.58,2.252) (2.6,2.254) (2.62,1.833) (2.64,1.724) (2.66,1.667) (2.68,1.809) (2.7,1.822) (2.72,1.835) (2.74,2.039) (2.76,1.956) (2.78,1.418) (2.8,1.460) (2.82,1.286) (2.84,1.351) (2.86,1.379) (2.88,1.281) (2.9,1.241) (2.92,1.305) (2.94,1.108) (2.96,1.048) (2.98,0.584) (3,0.572) (3.02,0.647) (3.04,0.494) (3.06,0.452) (3.08,0.492) (3.1,0.466) (3.12,0.473) (3.14,0.428) (3.16,-0.073) (3.18,-0.019) (3.2,0.015) (3.22,-0.014) (3.24,0.000) (3.26,0.071) (3.28,0.240) (3.3,0.246) (3.32,0.227) (3.34,0.209) (3.36,0.212) (3.38,0.301) (3.4,0.338) (3.42,0.315) (3.44,0.306) (3.46,0.296) (3.48,0.154) (3.5,0.251) (3.52,0.167) (3.54,0.120) (3.56,0.191) (3.58,0.078) (3.6,0.205) (3.62,0.542) (3.64,0.843) (3.66,0.754) (3.68,1.673) (3.7,1.792) (3.72,1.692) (3.74,1.573) (3.76,1.815) (3.78,2.318) (3.8,2.344) (3.82,2.330) (3.84,2.349) (3.86,2.356) (3.88,2.604) (3.9,2.557) (3.92,2.481) (3.94,2.462) (3.96,2.487) (3.98,2.439) (4,2.465) (4.02,2.474) (4.04,2.525) (4.06,2.911) (4.08,2.876) (4.1,2.806) (4.12,2.934) (4.14,2.986) (4.16,2.976) (4.18,2.946) (4.2,2.806) (4.22,2.855) (4.24,2.887) (4.26,2.880) (4.28,2.910) (4.3,2.856) (4.32,2.931) (4.34,2.940) (4.36,2.883) (4.38,2.865) (4.4,2.915) (4.42,2.937) (4.44,2.932) (4.46,3.037) (4.48,3.024) (4.5,3.026) (4.52,3.098) (4.54,3.133) (4.56,3.119) (4.58,3.111) (4.6,3.111) (4.62,3.113) (4.64,3.118) (4.66,3.193) (4.68,3.179) (4.7,3.181) (4.72,3.130) (4.74,3.125) (4.76,3.128) (4.78,3.125) (4.8,3.121) (4.82,3.094) (4.84,3.269) (4.86,3.302) (4.88,3.279) (4.9,3.216) (4.92,3.213) (4.94,3.194) (4.96,3.243) (4.98,3.219) (5,3.187) (5.02,3.185) (5.04,3.168) (5.06,3.136) (5.08,3.089) (5.1,3.107) (5.12,3.088) (5.14,3.121) (5.16,3.218) (5.18,3.242) (5.2,3.194)
\end{pspicture}

\vskip 0.6 cm

{\noindent \small Figure 22: average volatility (blue) and average correlation (red) based on the log-returns for 2001, both calculated in a moving window of 80 days and normalized so as to have mean two and standard deviation one.}

\vskip 0.4 cm

The figure shows a great increase in volatility just after September 11, followed by an increase in average correlation between the world stock market indices. This is expected from a crisis that was completely exogenous to the financial markets. A similar increase of both volatility and average correlation occur close to the beginning of the year, related with the burst of the dot-com bubble.

More illustrative is figure 23, which shows the average correlation and the average volatility, both calculated in a running window of 80 days, normalized so as to have mean 2 and standard deviation 1. The normalized average volatility is in blue, and the normalized average correlation is in red.

\begin{pspicture}(-2,0)(3.5,3)
\psset{xunit=2.5,yunit=0.6}
\psline{->}(0,0)(5.4,0) \psline{->}(0,0)(0,4) \rput(5.6,0){day} \rput(0.7,4){$<C>_n$,$<\text{vol.}>_n$}\scriptsize \psline(0.02,-0.2)(0.02,0.2) \rput(0.02,-0.6){01/02} \psline(0.46,-0.2)(0.46,0.2) \rput(0.46,-0.6){02/01} \psline(0.86,-0.2)(0.86,0.2) \rput(0.86,-0.6){03/01} \psline(1.3,-0.2)(1.3,0.2) \rput(1.3,-0.6){04/02} \psline(1.72,-0.2)(1.72,0.2) \rput(1.72,-0.6){05/01} \psline(2.18,-0.2)(2.18,0.2) \rput(2.18,-0.6){06/01} \psline(2.6,-0.2)(2.6,0.2) \rput(2.6,-0.6){07/02} \psline(3.04,-0.2)(3.04,0.2) \rput(3.04,-0.6){08/01} \psline(3.5,-0.2)(3.5,0.2) \rput(3.5,-0.6){09/03} \psline(3.9,-0.2)(3.9,0.2) \rput(3.9,-0.6){10/01} \psline(4.36,-0.2)(4.36,0.2) \rput(4.36,-0.6){11/01} \psline(4.8,-0.2)(4.8,0.2) \rput(4.8,-0.6){12/03} \psline(5.2,-0.2)(5.2,0.2) \rput(5.2,-0.6){12/31} \psline(-0.043,1)(0.043,1) \rput(-0.17,1){$1$} \psline(-0.043,2)(0.043,2) \rput(-0.17,2){$2$} \psline(-0.043,3)(0.043,3) \rput(-0.17,3){$3$}
\psline[linecolor=blue](1.62,2.116) (1.64,2.064) (1.66,2.034) (1.68,1.888) (1.7,1.959) (1.72,2.006) (1.74,2.009) (1.76,1.997) (1.78,2.000) (1.8,1.935) (1.82,1.938) (1.84,1.898) (1.86,1.803) (1.88,1.868) (1.9,1.850) (1.92,1.908) (1.94,1.926) (1.96,1.897) (1.98,2.004) (2,2.014) (2.02,2.067) (2.04,2.109) (2.06,2.086) (2.08,2.051) (2.1,2.033) (2.12,1.973) (2.14,1.973) (2.16,1.979) (2.18,1.982) (2.2,1.919) (2.22,1.939) (2.24,1.990) (2.26,1.920) (2.28,1.847) (2.3,1.742) (2.32,1.766) (2.34,1.859) (2.36,1.753) (2.38,1.794) (2.4,1.796) (2.42,1.809) (2.44,1.812) (2.46,1.830) (2.48,1.783) (2.5,1.786) (2.52,1.723) (2.54,1.702) (2.56,1.692) (2.58,1.697) (2.6,1.704) (2.62,1.519) (2.64,1.516) (2.66,1.412) (2.68,1.389) (2.7,1.421) (2.72,1.411) (2.74,1.416) (2.76,1.395) (2.78,1.229) (2.8,1.096) (2.82,0.982) (2.84,0.960) (2.86,0.934) (2.88,0.971) (2.9,0.965) (2.92,0.920) (2.94,0.794) (2.96,0.828) (2.98,0.666) (3,0.722) (3.02,0.731) (3.04,0.636) (3.06,0.610) (3.08,0.617) (3.1,0.641) (3.12,0.651) (3.14,0.638) (3.16,0.464) (3.18,0.518) (3.2,0.485) (3.22,0.426) (3.24,0.467) (3.26,0.495) (3.28,0.505) (3.3,0.514) (3.32,0.468) (3.34,0.490) (3.36,0.487) (3.38,0.429) (3.4,0.500) (3.42,0.502) (3.44,0.554) (3.46,0.542) (3.48,0.552) (3.5,0.549) (3.52,0.552) (3.54,0.560) (3.56,0.620) (3.58,0.619) (3.6,0.692) (3.62,0.729) (3.64,0.981) (3.66,1.047) (3.68,1.065) (3.7,1.315) (3.72,1.376) (3.74,1.362) (3.76,1.309) (3.78,1.553) (3.8,1.791) (3.82,2.029) (3.84,2.044) (3.86,2.097) (3.88,2.141) (3.9,2.310) (3.92,2.325) (3.94,2.248) (3.96,2.238) (3.98,2.330) (4,2.303) (4.02,2.270) (4.04,2.299) (4.06,2.406) (4.08,2.568) (4.1,2.591) (4.12,2.681) (4.14,2.703) (4.16,2.789) (4.18,2.842) (4.2,2.795) (4.22,2.858) (4.24,2.931) (4.26,2.941) (4.28,2.966) (4.3,2.940) (4.32,3.019) (4.34,3.125) (4.36,3.103) (4.38,3.060) (4.4,3.074) (4.42,3.154) (4.44,3.150) (4.46,3.117) (4.48,3.117) (4.5,3.101) (4.52,3.162) (4.54,3.265) (4.56,3.284) (4.58,3.295) (4.6,3.244) (4.62,3.298) (4.64,3.334) (4.66,3.318) (4.68,3.346) (4.7,3.337) (4.72,3.362) (4.74,3.382) (4.76,3.430) (4.78,3.344) (4.8,3.374) (4.82,3.387) (4.84,3.404) (4.86,3.569) (4.88,3.562) (4.9,3.469) (4.92,3.531) (4.94,3.510) (4.96,3.520) (4.98,3.620) (5,3.568) (5.02,3.613) (5.04,3.577) (5.06,3.562) (5.08,3.466) (5.1,3.509) (5.12,3.448) (5.14,3.425) (5.16,3.360) (5.18,3.353) (5.2,3.281)
\psline[linecolor=red](1.62,1.897) (1.64,1.944) (1.66,2.091) (1.68,1.844) (1.7,1.896) (1.72,1.900) (1.74,1.906) (1.76,1.931) (1.78,1.917) (1.8,1.841) (1.82,1.862) (1.84,1.880) (1.86,1.877) (1.88,1.873) (1.9,1.911) (1.92,1.969) (1.94,1.944) (1.96,2.067) (1.98,2.070) (2,2.048) (2.02,2.051) (2.04,2.049) (2.06,2.025) (2.08,2.031) (2.1,2.071) (2.12,2.042) (2.14,2.060) (2.16,2.056) (2.18,2.060) (2.2,2.114) (2.22,2.074) (2.24,2.082) (2.26,2.047) (2.28,2.020) (2.3,2.029) (2.32,2.037) (2.34,2.116) (2.36,2.063) (2.38,2.061) (2.4,2.080) (2.42,2.057) (2.44,2.061) (2.46,2.087) (2.48,2.126) (2.5,2.141) (2.52,2.190) (2.54,2.137) (2.56,2.110) (2.58,2.252) (2.6,2.254) (2.62,1.833) (2.64,1.724) (2.66,1.667) (2.68,1.809) (2.7,1.822) (2.72,1.835) (2.74,2.039) (2.76,1.956) (2.78,1.418) (2.8,1.460) (2.82,1.286) (2.84,1.351) (2.86,1.379) (2.88,1.281) (2.9,1.241) (2.92,1.305) (2.94,1.108) (2.96,1.048) (2.98,0.584) (3,0.572) (3.02,0.647) (3.04,0.494) (3.06,0.452) (3.08,0.492) (3.1,0.466) (3.12,0.473) (3.14,0.428) (3.16,-0.073) (3.18,-0.019) (3.2,0.015) (3.22,-0.014) (3.24,0.000) (3.26,0.071) (3.28,0.240) (3.3,0.246) (3.32,0.227) (3.34,0.209) (3.36,0.212) (3.38,0.301) (3.4,0.338) (3.42,0.315) (3.44,0.306) (3.46,0.296) (3.48,0.154) (3.5,0.251) (3.52,0.167) (3.54,0.120) (3.56,0.191) (3.58,0.078) (3.6,0.205) (3.62,0.542) (3.64,0.843) (3.66,0.754) (3.68,1.673) (3.7,1.792) (3.72,1.692) (3.74,1.573) (3.76,1.815) (3.78,2.318) (3.8,2.344) (3.82,2.330) (3.84,2.349) (3.86,2.356) (3.88,2.604) (3.9,2.557) (3.92,2.481) (3.94,2.462) (3.96,2.487) (3.98,2.439) (4,2.465) (4.02,2.474) (4.04,2.525) (4.06,2.911) (4.08,2.876) (4.1,2.806) (4.12,2.934) (4.14,2.986) (4.16,2.976) (4.18,2.946) (4.2,2.806) (4.22,2.855) (4.24,2.887) (4.26,2.880) (4.28,2.910) (4.3,2.856) (4.32,2.931) (4.34,2.940) (4.36,2.883) (4.38,2.865) (4.4,2.915) (4.42,2.937) (4.44,2.932) (4.46,3.037) (4.48,3.024) (4.5,3.026) (4.52,3.098) (4.54,3.133) (4.56,3.119) (4.58,3.111) (4.6,3.111) (4.62,3.113) (4.64,3.118) (4.66,3.193) (4.68,3.179) (4.7,3.181) (4.72,3.130) (4.74,3.125) (4.76,3.128) (4.78,3.125) (4.8,3.121) (4.82,3.094) (4.84,3.269) (4.86,3.302) (4.88,3.279) (4.9,3.216) (4.92,3.213) (4.94,3.194) (4.96,3.243) (4.98,3.219) (5,3.187) (5.02,3.185) (5.04,3.168) (5.06,3.136) (5.08,3.089) (5.1,3.107) (5.12,3.088) (5.14,3.121) (5.16,3.218) (5.18,3.242) (5.2,3.194)
\end{pspicture}

\vskip 0.6 cm

{\noindent \small Figure 23: average volatility (blue) and average correlation (red) based on the log-returns for 2001, both calculated in a moving window of 80 days and normalized so as to have mean two and standard deviation one.}

\vskip 0.4 cm

The covariance between the volatility and the average correlation is plotted in figure 24, calculated in a moving window of 30 days. One can readily identify a peak around September 11, but no peak related with the burst of the dot-com bubble, which was not a precisely defined event in time.

\begin{pspicture}(-2,-0.3)(3.5,3.7)
\psset{xunit=2.5,yunit=4}
\psline{->}(0,0)(5.4,0) \psline{->}(0,-0.2)(0,0.8) \rput(5.6,0){day} \rput(0.3,0.8){covar.}\scriptsize \psline(0.02,-0.025)(0.02,0.025) \rput(0.02,-0.075){01/03} \psline(0.46,-0.025)(0.46,0.025) \rput(0.46,-0.075){02/01} \psline(0.86,-0.025)(0.86,0.025) \rput(0.86,-0.075){03/01} \psline(1.3,-0.025)(1.3,0.025) \rput(1.3,-0.075){04/02} \psline(1.68,-0.025)(1.68,0.025) \rput(1.68,-0.075){05/02} \psline(2.12,-0.025)(2.12,0.025) \rput(2.12,-0.075){06/01} \psline(2.54,-0.025)(2.54,0.025) \rput(2.54,-0.075){07/02} \psline(2.98,-0.025)(2.98,0.025) \rput(2.98,-0.075){08/01} \psline(3.44,-0.025)(3.44,0.025) \rput(3.44,-0.075){09/03} \psline(3.84,-0.025)(3.84,0.025) \rput(3.84,-0.075){10/01} \psline(4.3,-0.025)(4.3,0.025) \rput(4.3,-0.075){11/01} \psline(4.74,-0.025)(4.74,0.025) \rput(4.74,-0.075){12/03} \psline(5.08,-0.025)(5.08,0.025) \rput(5.08,-0.075){12/28} \psline(-0.043,0.2)(0.043,0.2) \rput(-0.24,0.2){$0.002$} \psline(-0.043,0.4)(0.043,0.4) \rput(-0.24,0.4){$0.004$} \psline(-0.043,0.6)(0.043,0.6) \rput(-0.24,0.6){$0.006$}
\psline[linecolor=red](2.22,-0.0041) (2.24,-0.0024) (2.26,-0.0018) (2.28,-0.0008) (2.3,-0.0003) (2.32,0.0015) (2.34,0.0041) (2.36,0.0071) (2.38,0.0059) (2.4,0.0086) (2.42,0.0062) (2.44,0.0052) (2.46,0.0018) (2.48,0.0015) (2.5,0.0045) (2.52,0.0017) (2.54,0.0005) (2.56,0.0006) (2.58,-0.0002) (2.6,0.0007) (2.62,0.0052) (2.64,0.0060) (2.66,0.0034) (2.68,0.0032) (2.7,0.0023) (2.72,-0.0078) (2.74,-0.0056) (2.76,-0.0055) (2.78,-0.0082) (2.8,-0.0161) (2.82,-0.0023) (2.84,-0.0047) (2.86,0.0017) (2.88,0.0004) (2.9,-0.0152) (2.92,-0.0113) (2.94,-0.0013) (2.96,-0.0167) (2.98,-0.0185) (3,-0.0204) (3.02,-0.0384) (3.04,-0.0364) (3.06,-0.0234) (3.08,-0.0117) (3.1,0.0099) (3.12,0.0302) (3.14,0.0640) (3.16,0.0618) (3.18,0.0744) (3.2,0.0498) (3.22,0.0433) (3.24,0.0625) (3.26,0.0535) (3.28,0.0652) (3.3,0.0638) (3.32,0.0091) (3.34,0.0295) (3.36,0.0393) (3.38,0.0121) (3.4,0.0047) (3.42,0.0193) (3.44,0.0176) (3.46,0.0259) (3.48,0.0205) (3.5,-0.0142) (3.52,-0.0148) (3.54,-0.0010) (3.56,-0.0069) (3.58,-0.0139) (3.6,-0.0192) (3.62,-0.0261) (3.64,-0.0161) (3.66,0.0624) (3.68,0.0759) (3.7,0.0535) (3.72,0.2226) (3.74,0.2179) (3.76,0.1896) (3.78,0.1623) (3.8,0.3445) (3.82,0.4819) (3.84,0.6184) (3.86,0.5901) (3.88,0.5431) (3.9,0.4950) (3.92,0.5629) (3.94,0.5227) (3.96,0.4479) (3.98,0.3542) (4,0.3556) (4.02,0.2999) (4.04,0.2252) (4.06,0.1649) (4.08,0.1434) (4.1,0.1811) (4.12,0.0697) (4.14,0.0286) (4.16,-0.0574) (4.18,-0.1069) (4.2,-0.1184) (4.22,-0.1272) (4.24,-0.1480) (4.26,0.0134) (4.28,0.0009) (4.3,-0.0399) (4.32,0.0322) (4.34,0.0151) (4.36,-0.0195) (4.38,-0.0523) (4.4,-0.0271) (4.42,-0.0048) (4.44,0.0392) (4.46,0.0304) (4.48,0.0185) (4.5,0.0181) (4.52,0.0252) (4.54,0.0248) (4.56,0.0294) (4.58,0.0171) (4.6,0.0360) (4.62,0.0177) (4.64,0.0096) (4.66,-0.0008) (4.68,-0.0063) (4.7,-0.0031) (4.72,-0.0114) (4.74,-0.0127) (4.76,-0.0138) (4.78,-0.0113) (4.8,-0.0137) (4.82,-0.0177) (4.84,-0.0158) (4.86,-0.0092) (4.88,0.0054) (4.9,0.0038) (4.92,0.0032) (4.94,0.0076) (4.96,0.0119) (4.98,0.0112) (5,0.0097) (5.02,0.0038) (5.04,0.0119) (5.06,0.0070) (5.08,0.0054) (5.1,0.0109) (5.12,0.0093) (5.14,0.0115) (5.16,0.0143) (5.18,0.0137) (5.2,0.0159)
\end{pspicture}

\vskip 0.6 cm

\hskip 2 cm {\small Figure 24: covariance between volatility and average correlation as a function of time.}

\vskip 0.4 cm

\section{2008, Subprime Mortgage Crisis}

The last large financial crisis began in 2007, reached its peak in 2008, and is happening until now. This crisis was triggered by the default of a large number of mortgages in the USA. Subprimes are loans to borrowers who have low credit scores. Most of them had a small initial interest rate, adjustable for future payments, which led to many home foreclosures after the rates climbed substantially. Meanwhile, the loans were transformed in pools that were then resold to interested investors. Since the returns of such investments were high, a financial bubble was created, inflating the subprime mortgage market until the defaults started to pop up.

Because of their underestimation of risk, financial institutions worldwide lost trillions of dollars, and many of them declared banckrupcy. Because of that, credit lines tightened around the world, taking the financial crisis to the so called real economy. The world is yet to recover from this crisis, and many institutions are still to lose a good part of their assets in the following years.

Here we analyze the year 2008, which is considered the time when the subprime crisis reached its peak, marked by events like the Lehman Brothers' announcement of banckrupcy, and the liquidation of three of the largest investment banks in the USA. In our research, we add now 13 indices to the ones we used for 2001: IGBC from Colombia (Colo), BELEX 15 from Serbia (Serb), CROBEX from Croatia (Croa), SBI TOP from Slovenia (Slov), SASE 10 from Bosnia and Herzegovina (BoHe), MOSTE from Montenegro (Mont), MBI 10 from Macedonia (Mace), CSE from Cyprus (Cypr), Kwait SE Weigthed Index from Kwait (Kwai),  Bahrain All Share Index from Bahrain (Baha), ADX General Index from the United Arab Emirates (UAE), DSEI from Tanzania (Tanz), and FTSE/Namibia Overall from Namibia (Nami). So, we use a total of 92 indices, 4 from North America, 2 from Central America, 2 from the islands of the Atlantic, 6 from South America, 35 from Europe, 2 from Eurasia, 28 from Asia, 2 from Oceania, and 11 from Africa. The sample became larger mainly because of the partition of the former Yugoslavia into many countries.

Using the modified log-returns (\ref{padronized}) based on the closing indices from 01/02/2008 to 12/31/2008, we built a $92\times 92$ correlation matrix between those. This matrix has an average correlation $<C>=0.26$, standard deviation $\sigma =0.05$, and is based on $L=253$ days for the $M=92$ indices, which gives $Q=L/M=256/92\approx 2.78$.

The upper and lower bounds of the eigenvalues of the Mar\v{e}nko-Pastur distribution (\ref{dist}) are
\begin{equation}
\lambda_-=0.160 \ \ \text{and}\ \ \lambda_+=2.558 \ .
\end{equation}

The frequency distribution of the eigenvalues is displayed bellow (figure 25), plotted against the theoretical Mar\v{e}nko-Pastur distribution were it an infinite random matrix with mean zero and standard deviation 1. Figure 26 shows the eigenvalues of the correlation matrix in order of magnitude (the region associated with noise appears shaded).

\begin{pspicture}(-0.2,0)(3.5,4.5)
\psset{xunit=0.78,yunit=2}
\pspolygon*[linecolor=lightblue](0,0)(0,1.630)(0.2,1.630)(0.2,0)
\pspolygon*[linecolor=lightblue](0.2,0)(0.2,0.870)(0.4,0.870)(0.4,0)
\pspolygon*[linecolor=lightblue](0.4,0)(0.4,0.652)(0.6,0.652)(0.6,0)
\pspolygon*[linecolor=lightblue](0.6,0)(0.6,0.380)(0.8,0.380)(0.8,0)
\pspolygon*[linecolor=lightblue](0.8,0)(0.8,0.326)(1,0.326)(1,0)
\pspolygon*[linecolor=lightblue](1,0)(1,0.326)(1.2,0.326)(1.2,0)
\pspolygon*[linecolor=lightblue](1.2,0)(1.2,0.163)(1.4,0.163)(1.4,0)
\pspolygon*[linecolor=lightblue](1.4,0)(1.4,0.163)(1.6,0.163)(1.6,0)
\pspolygon*[linecolor=lightblue](1.6,0)(1.6,0.109)(1.8,0.109)(1.8,0)
\pspolygon*[linecolor=lightblue](1.8,0)(1.8,0.109)(2,0.109)(2,0)
\pspolygon*[linecolor=lightblue](2.2,0)(2.2,0.054)(2.4,0.054)(2.4,0)
\pspolygon*[linecolor=lightblue](2.4,0)(2.4,0.054)(2.6,0.054)(2.6,0)
\pspolygon*[linecolor=lightblue](2.8,0)(2.8,0.054)(3,0.054)(3,0)
\pspolygon*[linecolor=lightblue](7.2,0)(7.2,0.054)(7.4,0.054)(7.4,0)
\pspolygon*[linecolor=lightblue](9.2,0)(9.2,0.054)(9.4,0.054)(9.4,0)
\pspolygon[linecolor=blue](0,0)(0,1.630)(0.2,1.630)(0.2,0)
\pspolygon[linecolor=blue](0.2,0)(0.2,0.870)(0.4,0.870)(0.4,0)
\pspolygon[linecolor=blue](0.4,0)(0.4,0.652)(0.6,0.652)(0.6,0)
\pspolygon[linecolor=blue](0.6,0)(0.6,0.380)(0.8,0.380)(0.8,0)
\pspolygon[linecolor=blue](0.8,0)(0.8,0.326)(1,0.326)(1,0)
\pspolygon[linecolor=blue](1,0)(1,0.326)(1.2,0.326)(1.2,0)
\pspolygon[linecolor=blue](1.2,0)(1.2,0.163)(1.4,0.163)(1.4,0)
\pspolygon[linecolor=blue](1.4,0)(1.4,0.163)(1.6,0.163)(1.6,0)
\pspolygon[linecolor=blue](1.6,0)(1.6,0.109)(1.8,0.109)(1.8,0)
\pspolygon[linecolor=blue](1.8,0)(1.8,0.109)(2,0.109)(2,0)
\pspolygon[linecolor=blue](2.2,0)(2.2,0.054)(2.4,0.054)(2.4,0)
\pspolygon[linecolor=blue](2.4,0)(2.4,0.054)(2.6,0.054)(2.6,0)
\pspolygon[linecolor=blue](2.8,0)(2.8,0.054)(3,0.054)(3,0)
\pspolygon[linecolor=blue](7.2,0)(7.2,0.054)(7.4,0.054)(7.4,0)
\pspolygon[linecolor=blue](9.2,0)(9.2,0.054)(9.4,0.054)(9.4,0)
\psline{->}(0,0)(10,0) \psline[linecolor=white,linewidth=2pt](8.4,0)(8.6,0) \psline(8.3,-0.1)(8.5,0.1) \psline(8.5,-0.1)(8.7,0.1) \psline{->}(0,0)(0,2) \rput(10.4,0){$\lambda $} \rput(0.7,2){$\rho (\lambda )$} \scriptsize \psline(1,-0.05)(1,0.05) \rput(1,-0.15){1} \psline(2,-0.05)(2,0.05) \rput(2,-0.15){2} \psline(3,-0.05)(3,0.05) \rput(3,-0.15){3}  \psline(4,-0.05)(4,0.05) \rput(4,-0.15){4} \psline(5,-0.05)(5,0.05) \rput(5,-0.15){5} \psline(6,-0.05)(6,0.05) \rput(6,-0.15){6} \psline(7,-0.05)(7,0.05) \rput(7,-0.15){7} \psline(8,-0.05)(8,0.05) \rput(8,-0.15){8} \psline(9,-0.05)(9,0.05) \rput(9,-0.15){31} \psline(-0.1,0.4)(0.1,0.4) \rput(-0.5,0.4){$0.4$} \psline(-0.1,0.8)(0.1,0.8) \rput(-0.5,0.8){$0.8$} \psline(-0.1,1.2)(0.1,1.2) \rput(-0.5,1.2){$1.2$} \psline(-0.1,1.6)(0.1,1.6) \rput(-0.5,1.6){$1.6$}
\psplot[linecolor=red,plotpoints=500]{0.160}{2.558}{2.558 x sub x 0.160 sub mul 0.5 exp x -1 exp mul 0.443 mul} \small \rput(5,-0.4){Figure 25: frequency distribution of the eigenvalues of the} \rput(4.8,-0.6){correlation matrix for 2008. The theoretical distribution}  \rput(1.5,-0.8){is superimposed on it.}
\end{pspicture}
\begin{pspicture}(-5.8,0)(3.5,1.8)
\psset{xunit=0.78,yunit=2.2}
\pspolygon*[linecolor=lightgray](0.160,0)(0.160,0.8)(2.558,0.8)(2.558,0)
\psline{->}(0,0)(10,0) \psline[linecolor=white,linewidth=2pt](8.4,0)(8.6,0) \psline(8.3,-0.1)(8.5,0.1) \psline(8.5,-0.1)(8.7,0.1) \rput(10.4,0){$\lambda $} \scriptsize \psline(1,-0.05)(1,0.05) \rput(1,-0.15){1} \psline(2,-0.05)(2,0.05) \rput(2,-0.15){2} \psline(3,-0.05)(3,0.05) \rput(3,-0.15){3}  \psline(4,-0.05)(4,0.05) \rput(4,-0.15){4} \psline(5,-0.05)(5,0.05) \rput(5,-0.15){5} \psline(6,-0.05)(6,0.05) \rput(6,-0.15){6} \psline(7,-0.05)(7,0.05) \rput(7,-0.15){7} \psline(8,-0.05)(8,0.05) \rput(8,-0.15){8} \psline(9,-0.05)(9,0.05) \rput(9,-0.15){31}
\psline[linewidth=1pt](0.010,0)(0.010,0.5) \psline[linewidth=1pt](0.013,0)(0.013,0.5) \psline[linewidth=1pt](0.022,0)(0.022,0.5) \psline[linewidth=1pt](0.030,0)(0.030,0.5) \psline[linewidth=1pt](0.035,0)(0.035,0.5) \psline[linewidth=1pt](0.044,0)(0.044,0.5) \psline[linewidth=1pt](0.048,0)(0.048,0.5) \psline[linewidth=1pt](0.051,0)(0.051,0.5) \psline[linewidth=1pt](0.055,0)(0.055,0.5) \psline[linewidth=1pt](0.062,0)(0.062,0.5) \psline[linewidth=1pt](0.066,0)(0.066,0.5) \psline[linewidth=1pt](0.068,0)(0.068,0.5) \psline[linewidth=1pt](0.073,0)(0.073,0.5) \psline[linewidth=1pt](0.075,0)(0.075,0.5) \psline[linewidth=1pt](0.082,0)(0.082,0.5) \psline[linewidth=1pt](0.089,0)(0.089,0.5) \psline[linewidth=1pt](0.095,0)(0.095,0.5) \psline[linewidth=1pt](0.101,0)(0.101,0.5) \psline[linewidth=1pt](0.110,0)(0.110,0.5) \psline[linewidth=1pt](0.121,0)(0.121,0.5) \psline[linewidth=1pt](0.128,0)(0.128,0.5) \psline[linewidth=1pt](0.137,0)(0.137,0.5) \psline[linewidth=1pt](0.140,0)(0.140,0.5) \psline[linewidth=1pt](0.149,0)(0.149,0.5) \psline[linewidth=1pt](0.154,0)(0.154,0.5) \psline[linewidth=1pt](0.166,0)(0.166,0.5) \psline[linewidth=1pt](0.167,0)(0.167,0.5) \psline[linewidth=1pt](0.177,0)(0.177,0.5) \psline[linewidth=1pt](0.182,0)(0.182,0.5) \psline[linewidth=1pt](0.188,0)(0.188,0.5) \psline[linewidth=1pt](0.201,0)(0.201,0.5) \psline[linewidth=1pt](0.208,0)(0.208,0.5) \psline[linewidth=1pt](0.225,0)(0.225,0.5) \psline[linewidth=1pt](0.235,0)(0.235,0.5) \psline[linewidth=1pt](0.259,0)(0.259,0.5) \psline[linewidth=1pt](0.262,0)(0.262,0.5)
\psline[linewidth=1pt](0.281,0)(0.281,0.5) \psline[linewidth=1pt](0.291,0)(0.291,0.5)
\psline[linewidth=1pt](0.296,0)(0.296,0.5) \psline[linewidth=1pt](0.304,0)(0.304,0.5)
\psline[linewidth=1pt](0.331,0)(0.331,0.5) \psline[linewidth=1pt](0.339,0)(0.339,0.5) \psline[linewidth=1pt](0.343,0)(0.343,0.5) \psline[linewidth=1pt](0.367,0)(0.367,0.5) \psline[linewidth=1pt](0.371,0)(0.371,0.5) \psline[linewidth=1pt](0.395,0)(0.395,0.5) \psline[linewidth=1pt](0.403,0)(0.403,0.5) \psline[linewidth=1pt](0.433,0)(0.433,0.5) \psline[linewidth=1pt](0.446,0)(0.446,0.5) \psline[linewidth=1pt](0.457,0)(0.457,0.5) \psline[linewidth=1pt](0.471,0)(0.471,0.5) \psline[linewidth=1pt](0.494,0)(0.494,0.5) \psline[linewidth=1pt](0.502,0)(0.502,0.5) \psline[linewidth=1pt](0.527,0)(0.527,0.5) \psline[linewidth=1pt](0.540,0)(0.540,0.5) \psline[linewidth=1pt](0.575,0)(0.575,0.5) \psline[linewidth=1pt](0.577,0)(0.577,0.5) \psline[linewidth=1pt](0.587,0)(0.587,0.5) \psline[linewidth=1pt](0.648,0)(0.648,0.5) \psline[linewidth=1pt](0.658,0)(0.658,0.5) \psline[linewidth=1pt](0.663,0)(0.663,0.5) \psline[linewidth=1pt](0.717,0)(0.717,0.5) \psline[linewidth=1pt](0.746,0)(0.746,0.5) \psline[linewidth=1pt](0.769,0)(0.769,0.5) \psline[linewidth=1pt](0.787,0)(0.787,0.5) \psline[linewidth=1pt](0.822,0)(0.822,0.5) \psline[linewidth=1pt](0.833,0)(0.833,0.5) \psline[linewidth=1pt](0.865,0)(0.865,0.5) \psline[linewidth=1pt](0.877,0)(0.877,0.5) \psline[linewidth=1pt](0.922,0)(0.922,0.5) \psline[linewidth=1pt](0.988,0)(0.988,0.5) \psline[linewidth=1pt](1.027,0)(1.027,0.5) \psline[linewidth=1pt](1.040,0)(1.040,0.5) \psline[linewidth=1pt](1.060,0)(1.060,0.5) \psline[linewidth=1pt](1.108,0)(1.108,0.5) \psline[linewidth=1pt](1.164,0)(1.164,0.5)
\psline[linewidth=1pt](1.180,0)(1.180,0.5) \psline[linewidth=1pt](1.207,0)(1.207,0.5)
\psline[linewidth=1pt](1.236,0)(1.236,0.5) \psline[linewidth=1pt](1.372,0)(1.372,0.5)
\psline[linewidth=1pt](1.428,0)(1.428,0.5) \psline[linewidth=1pt](1.501,0)(1.501,0.5)
\psline[linewidth=1pt](1.545,0)(1.545,0.5) \psline[linewidth=1pt](1.638,0)(1.638,0.5)
\psline[linewidth=1pt](1.735,0)(1.735,0.5) \psline[linewidth=1pt](1.808,0)(1.808,0.5)
\psline[linewidth=1pt](1.917,0)(1.917,0.5) \psline[linewidth=1pt](2.233,0)(2.233,0.5)
\psline[linewidth=1pt](2.571,0)(2.571,0.5) \psline[linewidth=1pt](2.821,0)(2.821,0.5)
\psline[linewidth=1pt](7.277,0)(7.277,0.5) \psline[linewidth=1pt](9.284,0)(9.284,0.5)
\small \rput(4.9,-0.35){Figure 26: eigenvalues in order of magnitude. The} \rput(5.2,-0.55){shaded area corresponds to the eigenvalues predicted} \rput(2,-0.75){for a random matrix.}
\end{pspicture}

\vskip 2 cm

Note that the largest eigenvalue is even more out of scale than in previous crisis, what usually indicates a high level of correlation between the market indices and the presence of a powerful global market movement, although it is also influenced by the size of the sample of indices. We also have several eigenvalues that are bellow the minimum theoretical eingenvalue. The next picture (figure 27) shows eigenvector $e_{92}$, which corresponds to a combination of all indices in a market movement that explains about 34\% of the collective movement of all indices.

\begin{pspicture}(-0.1,0)(3.5,2.5)
\psset{xunit=0.187,yunit=5}
\pspolygon*[linecolor=lightblue](0.5,0)(0.5,0.089)(1.5,0.089)(1.5,0)
\pspolygon*[linecolor=lightblue](1.5,0)(1.5,0.082)(2.5,0.082)(2.5,0)
\pspolygon*[linecolor=lightblue](2.5,0)(2.5,0.105)(3.5,0.105)(3.5,0)
\pspolygon*[linecolor=lightblue](3.5,0)(3.5,0.113)(4.5,0.113)(4.5,0)
\pspolygon*[linecolor=lightblue](4.5,0)(4.5,0.025)(5.5,0.025)(5.5,0)
\pspolygon*[linecolor=lightblue](5.5,0)(5.5,0.023)(6.5,0.023)(6.5,0)
\pspolygon*[linecolor=lightblue](6.5,0)(6.5,0.008)(7.5,0.008)(7.5,0)
\pspolygon*[linecolor=lightblue](7.5,0)(7.5,0.033)(8.5,0.033)(8.5,0)
\pspolygon*[linecolor=lightblue](8.5,0)(8.5,0.118)(9.5,0.118)(9.5,0)
\pspolygon*[linecolor=lightblue](9.5,0)(9.5,0.117)(10.5,0.117)(10.5,0)
\pspolygon*[linecolor=lightblue](10.5,0)(10.5,0.119)(11.5,0.119)(11.5,0)
\pspolygon*[linecolor=lightblue](11.5,0)(11.5,0.126)(12.5,0.126)(12.5,0)
\pspolygon*[linecolor=lightblue](12.5,0)(12.5,0.025)(13.5,0.025)(13.5,0)
\pspolygon*[linecolor=lightblue](13.5,0)(13.5,0.133)(14.5,0.133)(14.5,0)
\pspolygon*[linecolor=lightblue](14.5,0)(14.5,0.155)(15.5,0.155)(15.5,0)
\pspolygon*[linecolor=lightblue](15.5,0)(15.5,0.127)(16.5,0.127)(16.5,0)
\pspolygon*[linecolor=lightblue](16.5,0)(16.5,0.154)(17.5,0.154)(17.5,0)
\pspolygon*[linecolor=lightblue](17.5,0)(17.5,0.147)(18.5,0.147)(18.5,0)
\pspolygon*[linecolor=lightblue](18.5,0)(18.5,0.149)(19.5,0.149)(19.5,0)
\pspolygon*[linecolor=lightblue](19.5,0)(19.5,0.154)(20.5,0.154)(20.5,0)
\pspolygon*[linecolor=lightblue](20.5,0)(20.5,0.154)(21.5,0.154)(21.5,0)
\pspolygon*[linecolor=lightblue](21.5,0)(21.5,0.020)(22.5,0.020)(22.5,0)
\pspolygon*[linecolor=lightblue](22.5,0)(22.5,0.145)(23.5,0.145)(23.5,0)
\pspolygon*[linecolor=lightblue](23.5,0)(23.5,0.152)(24.5,0.152)(24.5,0)
\pspolygon*[linecolor=lightblue](24.5,0)(24.5,0.148)(25.5,0.148)(25.5,0)
\pspolygon*[linecolor=lightblue](25.5,0)(25.5,0.148)(26.5,0.148)(26.5,0)
\pspolygon*[linecolor=lightblue](26.5,0)(26.5,0.158)(27.5,0.158)(27.5,0)
\pspolygon*[linecolor=lightblue](27.5,0)(27.5,0.147)(28.5,0.147)(28.5,0)
\pspolygon*[linecolor=lightblue](28.5,0)(28.5,0.145)(29.5,0.145)(29.5,0)
\pspolygon*[linecolor=lightred](29.5,0)(29.5,0.026)(30.5,0.026)(30.5,0)
\pspolygon*[linecolor=lightblue](30.5,0)(30.5,0.150)(31.5,0.150)(31.5,0)
\pspolygon*[linecolor=lightblue](31.5,0)(31.5,0.153)(32.5,0.153)(32.5,0)
\pspolygon*[linecolor=lightblue](32.5,0)(32.5,0.142)(33.5,0.142)(33.5,0)
\pspolygon*[linecolor=lightblue](33.5,0)(33.5,0.151)(34.5,0.151)(34.5,0)
\pspolygon*[linecolor=lightblue](34.5,0)(34.5,0.007)(35.5,0.007)(35.5,0)
\pspolygon*[linecolor=lightblue](35.5,0)(35.5,0.132)(36.5,0.132)(36.5,0)
\pspolygon*[linecolor=lightblue](36.5,0)(36.5,0.074)(37.5,0.074)(37.5,0)
\pspolygon*[linecolor=lightblue](37.5,0)(37.5,0.130)(38.5,0.130)(38.5,0)
\pspolygon*[linecolor=lightblue](38.5,0)(38.5,0.113)(39.5,0.113)(39.5,0)
\pspolygon*[linecolor=lightblue](39.5,0)(39.5,0.030)(40.5,0.030)(40.5,0)
\pspolygon*[linecolor=lightblue](40.5,0)(40.5,0.041)(41.5,0.041)(41.5,0)
\pspolygon*[linecolor=lightblue](41.5,0)(41.5,0.066)(42.5,0.066)(42.5,0)
\pspolygon*[linecolor=lightblue](42.5,0)(42.5,0.140)(43.5,0.140)(43.5,0)
\pspolygon*[linecolor=lightblue](43.5,0)(43.5,0.131)(44.5,0.131)(44.5,0)
\pspolygon*[linecolor=lightblue](44.5,0)(44.5,0.090)(45.5,0.090)(45.5,0)
\pspolygon*[linecolor=lightblue](45.5,0)(45.5,0.112)(46.5,0.112)(46.5,0)
\pspolygon*[linecolor=lightblue](46.5,0)(46.5,0.084)(47.5,0.084)(47.5,0)
\pspolygon*[linecolor=lightblue](47.5,0)(47.5,0.113)(48.5,0.113)(48.5,0)
\pspolygon*[linecolor=lightblue](48.5,0)(48.5,0.108)(49.5,0.108)(49.5,0)
\pspolygon*[linecolor=lightblue](49.5,0)(49.5,0.127)(50.5,0.127)(50.5,0)
\pspolygon*[linecolor=lightblue](50.5,0)(50.5,0.079)(51.5,0.079)(51.5,0)
\pspolygon*[linecolor=lightblue](51.5,0)(51.5,0.129)(52.5,0.129)(52.5,0)
\pspolygon*[linecolor=lightblue](52.5,0)(52.5,0.126)(53.5,0.126)(53.5,0)
\pspolygon*[linecolor=lightblue](53.5,0)(53.5,0.094)(54.5,0.094)(54.5,0)
\pspolygon*[linecolor=lightblue](54.5,0)(54.5,0.066)(55.5,0.066)(55.5,0)
\pspolygon*[linecolor=lightblue](55.5,0)(55.5,0.056)(56.5,0.056)(56.5,0)
\pspolygon*[linecolor=lightblue](56.5,0)(56.5,0.073)(57.5,0.073)(57.5,0)
\pspolygon*[linecolor=lightblue](57.5,0)(57.5,0.089)(58.5,0.089)(58.5,0)
\pspolygon*[linecolor=lightblue](58.5,0)(58.5,0.038)(59.5,0.038)(59.5,0)
\pspolygon*[linecolor=lightblue](59.5,0)(59.5,0.063)(60.5,0.063)(60.5,0)
\pspolygon*[linecolor=lightblue](60.5,0)(60.5,0.089)(61.5,0.089)(61.5,0)
\pspolygon*[linecolor=lightblue](61.5,0)(61.5,0.095)(62.5,0.095)(62.5,0)
\pspolygon*[linecolor=lightblue](62.5,0)(62.5,0.077)(63.5,0.077)(63.5,0)
\pspolygon*[linecolor=lightblue](63.5,0)(63.5,0.002)(64.5,0.002)(64.5,0)
\pspolygon*[linecolor=lightblue](64.5,0)(64.5,0.113)(65.5,0.113)(65.5,0)
\pspolygon*[linecolor=lightblue](65.5,0)(65.5,0.058)(66.5,0.058)(66.5,0)
\pspolygon*[linecolor=lightblue](66.5,0)(66.5,0.019)(67.5,0.019)(67.5,0)
\pspolygon*[linecolor=lightblue](67.5,0)(67.5,0.123)(68.5,0.123)(68.5,0)
\pspolygon*[linecolor=lightblue](68.5,0)(68.5,0.122)(69.5,0.122)(69.5,0)
\pspolygon*[linecolor=lightblue](69.5,0)(69.5,0.054)(70.5,0.054)(70.5,0)
\pspolygon*[linecolor=lightred](70.5,0)(70.5,0.006)(71.5,0.006)(71.5,0)
\pspolygon*[linecolor=lightblue](71.5,0)(71.5,0.100)(72.5,0.100)(72.5,0)
\pspolygon*[linecolor=lightblue](72.5,0)(72.5,0.115)(73.5,0.115)(73.5,0)
\pspolygon*[linecolor=lightblue](73.5,0)(73.5,0.119)(74.5,0.119)(74.5,0)
\pspolygon*[linecolor=lightblue](74.5,0)(74.5,0.038)(75.5,0.038)(75.5,0)
\pspolygon*[linecolor=lightblue](75.5,0)(75.5,0.095)(76.5,0.095)(76.5,0)
\pspolygon*[linecolor=lightblue](76.5,0)(76.5,0.133)(77.5,0.133)(77.5,0)
\pspolygon*[linecolor=lightblue](77.5,0)(77.5,0.108)(78.5,0.108)(78.5,0)
\pspolygon*[linecolor=lightblue](78.5,0)(78.5,0.108)(79.5,0.108)(79.5,0)
\pspolygon*[linecolor=lightblue](79.5,0)(79.5,0.118)(80.5,0.118)(80.5,0)
\pspolygon*[linecolor=lightblue](80.5,0)(80.5,0.094)(81.5,0.094)(81.5,0)
\pspolygon*[linecolor=lightblue](81.5,0)(81.5,0.071)(82.5,0.071)(82.5,0)
\pspolygon*[linecolor=lightblue](82.5,0)(82.5,0.057)(83.5,0.057)(83.5,0)
\pspolygon*[linecolor=lightblue](83.5,0)(83.5,0.094)(84.5,0.094)(84.5,0)
\pspolygon*[linecolor=lightblue](84.5,0)(84.5,0.010)(85.5,0.010)(85.5,0)
\pspolygon*[linecolor=lightred](85.5,0)(85.5,0.014)(86.5,0.014)(86.5,0)
\pspolygon*[linecolor=lightblue](86.5,0)(86.5,0.043)(87.5,0.043)(87.5,0)
\pspolygon*[linecolor=lightred](87.5,0)(87.5,0.021)(88.5,0.021)(88.5,0)
\pspolygon*[linecolor=lightblue](88.5,0)(88.5,0.127)(89.5,0.127)(89.5,0)
\pspolygon*[linecolor=lightred](89.5,0)(89.5,0.005)(90.5,0.005)(90.5,0)
\pspolygon*[linecolor=lightblue](90.5,0)(90.5,0.136)(91.5,0.136)(91.5,0)
\pspolygon*[linecolor=lightblue](91.5,0)(91.5,0.072)(92.5,0.072)(92.5,0)
\pspolygon[linecolor=blue](0.5,0)(0.5,0.089)(1.5,0.089)(1.5,0)
\pspolygon[linecolor=blue](1.5,0)(1.5,0.082)(2.5,0.082)(2.5,0)
\pspolygon[linecolor=blue](2.5,0)(2.5,0.105)(3.5,0.105)(3.5,0)
\pspolygon[linecolor=blue](3.5,0)(3.5,0.113)(4.5,0.113)(4.5,0)
\pspolygon[linecolor=blue](4.5,0)(4.5,0.025)(5.5,0.025)(5.5,0)
\pspolygon[linecolor=blue](5.5,0)(5.5,0.023)(6.5,0.023)(6.5,0)
\pspolygon[linecolor=blue](6.5,0)(6.5,0.008)(7.5,0.008)(7.5,0)
\pspolygon[linecolor=blue](7.5,0)(7.5,0.033)(8.5,0.033)(8.5,0)
\pspolygon[linecolor=blue](8.5,0)(8.5,0.118)(9.5,0.118)(9.5,0)
\pspolygon[linecolor=blue](9.5,0)(9.5,0.117)(10.5,0.117)(10.5,0)
\pspolygon[linecolor=blue](10.5,0)(10.5,0.119)(11.5,0.119)(11.5,0)
\pspolygon[linecolor=blue](11.5,0)(11.5,0.126)(12.5,0.126)(12.5,0)
\pspolygon[linecolor=blue](12.5,0)(12.5,0.025)(13.5,0.025)(13.5,0)
\pspolygon[linecolor=blue](13.5,0)(13.5,0.133)(14.5,0.133)(14.5,0)
\pspolygon[linecolor=blue](14.5,0)(14.5,0.155)(15.5,0.155)(15.5,0)
\pspolygon[linecolor=blue](15.5,0)(15.5,0.127)(16.5,0.127)(16.5,0)
\pspolygon[linecolor=blue](16.5,0)(16.5,0.154)(17.5,0.154)(17.5,0)
\pspolygon[linecolor=blue](17.5,0)(17.5,0.147)(18.5,0.147)(18.5,0)
\pspolygon[linecolor=blue](18.5,0)(18.5,0.149)(19.5,0.149)(19.5,0)
\pspolygon[linecolor=blue](19.5,0)(19.5,0.154)(20.5,0.154)(20.5,0)
\pspolygon[linecolor=blue](20.5,0)(20.5,0.154)(21.5,0.154)(21.5,0)
\pspolygon[linecolor=blue](21.5,0)(21.5,0.020)(22.5,0.020)(22.5,0)
\pspolygon[linecolor=blue](22.5,0)(22.5,0.145)(23.5,0.145)(23.5,0)
\pspolygon[linecolor=blue](23.5,0)(23.5,0.152)(24.5,0.152)(24.5,0)
\pspolygon[linecolor=blue](24.5,0)(24.5,0.148)(25.5,0.148)(25.5,0)
\pspolygon[linecolor=blue](25.5,0)(25.5,0.148)(26.5,0.148)(26.5,0)
\pspolygon[linecolor=blue](26.5,0)(26.5,0.158)(27.5,0.158)(27.5,0)
\pspolygon[linecolor=blue](27.5,0)(27.5,0.147)(28.5,0.147)(28.5,0)
\pspolygon[linecolor=blue](28.5,0)(28.5,0.145)(29.5,0.145)(29.5,0)
\pspolygon[linecolor=red](29.5,0)(29.5,0.026)(30.5,0.026)(30.5,0)
\pspolygon[linecolor=blue](30.5,0)(30.5,0.150)(31.5,0.150)(31.5,0)
\pspolygon[linecolor=blue](31.5,0)(31.5,0.153)(32.5,0.153)(32.5,0)
\pspolygon[linecolor=blue](32.5,0)(32.5,0.142)(33.5,0.142)(33.5,0)
\pspolygon[linecolor=blue](33.5,0)(33.5,0.151)(34.5,0.151)(34.5,0)
\pspolygon[linecolor=blue](34.5,0)(34.5,0.007)(35.5,0.007)(35.5,0)
\pspolygon[linecolor=blue](35.5,0)(35.5,0.132)(36.5,0.132)(36.5,0)
\pspolygon[linecolor=blue](36.5,0)(36.5,0.074)(37.5,0.074)(37.5,0)
\pspolygon[linecolor=blue](37.5,0)(37.5,0.130)(38.5,0.130)(38.5,0)
\pspolygon[linecolor=blue](38.5,0)(38.5,0.113)(39.5,0.113)(39.5,0)
\pspolygon[linecolor=blue](39.5,0)(39.5,0.030)(40.5,0.030)(40.5,0)
\pspolygon[linecolor=blue](40.5,0)(40.5,0.041)(41.5,0.041)(41.5,0)
\pspolygon[linecolor=blue](41.5,0)(41.5,0.066)(42.5,0.066)(42.5,0)
\pspolygon[linecolor=blue](42.5,0)(42.5,0.140)(43.5,0.140)(43.5,0)
\pspolygon[linecolor=blue](43.5,0)(43.5,0.131)(44.5,0.131)(44.5,0)
\pspolygon[linecolor=blue](44.5,0)(44.5,0.090)(45.5,0.090)(45.5,0)
\pspolygon[linecolor=blue](45.5,0)(45.5,0.112)(46.5,0.112)(46.5,0)
\pspolygon[linecolor=blue](46.5,0)(46.5,0.084)(47.5,0.084)(47.5,0)
\pspolygon[linecolor=blue](47.5,0)(47.5,0.113)(48.5,0.113)(48.5,0)
\pspolygon[linecolor=blue](48.5,0)(48.5,0.108)(49.5,0.108)(49.5,0)
\pspolygon[linecolor=blue](49.5,0)(49.5,0.127)(50.5,0.127)(50.5,0)
\pspolygon[linecolor=blue](50.5,0)(50.5,0.079)(51.5,0.079)(51.5,0)
\pspolygon[linecolor=blue](51.5,0)(51.5,0.129)(52.5,0.129)(52.5,0)
\pspolygon[linecolor=blue](52.5,0)(52.5,0.126)(53.5,0.126)(53.5,0)
\pspolygon[linecolor=blue](53.5,0)(53.5,0.094)(54.5,0.094)(54.5,0)
\pspolygon[linecolor=blue](54.5,0)(54.5,0.066)(55.5,0.066)(55.5,0)
\pspolygon[linecolor=blue](55.5,0)(55.5,0.056)(56.5,0.056)(56.5,0)
\pspolygon[linecolor=blue](56.5,0)(56.5,0.073)(57.5,0.073)(57.5,0)
\pspolygon[linecolor=blue](57.5,0)(57.5,0.089)(58.5,0.089)(58.5,0)
\pspolygon[linecolor=blue](58.5,0)(58.5,0.038)(59.5,0.038)(59.5,0)
\pspolygon[linecolor=blue](59.5,0)(59.5,0.063)(60.5,0.063)(60.5,0)
\pspolygon[linecolor=blue](60.5,0)(60.5,0.089)(61.5,0.089)(61.5,0)
\pspolygon[linecolor=blue](61.5,0)(61.5,0.095)(62.5,0.095)(62.5,0)
\pspolygon[linecolor=blue](62.5,0)(62.5,0.077)(63.5,0.077)(63.5,0)
\pspolygon[linecolor=blue](63.5,0)(63.5,0.002)(64.5,0.002)(64.5,0)
\pspolygon[linecolor=blue](64.5,0)(64.5,0.113)(65.5,0.113)(65.5,0)
\pspolygon[linecolor=blue](65.5,0)(65.5,0.058)(66.5,0.058)(66.5,0)
\pspolygon[linecolor=blue](66.5,0)(66.5,0.019)(67.5,0.019)(67.5,0)
\pspolygon[linecolor=blue](67.5,0)(67.5,0.123)(68.5,0.123)(68.5,0)
\pspolygon[linecolor=blue](68.5,0)(68.5,0.122)(69.5,0.122)(69.5,0)
\pspolygon[linecolor=blue](69.5,0)(69.5,0.054)(70.5,0.054)(70.5,0)
\pspolygon[linecolor=red](70.5,0)(70.5,0.006)(71.5,0.006)(71.5,0)
\pspolygon[linecolor=blue](71.5,0)(71.5,0.100)(72.5,0.100)(72.5,0)
\pspolygon[linecolor=blue](72.5,0)(72.5,0.115)(73.5,0.115)(73.5,0)
\pspolygon[linecolor=blue](73.5,0)(73.5,0.119)(74.5,0.119)(74.5,0)
\pspolygon[linecolor=blue](74.5,0)(74.5,0.038)(75.5,0.038)(75.5,0)
\pspolygon[linecolor=blue](75.5,0)(75.5,0.095)(76.5,0.095)(76.5,0)
\pspolygon[linecolor=blue](76.5,0)(76.5,0.133)(77.5,0.133)(77.5,0)
\pspolygon[linecolor=blue](77.5,0)(77.5,0.108)(78.5,0.108)(78.5,0)
\pspolygon[linecolor=blue](78.5,0)(78.5,0.108)(79.5,0.108)(79.5,0)
\pspolygon[linecolor=blue](79.5,0)(79.5,0.118)(80.5,0.118)(80.5,0)
\pspolygon[linecolor=blue](80.5,0)(80.5,0.094)(81.5,0.094)(81.5,0)
\pspolygon[linecolor=blue](81.5,0)(81.5,0.071)(82.5,0.071)(82.5,0)
\pspolygon[linecolor=blue](82.5,0)(82.5,0.057)(83.5,0.057)(83.5,0)
\pspolygon[linecolor=blue](83.5,0)(83.5,0.094)(84.5,0.094)(84.5,0)
\pspolygon[linecolor=blue](84.5,0)(84.5,0.010)(85.5,0.010)(85.5,0)
\pspolygon[linecolor=red](85.5,0)(85.5,0.014)(86.5,0.014)(86.5,0)
\pspolygon[linecolor=blue](86.5,0)(86.5,0.043)(87.5,0.043)(87.5,0)
\pspolygon[linecolor=red](87.5,0)(87.5,0.021)(88.5,0.021)(88.5,0)
\pspolygon[linecolor=blue](88.5,0)(88.5,0.127)(89.5,0.127)(89.5,0)
\pspolygon[linecolor=red](89.5,0)(89.5,0.005)(90.5,0.005)(90.5,0)
\pspolygon[linecolor=blue](90.5,0)(90.5,0.136)(91.5,0.136)(91.5,0)
\pspolygon[linecolor=blue](91.5,0)(91.5,0.072)(92.5,0.072)(92.5,0)
\psline{->}(0,0)(94,0) \psline{->}(0,0)(0,0.4) \rput(2,0.4){$e_{92}$} \scriptsize \psline(1,-0.02)(1,0.02) \rput(1,-0.06){S\&P} \psline(5,-0.02)(5,0.02) \rput(5,-0.06){Pana} \psline(10,-0.02)(10,0.02) \rput(10,-0.06){Arge} \psline(15,-0.02)(15,0.02) \rput(15,-0.06){UK} \psline(20,-0.02)(20,0.02) \rput(20,-0.06){Autr} \psline(25,-0.02)(25,0.02) \rput(25,-0.06){Luxe} \psline(30,-0.02)(30,0.02) \rput(30,-0.06){Icel} \psline(35,-0.02)(35,0.02) \rput(35,-0.06){Slok} \psline(40,-0.02)(40,0.02) \rput(40,-0.06){BoHe} \psline(45,-0.02)(45,0.02) \rput(45,-0.06){Bulg} \psline(49,-0.02)(49,0.02) \rput(49,-0.06){Russ} \psline(55,-0.02)(55,0.02) \rput(55,-0.06){Pale} \psline(60,-0.02)(60,0.02) \rput(60,-0.06){Bahr} \psline(65,-0.02)(65,0.02) \rput(65,-0.06){Indi} \psline(70,-0.02)(70,0.02) \rput(70,-0.06){Chin} \psline(75,-0.02)(75,0.02) \rput(75,-0.06){Viet} \psline(80,-0.02)(80,0.02) \rput(80,-0.06){Aust} \psline(85,-0.02)(85,0.02) \rput(85,-0.06){Ghan} \psline(90,-0.02)(90,0.02) \rput(90,-0.06){Bots} \scriptsize \psline(-0.28,0.1)(0.28,0.1) \rput(-1.6,0.1){$0.1$} \psline(-0.28,0.2)(0.28,0.2) \rput(-1.6,0.2){$0.2$} \psline(-0.28,0.3)(0.28,0.3) \rput(-1.6,0.3){$0.3$}
\end{pspicture}

\vskip 0.6 cm

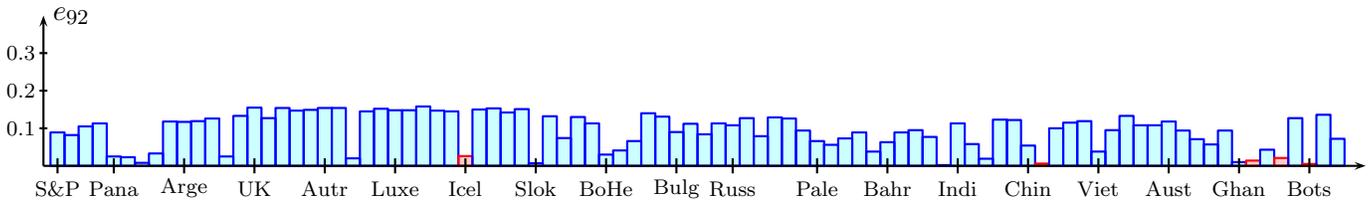
{\noindent \small Figure 27: contributions of the stock market indices to eigenvector $e_{92}$, corresponding to the largest eigenvalue of the correlation matrix. Blue bars indicate positive values, and red bars correspond to negative values. The indices are aligned in the following way: {\bf S\&P}, Nasd, Cana, Mexi, {\bf Pana}, CoRi, Berm, Jama, Braz, {\bf Arge}, Chil, Colo, Vene, Peru, {\bf UK}, Irel, Fran, Germ, Swit, {\bf Autr}, Ital, Malt, Belg, Neth, {\bf Luxe}, Swed, Denm, Finl, Norw, {\bf Icel}, Spai, Port, Gree, CzRe, {\bf Slok}, Hung, Serb, Croa, Slov, {\bf BoHe}, Mont, Mace, Pola, Roma, {\bf Bulg}, Esto, Latv, Lith, Ukra, {\bf Russ}, Kaza, Turk, Cypr, Isra, {\bf Pale}, Leba, Jord, SaAr, Kwai, {\bf Bahr}, Qata, UAE, Ohma, Paki, {\bf Indi}, SrLa, Bang, Japa, HoKo, {\bf Chin}, Mong, Taiw, SoKo, Thai, {\bf Viet}, Mala, Sing, Indo, Phil, {\bf Aust}, NeZe, Moro, Tuni, Egyp, {\bf Ghan}, Nige, Keny, Tanz, Nami, {\bf Bots}, SoAf, Maur.}

\vskip 0.4 cm

Indices with small negative contributions are those from Iceland, which suffered the effects of the crisis with greater impact than most of the other countries, Mongolia, Nigeria, Tanzania, and Botswana. Very small participations (less than $0.050$) are related with the indices from Central America, the Atlantic Islands, Venezuela, Malta, Slovakia, Bosnia and Herzegovina, Montenegro, Kwait, Pakistan, Bangladesh, Vietnam, Ghana, and Kenya. Indices with strong participation (greater than $0.100$) are those from Canada, Mexico, most South American countries, most of the European countries, Russia, Turkey, Cyprus, India, Japan, Hong Kong, Taiwan, South Korea, Thailand, Singapore, Indonesia, Philipines, Australia, Namibia, and South Africa. The surprise is the participations of the indices from the USA - S\&P 500 ($0.089$), and Nasdaq ($0.082$), which are lower than expected.

The following two figures show the relation between the average correlation and the volatility of the market index. Figure 28 shows the average correlation (in red) calculated in a moving window of 100 days, and the volatility of the market index (in blue), both normalized so as to have mean two and standard deviation one. Figure 29 shows the average correlation (in red) and the average volatility (in blue), both calculated in a moving window of 100 days, and normalized so as to have mean two and standard deviation one.

One can see that the period of high volatility seems to be preceded by a period of high correlation between the stock markets of the world. Figure 30 shows the evolution of the covariance between the mean correlation and the mean volatility, calculated in a moving window of 30 days.

\begin{pspicture}(-1.5,-0.2)(3.5,5.3)
\psset{xunit=2.5,yunit=0.8}
\psline{->}(0,0)(5.4,0) \psline{->}(0,0)(0,6) \rput(5.6,0){day} \rput(0.5,6){$<C>_n$,$\text{vol.}_n$}\scriptsize \psline(0.02,-0.1)(0.02,0.1) \rput(0.02,-0.3){01/02} \psline(0.46,-0.1)(0.46,0.1) \rput(0.46,-0.3){02/01} \psline(0.88,-0.1)(0.88,0.1) \rput(0.88,-0.3){03/03} \psline(1.26,-0.1)(1.26,0.1) \rput(1.26,-0.3){04/01} \psline(1.7,-0.1)(1.7,0.1) \rput(1.7,-0.3){05/02} \psline(2.12,-0.1)(2.12,0.1) \rput(2.12,-0.3){06/02} \psline(2.54,-0.1)(2.54,0.1) \rput(2.54,-0.3){07/01} \psline(3,-0.1)(3,0.1) \rput(3,-0.3){08/01} \psline(3.42,-0.1)(3.42,0.1) \rput(3.42,-0.3){09/01} \psline(3.86,-0.1)(3.86,0.1) \rput(3.86,-0.3){10/02} \psline(4.3,-0.1)(4.3,0.1) \rput(4.3,-0.3){11/03} \psline(4.7,-0.1)(4.7,0.1) \rput(4.7,-0.3){12/01} \psline(5.06,-0.1)(5.06,0.1) \rput(5.06,-0.3){12/30} \psline(-0.043,1)(0.043,1) \rput(-0.17,1){$1$} \psline(-0.043,2)(0.043,2) \rput(-0.17,2){$2$} \psline(-0.043,3)(0.043,3) \rput(-0.17,3){$3$} \psline(-0.043,4)(0.043,4) \rput(-0.17,4){$4$} \psline(-0.043,5)(0.043,5) \rput(-0.17,5){$5$}
\psline[linecolor=blue](2.02,1.665) (2.04,1.493) (2.06,1.200) (2.08,1.380) (2.1,1.201) (2.12,1.469) (2.14,1.366) (2.16,1.258) (2.18,1.505) (2.2,1.334) (2.22,1.475) (2.24,1.630) (2.26,1.865) (2.28,1.605) (2.3,1.115) (2.32,1.116) (2.34,1.185) (2.36,1.330) (2.38,1.653) (2.4,1.635) (2.42,1.681) (2.44,1.435) (2.46,1.645) (2.48,1.489) (2.5,2.004) (2.52,1.516) (2.54,1.139) (2.56,2.112) (2.58,1.463) (2.6,1.526) (2.62,1.410) (2.64,1.397) (2.66,1.873) (2.68,1.728) (2.7,1.548) (2.72,1.821) (2.74,1.101) (2.76,2.523) (2.78,1.175) (2.8,2.275) (2.82,1.402) (2.84,1.801) (2.86,1.503) (2.88,1.979) (2.9,1.439) (2.92,1.555) (2.94,1.334) (2.96,1.223) (2.98,2.026) (3,1.087) (3.02,1.729) (3.04,1.745) (3.06,1.451) (3.08,1.564) (3.1,1.270) (3.12,1.321) (3.14,1.224) (3.16,1.352) (3.18,1.662) (3.2,1.380) (3.22,1.087) (3.24,1.263) (3.26,2.046) (3.28,1.305) (3.3,1.371) (3.32,1.688) (3.34,1.334) (3.36,1.338) (3.38,1.213) (3.4,1.523) (3.42,1.374) (3.44,1.338) (3.46,1.135) (3.48,1.692) (3.5,2.211) (3.52,2.455) (3.54,2.387) (3.56,2.005) (3.58,1.842) (3.6,1.981) (3.62,1.489) (3.64,3.311) (3.66,3.154) (3.68,2.135) (3.7,1.840) (3.72,5.527) (3.74,1.197) (3.76,2.419) (3.78,1.200) (3.8,1.490) (3.82,1.912) (3.84,3.820) (3.86,1.342) (3.88,1.733) (3.9,1.388) (3.92,5.609) (3.94,2.804) (3.96,4.706) (3.98,1.290) (4,5.169) (4.02,5.107) (4.04,4.316) (4.06,4.329) (4.08,4.108) (4.1,1.144) (4.12,2.469) (4.14,1.108) (4.16,4.142) (4.18,2.685) (4.2,4.765) (4.22,3.495) (4.24,2.400) (4.26,3.742) (4.28,3.291) (4.3,2.249) (4.32,2.537) (4.34,3.521) (4.36,1.641) (4.38,4.015) (4.4,1.496) (4.42,1.766) (4.44,3.458) (4.46,3.018) (4.48,1.942) (4.5,1.235) (4.52,2.420) (4.54,2.213) (4.56,2.425) (4.58,3.680) (4.6,1.466) (4.62,3.310) (4.64,2.145) (4.66,1.178) (4.68,2.265) (4.7,1.618) (4.72,3.159) (4.74,1.484) (4.76,1.152) (4.78,1.337) (4.8,2.165) (4.82,3.984) (4.84,1.753) (4.86,1.332) (4.88,2.204) (4.9,1.458) (4.92,1.653) (4.94,1.444) (4.96,1.201) (4.98,1.676) (5,1.827) (5.02,1.222) (5.04,1.194) (5.06,1.543)
\psline[linecolor=red](2.02,1.561) (2.04,1.567) (2.06,1.567) (2.08,1.583) (2.1,1.590) (2.12,1.594) (2.14,1.601) (2.16,1.602) (2.18,1.618) (2.2,1.615) (2.22,1.610) (2.24,1.560) (2.26,1.567) (2.28,1.562) (2.3,1.265) (2.32,1.283) (2.34,1.362) (2.36,1.304) (2.38,1.287) (2.4,1.266) (2.42,1.271) (2.44,1.282) (2.46,1.294) (2.48,1.235) (2.5,1.207) (2.52,1.169) (2.54,1.188) (2.56,1.187) (2.58,1.190) (2.6,1.189) (2.62,1.137) (2.64,1.139) (2.66,1.115) (2.68,1.118) (2.7,1.097) (2.72,1.071) (2.74,1.132) (2.76,1.101) (2.78,1.133) (2.8,1.111) (2.82,1.105) (2.84,1.105) (2.86,1.134) (2.88,1.116) (2.9,1.087) (2.92,1.089) (2.94,1.065) (2.96,1.113) (2.98,1.096) (3,1.093) (3.02,1.109) (3.04,1.065) (3.06,1.055) (3.08,1.051) (3.1,0.878) (3.12,0.873) (3.14,0.892) (3.16,0.888) (3.18,0.830) (3.2,0.835) (3.22,0.837) (3.24,0.861) (3.26,0.864) (3.28,0.837) (3.3,0.806) (3.32,0.812) (3.34,0.808) (3.36,0.788) (3.38,0.803) (3.4,0.812) (3.42,0.807) (3.44,0.803) (3.46,0.789) (3.48,0.824) (3.5,0.894) (3.52,0.948) (3.54,0.941) (3.56,0.963) (3.58,0.998) (3.6,0.981) (3.62,1.253) (3.64,1.451) (3.66,1.418) (3.68,1.417) (3.7,2.162) (3.72,2.103) (3.74,2.137) (3.76,2.134) (3.78,2.125) (3.8,2.107) (3.82,2.198) (3.84,2.099) (3.86,2.063) (3.88,2.018) (3.9,2.538) (3.92,2.564) (3.94,2.869) (3.96,2.755) (3.98,3.022) (4,3.193) (4.02,3.238) (4.04,3.209) (4.06,3.309) (4.08,3.249) (4.1,3.231) (4.12,3.197) (4.14,3.276) (4.16,3.287) (4.18,3.419) (4.2,3.427) (4.22,3.354) (4.24,3.336) (4.26,3.345) (4.28,3.335) (4.3,3.325) (4.32,3.369) (4.34,3.329) (4.36,3.338) (4.38,3.314) (4.4,3.307) (4.42,3.341) (4.44,3.335) (4.46,3.310) (4.48,3.294) (4.5,3.293) (4.52,3.285) (4.54,3.246) (4.56,3.307) (4.58,3.274) (4.6,3.213) (4.62,3.211) (4.64,3.200) (4.66,3.216) (4.68,3.217) (4.7,3.210) (4.72,3.202) (4.74,3.191) (4.76,3.184) (4.78,3.168) (4.8,3.158) (4.82,3.151) (4.84,3.138) (4.86,3.139) (4.88,3.112) (4.9,3.115) (4.92,3.116) (4.94,3.113) (4.96,3.111) (4.98,3.094) (5,3.075) (5.02,3.071) (5.04,3.050) (5.06,3.058)
\end{pspicture}

\vskip 0.4 cm

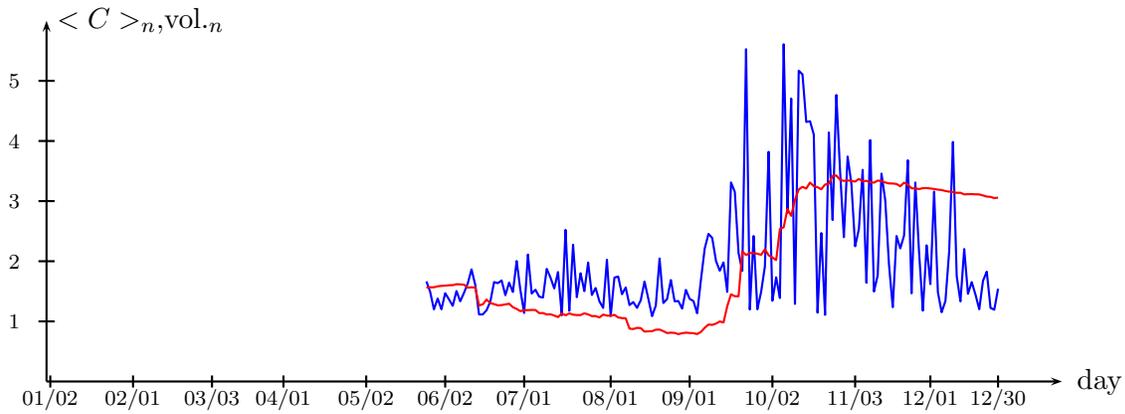
{\noindent \small Figure 28: volatility (blue) and average correlation (red) based on the log-returns for 2008, calculated in a moving window of 100 days, both normalized so as to have mean two and standard deviation one.}

\vskip 0.6 cm

\begin{pspicture}(-1.5,-0.2)(3.5,4)
\psset{xunit=2.5,yunit=0.8}
\psline{->}(0,0)(5.4,0) \psline{->}(0,0)(0,5) \rput(5.6,0){day} \rput(0.5,5){$<C>_n$,$\text{vol.}_n$}\scriptsize \psline(0.02,-0.1)(0.02,0.1) \rput(0.02,-0.3){01/02} \psline(0.46,-0.1)(0.46,0.1) \rput(0.46,-0.3){02/01} \psline(0.88,-0.1)(0.88,0.1) \rput(0.88,-0.3){03/03} \psline(1.26,-0.1)(1.26,0.1) \rput(1.26,-0.3){04/01} \psline(1.7,-0.1)(1.7,0.1) \rput(1.7,-0.3){05/02} \psline(2.12,-0.1)(2.12,0.1) \rput(2.12,-0.3){06/02} \psline(2.54,-0.1)(2.54,0.1) \rput(2.54,-0.3){07/01} \psline(3,-0.1)(3,0.1) \rput(3,-0.3){08/01} \psline(3.42,-0.1)(3.42,0.1) \rput(3.42,-0.3){09/01} \psline(3.86,-0.1)(3.86,0.1) \rput(3.86,-0.3){10/02} \psline(4.3,-0.1)(4.3,0.1) \rput(4.3,-0.3){11/03} \psline(4.7,-0.1)(4.7,0.1) \rput(4.7,-0.3){12/01} \psline(5.06,-0.1)(5.06,0.1) \rput(5.06,-0.3){12/30} \psline(-0.043,1)(0.043,1) \rput(-0.17,1){$1$} \psline(-0.043,2)(0.043,2) \rput(-0.17,2){$2$} \psline(-0.043,3)(0.043,3) \rput(-0.17,3){$3$} \psline(-0.043,4)(0.043,4) \rput(-0.17,4){$4$}
\psline[linecolor=blue](2.02,1.517) (2.04,1.521) (2.06,1.515) (2.08,1.509) (2.1,1.498) (2.12,1.505) (2.14,1.500) (2.16,1.492) (2.18,1.497) (2.2,1.504) (2.22,1.477) (2.24,1.447) (2.26,1.461) (2.28,1.460) (2.3,1.358) (2.32,1.336) (2.34,1.326) (2.36,1.266) (2.38,1.256) (2.4,1.242) (2.42,1.238) (2.44,1.249) (2.46,1.261) (2.48,1.236) (2.5,1.244) (2.52,1.216) (2.54,1.202) (2.56,1.216) (2.58,1.229) (2.6,1.234) (2.62,1.206) (2.64,1.206) (2.66,1.215) (2.68,1.217) (2.7,1.213) (2.72,1.224) (2.74,1.213) (2.76,1.244) (2.78,1.233) (2.8,1.255) (2.82,1.248) (2.84,1.271) (2.86,1.271) (2.88,1.277) (2.9,1.256) (2.92,1.256) (2.94,1.243) (2.96,1.236) (2.98,1.241) (3,1.213) (3.02,1.211) (3.04,1.215) (3.06,1.194) (3.08,1.202) (3.1,1.143) (3.12,1.123) (3.14,1.120) (3.16,1.106) (3.18,1.075) (3.2,1.081) (3.22,1.070) (3.24,1.068) (3.26,1.093) (3.28,1.073) (3.3,1.058) (3.32,1.075) (3.34,1.076) (3.36,1.066) (3.38,1.059) (3.4,1.067) (3.42,1.066) (3.44,1.061) (3.46,1.038) (3.48,1.049) (3.5,1.062) (3.52,1.107) (3.54,1.128) (3.56,1.156) (3.58,1.173) (3.6,1.203) (3.62,1.215) (3.64,1.280) (3.66,1.343) (3.68,1.370) (3.7,1.383) (3.72,1.510) (3.74,1.508) (3.76,1.551) (3.78,1.549) (3.8,1.560) (3.82,1.577) (3.84,1.662) (3.86,1.666) (3.88,1.675) (3.9,1.677) (3.92,1.822) (3.94,1.870) (3.96,1.969) (3.98,1.967) (4,2.103) (4.02,2.221) (4.04,2.318) (4.06,2.425) (4.08,2.518) (4.1,2.516) (4.12,2.550) (4.14,2.541) (4.16,2.640) (4.18,2.680) (4.2,2.798) (4.22,2.867) (4.24,2.893) (4.26,2.957) (4.28,3.015) (4.3,3.054) (4.32,3.102) (4.34,3.182) (4.36,3.193) (4.38,3.273) (4.4,3.269) (4.42,3.272) (4.44,3.341) (4.46,3.388) (4.48,3.403) (4.5,3.377) (4.52,3.408) (4.54,3.444) (4.56,3.455) (4.58,3.531) (4.6,3.529) (4.62,3.594) (4.64,3.619) (4.66,3.596) (4.68,3.614) (4.7,3.616) (4.72,3.662) (4.74,3.675) (4.76,3.628) (4.78,3.634) (4.8,3.630) (4.82,3.718) (4.84,3.717) (4.86,3.711) (4.88,3.719) (4.9,3.719) (4.92,3.723) (4.94,3.726) (4.96,3.726) (4.98,3.714) (5,3.739) (5.02,3.722) (5.04,3.703) (5.06,3.706)
\psline[linecolor=red](2.02,1.561) (2.04,1.567) (2.06,1.567) (2.08,1.583) (2.1,1.590) (2.12,1.594) (2.14,1.601) (2.16,1.602) (2.18,1.618) (2.2,1.615) (2.22,1.610) (2.24,1.560) (2.26,1.567) (2.28,1.562) (2.3,1.265) (2.32,1.283) (2.34,1.362) (2.36,1.304) (2.38,1.287) (2.4,1.266) (2.42,1.271) (2.44,1.282) (2.46,1.294) (2.48,1.235) (2.5,1.207) (2.52,1.169) (2.54,1.188) (2.56,1.187) (2.58,1.190) (2.6,1.189) (2.62,1.137) (2.64,1.139) (2.66,1.115) (2.68,1.118) (2.7,1.097) (2.72,1.071) (2.74,1.132) (2.76,1.101) (2.78,1.133) (2.8,1.111) (2.82,1.105) (2.84,1.105) (2.86,1.134) (2.88,1.116) (2.9,1.087) (2.92,1.089) (2.94,1.065) (2.96,1.113) (2.98,1.096) (3,1.093) (3.02,1.109) (3.04,1.065) (3.06,1.055) (3.08,1.051) (3.1,0.878) (3.12,0.873) (3.14,0.892) (3.16,0.888) (3.18,0.830) (3.2,0.835) (3.22,0.837) (3.24,0.861) (3.26,0.864) (3.28,0.837) (3.3,0.806) (3.32,0.812) (3.34,0.808) (3.36,0.788) (3.38,0.803) (3.4,0.812) (3.42,0.807) (3.44,0.803) (3.46,0.789) (3.48,0.824) (3.5,0.894) (3.52,0.948) (3.54,0.941) (3.56,0.963) (3.58,0.998) (3.6,0.981) (3.62,1.253) (3.64,1.451) (3.66,1.418) (3.68,1.417) (3.7,2.162) (3.72,2.103) (3.74,2.137) (3.76,2.134) (3.78,2.125) (3.8,2.107) (3.82,2.198) (3.84,2.099) (3.86,2.063) (3.88,2.018) (3.9,2.538) (3.92,2.564) (3.94,2.869) (3.96,2.755) (3.98,3.022) (4,3.193) (4.02,3.238) (4.04,3.209) (4.06,3.309) (4.08,3.249) (4.1,3.231) (4.12,3.197) (4.14,3.276) (4.16,3.287) (4.18,3.419) (4.2,3.427) (4.22,3.354) (4.24,3.336) (4.26,3.345) (4.28,3.335) (4.3,3.325) (4.32,3.369) (4.34,3.329) (4.36,3.338) (4.38,3.314) (4.4,3.307) (4.42,3.341) (4.44,3.335) (4.46,3.310) (4.48,3.294) (4.5,3.293) (4.52,3.285) (4.54,3.246) (4.56,3.307) (4.58,3.274) (4.6,3.213) (4.62,3.211) (4.64,3.200) (4.66,3.216) (4.68,3.217) (4.7,3.210) (4.72,3.202) (4.74,3.191) (4.76,3.184) (4.78,3.168) (4.8,3.158) (4.82,3.151) (4.84,3.138) (4.86,3.139) (4.88,3.112) (4.9,3.115) (4.92,3.116) (4.94,3.113) (4.96,3.111) (4.98,3.094) (5,3.075) (5.02,3.071) (5.04,3.050) (5.06,3.058)
\end{pspicture}

\vskip 0.4 cm

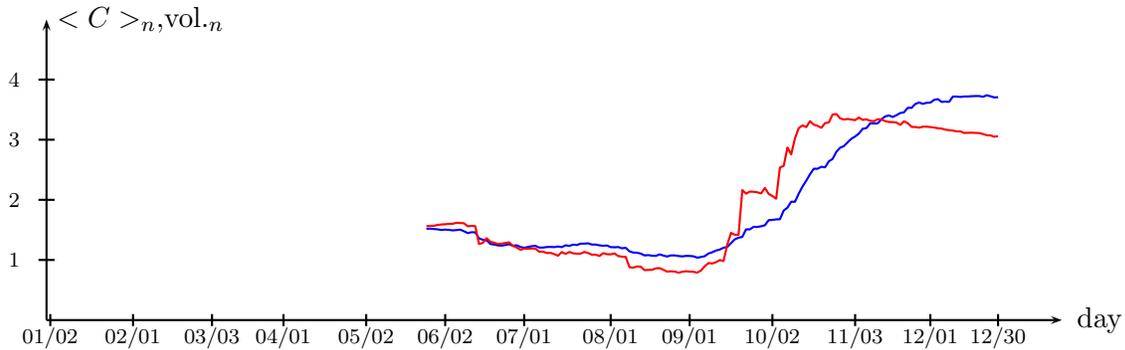
{\noindent \small Figure 29: average volatility (blue) and average correlation (red) based on the log-returns for 2008, both calculated in a moving window of 100 days and normalized so as to have mean two and standard deviation one.}

\vskip 0.2 cm

\vskip 0.6 cm

\begin{pspicture}(-1.5,-0.2)(3.5,4.9)
\psset{xunit=2.5,yunit=7}
\psline{->}(0,0)(5.4,0) \psline{->}(0,0)(0,0.7) \rput(5.6,0){day} \rput(0.3,0.7){covar.} \scriptsize \psline(0.02,-0.014)(0.02,0.014) \rput(0.02,-0.043){01/02} \psline(0.46,-0.014)(0.46,0.014) \rput(0.46,-0.043){02/01} \psline(0.88,-0.014)(0.88,0.014) \rput(0.88,-0.043){03/03} \psline(1.28,-0.014)(1.28,0.014) \rput(1.28,-0.043){04/01} \psline(1.72,-0.014)(1.72,0.014) \rput(1.72,-0.043){05/02} \psline(2.14,-0.014)(2.14,0.014) \rput(2.14,-0.043){06/02} \psline(2.56,-0.014)(2.56,0.014) \rput(2.56,-0.043){07/01} \psline(3.02,-0.014)(3.02,0.014) \rput(3.02,-0.043){08/01} \psline(3.44,-0.014)(3.44,0.014) \rput(3.44,-0.043){09/01} \psline(3.88,-0.014)(3.88,0.014) \rput(3.88,-0.043){10/01} \psline(4.34,-0.014)(4.34,0.014) \rput(4.34,-0.043){11/03} \psline(4.74,-0.014)(4.74,0.014) \rput(4.74,-0.043){12/01} \psline(5.16,-0.014)(5.16,0.014) \rput(5.16,-0.043){12/31} \psline(-0.043,0.2)(0.043,0.2) \rput(-0.17,0.2){$0.2$} \psline(-0.043,0.4)(0.043,0.4) \rput(-0.17,0.4){$0.4$} \psline(-0.043,0.6)(0.043,0.6) \rput(-0.17,0.6){$0.6$}
\psline[linecolor=red](2.62,-0.0073) (2.64,-0.0077) (2.66,-0.0072) (2.68,-0.0087) (2.7,-0.0096) (2.72,-0.0073) (2.74,-0.0093) (2.76,-0.0052) (2.78,-0.0085) (2.8,-0.0061) (2.82,-0.0067) (2.84,-0.0046) (2.86,-0.0060) (2.88,-0.0093) (2.9,-0.0106) (2.92,-0.0088) (2.94,-0.0069) (2.96,-0.0031) (2.98,-0.0011) (3,-0.0021) (3.02,-0.0011) (3.04,-0.0016) (3.06,-0.0010) (3.08,-0.0008) (3.1,-0.0002) (3.12,0.0012) (3.14,0.0035) (3.16,0.0074) (3.18,0.0069) (3.2,0.0066) (3.22,0.0082) (3.24,0.0120) (3.26,0.0143) (3.28,0.0105) (3.3,0.0115) (3.32,0.0126) (3.34,0.0109) (3.36,0.0142) (3.38,0.0113) (3.4,0.0148) (3.42,0.0106) (3.44,0.0114) (3.46,0.0102) (3.48,0.0113) (3.5,0.0069) (3.52,0.0065) (3.54,0.0074) (3.56,0.0095) (3.58,0.0129) (3.6,0.0106) (3.62,0.0152) (3.64,0.0130) (3.66,0.0443) (3.68,0.0712) (3.7,0.0792) (3.72,0.0837) (3.74,0.2230) (3.76,0.1952) (3.78,0.2116) (3.8,0.1861) (3.82,0.1681) (3.84,0.1587) (3.86,0.2037) (3.88,0.1879) (3.9,0.1711) (3.92,0.1357) (3.94,0.2648) (3.96,0.2769) (3.98,0.3542) (4,0.2776) (4.02,0.3952) (4.04,0.4889) (4.06,0.5231) (4.08,0.5382) (4.1,0.5488) (4.12,0.4623) (4.14,0.4384) (4.16,0.3656) (4.18,0.3736) (4.2,0.3217) (4.22,0.3342) (4.24,0.2818) (4.26,0.2821) (4.28,0.3068) (4.3,0.2724) (4.32,0.2324) (4.34,0.2912) (4.36,0.2523) (4.38,0.2184) (4.4,0.1753) (4.42,0.1106) (4.44,0.0640) (4.46,0.0928) (4.48,0.0278) (4.5,-0.0317) (4.52,-0.0765) (4.54,-0.0178) (4.56,-0.0199) (4.58,0.0112) (4.6,-0.0050) (4.62,0.0078) (4.64,0.0114) (4.66,0.0184) (4.68,0.0222) (4.7,0.0257) (4.72,0.0253) (4.74,0.0231) (4.76,0.0261) (4.78,0.0316) (4.8,0.0355) (4.82,0.0248) (4.84,0.0153) (4.86,0.0181) (4.88,0.0180) (4.9,0.0155) (4.92,0.0190) (4.94,0.0199) (4.96,0.0181) (4.98,0.0236) (5,0.0187) (5.02,0.0216) (5.04,0.0266) (5.06,0.0232)
\end{pspicture}

\vskip 0.4 cm

\hskip 2 cm {\noindent \small Figure 30: covariance between volatility and average correlation as a function of time.}

\vskip 0.4 cm

\section{Normality tests for the correlation matrix}

In this section, we make tests in order to check wether the elements of the correlation matrix exhibit a normal or close to normal probability distribution or not. A first analysis of the data might lead us to believe it does. Observe the following graphics, with the skewness and kurtosis of the probability distribution obtained by considering the elements of correlations matrix (except those of the diagonal) calculated over moving windows (the size of the windows vary for each of the years that were considered).

For 1987 (figures 31 and 32), the size of the running window is 30 days. Note that, near the Black Monday, which occurred in October, kurtosis dropped substantially, what seems to imply that the distribution of the coefficients of the correlation matrix approach that of a normal curve.

\begin{pspicture}(-2,-0.7)(3.5,4)
\psset{xunit=2.5,yunit=3}
\psline{->}(0,0)(5.4,0) \psline{->}(0,-0.2)(0,1.2) \rput(5.6,0){day} \rput(0.4,1.2){Skewness}\scriptsize \psline(0.02,-0.033)(0.02,0.033) \rput(0.02,-0.1){01/02} \psline(0.44,-0.033)(0.44,0.033) \rput(0.44,-0.1){02/02} \psline(0.84,-0.033)(0.84,0.033) \rput(0.84,-0.1){03/02} \psline(1.28,-0.033)(1.28,0.033) \rput(1.28,-0.1){04/01} \psline(1.7,-0.033)(1.7,0.033) \rput(1.7,-0.1){05/01} \psline(2.12,-0.033)(2.12,0.033) \rput(2.12,-0.1){06/01} \psline(2.56,-0.033)(2.56,0.033) \rput(2.56,-0.1){07/01} \psline(3.02,-0.033)(3.02,0.033) \rput(3.02,-0.1){08/03} \psline(3.44,-0.033)(3.44,0.033) \rput(3.44,-0.1){09/01} \psline(3.88,-0.033)(3.88,0.033) \rput(3.88,-0.1){10/01} \psline(4.32,-0.033)(4.32,0.033) \rput(4.32,-0.1){11/02} \psline(4.74,-0.033)(4.74,0.033) \rput(4.74,-0.1){12/01} \psline(5.16,-0.033)(5.16,0.033) \rput(5.16,-0.1){12/31} \psline(-0.043,0.2)(0.043,0.2) \rput(-0.17,0.2){$0.2$} \psline(-0.043,0.4)(0.043,0.4) \rput(-0.17,0.4){$0.4$} \psline(-0.043,0.6)(0.043,0.6) \rput(-0.17,0.6){$0.6$} \psline(-0.043,0.8)(0.043,0.8) \rput(-0.17,0.8){$0.8$} \psline(-0.043,1)(0.043,1) \rput(-0.17,1){$1$}
\psline[linecolor=blue](0.62,0.515) (0.64,0.525) (0.66,0.553) (0.68,0.592) (0.7,0.646) (0.72,0.655) (0.74,0.784) (0.76,0.877) (0.78,0.843) (0.8,0.872) (0.82,0.814) (0.84,0.714) (0.86,0.686) (0.88,0.707) (0.9,0.633) (0.92,0.617) (0.94,0.734) (0.96,0.698) (0.98,0.641) (1,0.539) (1.02,0.493) (1.04,0.466) (1.06,0.437) (1.08,0.446) (1.1,0.353) (1.12,0.386) (1.14,0.419) (1.16,0.367) (1.18,0.364) (1.2,0.289) (1.22,0.304) (1.24,0.481) (1.26,0.487) (1.28,0.272) (1.3,0.281) (1.32,0.279) (1.34,0.281) (1.36,0.202) (1.38,0.222) (1.4,0.275) (1.42,0.248) (1.44,0.184) (1.46,0.307) (1.48,0.292) (1.5,0.297) (1.52,0.315) (1.54,0.293) (1.56,0.305) (1.58,0.395) (1.6,0.469) (1.62,0.377) (1.64,0.373) (1.66,0.410) (1.68,0.397) (1.7,0.371) (1.72,0.358) (1.74,0.329) (1.76,0.362) (1.78,0.356) (1.8,0.339) (1.82,0.344) (1.84,0.307) (1.86,0.192) (1.88,0.248) (1.9,0.202) (1.92,0.240) (1.94,0.199) (1.96,0.157) (1.98,0.208) (2,0.275) (2.02,0.267) (2.04,0.192) (2.06,0.210) (2.08,0.223) (2.1,0.207) (2.12,0.287) (2.14,0.241) (2.16,0.249) (2.18,0.300) (2.2,0.323) (2.22,0.332) (2.24,0.230) (2.26,0.196) (2.28,0.192) (2.3,0.138) (2.32,0.012) (2.34,-0.095) (2.36,-0.029) (2.38,-0.027) (2.4,-0.018) (2.42,0.108) (2.44,0.051) (2.46,0.134) (2.48,0.210) (2.5,0.225) (2.52,0.383) (2.54,0.347) (2.56,0.325) (2.58,0.377) (2.6,0.260) (2.62,0.249) (2.64,0.164) (2.66,0.315) (2.68,0.230) (2.7,0.287) (2.72,0.279) (2.74,0.241) (2.76,0.229) (2.78,0.276) (2.8,0.353) (2.82,0.242) (2.84,0.219) (2.86,0.190) (2.88,0.146) (2.9,0.105) (2.92,0.124) (2.94,0.166) (2.96,0.177) (2.98,0.157) (3,0.206) (3.02,0.136) (3.04,0.139) (3.06,0.099) (3.08,0.171) (3.1,0.082) (3.12,0.094) (3.14,0.094) (3.16,0.075) (3.18,-0.019) (3.2,0.038) (3.22,0.044) (3.24,0.059) (3.26,0.114) (3.28,0.099) (3.3,0.151) (3.32,0.158) (3.34,0.212) (3.36,0.238) (3.38,0.263) (3.4,0.200) (3.42,0.220) (3.44,0.404) (3.46,0.441) (3.48,0.481) (3.5,0.525) (3.52,0.499) (3.54,0.478) (3.56,0.494) (3.58,0.458) (3.6,0.408) (3.62,0.356) (3.64,0.440) (3.66,0.508) (3.68,0.456) (3.7,0.405) (3.72,0.323) (3.74,0.329) (3.76,0.321) (3.78,0.263) (3.8,0.307) (3.82,0.328) (3.84,0.348) (3.86,0.464) (3.88,0.384) (3.9,0.422) (3.92,0.479) (3.94,0.398) (3.96,0.488) (3.98,0.503) (4,0.384) (4.02,0.369) (4.04,0.378) (4.06,0.321) (4.08,0.348) (4.1,0.276) (4.12,0.302) (4.14,0.349) (4.16,0.427) (4.18,0.444) (4.2,0.406) (4.22,0.440) (4.24,0.409) (4.26,0.396) (4.28,0.388) (4.3,0.383) (4.32,0.365) (4.34,0.341) (4.36,0.325) (4.38,0.315) (4.4,0.329) (4.42,0.255) (4.44,0.267) (4.46,0.294) (4.48,0.271) (4.5,0.267) (4.52,0.287) (4.54,0.322) (4.56,0.334) (4.58,0.323) (4.6,0.345) (4.62,0.123) (4.64,0.133) (4.66,0.081) (4.68,0.075) (4.7,0.094) (4.72,0.135) (4.74,0.474) (4.76,0.371) (4.78,0.295) (4.8,0.338) (4.82,0.450) (4.84,0.452) (4.86,0.453) (4.88,0.432) (4.9,0.421) (4.92,0.476) (4.94,0.432) (4.96,0.410) (4.98,0.375) (5,0.369) (5.02,0.449) (5.04,0.541) (5.06,0.507) (5.08,0.460) (5.1,0.428) (5.12,0.362)
\end{pspicture}

\vskip -0.1 cm

{\noindent \small Figure 31: skewness of the correlation matrix during the year 1987 for correlation matrices calculated over a running window of 30 days.}

\vskip 0.3 cm

\begin{pspicture}(-2,-0.3)(3.5,4.5)
\psset{xunit=2.5,yunit=0.6}
\psline{->}(0,0)(5.4,0) \psline{->}(0,0)(0,7) \rput(5.6,0){day} \rput(0.4,7){Kurtosis}\scriptsize \psline(0.02,-0.17)(0.02,0.17) \rput(0.02,-0.5){01/02} \psline(0.44,-0.17)(0.44,0.17) \rput(0.44,-0.5){02/02} \psline(0.84,-0.17)(0.84,0.17) \rput(0.84,-0.5){03/02} \psline(1.28,-0.17)(1.28,0.17) \rput(1.28,-0.5){04/01} \psline(1.7,-0.17)(1.7,0.17) \rput(1.7,-0.5){05/01} \psline(2.12,-0.17)(2.12,0.17) \rput(2.12,-0.5){06/01} \psline(2.56,-0.17)(2.56,0.17) \rput(2.56,-0.5){07/01} \psline(3.02,-0.17)(3.02,0.17) \rput(3.02,-0.5){08/03} \psline(3.44,-0.17)(3.44,0.17) \rput(3.44,-0.5){09/01} \psline(3.88,-0.17)(3.88,0.17) \rput(3.88,-0.5){10/01} \psline(4.32,-0.17)(4.32,0.17) \rput(4.32,-0.5){11/02} \psline(4.74,-0.17)(4.74,0.17) \rput(4.74,-0.5){12/01} \psline(5.16,-0.17)(5.16,0.17) \rput(5.16,-0.5){12/31} \psline(-0.043,1)(0.043,1) \rput(-0.1,1){$1$} \psline(-0.043,2)(0.043,2) \rput(-0.1,2){$2$} \psline(-0.043,3)(0.043,3) \rput(-0.1,3){$3$} \psline(-0.043,4)(0.043,4) \rput(-0.1,4){$4$} \psline(-0.043,5)(0.043,5) \rput(-0.1,5){$5$} \psline(-0.043,6)(0.043,6) \rput(-0.1,6){$6$}
\psline[linecolor=gray](0,3)(5.2,3)
\psline[linecolor=red](0.62,3.585) (0.64,3.682) (0.66,3.566) (0.68,3.702) (0.7,3.858) (0.72,3.908) (0.74,4.239) (0.76,4.382) (0.78,4.194) (0.8,4.528) (0.82,4.317) (0.84,4.347) (0.86,4.250) (0.88,4.323) (0.9,4.096) (0.92,4.089) (0.94,4.351) (0.96,4.318) (0.98,4.269) (1,3.975) (1.02,3.969) (1.04,4.081) (1.06,3.913) (1.08,4.305) (1.1,4.702) (1.12,4.799) (1.14,4.433) (1.16,4.461) (1.18,4.411) (1.2,4.379) (1.22,4.412) (1.24,4.672) (1.26,4.493) (1.28,4.393) (1.3,4.265) (1.32,4.217) (1.34,4.004) (1.36,4.303) (1.38,4.272) (1.4,4.477) (1.42,4.320) (1.44,3.978) (1.46,3.896) (1.48,3.875) (1.5,3.739) (1.52,3.711) (1.54,3.639) (1.56,3.688) (1.58,3.473) (1.6,3.796) (1.62,3.466) (1.64,3.478) (1.66,3.485) (1.68,3.257) (1.7,3.261) (1.72,3.079) (1.74,2.924) (1.76,3.000) (1.78,3.047) (1.8,3.033) (1.82,3.043) (1.84,2.978) (1.86,3.039) (1.88,3.127) (1.9,3.094) (1.92,3.301) (1.94,3.030) (1.96,3.050) (1.98,3.088) (2,3.021) (2.02,3.043) (2.04,2.981) (2.06,3.040) (2.08,2.982) (2.1,3.042) (2.12,3.179) (2.14,3.168) (2.16,3.080) (2.18,3.092) (2.2,3.059) (2.22,2.964) (2.24,3.255) (2.26,3.215) (2.28,3.144) (2.3,3.206) (2.32,3.203) (2.34,3.236) (2.36,3.297) (2.38,3.345) (2.4,3.305) (2.42,3.279) (2.44,3.428) (2.46,3.367) (2.48,3.209) (2.5,3.292) (2.52,3.706) (2.54,3.352) (2.56,3.363) (2.58,3.380) (2.6,3.436) (2.62,3.324) (2.64,3.157) (2.66,3.236) (2.68,2.959) (2.7,3.010) (2.72,3.137) (2.74,3.033) (2.76,3.126) (2.78,3.294) (2.8,3.399) (2.82,3.152) (2.84,3.116) (2.86,3.119) (2.88,3.204) (2.9,3.155) (2.92,3.003) (2.94,2.923) (2.96,2.994) (2.98,3.005) (3,2.975) (3.02,2.878) (3.04,2.818) (3.06,2.885) (3.08,2.939) (3.1,2.922) (3.12,2.949) (3.14,2.960) (3.16,2.813) (3.18,3.080) (3.2,2.966) (3.22,2.947) (3.24,3.034) (3.26,3.040) (3.28,3.019) (3.3,3.170) (3.32,3.217) (3.34,3.081) (3.36,3.037) (3.38,3.043) (3.4,3.206) (3.42,3.369) (3.44,3.432) (3.46,3.616) (3.48,3.603) (3.5,3.571) (3.52,3.703) (3.54,3.607) (3.56,3.871) (3.58,3.982) (3.6,4.160) (3.62,3.956) (3.64,4.055) (3.66,4.216) (3.68,4.388) (3.7,4.550) (3.72,4.419) (3.74,4.067) (3.76,3.962) (3.78,4.146) (3.8,3.950) (3.82,3.966) (3.84,3.834) (3.86,3.832) (3.88,3.437) (3.9,3.734) (3.92,3.785) (3.94,3.661) (3.96,3.949) (3.98,3.896) (4,3.568) (4.02,3.595) (4.04,3.416) (4.06,3.278) (4.08,3.435) (4.1,1.969) (4.12,1.969) (4.14,1.973) (4.16,2.184) (4.18,2.217) (4.2,2.271) (4.22,2.339) (4.24,2.334) (4.26,2.422) (4.28,2.470) (4.3,2.386) (4.32,2.364) (4.34,2.382) (4.36,2.400) (4.38,2.386) (4.4,2.365) (4.42,2.342) (4.44,2.381) (4.46,2.351) (4.48,2.357) (4.5,2.310) (4.52,2.290) (4.54,2.297) (4.56,2.296) (4.58,2.309) (4.6,2.327) (4.62,2.315) (4.64,2.318) (4.66,2.434) (4.68,2.430) (4.7,2.422) (4.72,2.457) (4.74,2.641) (4.76,2.476) (4.78,2.315) (4.8,2.374) (4.82,2.569) (4.84,2.482) (4.86,2.493) (4.88,2.507) (4.9,2.505) (4.92,2.513) (4.94,2.472) (4.96,2.555) (4.98,2.544) (5,2.508) (5.02,2.465) (5.04,2.723) (5.06,2.638) (5.08,2.734) (5.1,2.637) (5.12,2.564)
\end{pspicture}

\vskip 0.3 cm

{\noindent \small Figure 32: kurtosis of the correlation matrix during the year 1987 for correlation matrices calculated over a running window of 30 days.}

\vskip 0.3 cm

For 1998, the running window has 70 days. The skewness (figure 33) remains nearly constant for most of the time, and the kurtosis (figure 34) of the same distribution stays near 3 during the same period.

\begin{pspicture}(-2,-0.4)(3.5,3.6)
\psset{xunit=2.5,yunit=3}
\psline{->}(0,0)(5.6,0) \psline{->}(0,0)(0,1) \rput(5.8,0){day} \rput(0.4,1){Skewness} \scriptsize \psline(0.02,-0.033)(0.02,0.033) \rput(0.02,-0.1){01/02} \psline(0.44,-0.033)(0.44,0.033) \rput(0.44,-0.1){02/02} \psline(0.84,-0.033)(0.84,0.033) \rput(0.84,-0.1){03/02} \psline(1.28,-0.033)(1.28,0.033) \rput(1.28,-0.1){04/01} \psline(1.7,-0.033)(1.7,0.033) \rput(1.7,-0.1){05/04} \psline(2.1,-0.033)(2.1,0.033) \rput(2.1,-0.1){06/01} \psline(2.54,-0.033)(2.54,0.033) \rput(2.54,-0.1){07/01} \psline(3,-0.033)(3,0.033) \rput(3,-0.1){08/03} \psline(3.42,-0.033)(3.42,0.033) \rput(3.42,-0.1){09/01} \psline(3.86,-0.033)(3.86,0.033) \rput(3.86,-0.1){10/01} \psline(4.3,-0.033)(4.3,0.033) \rput(4.3,-0.1){11/02} \psline(4.72,-0.033)(4.72,0.033) \rput(4.72,-0.1){12/01} \psline(5.14,-0.033)(5.14,0.033) \rput(5.13,-0.1){12/31} \psline(-0.04,0.2)(0.04,0.2) \rput(-0.15,0.2){$0.2$} \psline(-0.04,0.4)(0.04,0.4) \rput(-0.15,0.4){$0.4$} \psline(-0.04,0.6)(0.04,0.6) \rput(-0.15,0.6){$0.6$} \psline(-0.04,0.8)(0.04,0.8) \rput(-0.15,0.8){$0.8$} %
\psline[linecolor=blue](1.42,0.412) (1.44,0.395) (1.46,0.362) (1.48,0.391) (1.5,0.417) (1.52,0.447) (1.54,0.499) (1.56,0.390) (1.58,0.417) (1.6,0.450) (1.62,0.540) (1.64,0.588) (1.66,0.622) (1.68,0.648) (1.7,0.684) (1.72,0.715) (1.74,0.671) (1.76,0.673) (1.78,0.668) (1.8,0.680) (1.82,0.690) (1.84,0.669) (1.86,0.780) (1.88,0.751) (1.9,0.715) (1.92,0.728) (1.94,0.738) (1.96,0.727) (1.98,0.738) (2,0.735) (2.02,0.724) (2.04,0.629) (2.06,0.632) (2.08,0.635) (2.1,0.560) (2.12,0.581) (2.14,0.609) (2.16,0.621) (2.18,0.634) (2.2,0.632) (2.22,0.638) (2.24,0.593) (2.26,0.580) (2.28,0.608) (2.3,0.558) (2.32,0.606) (2.34,0.555) (2.36,0.613) (2.38,0.602) (2.4,0.621) (2.42,0.618) (2.44,0.638) (2.46,0.660) (2.48,0.656) (2.5,0.667) (2.52,0.678) (2.54,0.678) (2.56,0.681) (2.58,0.672) (2.6,0.681) (2.62,0.669) (2.64,0.678) (2.66,0.678) (2.68,0.678) (2.7,0.678) (2.72,0.621) (2.74,0.589) (2.76,0.577) (2.78,0.578) (2.8,0.568) (2.82,0.581) (2.84,0.564) (2.86,0.553) (2.88,0.566) (2.9,0.565) (2.92,0.570) (2.94,0.575) (2.96,0.578) (2.98,0.596) (3,0.566) (3.02,0.565) (3.04,0.546) (3.06,0.539) (3.08,0.594) (3.1,0.615) (3.12,0.520) (3.14,0.553) (3.16,0.549) (3.18,0.538) (3.2,0.539) (3.22,0.557) (3.24,0.555) (3.26,0.589) (3.28,0.568) (3.3,0.555) (3.32,0.583) (3.34,0.591) (3.36,0.560) (3.38,0.538) (3.4,0.551) (3.42,0.601) (3.44,0.583) (3.46,0.591) (3.48,0.604) (3.5,0.497) (3.52,0.540) (3.54,0.539) (3.56,0.586) (3.58,0.600) (3.6,0.610) (3.62,0.655) (3.64,0.682) (3.66,0.719) (3.68,0.729) (3.7,0.743) (3.72,0.749) (3.74,0.727) (3.76,0.752) (3.78,0.724) (3.8,0.719) (3.82,0.719) (3.84,0.713) (3.86,0.704) (3.88,0.698) (3.9,0.681) (3.92,0.676) (3.94,0.692) (3.96,0.728) (3.98,0.726) (4,0.721) (4.02,0.720) (4.04,0.715) (4.06,0.704) (4.08,0.704) (4.1,0.701) (4.12,0.693) (4.14,0.709) (4.16,0.717) (4.18,0.711) (4.2,0.710) (4.22,0.706) (4.24,0.707) (4.26,0.710) (4.28,0.699) (4.3,0.686) (4.32,0.693) (4.34,0.695) (4.36,0.718) (4.38,0.711) (4.4,0.712) (4.42,0.717) (4.44,0.722) (4.46,0.724) (4.48,0.724) (4.5,0.694) (4.52,0.703) (4.54,0.725) (4.56,0.708) (4.58,0.704) (4.6,0.687) (4.62,0.691) (4.64,0.668) (4.66,0.664) (4.68,0.665) (4.7,0.659) (4.72,0.687) (4.74,0.683) (4.76,0.668) (4.78,0.691) (4.8,0.706) (4.82,0.712) (4.84,0.715) (4.86,0.718) (4.88,0.720) (4.9,0.688) (4.92,0.732) (4.94,0.728) (4.96,0.726) (4.98,0.754) (5,0.750) (5.02,0.745) (5.04,0.715) (5.06,0.707) (5.08,0.730) (5.1,0.732) (5.12,0.758) (5.14,0.764)
\end{pspicture}

\vskip 0.1 cm

{\noindent \small Figure 33: skewness of the correlation matrix during the year 1998 for correlation matrices calculated over a running window of 70 days.}

\vskip 0.3 cm

\begin{pspicture}(-2,-0.3)(3.5,4.2)
\psset{xunit=2.5,yunit=1}
\psline{->}(-0.2,0)(5.4,0) \psline{->}(-0.2,0)(-0.2,4) \rput(5.6,0){day} \rput(0.2,4){Kurtosis}\scriptsize \psline(0.02,-0.1)(0.02,0.1) \rput(0.02,-0.3){01/02} \psline(0.44,-0.1)(0.44,0.1) \rput(0.44,-0.3){02/02} \psline(0.84,-0.1)(0.84,0.1) \rput(0.84,-0.3){03/02} \psline(1.28,-0.1)(1.28,0.1) \rput(1.28,-0.3){04/01} \psline(1.7,-0.1)(1.7,0.1) \rput(1.7,-0.3){05/04} \psline(2.1,-0.1)(2.1,0.1) \rput(2.1,-0.3){06/01} \psline(2.54,-0.1)(2.54,0.1) \rput(2.54,-0.3){07/01} \psline(3,-0.1)(3,0.1) \rput(3,-0.3){08/03} \psline(3.42,-0.1)(3.42,0.1) \rput(3.42,-0.3){09/01} \psline(3.86,-0.1)(3.86,0.1) \rput(3.86,-0.3){10/01} \psline(4.3,-0.1)(4.3,0.1) \rput(4.3,-0.3){11/02} \psline(4.72,-0.1)(4.72,0.1) \rput(4.72,-0.3){12/01} \psline(5.14,-0.1)(5.14,0.1) \rput(5.13,-0.3){12/31} \psline(-0.243,1)(-0.157,1) \rput(-0.4,1){$1$} \psline(-0.243,2)(-0.157,2) \rput(-0.4,2){$2$} \psline(-0.243,3)(-0.157,3) \rput(-0.4,3){$3$}
\psline[linecolor=gray](-0.2,3)(5.3,3)
\psline[linecolor=red](1.42,2.677) (1.44,2.673) (1.46,2.633) (1.48,2.587) (1.5,2.621) (1.52,2.714) (1.54,2.923) (1.56,2.989) (1.58,3.104) (1.6,3.112) (1.62,3.042) (1.64,3.214) (1.66,3.321) (1.68,3.312) (1.7,3.351) (1.72,3.475) (1.74,3.408) (1.76,3.365) (1.78,3.355) (1.8,3.425) (1.82,3.418) (1.84,3.327) (1.86,3.664) (1.88,3.606) (1.9,3.537) (1.92,3.572) (1.94,3.594) (1.96,3.573) (1.98,3.571) (2,3.651) (2.02,3.696) (2.04,3.567) (2.06,3.574) (2.08,3.598) (2.1,3.410) (2.12,3.396) (2.14,3.454) (2.16,3.487) (2.18,3.512) (2.2,3.526) (2.22,3.542) (2.24,3.431) (2.26,3.374) (2.28,3.452) (2.3,3.223) (2.32,3.317) (2.34,3.194) (2.36,3.294) (2.38,3.333) (2.4,3.378) (2.42,3.376) (2.44,3.355) (2.46,3.446) (2.48,3.472) (2.5,3.473) (2.52,3.485) (2.54,3.435) (2.56,3.453) (2.58,3.491) (2.6,3.500) (2.62,3.441) (2.64,3.445) (2.66,3.457) (2.68,3.448) (2.7,3.442) (2.72,3.385) (2.74,3.326) (2.76,3.335) (2.78,3.328) (2.8,3.323) (2.82,3.337) (2.84,3.282) (2.86,3.252) (2.88,3.278) (2.9,3.250) (2.92,3.243) (2.94,3.233) (2.96,3.221) (2.98,3.253) (3,3.154) (3.02,3.163) (3.04,3.158) (3.06,3.118) (3.08,3.241) (3.1,3.321) (3.12,2.984) (3.14,3.023) (3.16,3.003) (3.18,2.977) (3.2,3.007) (3.22,2.987) (3.24,2.963) (3.26,2.980) (3.28,2.856) (3.3,2.861) (3.32,2.907) (3.34,2.856) (3.36,2.543) (3.38,2.542) (3.4,2.582) (3.42,2.698) (3.44,2.600) (3.46,2.615) (3.48,2.659) (3.5,2.551) (3.52,2.614) (3.54,2.618) (3.56,2.608) (3.58,2.638) (3.6,2.618) (3.62,2.754) (3.64,2.745) (3.66,2.760) (3.68,2.819) (3.7,2.848) (3.72,2.790) (3.74,2.680) (3.76,2.751) (3.78,2.714) (3.8,2.694) (3.82,2.695) (3.84,2.699) (3.86,2.609) (3.88,2.619) (3.9,2.599) (3.92,2.639) (3.94,2.661) (3.96,2.752) (3.98,2.738) (4,2.701) (4.02,2.702) (4.04,2.716) (4.06,2.699) (4.08,2.684) (4.1,2.686) (4.12,2.689) (4.14,2.730) (4.16,2.737) (4.18,2.723) (4.2,2.727) (4.22,2.732) (4.24,2.742) (4.26,2.741) (4.28,2.725) (4.3,2.700) (4.32,2.722) (4.34,2.728) (4.36,2.754) (4.38,2.765) (4.4,2.768) (4.42,2.782) (4.44,2.786) (4.46,2.793) (4.48,2.798) (4.5,2.737) (4.52,2.750) (4.54,2.819) (4.56,2.788) (4.58,2.787) (4.6,2.765) (4.62,2.769) (4.64,2.738) (4.66,2.741) (4.68,2.739) (4.7,2.769) (4.72,2.790) (4.74,2.787) (4.76,2.837) (4.78,3.010) (4.8,3.055) (4.82,3.096) (4.84,3.080) (4.86,3.128) (4.88,3.137) (4.9,3.098) (4.92,3.155) (4.94,3.149) (4.96,3.133) (4.98,3.259) (5,3.232) (5.02,3.248) (5.04,3.138) (5.06,3.110) (5.08,3.251) (5.1,3.190) (5.12,3.243) (5.14,3.267)
\end{pspicture}

\vskip 0.3 cm

{\noindent \small Figure 34: kurtosis of the correlation matrix during the year 1998 for correlation matrices calculated over a running window of 70 days.}

\vskip 0.3 cm

The next two graphics show the skewness (figure 35) and the kurtosis (figure 36) for the elements of the correlation matrix, except its diagonal, calculated in a running window of 80 days for 2001. Both skewness and kurtosis present a peak in September 11, but otherwise remain nearly constant throughout the period.

\begin{pspicture}(-2,-0.5)(3.5,3.2)
\psset{xunit=2.5,yunit=1.7}
\psline{->}(0,0)(5.4,0) \psline{->}(0,-0.2)(0,1.6) \rput(5.6,0){day} \rput(0.4,1.6){Skewness}\scriptsize \psline(0.02,-0.0588)(0.02,0.0588) \rput(0.02,-0.176){01/02} \psline(0.46,-0.0588)(0.46,0.0588) \rput(0.46,-0.176){02/01} \psline(0.86,-0.0588)(0.86,0.0588) \rput(0.86,-0.176){03/01} \psline(1.3,-0.0588)(1.3,0.0588) \rput(1.3,-0.176){04/02} \psline(1.72,-0.0588)(1.72,0.0588) \rput(1.72,-0.176){05/01} \psline(2.18,-0.0588)(2.18,0.0588) \rput(2.18,-0.176){06/01} \psline(2.6,-0.0588)(2.6,0.0588) \rput(2.6,-0.176){07/02} \psline(3.04,-0.0588)(3.04,0.0588) \rput(3.04,-0.176){08/01} \psline(3.5,-0.0588)(3.5,0.0588) \rput(3.5,-0.176){09/03} \psline(3.9,-0.0588)(3.9,0.0588) \rput(3.9,-0.176){10/01} \psline(4.36,-0.0588)(4.36,0.0588) \rput(4.36,-0.176){11/01} \psline(4.8,-0.0588)(4.8,0.0588) \rput(4.8,-0.176){12/03} \psline(5.2,-0.0588)(5.2,0.0588) \rput(5.2,-0.176){12/31} \psline(-0.043,0.4)(0.043,0.4) \rput(-0.17,0.4){$0.4$} \psline(-0.043,0.8)(0.043,0.8) \rput(-0.17,0.8){$0.8$} \psline(-0.043,1.2)(0.043,1.2) \rput(-0.17,1.2){$1.2$} %
\psline[linecolor=blue](1.62,0.992) (1.64,0.994) (1.66,0.990) (1.68,1.044) (1.7,1.025) (1.72,1.023) (1.74,1.030) (1.76,1.022) (1.78,1.013) (1.8,1.026) (1.82,1.030) (1.84,1.025) (1.86,1.032) (1.88,1.042) (1.9,1.038) (1.92,1.053) (1.94,1.061) (1.96,1.026) (1.98,1.026) (2,1.029) (2.02,1.014) (2.04,1.020) (2.06,1.013) (2.08,1.006) (2.1,0.997) (2.12,1.015) (2.14,0.993) (2.16,1.006) (2.18,1.003) (2.2,1.003) (2.22,1.014) (2.24,1.013) (2.26,1.025) (2.28,1.023) (2.3,1.024) (2.32,1.025) (2.34,1.024) (2.36,1.032) (2.38,1.027) (2.4,0.998) (2.42,1.010) (2.44,0.995) (2.46,0.988) (2.48,0.970) (2.5,0.975) (2.52,0.972) (2.54,0.971) (2.56,0.974) (2.58,0.951) (2.6,0.940) (2.62,0.934) (2.64,0.948) (2.66,0.959) (2.68,0.947) (2.7,0.933) (2.72,0.943) (2.74,0.938) (2.76,0.927) (2.78,0.897) (2.8,0.857) (2.82,0.815) (2.84,0.796) (2.86,0.798) (2.88,0.845) (2.9,0.880) (2.92,0.877) (2.94,0.856) (2.96,0.878) (2.98,0.896) (3,0.891) (3.02,0.864) (3.04,0.865) (3.06,0.856) (3.08,0.884) (3.1,0.886) (3.12,0.879) (3.14,0.888) (3.16,0.935) (3.18,0.950) (3.2,0.975) (3.22,0.989) (3.24,0.998) (3.26,1.017) (3.28,1.039) (3.3,1.014) (3.32,1.024) (3.34,1.026) (3.36,1.021) (3.38,1.003) (3.4,0.979) (3.42,0.998) (3.44,1.004) (3.46,1.017) (3.48,1.019) (3.5,1.019) (3.52,1.048) (3.54,1.054) (3.56,1.075) (3.58,1.109) (3.6,1.088) (3.62,1.304) (3.64,1.276) (3.66,1.281) (3.68,1.213) (3.7,1.137) (3.72,1.156) (3.74,1.158) (3.76,1.170) (3.78,1.098) (3.8,1.108) (3.82,1.099) (3.84,1.079) (3.86,1.079) (3.88,1.048) (3.9,1.058) (3.92,1.072) (3.94,1.071) (3.96,1.063) (3.98,1.066) (4,1.067) (4.02,1.064) (4.04,1.047) (4.06,0.990) (4.08,0.994) (4.1,0.992) (4.12,0.964) (4.14,0.957) (4.16,0.953) (4.18,0.950) (4.2,0.955) (4.22,0.953) (4.24,0.960) (4.26,0.975) (4.28,0.975) (4.3,0.966) (4.32,0.956) (4.34,0.956) (4.36,0.963) (4.38,0.963) (4.4,0.974) (4.42,0.966) (4.44,0.964) (4.46,0.965) (4.48,0.968) (4.5,0.977) (4.52,0.980) (4.54,0.978) (4.56,0.980) (4.58,0.975) (4.6,0.971) (4.62,0.966) (4.64,0.969) (4.66,0.969) (4.68,0.970) (4.7,0.950) (4.72,0.959) (4.74,0.952) (4.76,0.946) (4.78,0.947) (4.8,0.945) (4.82,0.938) (4.84,0.927) (4.86,0.928) (4.88,0.932) (4.9,0.929) (4.92,0.925) (4.94,0.926) (4.96,0.928) (4.98,0.928) (5,0.919) (5.02,0.912) (5.04,0.918) (5.06,0.919) (5.08,0.945) (5.1,0.935) (5.12,0.925) (5.14,0.919) (5.16,0.916) (5.18,0.919) (5.2,0.914)
\end{pspicture}

\vskip 0.1 cm

{\noindent \small Figure 35: skewness of the correlation matrix during the year 2001 for correlation matrices calculated over a running window of 80 days.}

\vskip 0.3 cm

\begin{pspicture}(-2,-0.2)(3.5,3.7)
\psset{xunit=2.5,yunit=0.45}
\psline{->}(0,0)(5.4,0) \psline{->}(0,0)(0,8) \rput(5.6,0){day} \rput(0.4,8){Kurtosis}\scriptsize \psline(0.02,-0.22)(0.02,0.22) \rput(0.02,-0.7){01/02} \psline(0.46,-0.22)(0.46,0.22) \rput(0.46,-0.7){02/01} \psline(0.86,-0.22)(0.86,0.22) \rput(0.86,-0.7){03/01} \psline(1.3,-0.22)(1.3,0.22) \rput(1.3,-0.7){04/02} \psline(1.72,-0.22)(1.72,0.22) \rput(1.72,-0.7){05/01} \psline(2.18,-0.22)(2.18,0.22) \rput(2.18,-0.7){06/01} \psline(2.6,-0.22)(2.6,0.22) \rput(2.6,-0.7){07/02} \psline(3.04,-0.22)(3.04,0.22) \rput(3.04,-0.7){08/01} \psline(3.5,-0.22)(3.5,0.22) \rput(3.5,-0.7){09/03} \psline(3.9,-0.22)(3.9,0.22) \rput(3.9,-0.7){10/01} \psline(4.36,-0.22)(4.36,0.22) \rput(4.36,-0.7){11/01} \psline(4.8,-0.22)(4.8,0.22) \rput(4.8,-0.7){12/03} \psline(5.2,-0.22)(5.2,0.22) \rput(5.2,-0.7){12/31} \psline(-0.043,2)(0.043,2) \rput(-0.15,2){$2$} \psline(-0.043,4)(0.043,4) \rput(-0.15,4){$4$} \psline(-0.043,6)(0.043,6) \rput(-0.15,6){$6$}
\psline[linecolor=gray](0,3)(5.2,3)
\psline[linecolor=red](1.62,4.312) (1.64,4.296) (1.66,4.169) (1.68,4.321) (1.7,4.263) (1.72,4.249) (1.74,4.239) (1.76,4.225) (1.78,4.199) (1.8,4.232) (1.82,4.223) (1.84,4.225) (1.86,4.285) (1.88,4.297) (1.9,4.266) (1.92,4.252) (1.94,4.261) (1.96,4.159) (1.98,4.152) (2,4.130) (2.02,4.101) (2.04,4.104) (2.06,4.095) (2.08,4.061) (2.1,4.050) (2.12,4.075) (2.14,4.013) (2.16,4.039) (2.18,4.040) (2.2,4.047) (2.22,4.085) (2.24,4.078) (2.26,4.104) (2.28,4.093) (2.3,4.065) (2.32,4.079) (2.34,4.079) (2.36,4.127) (2.38,4.114) (2.4,4.019) (2.42,4.056) (2.44,4.014) (2.46,4.002) (2.48,3.968) (2.5,3.979) (2.52,3.960) (2.54,3.952) (2.56,3.974) (2.58,3.929) (2.6,3.936) (2.62,4.054) (2.64,4.064) (2.66,4.152) (2.68,4.098) (2.7,4.047) (2.72,4.036) (2.74,3.976) (2.76,3.964) (2.78,3.981) (2.8,4.000) (2.82,3.974) (2.84,3.924) (2.86,3.894) (2.88,4.015) (2.9,4.111) (2.92,4.088) (2.94,4.153) (2.96,4.163) (2.98,4.334) (3,4.466) (3.02,4.355) (3.04,4.360) (3.06,4.357) (3.08,4.376) (3.1,4.373) (3.12,4.355) (3.14,4.394) (3.16,4.848) (3.18,4.874) (3.2,4.965) (3.22,5.055) (3.24,5.055) (3.26,5.071) (3.28,5.163) (3.3,5.118) (3.32,5.233) (3.34,5.230) (3.36,5.213) (3.38,5.145) (3.4,5.054) (3.42,5.072) (3.44,5.073) (3.46,5.076) (3.48,5.161) (3.5,5.103) (3.52,5.202) (3.54,5.282) (3.56,5.218) (3.58,5.319) (3.6,5.266) (3.62,5.789) (3.64,5.539) (3.66,5.501) (3.68,5.221) (3.7,5.103) (3.72,5.182) (3.74,5.265) (3.76,5.024) (3.78,4.655) (3.8,4.624) (3.82,4.642) (3.84,4.571) (3.86,4.574) (3.88,4.418) (3.9,4.456) (3.92,4.496) (3.94,4.461) (3.96,4.340) (3.98,4.356) (4,4.349) (4.02,4.388) (4.04,4.333) (4.06,4.127) (4.08,4.153) (4.1,4.153) (4.12,4.096) (4.14,4.076) (4.16,4.031) (4.18,4.038) (4.2,4.048) (4.22,4.009) (4.24,4.003) (4.26,4.060) (4.28,4.040) (4.3,4.022) (4.32,3.967) (4.34,3.935) (4.36,3.950) (4.38,3.963) (4.4,3.944) (4.42,3.890) (4.44,3.880) (4.46,3.856) (4.48,3.882) (4.5,3.921) (4.52,3.889) (4.54,3.876) (4.56,3.879) (4.58,3.866) (4.6,3.865) (4.62,3.861) (4.64,3.864) (4.66,3.860) (4.68,3.863) (4.7,3.829) (4.72,3.853) (4.74,3.821) (4.76,3.823) (4.78,3.829) (4.8,3.833) (4.82,3.801) (4.84,3.719) (4.86,3.715) (4.88,3.715) (4.9,3.710) (4.92,3.701) (4.94,3.715) (4.96,3.707) (4.98,3.727) (5,3.747) (5.02,3.753) (5.04,3.775) (5.06,3.810) (5.08,3.912) (5.1,3.894) (5.12,3.886) (5.14,3.873) (5.16,3.889) (5.18,3.865) (5.2,3.903)
\end{pspicture}

\vskip 0.3 cm

{\noindent \small Figure 36: kurtosis of the correlation matrix during the year 2001 for correlation matrices calculated over a running window of 80 days.}

\vskip 0.3 cm

Figures 37 and 38 show the skewness and the kurtosis for the elements of the correlation matrix, except its diagonal, calculated in a running window of 100 days, for 2008. Note that the skewness becomes negative for the time after the beginning of the crisis, something that didn't happen in the previous cases. The kurtosis drops to values bellow 3 for the period of crisis.

\vskip 0.6 cm

\begin{pspicture}(-2,-0.1)(3.5,2.2)
\psset{xunit=2.5,yunit=1.5}
\psline{->}(0,0)(5.4,0) \psline{->}(0,-0.2)(0,1.6) \rput(5.6,0){day} \rput(0.4,1.6){Skewness}\scriptsize \psline(0.02,-0.067)(0.02,0.067) \rput(0.02,-0.2){01/02} \psline(0.46,-0.067)(0.46,0.067) \rput(0.46,-0.2){02/01} \psline(0.88,-0.067)(0.88,0.067) \rput(0.88,-0.2){03/03} \psline(1.26,-0.067)(1.26,0.067) \rput(1.26,-0.2){04/01} \psline(1.7,-0.067)(1.7,0.067) \rput(1.7,-0.2){05/02} \psline(2.12,-0.067)(2.12,0.067) \rput(2.12,-0.2){06/02} \psline(2.54,-0.067)(2.54,0.067) \rput(2.54,-0.2){07/01} \psline(3,-0.067)(3,0.067) \rput(3,-0.2){08/01} \psline(3.42,-0.067)(3.42,0.067) \rput(3.42,-0.2){09/01} \psline(3.86,-0.067)(3.86,0.067) \rput(3.86,-0.2){10/02} \psline(4.3,-0.067)(4.3,0.067) \rput(4.3,-0.2){11/03} \psline(4.7,-0.067)(4.7,0.067) \rput(4.7,-0.2){12/01} \psline(5.06,-0.067)(5.06,0.067) \rput(5.06,-0.2){12/30} \psline(-0.043,0.4)(0.043,0.4) \rput(-0.17,0.4){$0.4$} \psline(-0.043,0.8)(0.043,0.8) \rput(-0.17,0.8){$0.8$} \psline(-0.043,1.2)(0.043,1.2)
\psline[linecolor=blue](2.02,0.866) (2.04,0.868) (2.06,0.868) (2.08,0.864) (2.1,0.861) (2.12,0.859) (2.14,0.855) (2.16,0.859) (2.18,0.851) (2.2,0.856) (2.22,0.838) (2.24,0.882) (2.26,0.873) (2.28,0.883) (2.3,0.906) (2.32,0.971) (2.34,0.904) (2.36,0.902) (2.38,0.910) (2.4,0.947) (2.42,0.942) (2.44,0.932) (2.46,0.926) (2.48,0.936) (2.5,0.946) (2.52,0.940) (2.54,0.940) (2.56,0.932) (2.58,0.928) (2.6,0.936) (2.62,0.944) (2.64,0.947) (2.66,0.965) (2.68,0.970) (2.7,0.998) (2.72,1.011) (2.74,0.933) (2.76,0.955) (2.78,0.967) (2.8,0.981) (2.82,0.964) (2.84,0.955) (2.86,0.917) (2.88,0.941) (2.9,0.943) (2.92,0.931) (2.94,0.929) (2.96,0.919) (2.98,0.933) (3,0.947) (3.02,0.951) (3.04,0.969) (3.06,0.989) (3.08,0.993) (3.1,1.054) (3.12,1.054) (3.14,1.042) (3.16,1.053) (3.18,1.180) (3.2,1.173) (3.22,1.169) (3.24,1.166) (3.26,1.170) (3.28,1.153) (3.3,1.186) (3.32,1.188) (3.34,1.190) (3.36,1.185) (3.38,1.184) (3.4,1.179) (3.42,1.180) (3.44,1.186) (3.46,1.188) (3.48,1.180) (3.5,1.171) (3.52,1.127) (3.54,1.100) (3.56,1.116) (3.58,1.109) (3.6,1.105) (3.62,0.909) (3.64,0.887) (3.66,0.875) (3.68,0.833) (3.7,0.549) (3.72,0.575) (3.74,0.564) (3.76,0.564) (3.78,0.564) (3.8,0.573) (3.82,0.596) (3.84,0.640) (3.86,0.646) (3.88,0.654) (3.9,0.464) (3.92,0.476) (3.94,0.331) (3.96,0.381) (3.98,0.294) (4,0.249) (4.02,0.031) (4.04,0.089) (4.06,0.068) (4.08,0.058) (4.1,0.064) (4.12,0.078) (4.14,0.063) (4.16,0.045) (4.18,0.003) (4.2,-0.007) (4.22,-0.010) (4.24,-0.017) (4.26,-0.038) (4.28,-0.050) (4.3,-0.060) (4.32,-0.078) (4.34,-0.058) (4.36,-0.090) (4.38,-0.074) (4.4,-0.068) (4.42,-0.095) (4.44,-0.100) (4.46,-0.094) (4.48,-0.084) (4.5,-0.077) (4.52,-0.074) (4.54,-0.063) (4.56,-0.065) (4.58,-0.051) (4.6,0.003) (4.62,0.003) (4.64,0.004) (4.66,-0.002) (4.68,-0.014) (4.7,0.008) (4.72,0.004) (4.74,0.012) (4.76,0.014) (4.78,0.023) (4.8,0.017) (4.82,0.014) (4.84,0.019) (4.86,0.023) (4.88,0.027) (4.9,0.019) (4.92,0.015) (4.94,0.016) (4.96,0.014) (4.98,0.015) (5,0.030) (5.02,0.028) (5.04,0.040) (5.06,0.031)
\end{pspicture}

\vskip 0.4 cm

{\noindent \small Figure 37: skewness of the correlation matrix during the year 2008 for correlation matrices calculated over a running window of 100 days.}

\vskip 0.3 cm

\begin{pspicture}(-1.5,-0.2)(3.5,3)
\psset{xunit=2.5,yunit=0.5}
\psline{->}(0,0)(5.4,0) \psline{->}(0,0)(0,6) \rput(5.6,0){day} \rput(0.4,6){Kurtosis} \scriptsize \psline(0.02,-0.2)(0.02,0.2) \rput(0.02,-0.6){01/02} \psline(0.46,-0.2)(0.46,0.2) \rput(0.46,-0.6){02/01} \psline(0.88,-0.2)(0.88,0.2) \rput(0.88,-0.6){03/03} \psline(1.26,-0.2)(1.26,0.2) \rput(1.26,-0.6){04/01} \psline(1.7,-0.2)(1.7,0.2) \rput(1.7,-0.6){05/02} \psline(2.12,-0.2)(2.12,0.2) \rput(2.12,-0.6){06/02} \psline(2.54,-0.2)(2.54,0.2) \rput(2.54,-0.6){07/01} \psline(3,-0.2)(3,0.2) \rput(3,-0.6){08/01} \psline(3.42,-0.2)(3.42,0.2) \rput(3.42,-0.6){09/01} \psline(3.86,-0.2)(3.86,0.2) \rput(3.86,-0.6){10/02} \psline(4.3,-0.2)(4.3,0.2) \rput(4.3,-0.6){11/03} \psline(4.7,-0.2)(4.7,0.2) \rput(4.7,-0.6){12/01} \psline(5.06,-0.2)(5.06,0.2) \rput(5.06,-0.6){12/30} \psline(-0.043,1)(0.043,1) \rput(-0.15,1){$1$} \psline(-0.043,2)(0.043,2) \rput(-0.15,2){$2$} \psline(-0.043,3)(0.043,3) \rput(-0.15,3){$3$} \psline(-0.043,4)(0.043,4) \rput(-0.15,4){$4$} \psline(-0.043,5)(0.043,5) \rput(-0.15,5){$4$}
\psline[linecolor=gray](0,3)(5.2,3)
\psline[linecolor=red](2.02,3.199) (2.04,3.197) (2.06,3.198) (2.08,3.200) (2.1,3.203) (2.12,3.196) (2.14,3.201) (2.16,3.195) (2.18,3.189) (2.2,3.186) (2.22,3.169) (2.24,3.257) (2.26,3.230) (2.28,3.246) (2.3,3.462) (2.32,3.642) (2.34,3.517) (2.36,3.612) (2.38,3.636) (2.4,3.727) (2.42,3.771) (2.44,3.719) (2.46,3.686) (2.48,3.731) (2.5,3.744) (2.52,3.792) (2.54,3.732) (2.56,3.748) (2.58,3.775) (2.6,3.795) (2.62,3.847) (2.64,3.859) (2.66,3.874) (2.68,3.907) (2.7,3.971) (2.72,4.016) (2.74,3.771) (2.76,3.828) (2.78,3.781) (2.8,3.876) (2.82,3.828) (2.84,3.800) (2.86,3.707) (2.88,3.769) (2.9,3.805) (2.92,3.794) (2.94,3.772) (2.96,3.679) (2.98,3.705) (3,3.723) (3.02,3.768) (3.04,3.857) (3.06,3.888) (3.08,3.903) (3.1,4.222) (3.12,4.278) (3.14,4.245) (3.16,4.259) (3.18,4.677) (3.2,4.661) (3.22,4.647) (3.24,4.613) (3.26,4.603) (3.28,4.567) (3.3,4.638) (3.32,4.616) (3.34,4.606) (3.36,4.617) (3.38,4.632) (3.4,4.618) (3.42,4.624) (3.44,4.644) (3.46,4.643) (3.48,4.580) (3.5,4.575) (3.52,4.369) (3.54,4.284) (3.56,4.280) (3.58,4.242) (3.6,4.187) (3.62,3.852) (3.64,3.690) (3.66,3.649) (3.68,3.633) (3.7,2.849) (3.72,2.925) (3.74,2.885) (3.76,2.875) (3.78,2.861) (3.8,2.852) (3.82,2.806) (3.84,2.881) (3.86,2.920) (3.88,2.996) (3.9,2.725) (3.92,2.764) (3.94,2.644) (3.96,2.675) (3.98,2.572) (4,2.501) (4.02,2.500) (4.04,2.418) (4.06,2.394) (4.08,2.419) (4.1,2.399) (4.12,2.416) (4.14,2.388) (4.16,2.377) (4.18,2.354) (4.2,2.385) (4.22,2.350) (4.24,2.324) (4.26,2.314) (4.28,2.293) (4.3,2.273) (4.32,2.313) (4.34,2.316) (4.36,2.348) (4.38,2.338) (4.4,2.337) (4.42,2.341) (4.44,2.319) (4.46,2.325) (4.48,2.333) (4.5,2.343) (4.52,2.353) (4.54,2.326) (4.56,2.327) (4.58,2.342) (4.6,2.354) (4.62,2.341) (4.64,2.340) (4.66,2.318) (4.68,2.320) (4.7,2.310) (4.72,2.315) (4.74,2.319) (4.76,2.334) (4.78,2.352) (4.8,2.346) (4.82,2.337) (4.84,2.338) (4.86,2.345) (4.88,2.348) (4.9,2.343) (4.92,2.339) (4.94,2.349) (4.96,2.350) (4.98,2.365) (5,2.365) (5.02,2.367) (5.04,2.380) (5.06,2.374)
\end{pspicture}

\vskip 0.3 cm

{\noindent \small Figure 38: kurtosis of the correlation matrix during the year 2008 for correlation matrices calculated over a running window of 100 days.}

\vskip 0.3 cm

Since a perfect normal distribution would have skewness zero and kurtosis 3, we may see that the distribution of the elements of the correlation matrix on international indices in periods of crisis are not normal, although in the case of 1987 and 2008, it seems to be the case. This assumption is contradicted if one plots the probability distributions of the correlation matrix every two months (figures 39, 40, 41, and 42). During the months of highest volatility of each crisis (October for 1987, August for 1998, September for 2001, and October for 2008), the probability distribution deviate somewhat from a normal distribution.

One can see that the probability distributions for 2008 are less strongly peaked than for the other years, but this is mainly caused by the inclusion of a large number of weakly correlated indices. One can also notice that, in the months of crises, the average correlation grows, but the correlation gets more evenly distributed among the possible spectrum.

\vskip 0.3 cm

\begin{minipage}{8 cm}
\begin{pspicture}(-5.6,-2.9)(1,2.2)
\psset{xunit=1,yunit=1} \scriptsize
\pstThreeDLine[linecolor=blue](0,-0.35,0)(0,-0.35,0.395)(0,-0.25,0.395)(0,-0.25,0)(0,-0.35,0)
\pstThreeDLine[linecolor=blue](0,-0.25,0)(0,-0.25,0.949)(0,-0.15,0.949)(0,-0.15,0)(0,-0.25,0)
\pstThreeDLine[linecolor=blue](0,-0.15,0)(0,-0.15,1.700)(0,-0.05,1.700)(0,-0.05,0)(0,-0.15,0)
\pstThreeDLine[linecolor=blue](0,-0.05,0)(0,-0.05,1.897)(0,0.05,1.897)(0,0.05,0)(0,-0.05,0)
\pstThreeDLine[linecolor=blue](0,0.05,0)(0,0.05,2.332)(0,0.15,2.332)(0,0.15,0)(0,0.05,0)
\pstThreeDLine[linecolor=blue](0,0.15,0)(0,0.15,1.344)(0,0.25,1.344)(0,0.25,0)(0,0.15,0)
\pstThreeDLine[linecolor=blue](0,0.25,0)(0,0.25,0.672)(0,0.35,0.672)(0,0.35,0)(0,0.25,0)
\pstThreeDLine[linecolor=blue](0,0.35,0)(0,0.35,0.356)(0,0.45,0.356)(0,0.45,0)(0,0.35,0)
\pstThreeDLine[linecolor=blue](0,0.45,0)(0,0.45,0.158)(0,0.55,0.158)(0,0.55,0)(0,0.45,0)
\pstThreeDLine[linecolor=blue](0,0.55,0)(0,0.55,0.040)(0,0.65,0.040)(0,0.65,0)(0,0.55,0)
\pstThreeDLine[linecolor=blue](0,0.65,0)(0,0.65,0.119)(0,0.75,0.119)(0,0.75,0)(0,0.65,0)
\pstThreeDLine[linecolor=blue](0,0.75,0)(0,0.75,0.040)(0,0.85,0.040)(0,0.85,0)(0,0.75,0)
\pstThreeDPut(0,1.7,0){Jan/Feb}
\pstThreeDLine[linecolor=blue](1,-0.45,0)(1,-0.45,0.119)(1,-0.35,0.119)(1,-0.35,0)(1,-0.45,0)
\pstThreeDLine[linecolor=blue](1,-0.35,0)(1,-0.35,0.277)(1,-0.25,0.277)(1,-0.25,0)(1,-0.35,0)
\pstThreeDLine[linecolor=blue](1,-0.25,0)(1,-0.25,0.672)(1,-0.15,0.672)(1,-0.15,0)(1,-0.25,0)
\pstThreeDLine[linecolor=blue](1,-0.15,0)(1,-0.15,1.107)(1,-0.05,1.107)(1,-0.05,0)(1,-0.15,0)
\pstThreeDLine[linecolor=blue](1,-0.05,0)(1,-0.05,2.569)(1,0.05,2.569)(1,0.05,0)(1,-0.05,0)
\pstThreeDLine[linecolor=blue](1,0.05,0)(1,0.05,1.858)(1,0.15,1.858)(1,0.15,0)(1,0.05,0)
\pstThreeDLine[linecolor=blue](1,0.15,0)(1,0.15,1.621)(1,0.25,1.621)(1,0.25,0)(1,0.15,0)
\pstThreeDLine[linecolor=blue](1,0.25,0)(1,0.25,1.067)(1,0.35,1.067)(1,0.35,0)(1,0.25,0)
\pstThreeDLine[linecolor=blue](1,0.35,0)(1,0.35,0.277)(1,0.45,0.277)(1,0.45,0)(1,0.35,0)
\pstThreeDLine[linecolor=blue](1,0.45,0)(1,0.45,0.316)(1,0.55,0.316)(1,0.55,0)(1,0.45,0)
\pstThreeDLine[linecolor=blue](1,0.65,0)(1,0.65,0.040)(1,0.75,0.040)(1,0.75,0)(1,0.65,0)
\pstThreeDLine[linecolor=blue](1,0.75,0)(1,0.75,0.040)(1,0.85,0.040)(1,0.85,0)(1,0.75,0)
\pstThreeDLine[linecolor=blue](1,0.85,0)(1,0.85,0.040)(1,0.95,0.040)(1,0.95,0)(1,0.85,0)
\pstThreeDPut(1,1.7,0){Mar/Apr}
\pstThreeDLine[linecolor=blue](2,-0.35,0)(2,-0.35,0.158)(2,-0.25,0.158)(2,-0.25,0)(2,-0.35,0)
\pstThreeDLine[linecolor=blue](2,-0.25,0)(2,-0.25,0.988)(2,-0.15,0.988)(2,-0.15,0)(2,-0.25,0)
\pstThreeDLine[linecolor=blue](2,-0.15,0)(2,-0.15,1.304)(2,-0.05,1.304)(2,-0.05,0)(2,-0.15,0)
\pstThreeDLine[linecolor=blue](2,-0.05,0)(2,-0.05,1.937)(2,0.05,1.937)(2,0.05,0)(2,-0.05,0)
\pstThreeDLine[linecolor=blue](2,0.05,0)(2,0.05,2.332)(2,0.15,2.332)(2,0.15,0)(2,0.05,0)
\pstThreeDLine[linecolor=blue](2,0.15,0)(2,0.15,1.660)(2,0.25,1.660)(2,0.25,0)(2,0.15,0)
\pstThreeDLine[linecolor=blue](2,0.25,0)(2,0.25,1.186)(2,0.35,1.186)(2,0.35,0)(2,0.25,0)
\pstThreeDLine[linecolor=blue](2,0.35,0)(2,0.35,0.277)(2,0.45,0.277)(2,0.45,0)(2,0.35,0)
\pstThreeDLine[linecolor=blue](2,0.45,0)(2,0.45,0.079)(2,0.55,0.079)(2,0.55,0)(2,0.45,0)
\pstThreeDLine[linecolor=blue](2,0.65,0)(2,0.65,0.040)(2,0.75,0.040)(2,0.75,0)(2,0.65,0)
\pstThreeDLine[linecolor=blue](2,0.75,0)(2,0.75,0.040)(2,0.85,0.040)(2,0.85,0)(2,0.75,0)
\pstThreeDPut(2,1.7,0){May/Jun}
\pstThreeDLine[linecolor=blue](3,-0.45,0)(3,-0.45,0.040)(3,-0.35,0.040)(3,-0.35,0)(3,-0.45,0)
\pstThreeDLine[linecolor=blue](3,-0.35,0)(3,-0.35,0.316)(3,-0.25,0.316)(3,-0.25,0)(3,-0.35,0)
\pstThreeDLine[linecolor=blue](3,-0.25,0)(3,-0.25,0.553)(3,-0.15,0.553)(3,-0.15,0)(3,-0.25,0)
\pstThreeDLine[linecolor=blue](3,-0.15,0)(3,-0.15,1.344)(3,-0.05,1.344)(3,-0.05,0)(3,-0.15,0)
\pstThreeDLine[linecolor=blue](3,-0.05,0)(3,-0.05,1.344)(3,0.05,1.344)(3,0.05,0)(3,-0.05,0)
\pstThreeDLine[linecolor=blue](3,0.05,0)(3,0.05,2.332)(3,0.15,2.332)(3,0.15,0)(3,0.05,0)
\pstThreeDLine[linecolor=blue](3,0.15,0)(3,0.15,2.213)(3,0.25,2.213)(3,0.25,0)(3,0.15,0)
\pstThreeDLine[linecolor=blue](3,0.25,0)(3,0.25,1.186)(3,0.35,1.186)(3,0.35,0)(3,0.25,0)
\pstThreeDLine[linecolor=blue](3,0.35,0)(3,0.35,0.395)(3,0.45,0.395)(3,0.45,0)(3,0.35,0)
\pstThreeDLine[linecolor=blue](3,0.45,0)(3,0.45,0.079)(3,0.55,0.079)(3,0.55,0)(3,0.45,0)
\pstThreeDLine[linecolor=blue](3,0.55,0)(3,0.55,0.119)(3,0.65,0.119)(3,0.65,0)(3,0.55,0)
\pstThreeDLine[linecolor=blue](3,0.65,0)(3,0.65,0.040)(3,0.75,0.040)(3,0.75,0)(3,0.65,0)
\pstThreeDLine[linecolor=blue](3,0.85,0)(3,0.85,0.040)(3,0.95,0.040)(3,0.95,0)(3,0.85,0)
\pstThreeDPut(3,1.7,0){Jul/Aug}
\pstThreeDLine[linecolor=blue](4,-0.25,0)(4,-0.25,0.237)(4,-0.15,0.237)(4,-0.15,0)(4,-0.25,0)
\pstThreeDLine[linecolor=blue](4,-0.15,0)(4,-0.15,0.672)(4,-0.05,0.672)(4,-0.05,0)(4,-0.15,0)
\pstThreeDLine[linecolor=blue](4,-0.05,0)(4,-0.05,1.067)(4,0.05,1.067)(4,0.05,0)(4,-0.05,0)
\pstThreeDLine[linecolor=blue](4,0.05,0)(4,0.05,1.462)(4,0.15,1.462)(4,0.15,0)(4,0.05,0)
\pstThreeDLine[linecolor=blue](4,0.15,0)(4,0.15,1.542)(4,0.25,1.542)(4,0.25,0)(4,0.15,0)
\pstThreeDLine[linecolor=blue](4,0.25,0)(4,0.25,1.462)(4,0.35,1.462)(4,0.35,0)(4,0.25,0)
\pstThreeDLine[linecolor=blue](4,0.35,0)(4,0.35,0.751)(4,0.45,0.751)(4,0.45,0)(4,0.35,0)
\pstThreeDLine[linecolor=blue](4,0.45,0)(4,0.45,0.474)(4,0.55,0.474)(4,0.55,0)(4,0.45,0)
\pstThreeDLine[linecolor=blue](4,0.55,0)(4,0.55,0.870)(4,0.65,0.870)(4,0.65,0)(4,0.55,0)
\pstThreeDLine[linecolor=blue](4,0.65,0)(4,0.65,0.632)(4,0.75,0.632)(4,0.75,0)(4,0.65,0)
\pstThreeDLine[linecolor=blue](4,0.75,0)(4,0.75,0.474)(4,0.85,0.474)(4,0.85,0)(4,0.75,0)
\pstThreeDLine[linecolor=blue](4,0.85,0)(4,0.85,0.158)(4,0.95,0.158)(4,0.95,0)(4,0.75,0)
\pstThreeDLine[linecolor=blue](4,0.95,0)(4,0.95,0.198)(4,1.05,0.198)(4,1.05,0)(4,0.75,0)
\pstThreeDPut(4,1.7,0){Sept/Oct}
\pstThreeDLine[linecolor=blue](5,-0.55,0)(5,-0.55,0.040)(5,-0.45,0.040)(5,-0.45,0)(5,-0.55,0)
\pstThreeDLine[linecolor=blue](5,-0.45,0)(5,-0.45,0.079)(5,-0.35,0.079)(5,-0.35,0)(5,-0.45,0)
\pstThreeDLine[linecolor=blue](5,-0.35,0)(5,-0.35,0.158)(5,-0.25,0.158)(5,-0.25,0)(5,-0.35,0)
\pstThreeDLine[linecolor=blue](5,-0.25,0)(5,-0.25,0.751)(5,-0.15,0.751)(5,-0.15,0)(5,-0.25,0)
\pstThreeDLine[linecolor=blue](5,-0.15,0)(5,-0.15,1.621)(5,-0.05,1.621)(5,-0.05,0)(5,-0.15,0)
\pstThreeDLine[linecolor=blue](5,-0.05,0)(5,-0.05,1.146)(5,0.05,1.146)(5,0.05,0)(5,-0.05,0)
\pstThreeDLine[linecolor=blue](5,0.05,0)(5,0.05,1.344)(5,0.15,1.344)(5,0.15,0)(5,0.05,0)
\pstThreeDLine[linecolor=blue](5,0.15,0)(5,0.15,1.304)(5,0.25,1.304)(5,0.25,0)(5,0.15,0)
\pstThreeDLine[linecolor=blue](5,0.25,0)(5,0.25,0.949)(5,0.35,0.949)(5,0.35,0)(5,0.25,0)
\pstThreeDLine[linecolor=blue](5,0.35,0)(5,0.35,0.672)(5,0.45,0.672)(5,0.45,0)(5,0.35,0)
\pstThreeDLine[linecolor=blue](5,0.45,0)(5,0.45,0.672)(5,0.55,0.672)(5,0.55,0)(5,0.45,0)
\pstThreeDLine[linecolor=blue](5,0.55,0)(5,0.55,0.514)(5,0.65,0.514)(5,0.65,0)(5,0.55,0)
\pstThreeDLine[linecolor=blue](5,0.65,0)(5,0.65,0.514)(5,0.75,0.514)(5,0.75,0)(5,0.65,0)
\pstThreeDLine[linecolor=blue](5,0.75,0)(5,0.75,0.119)(5,0.85,0.119)(5,0.85,0)(5,0.75,0)
\pstThreeDLine[linecolor=blue](5,0.85,0)(5,0.85,0.040)(5,0.95,0.040)(5,0.95,0)(5,0.85,0)
\pstThreeDPut(5,1.7,0){Nov/Dec}
\pstThreeDLine[linecolor=red](6,-0.15,0)(6,-0.15,0.119)(6,-0.05,0.119)(6,-0.05,0)(6,-0.15,0)
\pstThreeDLine[linecolor=red](6,-0.05,0)(6,-0.05,2.174)(6,0.05,2.174)(6,0.05,0)(6,-0.05,0)
\pstThreeDLine[linecolor=red](6,0.05,0)(6,0.05,2.806)(6,0.15,2.806)(6,0.15,0)(6,0.05,0)
\pstThreeDLine[linecolor=red](6,0.15,0)(6,0.15,1.858)(6,0.25,1.858)(6,0.25,0)(6,0.15,0)
\pstThreeDLine[linecolor=red](6,0.25,0)(6,0.25,0.791)(6,0.35,0.791)(6,0.35,0)(6,0.25,0)
\pstThreeDLine[linecolor=red](6,0.35,0)(6,0.35,0.514)(6,0.45,0.514)(6,0.45,0)(6,0.35,0)
\pstThreeDLine[linecolor=red](6,0.45,0)(6,0.45,0.751)(6,0.55,0.751)(6,0.55,0)(6,0.45,0)
\pstThreeDLine[linecolor=red](6,0.55,0)(6,0.55,0.514)(6,0.65,0.514)(6,0.65,0)(6,0.55,0)
\pstThreeDLine[linecolor=red](6,0.65,0)(6,0.65,0.198)(6,0.75,0.198)(6,0.75,0)(6,0.65,0)
\pstThreeDLine[linecolor=red](6,0.75,0)(6,0.75,0.237)(6,0.85,0.237)(6,0.85,0)(6,0.75,0)
\pstThreeDLine[linecolor=red](6,0.85,0)(6,0.85,0.040)(6,0.95,0.040)(6,0.95,0)(6,0.85,0)
\pstThreeDPut(6,1.7,0){1987}
\end{pspicture}

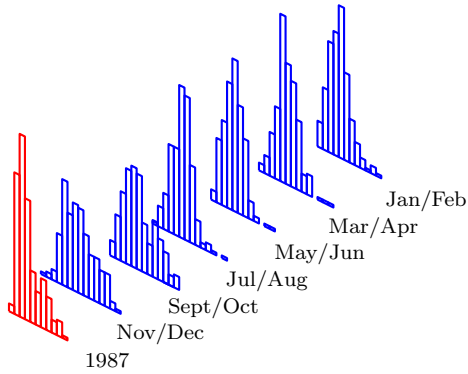
{\noindent \small Figure 39: probability distributions of the correlation matrix calculated every two months in 1987. The probability distribution for the data for the whole year appears last, in red.}
\end{minipage}
\hskip 1.3 cm \begin{minipage}{8 cm}
\begin{pspicture}(-5.6,-2.9)(1,2.2)
\psset{xunit=1,yunit=1} \scriptsize
\pstThreeDLine[linecolor=blue](0,-0.45,0)(0,-0.45,0.020)(0,-0.35,0.020)(0,-0.35,0)(0,-0.35,0)
\pstThreeDLine[linecolor=blue](0,-0.35,0)(0,-0.35,0.128)(0,-0.25,0.128)(0,-0.25,0)(0,-0.35,0)
\pstThreeDLine[linecolor=blue](0,-0.25,0)(0,-0.25,0.251)(0,-0.15,0.251)(0,-0.15,0)(0,-0.25,0)
\pstThreeDLine[linecolor=blue](0,-0.15,0)(0,-0.15,0.681)(0,-0.05,0.681)(0,-0.05,0)(0,-0.15,0)
\pstThreeDLine[linecolor=blue](0,-0.05,0)(0,-0.05,1.219)(0,0.05,1.219)(0,0.05,0)(0,-0.05,0)
\pstThreeDLine[linecolor=blue](0,0.05,0)(0,0.05,1.654)(0,0.15,1.654)(0,0.15,0)(0,0.05,0)
\pstThreeDLine[linecolor=blue](0,0.15,0)(0,0.15,1.500)(0,0.25,1.500)(0,0.25,0)(0,0.15,0)
\pstThreeDLine[linecolor=blue](0,0.25,0)(0,0.25,1.239)(0,0.35,1.239)(0,0.35,0)(0,0.25,0)
\pstThreeDLine[linecolor=blue](0,0.35,0)(0,0.35,1.111)(0,0.45,1.111)(0,0.45,0)(0,0.35,0)
\pstThreeDLine[linecolor=blue](0,0.45,0)(0,0.45,1.019)(0,0.55,1.019)(0,0.55,0)(0,0.45,0)
\pstThreeDLine[linecolor=blue](0,0.55,0)(0,0.55,0.676)(0,0.65,0.676)(0,0.65,0)(0,0.55,0)
\pstThreeDLine[linecolor=blue](0,0.65,0)(0,0.65,0.328)(0,0.75,0.328)(0,0.75,0)(0,0.65,0)
\pstThreeDLine[linecolor=blue](0,0.75,0)(0,0.75,0.164)(0,0.85,0.164)(0,0.85,0)(0,0.75,0)
\pstThreeDLine[linecolor=blue](0,0.85,0)(0,0.85,0.010)(0,0.95,0.010)(0,0.95,0)(0,0.75,0)
\pstThreeDPut(0,1.7,0){Jan/Feb}
\pstThreeDLine[linecolor=blue](1,-0.45,0)(1,-0.45,0.020)(1,-0.35,0.020)(1,-0.35,0)(1,-0.45,0)
\pstThreeDLine[linecolor=blue](1,-0.35,0)(1,-0.35,0.072)(1,-0.25,0.072)(1,-0.25,0)(1,-0.35,0)
\pstThreeDLine[linecolor=blue](1,-0.25,0)(1,-0.25,0.312)(1,-0.15,0.312)(1,-0.15,0)(1,-0.25,0)
\pstThreeDLine[linecolor=blue](1,-0.15,0)(1,-0.15,1.106)(1,-0.05,1.106)(1,-0.05,0)(1,-0.15,0)
\pstThreeDLine[linecolor=blue](1,-0.05,0)(1,-0.05,1.736)(1,0.05,1.736)(1,0.05,0)(1,-0.05,0)
\pstThreeDLine[linecolor=blue](1,0.05,0)(1,0.05,1.946)(1,0.15,1.946)(1,0.15,0)(1,0.05,0)
\pstThreeDLine[linecolor=blue](1,0.15,0)(1,0.15,1.792)(1,0.25,1.792)(1,0.25,0)(1,0.15,0)
\pstThreeDLine[linecolor=blue](1,0.25,0)(1,0.25,1.157)(1,0.35,1.157)(1,0.35,0)(1,0.25,0)
\pstThreeDLine[linecolor=blue](1,0.35,0)(1,0.35,0.773)(1,0.45,0.773)(1,0.45,0)(1,0.35,0)
\pstThreeDLine[linecolor=blue](1,0.45,0)(1,0.45,0.512)(1,0.55,0.512)(1,0.55,0)(1,0.45,0)
\pstThreeDLine[linecolor=blue](1,0.55,0)(1,0.55,0.302)(1,0.65,0.302)(1,0.65,0)(1,0.45,0)
\pstThreeDLine[linecolor=blue](1,0.65,0)(1,0.65,0.179)(1,0.75,0.179)(1,0.75,0)(1,0.65,0)
\pstThreeDLine[linecolor=blue](1,0.75,0)(1,0.75,0.082)(1,0.85,0.082)(1,0.85,0)(1,0.75,0)
\pstThreeDLine[linecolor=blue](1,0.85,0)(1,0.85,0.010)(1,0.95,0.010)(1,0.95,0)(1,0.85,0)
\pstThreeDPut(1,1.7,0){Mar/Apr}
\pstThreeDLine[linecolor=blue](2,-0.45,0)(2,-0.45,0.010)(2,-0.35,0.010)(2,-0.35,0)(2,-0.45,0)
\pstThreeDLine[linecolor=blue](2,-0.35,0)(2,-0.35,0.077)(2,-0.25,0.077)(2,-0.25,0)(2,-0.35,0)
\pstThreeDLine[linecolor=blue](2,-0.25,0)(2,-0.25,0.261)(2,-0.15,0.261)(2,-0.15,0)(2,-0.25,0)
\pstThreeDLine[linecolor=blue](2,-0.15,0)(2,-0.15,0.789)(2,-0.05,0.789)(2,-0.05,0)(2,-0.15,0)
\pstThreeDLine[linecolor=blue](2,-0.05,0)(2,-0.05,1.239)(2,0.05,1.239)(2,0.05,0)(2,-0.05,0)
\pstThreeDLine[linecolor=blue](2,0.05,0)(2,0.05,1.833)(2,0.15,1.833)(2,0.15,0)(2,0.05,0)
\pstThreeDLine[linecolor=blue](2,0.15,0)(2,0.15,1.874)(2,0.25,1.874)(2,0.25,0)(2,0.15,0)
\pstThreeDLine[linecolor=blue](2,0.25,0)(2,0.25,1.521)(2,0.35,1.521)(2,0.35,0)(2,0.25,0)
\pstThreeDLine[linecolor=blue](2,0.35,0)(2,0.35,0.891)(2,0.45,0.891)(2,0.45,0)(2,0.35,0)
\pstThreeDLine[linecolor=blue](2,0.45,0)(2,0.45,0.691)(2,0.55,0.691)(2,0.55,0)(2,0.45,0)
\pstThreeDLine[linecolor=blue](2,0.55,0)(2,0.55,0.364)(2,0.65,0.364)(2,0.65,0)(2,0.55,0)
\pstThreeDLine[linecolor=blue](2,0.65,0)(2,0.65,0.246)(2,0.75,0.246)(2,0.75,0)(2,0.65,0)
\pstThreeDLine[linecolor=blue](2,0.75,0)(2,0.75,0.138)(2,0.85,0.138)(2,0.85,0)(2,0.75,0)
\pstThreeDLine[linecolor=blue](2,0.85,0)(2,0.85,0.067)(2,0.95,0.067)(2,0.95,0)(2,0.85,0)
\pstThreeDPut(2,1.7,0){May/Jun}
\pstThreeDLine[linecolor=blue](3,-0.45,0)(3,-0.45,0.010)(3,-0.35,0.010)(3,-0.35,0)(3,-0.45,0)
\pstThreeDLine[linecolor=blue](3,-0.35,0)(3,-0.35,0.082)(3,-0.25,0.082)(3,-0.25,0)(3,-0.35,0)
\pstThreeDLine[linecolor=blue](3,-0.25,0)(3,-0.25,0.312)(3,-0.15,0.312)(3,-0.15,0)(3,-0.25,0)
\pstThreeDLine[linecolor=blue](3,-0.15,0)(3,-0.15,0.870)(3,-0.05,0.870)(3,-0.05,0)(3,-0.15,0)
\pstThreeDLine[linecolor=blue](3,-0.05,0)(3,-0.05,1.285)(3,0.05,1.285)(3,0.05,0)(3,-0.05,0)
\pstThreeDLine[linecolor=blue](3,0.05,0)(3,0.05,1.628)(3,0.15,1.628)(3,0.15,0)(3,0.05,0)
\pstThreeDLine[linecolor=blue](3,0.15,0)(3,0.15,1.434)(3,0.25,1.434)(3,0.25,0)(3,0.15,0)
\pstThreeDLine[linecolor=blue](3,0.25,0)(3,0.25,1.019)(3,0.35,1.019)(3,0.35,0)(3,0.25,0)
\pstThreeDLine[linecolor=blue](3,0.35,0)(3,0.35,0.722)(3,0.45,0.722)(3,0.45,0)(3,0.35,0)
\pstThreeDLine[linecolor=blue](3,0.45,0)(3,0.45,0.691)(3,0.55,0.691)(3,0.55,0)(3,0.45,0)
\pstThreeDLine[linecolor=blue](3,0.55,0)(3,0.55,0.829)(3,0.65,0.829)(3,0.65,0)(3,0.55,0)
\pstThreeDLine[linecolor=blue](3,0.65,0)(3,0.65,0.538)(3,0.75,0.538)(3,0.75,0)(3,0.65,0)
\pstThreeDLine[linecolor=blue](3,0.75,0)(3,0.75,0.369)(3,0.85,0.369)(3,0.85,0)(3,0.75,0)
\pstThreeDLine[linecolor=blue](3,0.85,0)(3,0.85,0.205)(3,0.95,0.205)(3,0.95,0)(3,0.85,0)
\pstThreeDLine[linecolor=blue](3,0.95,0)(3,0.95,0.005)(3,1.05,0.005)(3,1.05,0)(3,0.95,0)
\pstThreeDPut(3,1.7,0){Jul/Aug}
\pstThreeDLine[linecolor=blue](4,-0.45,0)(4,-0.45,0.041)(4,-0.35,0.041)(4,-0.35,0)(4,-0.45,0)
\pstThreeDLine[linecolor=blue](4,-0.35,0)(4,-0.35,0.179)(4,-0.25,0.179)(4,-0.25,0)(4,-0.35,0)
\pstThreeDLine[linecolor=blue](4,-0.25,0)(4,-0.25,0.379)(4,-0.15,0.379)(4,-0.15,0)(4,-0.25,0)
\pstThreeDLine[linecolor=blue](4,-0.15,0)(4,-0.15,0.927)(4,-0.05,0.927)(4,-0.05,0)(4,-0.15,0)
\pstThreeDLine[linecolor=blue](4,-0.05,0)(4,-0.05,1.244)(4,0.05,1.244)(4,0.05,0)(4,-0.05,0)
\pstThreeDLine[linecolor=blue](4,0.05,0)(4,0.05,1.664)(4,0.15,1.664)(4,0.15,0)(4,0.05,0)
\pstThreeDLine[linecolor=blue](4,0.15,0)(4,0.15,1.439)(4,0.25,1.439)(4,0.25,0)(4,0.15,0)
\pstThreeDLine[linecolor=blue](4,0.25,0)(4,0.25,1.096)(4,0.35,1.096)(4,0.35,0)(4,0.25,0)
\pstThreeDLine[linecolor=blue](4,0.35,0)(4,0.35,0.927)(4,0.45,0.927)(4,0.45,0)(4,0.35,0)
\pstThreeDLine[linecolor=blue](4,0.45,0)(4,0.45,0.783)(4,0.55,0.783)(4,0.55,0)(4,0.45,0)
\pstThreeDLine[linecolor=blue](4,0.55,0)(4,0.55,0.440)(4,0.65,0.440)(4,0.65,0)(4,0.55,0)
\pstThreeDLine[linecolor=blue](4,0.65,0)(4,0.65,0.358)(4,0.75,0.358)(4,0.75,0)(4,0.65,0)
\pstThreeDLine[linecolor=blue](4,0.75,0)(4,0.75,0.338)(4,0.85,0.338)(4,0.85,0)(4,0.75,0)
\pstThreeDLine[linecolor=blue](4,0.85,0)(4,0.85,0.174)(4,0.95,0.174)(4,0.95,0)(4,0.75,0)
\pstThreeDLine[linecolor=blue](4,0.95,0)(4,0.95,0.010)(4,1.05,0.010)(4,1.05,0)(4,0.75,0)
\pstThreeDPut(4,1.7,0){Sept/Oct}
\pstThreeDLine[linecolor=blue](5,-0.45,0)(5,-0.45,0.031)(5,-0.35,0.031)(5,-0.35,0)(5,-0.45,0)
\pstThreeDLine[linecolor=blue](5,-0.35,0)(5,-0.35,0.123)(5,-0.25,0.123)(5,-0.25,0)(5,-0.35,0)
\pstThreeDLine[linecolor=blue](5,-0.25,0)(5,-0.25,0.538)(5,-0.15,0.538)(5,-0.15,0)(5,-0.25,0)
\pstThreeDLine[linecolor=blue](5,-0.15,0)(5,-0.15,1.111)(5,-0.05,1.111)(5,-0.05,0)(5,-0.15,0)
\pstThreeDLine[linecolor=blue](5,-0.05,0)(5,-0.05,1.695)(5,0.05,1.695)(5,0.05,0)(5,-0.05,0)
\pstThreeDLine[linecolor=blue](5,0.05,0)(5,0.05,1.787)(5,0.15,1.787)(5,0.15,0)(5,0.05,0)
\pstThreeDLine[linecolor=blue](5,0.15,0)(5,0.15,1.377)(5,0.25,1.377)(5,0.25,0)(5,0.15,0)
\pstThreeDLine[linecolor=blue](5,0.25,0)(5,0.25,1.331)(5,0.35,1.331)(5,0.35,0)(5,0.25,0)
\pstThreeDLine[linecolor=blue](5,0.35,0)(5,0.35,0.824)(5,0.45,0.824)(5,0.45,0)(5,0.35,0)
\pstThreeDLine[linecolor=blue](5,0.45,0)(5,0.45,0.445)(5,0.55,0.445)(5,0.55,0)(5,0.45,0)
\pstThreeDLine[linecolor=blue](5,0.55,0)(5,0.55,0.394)(5,0.65,0.394)(5,0.65,0)(5,0.55,0)
\pstThreeDLine[linecolor=blue](5,0.65,0)(5,0.65,0.215)(5,0.75,0.215)(5,0.75,0)(5,0.65,0)
\pstThreeDLine[linecolor=blue](5,0.75,0)(5,0.75,0.067)(5,0.85,0.067)(5,0.85,0)(5,0.75,0)
\pstThreeDLine[linecolor=blue](5,0.85,0)(5,0.85,0.061)(5,0.95,0.061)(5,0.95,0)(5,0.85,0)
\pstThreeDPut(5,1.7,0){Nov/Dec}
\pstThreeDLine[linecolor=red](6,-0.25,0)(6,-0.25,0.005)(6,-0.15,0.005)(6,-0.15,0)(6,-0.25,0)
\pstThreeDLine[linecolor=red](6,-0.15,0)(6,-0.15,0.210)(6,-0.05,0.210)(6,-0.05,0)(6,-0.15,0)
\pstThreeDLine[linecolor=red](6,-0.05,0)(6,-0.05,1.971)(6,0.05,1.971)(6,0.05,0)(6,-0.05,0)
\pstThreeDLine[linecolor=red](6,0.05,0)(6,0.05,2.785)(6,0.15,2.785)(6,0.15,0)(6,0.05,0)
\pstThreeDLine[linecolor=red](6,0.15,0)(6,0.15,1.603)(6,0.25,1.603)(6,0.25,0)(6,0.15,0)
\pstThreeDLine[linecolor=red](6,0.25,0)(6,0.25,0.993)(6,0.35,0.993)(6,0.35,0)(6,0.25,0)
\pstThreeDLine[linecolor=red](6,0.35,0)(6,0.35,0.983)(6,0.45,0.983)(6,0.45,0)(6,0.35,0)
\pstThreeDLine[linecolor=red](6,0.45,0)(6,0.45,0.538)(6,0.55,0.538)(6,0.55,0)(6,0.45,0)
\pstThreeDLine[linecolor=red](6,0.55,0)(6,0.55,0.343)(6,0.65,0.343)(6,0.65,0)(6,0.55,0)
\pstThreeDLine[linecolor=red](6,0.65,0)(6,0.65,0.312)(6,0.75,0.312)(6,0.75,0)(6,0.65,0)
\pstThreeDLine[linecolor=red](6,0.75,0)(6,0.75,0.215)(6,0.85,0.215)(6,0.85,0)(6,0.75,0)
\pstThreeDLine[linecolor=red](6,0.85,0)(6,0.85,0.041)(6,0.95,0.041)(6,0.95,0)(6,0.85,0)
\pstThreeDPut(6,1.7,0){1998}
\end{pspicture}

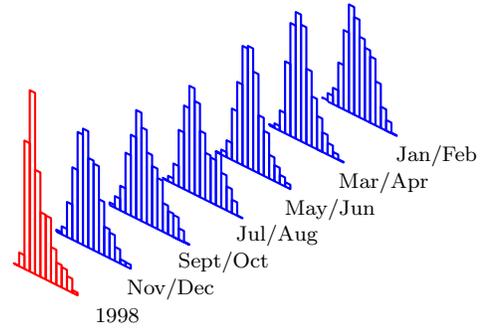
{\noindent \small Figure 40: probability distributions of the correlation matrix calculated every two months in 1998. The probability distribution for the data for the whole year appears last, in red.}
\end{minipage}

\vskip 1 cm

\begin{minipage}{8 cm}
\begin{pspicture}(-5.6,-2.9)(1,1)
\psset{xunit=1,yunit=1} \scriptsize
\pstThreeDLine[linecolor=blue](0,-0.65,0)(0,-0.65,0.006)(0,-0.55,0.006)(0,-0.55,0)(0,-0.65,0)
\pstThreeDLine[linecolor=blue](0,-0.55,0)(0,-0.55,0.006)(0,-0.45,0.006)(0,-0.45,0)(0,-0.55,0)
\pstThreeDLine[linecolor=blue](0,-0.45,0)(0,-0.45,0.052)(0,-0.35,0.052)(0,-0.35,0)(0,-0.45,0)
\pstThreeDLine[linecolor=blue](0,-0.35,0)(0,-0.35,0.156)(0,-0.25,0.156)(0,-0.25,0)(0,-0.35,0)
\pstThreeDLine[linecolor=blue](0,-0.25,0)(0,-0.25,0.526)(0,-0.15,0.526)(0,-0.15,0)(0,-0.25,0)
\pstThreeDLine[linecolor=blue](0,-0.15,0)(0,-0.15,1.152)(0,-0.05,1.152)(0,-0.05,0)(0,-0.15,0)
\pstThreeDLine[linecolor=blue](0,-0.05,0)(0,-0.05,1.941)(0,0.05,1.941)(0,0.05,0)(0,-0.05,0)
\pstThreeDLine[linecolor=blue](0,0.05,0)(0,0.05,2.061)(0,0.15,2.061)(0,0.15,0)(0,0.05,0)
\pstThreeDLine[linecolor=blue](0,0.15,0)(0,0.15,1.876)(0,0.25,1.876)(0,0.25,0)(0,0.15,0)
\pstThreeDLine[linecolor=blue](0,0.25,0)(0,0.25,1.155)(0,0.35,1.155)(0,0.35,0)(0,0.25,0)
\pstThreeDLine[linecolor=blue](0,0.35,0)(0,0.35,0.503)(0,0.45,0.503)(0,0.45,0)(0,0.35,0)
\pstThreeDLine[linecolor=blue](0,0.45,0)(0,0.45,0.260)(0,0.55,0.260)(0,0.55,0)(0,0.45,0)
\pstThreeDLine[linecolor=blue](0,0.55,0)(0,0.55,0.146)(0,0.65,0.146)(0,0.65,0)(0,0.55,0)
\pstThreeDLine[linecolor=blue](0,0.65,0)(0,0.65,0.084)(0,0.75,0.084)(0,0.75,0)(0,0.65,0)
\pstThreeDLine[linecolor=blue](0,0.75,0)(0,0.75,0.058)(0,0.85,0.058)(0,0.85,0)(0,0.75,0)
\pstThreeDLine[linecolor=blue](0,0.85,0)(0,0.85,0.013)(0,0.95,0.013)(0,0.95,0)(0,0.85,0)
\pstThreeDLine[linecolor=blue](0,0.95,0)(0,0.95,0.003)(0,1.05,0.003)(0,1.05,0)(0,0.95,0)
\pstThreeDPut(0,1.7,0){Jan/Feb}
\pstThreeDLine[linecolor=blue](1,-0.55,0)(1,-0.55,0.003)(1,-0.45,0.003)(1,-0.45,0)(1,-0.55,0)
\pstThreeDLine[linecolor=blue](1,-0.45,0)(1,-0.45,0.007)(1,-0.35,0.007)(1,-0.35,0)(1,-0.45,0)
\pstThreeDLine[linecolor=blue](1,-0.35,0)(1,-0.35,0.090)(1,-0.25,0.090)(1,-0.25,0)(1,-0.35,0)
\pstThreeDLine[linecolor=blue](1,-0.25,0)(1,-0.25,0.296)(1,-0.15,0.296)(1,-0.15,0)(1,-0.25,0)
\pstThreeDLine[linecolor=blue](1,-0.15,0)(1,-0.15,0.842)(1,-0.05,0.842)(1,-0.05,0)(1,-0.15,0)
\pstThreeDLine[linecolor=blue](1,-0.05,0)(1,-0.05,1.652)(1,0.05,1.652)(1,0.05,0)(1,-0.05,0)
\pstThreeDLine[linecolor=blue](1,0.05,0)(1,0.05,1.968)(1,0.15,1.968)(1,0.15,0)(1,0.05,0)
\pstThreeDLine[linecolor=blue](1,0.15,0)(1,0.15,1.752)(1,0.25,1.752)(1,0.25,0)(1,0.15,0)
\pstThreeDLine[linecolor=blue](1,0.25,0)(1,0.25,1.146)(1,0.35,1.146)(1,0.35,0)(1,0.25,0)
\pstThreeDLine[linecolor=blue](1,0.35,0)(1,0.35,0.713)(1,0.45,0.713)(1,0.45,0)(1,0.35,0)
\pstThreeDLine[linecolor=blue](1,0.45,0)(1,0.45,0.500)(1,0.55,0.500)(1,0.55,0)(1,0.45,0)
\pstThreeDLine[linecolor=blue](1,0.55,0)(1,0.55,0.466)(1,0.65,0.466)(1,0.65,0)(1,0.55,0)
\pstThreeDLine[linecolor=blue](1,0.65,0)(1,0.65,0.300)(1,0.75,0.300)(1,0.75,0)(1,0.65,0)
\pstThreeDLine[linecolor=blue](1,0.75,0)(1,0.75,0.163)(1,0.85,0.163)(1,0.85,0)(1,0.75,0)
\pstThreeDLine[linecolor=blue](1,0.85,0)(1,0.85,0.093)(1,0.95,0.093)(1,0.95,0)(1,0.85,0)
\pstThreeDLine[linecolor=blue](1,0.95,0)(1,0.95,0.010)(1,1.05,0.010)(1,1.05,0)(1,0.95,0)
\pstThreeDPut(1,1.7,0){Mar/Apr}
\pstThreeDLine[linecolor=blue](2,-0.55,0)(2,-0.55,0.003)(2,-0.45,0.003)(2,-0.45,0)(2,-0.55,0)
\pstThreeDLine[linecolor=blue](2,-0.45,0)(2,-0.45,0.016)(2,-0.35,0.016)(2,-0.35,0)(2,-0.45,0)
\pstThreeDLine[linecolor=blue](2,-0.35,0)(2,-0.35,0.123)(2,-0.25,0.123)(2,-0.25,0)(2,-0.35,0)
\pstThreeDLine[linecolor=blue](2,-0.25,0)(2,-0.25,0.487)(2,-0.15,0.487)(2,-0.15,0)(2,-0.25,0)
\pstThreeDLine[linecolor=blue](2,-0.15,0)(2,-0.15,1.207)(2,-0.05,1.207)(2,-0.05,0)(2,-0.15,0)
\pstThreeDLine[linecolor=blue](2,-0.05,0)(2,-0.05,1.960)(2,0.05,1.960)(2,0.05,0)(2,-0.05,0)
\pstThreeDLine[linecolor=blue](2,0.05,0)(2,0.05,2.103)(2,0.15,2.103)(2,0.15,0)(2,0.05,0)
\pstThreeDLine[linecolor=blue](2,0.15,0)(2,0.15,1.873)(2,0.25,1.873)(2,0.25,0)(2,0.15,0)
\pstThreeDLine[linecolor=blue](2,0.25,0)(2,0.25,1.130)(2,0.35,1.130)(2,0.35,0)(2,0.25,0)
\pstThreeDLine[linecolor=blue](2,0.35,0)(2,0.35,0.610)(2,0.45,0.610)(2,0.45,0)(2,0.35,0)
\pstThreeDLine[linecolor=blue](2,0.45,0)(2,0.45,0.269)(2,0.55,0.269)(2,0.55,0)(2,0.45,0)
\pstThreeDLine[linecolor=blue](2,0.55,0)(2,0.55,0.133)(2,0.65,0.133)(2,0.65,0)(2,0.55,0)
\pstThreeDLine[linecolor=blue](2,0.65,0)(2,0.65,0.068)(2,0.75,0.068)(2,0.75,0)(2,0.65,0)
\pstThreeDLine[linecolor=blue](2,0.75,0)(2,0.75,0.010)(2,0.85,0.010)(2,0.85,0)(2,0.75,0)
\pstThreeDLine[linecolor=blue](2,0.85,0)(2,0.85,0.006)(2,0.95,0.006)(2,0.95,0)(2,0.85,0)
\pstThreeDPut(2,1.7,0){May/Jun}
\pstThreeDLine[linecolor=blue](3,-0.45,0)(3,-0.45,0.029)(3,-0.35,0.029)(3,-0.35,0)(3,-0.45,0)
\pstThreeDLine[linecolor=blue](3,-0.35,0)(3,-0.35,0.107)(3,-0.25,0.107)(3,-0.25,0)(3,-0.35,0)
\pstThreeDLine[linecolor=blue](3,-0.25,0)(3,-0.25,0.471)(3,-0.15,0.471)(3,-0.15,0)(3,-0.25,0)
\pstThreeDLine[linecolor=blue](3,-0.15,0)(3,-0.15,1.178)(3,-0.05,1.178)(3,-0.05,0)(3,-0.15,0)
\pstThreeDLine[linecolor=blue](3,-0.05,0)(3,-0.05,1.934)(3,0.05,1.934)(3,0.05,0)(3,-0.05,0)
\pstThreeDLine[linecolor=blue](3,0.05,0)(3,0.05,2.256)(3,0.15,2.256)(3,0.15,0)(3,0.05,0)
\pstThreeDLine[linecolor=blue](3,0.15,0)(3,0.15,1.905)(3,0.25,1.905)(3,0.25,0)(3,0.15,0)
\pstThreeDLine[linecolor=blue](3,0.25,0)(3,0.25,0.977)(3,0.35,0.977)(3,0.35,0)(3,0.25,0)
\pstThreeDLine[linecolor=blue](3,0.35,0)(3,0.35,0.532)(3,0.45,0.532)(3,0.45,0)(3,0.35,0)
\pstThreeDLine[linecolor=blue](3,0.45,0)(3,0.45,0.240)(3,0.55,0.240)(3,0.55,0)(3,0.45,0)
\pstThreeDLine[linecolor=blue](3,0.55,0)(3,0.55,0.208)(3,0.65,0.208)(3,0.65,0)(3,0.55,0)
\pstThreeDLine[linecolor=blue](3,0.65,0)(3,0.65,0.065)(3,0.75,0.065)(3,0.75,0)(3,0.65,0)
\pstThreeDLine[linecolor=blue](3,0.75,0)(3,0.75,0.062)(3,0.85,0.062)(3,0.85,0)(3,0.75,0)
\pstThreeDLine[linecolor=blue](3,0.85,0)(3,0.85,0.032)(3,0.95,0.032)(3,0.95,0)(3,0.85,0)
\pstThreeDLine[linecolor=blue](3,0.95,0)(3,0.95,0.003)(3,1.05,0.003)(3,1.05,0)(3,0.95,0)
\pstThreeDPut(3,1.7,0){Jul/Aug}
\pstThreeDLine[linecolor=blue](4,-0.45,0)(4,-0.45,0.006)(4,-0.35,0.006)(4,-0.35,0)(4,-0.45,0)
\pstThreeDLine[linecolor=blue](4,-0.35,0)(4,-0.35,0.094)(4,-0.25,0.094)(4,-0.25,0)(4,-0.35,0)
\pstThreeDLine[linecolor=blue](4,-0.25,0)(4,-0.25,0.354)(4,-0.15,0.354)(4,-0.15,0)(4,-0.25,0)
\pstThreeDLine[linecolor=blue](4,-0.15,0)(4,-0.15,0.837)(4,-0.05,0.837)(4,-0.05,0)(4,-0.15,0)
\pstThreeDLine[linecolor=blue](4,-0.05,0)(4,-0.05,1.522)(4,0.05,1.522)(4,0.05,0)(4,-0.05,0)
\pstThreeDLine[linecolor=blue](4,0.05,0)(4,0.05,1.762)(4,0.15,1.762)(4,0.15,0)(4,0.05,0)
\pstThreeDLine[linecolor=blue](4,0.15,0)(4,0.15,1.668)(4,0.25,1.668)(4,0.25,0)(4,0.15,0)
\pstThreeDLine[linecolor=blue](4,0.25,0)(4,0.25,1.240)(4,0.35,1.240)(4,0.35,0)(4,0.25,0)
\pstThreeDLine[linecolor=blue](4,0.35,0)(4,0.35,0.902)(4,0.45,0.902)(4,0.45,0)(4,0.35,0)
\pstThreeDLine[linecolor=blue](4,0.45,0)(4,0.45,0.652)(4,0.55,0.652)(4,0.55,0)(4,0.45,0)
\pstThreeDLine[linecolor=blue](4,0.55,0)(4,0.55,0.493)(4,0.65,0.493)(4,0.65,0)(4,0.55,0)
\pstThreeDLine[linecolor=blue](4,0.65,0)(4,0.65,0.260)(4,0.75,0.260)(4,0.75,0)(4,0.65,0)
\pstThreeDLine[linecolor=blue](4,0.75,0)(4,0.75,0.094)(4,0.85,0.094)(4,0.85,0)(4,0.75,0)
\pstThreeDLine[linecolor=blue](4,0.85,0)(4,0.85,0.084)(4,0.95,0.084)(4,0.95,0)(4,0.75,0)
\pstThreeDLine[linecolor=blue](4,0.95,0)(4,0.95,0.029)(4,1.05,0.029)(4,1.05,0)(4,0.75,0)
\pstThreeDPut(4,1.7,0){Sept/Oct}
\pstThreeDLine[linecolor=blue](5,-0.45,0)(5,-0.45,0.019)(5,-0.35,0.019)(5,-0.35,0)(5,-0.45,0)
\pstThreeDLine[linecolor=blue](5,-0.35,0)(5,-0.35,0.169)(5,-0.25,0.169)(5,-0.25,0)(5,-0.35,0)
\pstThreeDLine[linecolor=blue](5,-0.25,0)(5,-0.25,0.555)(5,-0.15,0.555)(5,-0.15,0)(5,-0.25,0)
\pstThreeDLine[linecolor=blue](5,-0.15,0)(5,-0.15,1.165)(5,-0.05,1.165)(5,-0.05,0)(5,-0.15,0)
\pstThreeDLine[linecolor=blue](5,-0.05,0)(5,-0.05,1.960)(5,0.05,1.960)(5,0.05,0)(5,-0.05,0)
\pstThreeDLine[linecolor=blue](5,0.05,0)(5,0.05,2.048)(5,0.15,2.048)(5,0.15,0)(5,0.05,0)
\pstThreeDLine[linecolor=blue](5,0.15,0)(5,0.15,1.639)(5,0.25,1.639)(5,0.25,0)(5,0.15,0)
\pstThreeDLine[linecolor=blue](5,0.25,0)(5,0.25,0.948)(5,0.35,0.948)(5,0.35,0)(5,0.25,0)
\pstThreeDLine[linecolor=blue](5,0.35,0)(5,0.35,0.636)(5,0.45,0.636)(5,0.45,0)(5,0.35,0)
\pstThreeDLine[linecolor=blue](5,0.45,0)(5,0.45,0.428)(5,0.55,0.428)(5,0.55,0)(5,0.45,0)
\pstThreeDLine[linecolor=blue](5,0.55,0)(5,0.55,0.224)(5,0.65,0.224)(5,0.65,0)(5,0.55,0)
\pstThreeDLine[linecolor=blue](5,0.65,0)(5,0.65,0.078)(5,0.75,0.078)(5,0.75,0)(5,0.65,0)
\pstThreeDLine[linecolor=blue](5,0.75,0)(5,0.75,0.055)(5,0.85,0.055)(5,0.85,0)(5,0.75,0)
\pstThreeDLine[linecolor=blue](5,0.85,0)(5,0.85,0.058)(5,0.95,0.058)(5,0.95,0)(5,0.95,0)
\pstThreeDLine[linecolor=blue](5,0.95,0)(5,0.95,0.016)(5,1.05,0.016)(5,1.05,0)(5,0.75,0)
\pstThreeDPut(5,1.7,0){Nov/Dec}
\pstThreeDLine[linecolor=red](6,-0.15,0)(6,-0.15,0.279)(6,-0.05,0.279)(6,-0.05,0)(6,-0.15,0)
\pstThreeDLine[linecolor=red](6,-0.05,0)(6,-0.05,2.379)(6,0.05,2.379)(6,0.05,0)(6,-0.05,0)
\pstThreeDLine[linecolor=red](6,0.05,0)(6,0.05,3.658)(6,0.15,3.658)(6,0.15,0)(6,0.05,0)
\pstThreeDLine[linecolor=red](6,0.15,0)(6,0.15,1.516)(6,0.25,1.516)(6,0.25,0)(6,0.15,0)
\pstThreeDLine[linecolor=red](6,0.25,0)(6,0.25,0.824)(6,0.35,0.824)(6,0.35,0)(6,0.25,0)
\pstThreeDLine[linecolor=red](6,0.35,0)(6,0.35,0.639)(6,0.45,0.639)(6,0.45,0)(6,0.35,0)
\pstThreeDLine[linecolor=red](6,0.45,0)(6,0.45,0.338)(6,0.55,0.338)(6,0.55,0)(6,0.45,0)
\pstThreeDLine[linecolor=red](6,0.55,0)(6,0.55,0.201)(6,0.65,0.201)(6,0.65,0)(6,0.55,0)
\pstThreeDLine[linecolor=red](6,0.65,0)(6,0.65,0.052)(6,0.75,0.052)(6,0.75,0)(6,0.65,0)
\pstThreeDLine[linecolor=red](6,0.75,0)(6,0.75,0.068)(6,0.85,0.068)(6,0.85,0)(6,0.75,0)
\pstThreeDLine[linecolor=red](6,0.85,0)(6,0.85,0.045)(6,0.95,0.045)(6,0.95,0)(6,0.85,0)
\pstThreeDPut(6,1.7,0){2001}
\end{pspicture}

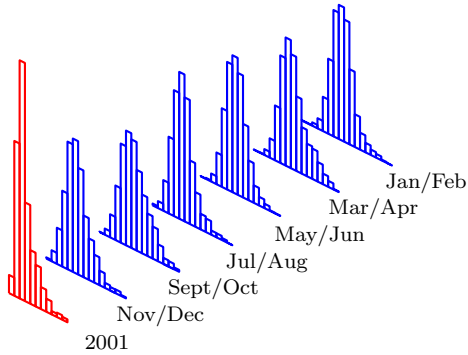
{\noindent \small Figure 41: probability distributions of the correlation matrix calculated every two months in 2001. The probability distribution for the data for the whole year appears last, in red.}
\end{minipage}
\hskip 1.3 cm \begin{minipage}{8 cm}
\begin{pspicture}(-5.6,-2.9)(1,1)
\psset{xunit=1,yunit=1} \scriptsize
\pstThreeDLine[linecolor=blue](0,-0.55,0)(0,-0.55,0.002)(0,-0.45,0.002)(0,-0.45,0)(0,-0.55,0)
\pstThreeDLine[linecolor=blue](0,-0.45,0)(0,-0.45,0.010)(0,-0.35,0.010)(0,-0.35,0)(0,-0.45,0)
\pstThreeDLine[linecolor=blue](0,-0.35,0)(0,-0.35,0.072)(0,-0.25,0.072)(0,-0.25,0)(0,-0.35,0)
\pstThreeDLine[linecolor=blue](0,-0.25,0)(0,-0.25,0.248)(0,-0.15,0.248)(0,-0.15,0)(0,-0.25,0)
\pstThreeDLine[linecolor=blue](0,-0.15,0)(0,-0.15,0.604)(0,-0.05,0.604)(0,-0.05,0)(0,-0.15,0)
\pstThreeDLine[linecolor=blue](0,-0.05,0)(0,-0.05,1.292)(0,0.05,1.292)(0,0.05,0)(0,-0.05,0)
\pstThreeDLine[linecolor=blue](0,0.05,0)(0,0.05,1.734)(0,0.15,1.734)(0,0.15,0)(0,0.05,0)
\pstThreeDLine[linecolor=blue](0,0.15,0)(0,0.15,1.780)(0,0.25,1.780)(0,0.25,0)(0,0.15,0)
\pstThreeDLine[linecolor=blue](0,0.25,0)(0,0.25,1.187)(0,0.35,1.187)(0,0.35,0)(0,0.25,0)
\pstThreeDLine[linecolor=blue](0,0.35,0)(0,0.35,0.817)(0,0.45,0.817)(0,0.45,0)(0,0.35,0)
\pstThreeDLine[linecolor=blue](0,0.45,0)(0,0.45,0.577)(0,0.55,0.577)(0,0.55,0)(0,0.45,0)
\pstThreeDLine[linecolor=blue](0,0.55,0)(0,0.55,0.561)(0,0.65,0.561)(0,0.65,0)(0,0.55,0)
\pstThreeDLine[linecolor=blue](0,0.65,0)(0,0.65,0.461)(0,0.75,0.461)(0,0.75,0)(0,0.65,0)
\pstThreeDLine[linecolor=blue](0,0.75,0)(0,0.75,0.449)(0,0.85,0.449)(0,0.85,0)(0,0.75,0)
\pstThreeDLine[linecolor=blue](0,0.85,0)(0,0.85,0.177)(0,0.95,0.177)(0,0.95,0)(0,0.85,0)
\pstThreeDLine[linecolor=blue](0,0.95,0)(0,0.95,0.048)(0,1.05,0.048)(0,1.05,0)(0,0.95,0)
\pstThreeDPut(0,1.7,0){Jan/Feb}
\pstThreeDLine[linecolor=blue](1,-0.65,0)(1,-0.65,0.002)(1,-0.55,0.002)(1,-0.55,0)(1,-0.65,0)
\pstThreeDLine[linecolor=blue](1,-0.45,0)(1,-0.45,0.029)(1,-0.35,0.029)(1,-0.35,0)(1,-0.45,0)
\pstThreeDLine[linecolor=blue](1,-0.35,0)(1,-0.35,0.189)(1,-0.25,0.189)(1,-0.25,0)(1,-0.35,0)
\pstThreeDLine[linecolor=blue](1,-0.25,0)(1,-0.25,0.444)(1,-0.15,0.444)(1,-0.15,0)(1,-0.25,0)
\pstThreeDLine[linecolor=blue](1,-0.15,0)(1,-0.15,0.932)(1,-0.05,0.932)(1,-0.05,0)(1,-0.15,0)
\pstThreeDLine[linecolor=blue](1,-0.05,0)(1,-0.05,1.297)(1,0.05,1.297)(1,0.05,0)(1,-0.05,0)
\pstThreeDLine[linecolor=blue](1,0.05,0)(1,0.05,1.474)(1,0.15,1.474)(1,0.15,0)(1,0.05,0)
\pstThreeDLine[linecolor=blue](1,0.15,0)(1,0.15,1.271)(1,0.25,1.271)(1,0.25,0)(1,0.15,0)
\pstThreeDLine[linecolor=blue](1,0.25,0)(1,0.25,1.211)(1,0.35,1.211)(1,0.35,0)(1,0.25,0)
\pstThreeDLine[linecolor=blue](1,0.35,0)(1,0.35,1.018)(1,0.45,1.018)(1,0.45,0)(1,0.35,0)
\pstThreeDLine[linecolor=blue](1,0.45,0)(1,0.45,0.896)(1,0.55,0.896)(1,0.55,0)(1,0.45,0)
\pstThreeDLine[linecolor=blue](1,0.55,0)(1,0.55,0.492)(1,0.65,0.492)(1,0.65,0)(1,0.55,0)
\pstThreeDLine[linecolor=blue](1,0.65,0)(1,0.65,0.301)(1,0.75,0.301)(1,0.75,0)(1,0.65,0)
\pstThreeDLine[linecolor=blue](1,0.75,0)(1,0.75,0.275)(1,0.85,0.275)(1,0.85,0)(1,0.75,0)
\pstThreeDLine[linecolor=blue](1,0.85,0)(1,0.85,0.119)(1,0.95,0.119)(1,0.95,0)(1,0.85,0)
\pstThreeDLine[linecolor=blue](1,0.95,0)(1,0.95,0.050)(1,1.05,0.050)(1,1.05,0)(1,0.95,0)
\pstThreeDPut(1,1.7,0){Mar/Apr}
\pstThreeDLine[linecolor=blue](2,-0.55,0)(2,-0.55,0.005)(2,-0.45,0.005)(2,-0.45,0)(2,-0.55,0)
\pstThreeDLine[linecolor=blue](2,-0.45,0)(2,-0.45,0.022)(2,-0.35,0.022)(2,-0.35,0)(2,-0.45,0)
\pstThreeDLine[linecolor=blue](2,-0.35,0)(2,-0.35,0.136)(2,-0.25,0.136)(2,-0.25,0)(2,-0.35,0)
\pstThreeDLine[linecolor=blue](2,-0.25,0)(2,-0.25,0.452)(2,-0.15,0.452)(2,-0.15,0)(2,-0.25,0)
\pstThreeDLine[linecolor=blue](2,-0.15,0)(2,-0.15,1.006)(2,-0.05,1.006)(2,-0.05,0)(2,-0.15,0)
\pstThreeDLine[linecolor=blue](2,-0.05,0)(2,-0.05,1.742)(2,0.05,1.742)(2,0.05,0)(2,-0.05,0)
\pstThreeDLine[linecolor=blue](2,0.05,0)(2,0.05,2.052)(2,0.15,2.052)(2,0.15,0)(2,0.05,0)
\pstThreeDLine[linecolor=blue](2,0.15,0)(2,0.15,1.624)(2,0.25,1.624)(2,0.25,0)(2,0.15,0)
\pstThreeDLine[linecolor=blue](2,0.25,0)(2,0.25,1.214)(2,0.35,1.214)(2,0.35,0)(2,0.25,0)
\pstThreeDLine[linecolor=blue](2,0.35,0)(2,0.35,0.719)(2,0.45,0.719)(2,0.45,0)(2,0.35,0)
\pstThreeDLine[linecolor=blue](2,0.45,0)(2,0.45,0.401)(2,0.55,0.401)(2,0.55,0)(2,0.45,0)
\pstThreeDLine[linecolor=blue](2,0.55,0)(2,0.55,0.299)(2,0.65,0.299)(2,0.65,0)(2,0.55,0)
\pstThreeDLine[linecolor=blue](2,0.65,0)(2,0.65,0.201)(2,0.75,0.201)(2,0.75,0)(2,0.65,0)
\pstThreeDLine[linecolor=blue](2,0.75,0)(2,0.75,0.081)(2,0.85,0.081)(2,0.85,0)(2,0.75,0)
\pstThreeDLine[linecolor=blue](2,0.85,0)(2,0.85,0.041)(2,0.95,0.041)(2,0.95,0)(2,0.85,0)
\pstThreeDLine[linecolor=blue](2,0.95,0)(2,0.95,0.007)(2,1.05,0.007)(2,1.05,0)(2,0.95,0)
\pstThreeDPut(2,1.7,0){May/Jun}
\pstThreeDLine[linecolor=blue](3,-0.45,0)(3,-0.45,0.005)(3,-0.35,0.005)(3,-0.35,0)(3,-0.45,0)
\pstThreeDLine[linecolor=blue](3,-0.35,0)(3,-0.35,0.136)(3,-0.25,0.136)(3,-0.25,0)(3,-0.35,0)
\pstThreeDLine[linecolor=blue](3,-0.25,0)(3,-0.25,0.363)(3,-0.15,0.363)(3,-0.15,0)(3,-0.25,0)
\pstThreeDLine[linecolor=blue](3,-0.15,0)(3,-0.15,0.934)(3,-0.05,0.934)(3,-0.05,0)(3,-0.15,0)
\pstThreeDLine[linecolor=blue](3,-0.05,0)(3,-0.05,1.687)(3,0.05,1.687)(3,0.05,0)(3,-0.05,0)
\pstThreeDLine[linecolor=blue](3,0.05,0)(3,0.05,1.854)(3,0.15,1.854)(3,0.15,0)(3,0.05,0)
\pstThreeDLine[linecolor=blue](3,0.15,0)(3,0.15,1.713)(3,0.25,1.713)(3,0.25,0)(3,0.15,0)
\pstThreeDLine[linecolor=blue](3,0.25,0)(3,0.25,1.192)(3,0.35,1.192)(3,0.35,0)(3,0.25,0)
\pstThreeDLine[linecolor=blue](3,0.35,0)(3,0.35,0.755)(3,0.45,0.755)(3,0.45,0)(3,0.35,0)
\pstThreeDLine[linecolor=blue](3,0.45,0)(3,0.45,0.604)(3,0.55,0.604)(3,0.55,0)(3,0.45,0)
\pstThreeDLine[linecolor=blue](3,0.55,0)(3,0.55,0.323)(3,0.65,0.323)(3,0.65,0)(3,0.55,0)
\pstThreeDLine[linecolor=blue](3,0.65,0)(3,0.65,0.184)(3,0.75,0.184)(3,0.75,0)(3,0.65,0)
\pstThreeDLine[linecolor=blue](3,0.75,0)(3,0.75,0.127)(3,0.85,0.127)(3,0.85,0)(3,0.75,0)
\pstThreeDLine[linecolor=blue](3,0.85,0)(3,0.85,0.098)(3,0.95,0.098)(3,0.95,0)(3,0.85,0)
\pstThreeDLine[linecolor=blue](3,0.95,0)(3,0.95,0.026)(3,1.05,0.026)(3,1.05,0)(3,0.95,0)
\pstThreeDPut(3,1.7,0){Jul/Aug}
\pstThreeDLine[linecolor=blue](4,-0.55,0)(4,-0.55,0.005)(4,-0.45,0.005)(4,-0.45,0)(4,-0.55,0)
\pstThreeDLine[linecolor=blue](4,-0.45,0)(4,-0.45,0.007)(4,-0.35,0.007)(4,-0.35,0)(4,-0.45,0)
\pstThreeDLine[linecolor=blue](4,-0.35,0)(4,-0.35,0.050)(4,-0.25,0.050)(4,-0.25,0)(4,-0.35,0)
\pstThreeDLine[linecolor=blue](4,-0.25,0)(4,-0.25,0.165)(4,-0.15,0.165)(4,-0.15,0)(4,-0.25,0)
\pstThreeDLine[linecolor=blue](4,-0.15,0)(4,-0.15,0.380)(4,-0.05,0.380)(4,-0.05,0)(4,-0.15,0)
\pstThreeDLine[linecolor=blue](4,-0.05,0)(4,-0.05,0.736)(4,0.05,0.736)(4,0.05,0)(4,-0.05,0)
\pstThreeDLine[linecolor=blue](4,0.05,0)(4,0.05,0.788)(4,0.15,0.788)(4,0.15,0)(4,0.05,0)
\pstThreeDLine[linecolor=blue](4,0.15,0)(4,0.15,0.788)(4,0.25,0.788)(4,0.25,0)(4,0.15,0)
\pstThreeDLine[linecolor=blue](4,0.25,0)(4,0.25,0.841)(4,0.35,0.841)(4,0.35,0)(4,0.25,0)
\pstThreeDLine[linecolor=blue](4,0.35,0)(4,0.35,1.135)(4,0.45,1.135)(4,0.45,0)(4,0.35,0)
\pstThreeDLine[linecolor=blue](4,0.45,0)(4,0.45,1.462)(4,0.55,1.462)(4,0.55,0)(4,0.45,0)
\pstThreeDLine[linecolor=blue](4,0.55,0)(4,0.55,1.488)(4,0.65,1.488)(4,0.65,0)(4,0.55,0)
\pstThreeDLine[linecolor=blue](4,0.65,0)(4,0.65,1.240)(4,0.75,1.240)(4,0.75,0)(4,0.65,0)
\pstThreeDLine[linecolor=blue](4,0.75,0)(4,0.75,0.602)(4,0.85,0.602)(4,0.85,0)(4,0.75,0)
\pstThreeDLine[linecolor=blue](4,0.85,0)(4,0.85,0.229)(4,0.95,0.229)(4,0.95,0)(4,0.75,0)
\pstThreeDLine[linecolor=blue](4,0.95,0)(4,0.95,0.084)(4,1.05,0.084)(4,1.05,0)(4,0.75,0)
\pstThreeDPut(4,1.7,0){Sept/Oct}
\pstThreeDLine[linecolor=blue](5,-0.65,0)(5,-0.65,0.002)(5,-0.55,0.002)(5,-0.55,0)(5,-0.65,0)
\pstThreeDLine[linecolor=blue](5,-0.55,0)(5,-0.55,0.002)(5,-0.45,0.002)(5,-0.45,0)(5,-0.55,0)
\pstThreeDLine[linecolor=blue](5,-0.45,0)(5,-0.45,0.041)(5,-0.35,0.041)(5,-0.35,0)(5,-0.45,0)
\pstThreeDLine[linecolor=blue](5,-0.35,0)(5,-0.35,0.141)(5,-0.25,0.141)(5,-0.25,0)(5,-0.35,0)
\pstThreeDLine[linecolor=blue](5,-0.25,0)(5,-0.25,0.368)(5,-0.15,0.368)(5,-0.15,0)(5,-0.25,0)
\pstThreeDLine[linecolor=blue](5,-0.15,0)(5,-0.15,0.702)(5,-0.05,0.702)(5,-0.05,0)(5,-0.15,0)
\pstThreeDLine[linecolor=blue](5,-0.05,0)(5,-0.05,1.094)(5,0.05,1.094)(5,0.05,0)(5,-0.05,0)
\pstThreeDLine[linecolor=blue](5,0.05,0)(5,0.05,1.405)(5,0.15,1.405)(5,0.15,0)(5,0.05,0)
\pstThreeDLine[linecolor=blue](5,0.15,0)(5,0.15,1.316)(5,0.25,1.316)(5,0.25,0)(5,0.15,0)
\pstThreeDLine[linecolor=blue](5,0.25,0)(5,0.25,1.180)(5,0.35,1.180)(5,0.35,0)(5,0.25,0)
\pstThreeDLine[linecolor=blue](5,0.35,0)(5,0.35,1.216)(5,0.45,1.216)(5,0.45,0)(5,0.35,0)
\pstThreeDLine[linecolor=blue](5,0.45,0)(5,0.45,0.910)(5,0.55,0.910)(5,0.55,0)(5,0.45,0)
\pstThreeDLine[linecolor=blue](5,0.55,0)(5,0.55,0.702)(5,0.65,0.702)(5,0.65,0)(5,0.55,0)
\pstThreeDLine[linecolor=blue](5,0.65,0)(5,0.65,0.466)(5,0.75,0.466)(5,0.75,0)(5,0.65,0)
\pstThreeDLine[linecolor=blue](5,0.75,0)(5,0.75,0.244)(5,0.85,0.244)(5,0.85,0)(5,0.75,0)
\pstThreeDLine[linecolor=blue](5,0.85,0)(5,0.85,0.119)(5,0.95,0.119)(5,0.95,0)(5,0.95,0)
\pstThreeDLine[linecolor=blue](5,0.95,0)(5,0.95,0.091)(5,1.05,0.091)(5,1.05,0)(5,0.75,0)
\pstThreeDPut(5,1.7,0){Nov/Dec}
\pstThreeDLine[linecolor=red](6,-0.35,0)(6,-0.35,0.002)(6,-0.25,0.002)(6,-0.25,0)(6,-0.35,0)
\pstThreeDLine[linecolor=red](6,-0.25,0)(6,-0.25,0.022)(6,-0.15,0.022)(6,-0.15,0)(6,-0.25,0)
\pstThreeDLine[linecolor=red](6,-0.15,0)(6,-0.15,0.179)(6,-0.05,0.179)(6,-0.05,0)(6,-0.15,0)
\pstThreeDLine[linecolor=red](6,-0.05,0)(6,-0.05,0.905)(6,0.05,0.905)(6,0.05,0)(6,-0.05,0)
\pstThreeDLine[linecolor=red](6,0.05,0)(6,0.05,1.761)(6,0.15,1.761)(6,0.15,0)(6,0.05,0)
\pstThreeDLine[linecolor=red](6,0.15,0)(6,0.15,1.759)(6,0.25,1.759)(6,0.25,0)(6,0.15,0)
\pstThreeDLine[linecolor=red](6,0.25,0)(6,0.25,1.491)(6,0.35,1.491)(6,0.35,0)(6,0.25,0)
\pstThreeDLine[linecolor=red](6,0.35,0)(6,0.35,1.340)(6,0.45,1.340)(6,0.45,0)(6,0.35,0)
\pstThreeDLine[linecolor=red](6,0.45,0)(6,0.45,1.087)(6,0.55,1.087)(6,0.55,0)(6,0.45,0)
\pstThreeDLine[linecolor=red](6,0.55,0)(6,0.55,0.724)(6,0.65,0.724)(6,0.65,0)(6,0.55,0)
\pstThreeDLine[linecolor=red](6,0.65,0)(6,0.65,0.542)(6,0.75,0.542)(6,0.75,0)(6,0.65,0)
\pstThreeDLine[linecolor=red](6,0.75,0)(6,0.75,0.194)(6,0.85,0.194)(6,0.85,0)(6,0.75,0)
\pstThreeDLine[linecolor=red](6,0.85,0)(6,0.85,0.127)(6,0.95,0.127)(6,0.95,0)(6,0.85,0)
\pstThreeDLine[linecolor=red](6,0.95,0)(6,0.95,0.038)(6,1.05,0.038)(6,1.05,0)(6,0.95,0)
\pstThreeDPut(6,1.7,0){2008}
\end{pspicture}

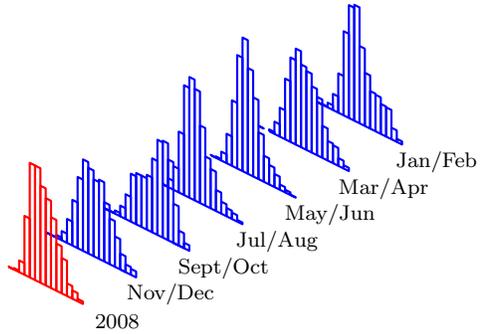
{\noindent \small Figure 42: probability distributions of the correlation matrix calculated every two months in 2008. The probability distribution for the data for the whole year appears last, in red.}
\end{minipage}

\vskip 0.3 cm

Our claim that the probability distributions are far from Gaussian during periods of high volatility may be substantiated by using two tests for normality of those distributions. The Jarque-Bera test \cite{JB} is based on the formula
\begin{equation}
JB=\frac{N}{6}\left[ s^2+\frac{(k-3)^2}{4}\right] \ ,
\end{equation}
where $N$ is the size of the sample, $s$ is its skewness, and $k$ is its kurtosis. The Lilliefors test \cite{Lilli}, a variant of the Kolmogorov-Smirnov test, is based on the formula
\begin{equation}
L={\rm max}\left| E(x)-N(x)\right| \ ,
\end{equation}
where $E(x)$ is the cumulative distribution function estimated from the data and $N(x)$ is the cumulative distribution function of a normal distribution with the same mean and standard deviation as the data.

The Jarque-Bera test rejects the null hypothesis that the distribution is normal at the $5\% $ significance level for all months of the years we have studied. The Lilliefors test rejects the null hypothesis that the distribution is normal at the $5\% $ significance level for all months except March/April and May/June, 1987. When applied to the whole years of data, both tests strongly reject the hypothesis that the distribution of the correlation matrix is similar to a normal distribution.

\section{Gauging the results}

As we commented in the introduction of this article, one of the major concerns when dealing with data from stock markets all over the world is that most of them do not operate at the same hours. This leads to incorrections when one tries to study the correlations between them. Another source of concern is that sometimes the correlations between markets may not be measured correctly by the Pearson correlation coefficient, since it is better suited for linear correlation, which may not be the case. Other correlation coefficients, like Spearman's or Kendall's rank correlation coefficients, may capture relations which are not seen using Pearson's correlation.

In order to gauge the effect of these two possible problems, we did two aditional analysis of the data. In the first one, we phased the data of Eastern markets (from Russia to the east) so that the data of Western stock markets were compared with data from the next day of Eastern markets. In the second one, we switched to Spearman's correlation whenever Pearson's correlation was used. We did all the calculations again for both cases and compared the results with the ones previously obtained. An account of the comparisons is given now for the four crises being considered.

For 1987 with phased data, the average correlation becomes $<C>=0.15$, slightly smaller than the value $<C>=0.16$ for the unphased data. Using Spearman's correlation, we obtain $<C>_S=0.07$ (remember it is a different type of correlation, and so it should not be compared numerically with the Pearson correlation). For the phased data, the maximum eingenvalue, which was $\lambda _{\text{max}}=6.500$, becomes $\lambda _{\text{max}}=6.135$, and for Spearman's correlation, it becomes $\lambda _{\text{max}}=3.977$.

While for the original data Indonesia had a substantial negative participation in the eigenvector with the highest eigenvalue, no index has relevant negative participation for the phased data, and Taiwan and South Korea increase their participation, although Hong Kong decreases its own. For the eigenvector obtained with Spearman's correlation, Brazil, Finland, Bangladesh, and Taiwan acquire small negative participations, Sri Lanka and Indonesia maintaining their negative coefficients.

For the phased data, there is nearly no change in the relations between average correlation and volatility, or between average correlation and average volatility, and the skewness and kurtosis of the probability distributions for the correlation matrix are also very similar. For the results obtained using Spearman's correlation, the agreement between the average correlation and average volatility is much greater for the data concerned with the begining of the crisis.

For 1998, the average correlation $<C>=0.17$ drops to $<C>=0.15$ for the phased data and is given by $<C>_S=0.16$ for the data related with Spearman's correlation. The maximum eigenvalue goes from $\lambda _{\text{max}}=16.897$ to $\lambda _{\text{max}}=15.511$ (phased data) and $\lambda _{\text{max}}=16.022$ (Spearman's correlation). The participation of the Asian markets in the eigenvector corresponding to the largest eigenvalue grows for phased data and keeps essentially the same for Spearman's correlation. There are no substantial changes between average correlation and volatility and average volatility calculated in a moving window, nor in the skewness and kurtosis of the probability distribution of the off-diagonal elements of the correlation matrix, although for Spearman's correlation, the average correlation and the average volatility are slightly more connected.

For 2001, the average correlation $<C>=0.11$ remains $<C>=0.11$ for the phased data, and it is $<C>_S=0.07$ for the data obtained using Spearman's correlation. The maximum eigenvalue goes from $\lambda _{\text{max}}=15.284$ to $\lambda _{\text{max}}=15.052$ (phased data) and $\lambda _{\text{max}}=10.577$ (Spearman's correlation). For the phased data, the number of participating Asian countries clearly grows, and the average participation, including those of some Western countries, also grows, but not substantially. For Spearman's correlation, the participations of indices in the market mode do not change substantially. There are no substantial changes to the skewness and kurtosis of the probability distribution of the off-diagonal elements of the correlation matrix if we use phased data or Spearman's correlation. For Spearman's correlation, the relation between average correlation and average volatility is even clearer.

For 2008, the average correlation $<C>=0.26$ drops to $<C>=0.21$ for the phased data and is $<C>_S=0.22$ for the data related with Spearman's correlation. The maximum eigenvalue goes from $\lambda _{\text{max}}=31.284$ to $\lambda _{\text{max}}=26.761$ (phased data) and $\lambda _{\text{max}}=26.587$ (Spearman's correlation). The participation of Asian and African markets increase slightly in the eigenvector corresponding to the largest eigenvalue for the case of phased data. For Spearman's correlation, participations do not change significantly. The relation between average correlation and average volatility becomes stronger using phased data, and increases drastically for Spearman's correlation. There is nearly no change in the skewness and kurtosis of the probability distribution of the off-diagonal elements of the correlation matrix for phased data, but for Spearman's correlation the skewness and kurtosis curves become smoother.

What we may conclude from this analysis is that the use of phased data gives occasional better results, but in general makes the average correlation between indices lower. So, we don't really have compelling reasons to use phased data. Now, for the calculations using Spearman's rank correlation, the agreement between average correlation and average volatility increases, sometimes drastically, as may be seen by comparing figures 43, 44, 45, and 46 (shown next) with figures 11, 17, 23, and 29, respectively.

\begin{pspicture}(-2,-0.3)(3.5,3)
\psset{xunit=2.5,yunit=0.4}
\psline{->}(0,0)(5.4,0) \psline{->}(0,0)(0,6) \rput(5.6,0){day} \rput(0.7,6){$<C>_n$,$<\text{vol}>_n$}\scriptsize \psline(0.02,-0.2)(0.02,0.2) \rput(0.02,-0.6){01/02} \psline(0.44,-0.2)(0.44,0.2) \rput(0.44,-0.6){02/02} \psline(0.84,-0.2)(0.84,0.2) \rput(0.84,-0.6){03/02} \psline(1.28,-0.2)(1.28,0.2) \rput(1.28,-0.6){04/01} \psline(1.7,-0.2)(1.7,0.2) \rput(1.7,-0.6){05/01} \psline(2.12,-0.2)(2.12,0.2) \rput(2.12,-0.6){06/01} \psline(2.56,-0.2)(2.56,0.2) \rput(2.56,-0.6){07/01} \psline(3.02,-0.2)(3.02,0.2) \rput(3.02,-0.6){08/03} \psline(3.44,-0.2)(3.44,0.2) \rput(3.44,-0.6){09/01} \psline(3.88,-0.2)(3.88,0.2) \rput(3.88,-0.6){10/01} \psline(4.32,-0.2)(4.32,0.2) \rput(4.32,-0.6){11/02} \psline(4.74,-0.2)(4.74,0.2) \rput(4.74,-0.6){12/01} \psline(5.16,-0.2)(5.16,0.2) \rput(5.16,-0.6){12/31} \psline(-0.043,2)(0.043,2) \rput(-0.17,2){$2$} \psline(-0.043,4)(0.043,4) \rput(-0.17,4){$4$}
\psline[linecolor=blue](0.62,1.441) (0.64,1.458) (0.66,1.420) (0.68,1.427) (0.7,1.423) (0.72,1.450) (0.74,1.467) (0.76,1.474) (0.78,1.466) (0.8,1.487) (0.82,1.445) (0.84,1.458) (0.86,1.459) (0.88,1.455) (0.9,1.481) (0.92,1.482) (0.94,1.479) (0.96,1.438) (0.98,1.427) (1,1.421) (1.02,1.415) (1.04,1.401) (1.06,1.406) (1.08,1.400) (1.1,1.404) (1.12,1.391) (1.14,1.410) (1.16,1.427) (1.18,1.417) (1.2,1.400) (1.22,1.397) (1.24,1.396) (1.26,1.455) (1.28,1.437) (1.3,1.418) (1.32,1.420) (1.34,1.431) (1.36,1.457) (1.38,1.474) (1.4,1.457) (1.42,1.469) (1.44,1.479) (1.46,1.520) (1.48,1.567) (1.5,1.580) (1.52,1.601) (1.54,1.604) (1.56,1.604) (1.58,1.616) (1.6,1.612) (1.62,1.621) (1.64,1.681) (1.66,1.680) (1.68,1.680) (1.7,1.694) (1.72,1.673) (1.74,1.667) (1.76,1.677) (1.78,1.670) (1.8,1.688) (1.82,1.703) (1.84,1.701) (1.86,1.651) (1.88,1.648) (1.9,1.654) (1.92,1.653) (1.94,1.641) (1.96,1.647) (1.98,1.645) (2,1.646) (2.02,1.633) (2.04,1.623) (2.06,1.577) (2.08,1.544) (2.1,1.523) (2.12,1.494) (2.14,1.501) (2.16,1.525) (2.18,1.520) (2.2,1.530) (2.22,1.508) (2.24,1.443) (2.26,1.426) (2.28,1.439) (2.3,1.429) (2.32,1.446) (2.34,1.436) (2.36,1.433) (2.38,1.430) (2.4,1.420) (2.42,1.403) (2.44,1.388) (2.46,1.378) (2.48,1.381) (2.5,1.391) (2.52,1.378) (2.54,1.378) (2.56,1.317) (2.58,1.362) (2.6,1.399) (2.62,1.400) (2.64,1.402) (2.66,1.410) (2.68,1.408) (2.7,1.413) (2.72,1.437) (2.74,1.449) (2.76,1.439) (2.78,1.456) (2.8,1.482) (2.82,1.482) (2.84,1.496) (2.86,1.527) (2.88,1.539) (2.9,1.517) (2.92,1.498) (2.94,1.534) (2.96,1.542) (2.98,1.568) (3,1.569) (3.02,1.578) (3.04,1.574) (3.06,1.595) (3.08,1.588) (3.1,1.574) (3.12,1.586) (3.14,1.616) (3.16,1.617) (3.18,1.558) (3.2,1.517) (3.22,1.532) (3.24,1.570) (3.26,1.592) (3.28,1.601) (3.3,1.592) (3.32,1.559) (3.34,1.548) (3.36,1.548) (3.38,1.518) (3.4,1.487) (3.42,1.492) (3.44,1.487) (3.46,1.476) (3.48,1.461) (3.5,1.467) (3.52,1.498) (3.54,1.443) (3.56,1.419) (3.58,1.412) (3.6,1.420) (3.62,1.419) (3.64,1.455) (3.66,1.448) (3.68,1.466) (3.7,1.483) (3.72,1.466) (3.74,1.439) (3.76,1.467) (3.78,1.484) (3.8,1.501) (3.82,1.514) (3.84,1.464) (3.86,1.453) (3.88,1.479) (3.9,1.485) (3.92,1.503) (3.94,1.529) (3.96,1.529) (3.98,1.533) (4,1.556) (4.02,1.572) (4.04,1.572) (4.06,1.582) (4.08,1.646) (4.1,1.726) (4.12,2.237) (4.14,2.577) (4.16,2.856) (4.18,3.002) (4.2,3.048) (4.22,3.505) (4.24,3.491) (4.26,3.663) (4.28,3.650) (4.3,3.849) (4.32,3.866) (4.34,3.972) (4.36,4.099) (4.38,4.113) (4.4,4.102) (4.42,4.221) (4.44,4.387) (4.46,4.452) (4.48,4.626) (4.5,4.645) (4.52,4.688) (4.54,4.755) (4.56,4.741) (4.58,4.812) (4.6,4.838) (4.62,4.875) (4.64,4.925) (4.66,4.897) (4.68,4.856) (4.7,4.785) (4.72,4.431) (4.74,4.102) (4.76,3.813) (4.78,3.704) (4.8,3.741) (4.82,3.274) (4.84,3.341) (4.86,3.189) (4.88,3.187) (4.9,2.999) (4.92,3.033) (4.94,2.971) (4.96,2.885) (4.98,2.859) (5,2.886) (5.02,2.837) (5.04,2.684) (5.06,2.649) (5.08,2.459) (5.1,2.546) (5.12,2.514)
\psline[linecolor=red](0.62,1.165) (0.64,1.149) (0.66,1.181) (0.68,1.125) (0.7,1.130) (0.72,1.152) (0.74,1.153) (0.76,1.055) (0.78,1.023) (0.8,1.016) (0.82,1.115) (0.84,1.125) (0.86,1.134) (0.88,1.123) (0.9,1.092) (0.92,1.119) (0.94,1.132) (0.96,0.986) (0.98,1.012) (1,1.028) (1.02,1.026) (1.04,1.090) (1.06,1.092) (1.08,1.086) (1.1,1.075) (1.12,1.071) (1.14,1.179) (1.16,1.189) (1.18,0.996) (1.2,1.011) (1.22,0.967) (1.24,1.036) (1.26,1.045) (1.28,1.139) (1.3,1.155) (1.32,1.141) (1.34,1.261) (1.36,1.306) (1.38,1.293) (1.4,1.328) (1.42,1.351) (1.44,1.229) (1.46,1.219) (1.48,1.233) (1.5,1.305) (1.52,1.318) (1.54,1.271) (1.56,1.281) (1.58,1.320) (1.6,1.416) (1.62,1.665) (1.64,1.617) (1.66,1.550) (1.68,1.518) (1.7,1.521) (1.72,1.547) (1.74,1.545) (1.76,1.469) (1.78,1.469) (1.8,1.409) (1.82,1.438) (1.84,1.457) (1.86,1.418) (1.88,1.440) (1.9,1.563) (1.92,1.554) (1.94,1.881) (1.96,1.752) (1.98,1.714) (2,1.738) (2.02,1.727) (2.04,1.720) (2.06,1.750) (2.08,1.759) (2.1,1.771) (2.12,1.730) (2.14,1.781) (2.16,1.707) (2.18,1.706) (2.2,1.716) (2.22,1.661) (2.24,1.423) (2.26,1.467) (2.28,1.480) (2.3,1.503) (2.32,1.574) (2.34,1.569) (2.36,1.492) (2.38,1.474) (2.4,1.443) (2.42,1.429) (2.44,1.364) (2.46,1.334) (2.48,1.399) (2.5,1.362) (2.52,1.223) (2.54,1.241) (2.56,1.085) (2.58,1.112) (2.6,1.092) (2.62,1.180) (2.64,1.207) (2.66,1.193) (2.68,1.179) (2.7,1.215) (2.72,1.225) (2.74,1.113) (2.76,1.151) (2.78,1.184) (2.8,1.171) (2.82,1.273) (2.84,1.555) (2.86,1.582) (2.88,1.566) (2.9,1.563) (2.92,1.751) (2.94,1.754) (2.96,1.788) (2.98,1.786) (3,1.789) (3.02,1.804) (3.04,1.819) (3.06,1.788) (3.08,1.783) (3.1,1.784) (3.12,1.731) (3.14,1.731) (3.16,1.692) (3.18,1.603) (3.2,1.687) (3.22,1.991) (3.24,2.099) (3.26,2.080) (3.28,2.110) (3.3,2.093) (3.32,2.028) (3.34,2.019) (3.36,2.033) (3.38,2.027) (3.4,1.983) (3.42,1.972) (3.44,1.911) (3.46,1.824) (3.48,1.808) (3.5,1.743) (3.52,1.728) (3.54,1.519) (3.56,1.550) (3.58,1.508) (3.6,1.574) (3.62,1.691) (3.64,1.713) (3.66,1.670) (3.68,1.774) (3.7,1.777) (3.72,1.806) (3.74,1.740) (3.76,1.733) (3.78,1.819) (3.8,1.839) (3.82,1.821) (3.84,1.680) (3.86,1.662) (3.88,1.723) (3.9,1.743) (3.92,1.759) (3.94,1.802) (3.96,1.757) (3.98,1.686) (4,1.651) (4.02,1.622) (4.04,1.731) (4.06,1.954) (4.08,2.073) (4.1,2.581) (4.12,3.015) (4.14,3.089) (4.16,3.114) (4.18,3.282) (4.2,3.885) (4.22,3.743) (4.24,3.908) (4.26,3.789) (4.28,4.078) (4.3,3.966) (4.32,4.129) (4.34,4.293) (4.36,4.285) (4.38,4.228) (4.4,4.312) (4.42,4.522) (4.44,4.376) (4.46,4.343) (4.48,4.184) (4.5,4.132) (4.52,4.089) (4.54,4.085) (4.56,4.065) (4.58,3.982) (4.6,4.075) (4.62,4.063) (4.64,4.084) (4.66,4.067) (4.68,4.078) (4.7,4.036) (4.72,3.826) (4.74,3.696) (4.76,3.654) (4.78,3.809) (4.8,3.779) (4.82,3.484) (4.84,3.716) (4.86,3.602) (4.88,3.702) (4.9,3.267) (4.92,3.313) (4.94,3.323) (4.96,3.263) (4.98,3.382) (5,3.418) (5.02,3.154) (5.04,2.776) (5.06,2.860) (5.08,2.900) (5.1,2.964) (5.12,2.955)
\end{pspicture}

\vskip 0.1 cm

{\noindent \small Figure 43: average volatility (blue) and average correlation (red) based on the log-returns for 1987, calculated in a moving window of 30 days and normalized so as to have mean two and standard deviation one, using Spearman's rank correlation.}

\vskip 0.7 cm

\begin{pspicture}(-1.5,-0.7)(3.5,2)
\psset{xunit=2.7,yunit=0.55}
\psline{->}(0,0)(5.6,0) \psline{->}(0,-1)(0,4) \rput(5.8,0){day} \rput(0.5,4){$<C>_n$,$\text{vol.}_n$}\scriptsize \psline(0.02,-0.18)(0.02,0.18) \rput(0.02,-0.54){01/02} \psline(0.44,-0.18)(0.44,0.18) \rput(0.44,-0.54){02/02} \psline(0.84,-0.18)(0.84,0.18) \rput(0.84,-0.54){03/02} \psline(1.28,-0.18)(1.28,0.18) \rput(1.28,-0.54){04/01} \psline(1.7,-0.18)(1.7,0.18) \rput(1.7,-0.54){05/04} \psline(2.1,-0.18)(2.1,0.18) \rput(2.1,-0.54){06/01} \psline(2.54,-0.18)(2.54,0.18) \rput(2.54,-0.54){07/01} \psline(3,-0.18)(3,0.18) \rput(3,-0.54){08/03} \psline(3.42,-0.18)(3.42,0.18) \rput(3.42,-0.54){09/01} \psline(3.86,-0.18)(3.86,0.18) \rput(3.86,-0.54){10/01} \psline(4.3,-0.18)(4.3,0.18) \rput(4.3,-0.54){11/02} \psline(4.72,-0.18)(4.72,0.18) \rput(4.72,-0.54){12/01} \psline(5.14,-0.18)(5.14,0.18) \rput(5.13,-0.54){12/31} \psline(-0.037,1)(0.037,1) \rput(-0.111,1){$1$} \psline(-0.037,2)(0.037,2) \rput(-0.111,2){$2$} \psline(-0.037,3)(0.037,3) \rput(-0.111,3){$3$}
\psline[linecolor=blue](1.42,1.204) (1.44,1.176) (1.46,1.169) (1.48,1.159) (1.5,1.153) (1.52,1.104) (1.54,1.043) (1.56,0.937) (1.58,0.867) (1.6,0.845) (1.62,0.852) (1.64,0.898) (1.66,0.860) (1.68,0.851) (1.7,0.889) (1.72,0.894) (1.74,0.913) (1.76,0.931) (1.78,0.935) (1.8,0.915) (1.82,0.919) (1.84,0.921) (1.86,0.852) (1.88,0.845) (1.9,0.821) (1.92,0.849) (1.94,0.848) (1.96,0.859) (1.98,0.871) (2,0.866) (2.02,0.835) (2.04,0.864) (2.06,0.954) (2.08,0.918) (2.1,0.908) (2.12,0.962) (2.14,0.963) (2.16,0.962) (2.18,0.956) (2.2,0.951) (2.22,0.945) (2.24,0.941) (2.26,0.955) (2.28,0.998) (2.3,1.021) (2.32,1.067) (2.34,1.023) (2.36,1.096) (2.38,1.086) (2.4,1.080) (2.42,1.106) (2.44,1.108) (2.46,1.113) (2.48,1.119) (2.5,1.125) (2.52,1.115) (2.54,1.083) (2.56,1.114) (2.58,1.133) (2.6,1.129) (2.62,1.128) (2.64,1.121) (2.66,1.120) (2.68,1.126) (2.7,1.135) (2.72,1.139) (2.74,1.176) (2.76,1.181) (2.78,1.178) (2.8,1.203) (2.82,1.201) (2.84,1.216) (2.86,1.235) (2.88,1.257) (2.9,1.247) (2.92,1.284) (2.94,1.278) (2.96,1.288) (2.98,1.315) (3,1.306) (3.02,1.343) (3.04,1.272) (3.06,1.311) (3.08,1.344) (3.1,1.302) (3.12,1.311) (3.14,1.424) (3.16,1.435) (3.18,1.462) (3.2,1.494) (3.22,1.510) (3.24,1.532) (3.26,1.541) (3.28,1.578) (3.3,1.700) (3.32,1.667) (3.34,1.685) (3.36,1.751) (3.38,1.917) (3.4,1.992) (3.42,2.054) (3.44,2.079) (3.46,2.081) (3.48,2.154) (3.5,2.169) (3.52,2.221) (3.54,2.265) (3.56,2.282) (3.58,2.429) (3.6,2.436) (3.62,2.498) (3.64,2.487) (3.66,2.450) (3.68,2.530) (3.7,2.492) (3.72,2.520) (3.74,2.585) (3.76,2.608) (3.78,2.628) (3.8,2.664) (3.82,2.690) (3.84,2.675) (3.86,2.741) (3.88,2.890) (3.9,2.931) (3.92,2.991) (3.94,3.056) (3.96,3.030) (3.98,3.093) (4,3.151) (4.02,3.288) (4.04,3.304) (4.06,3.323) (4.08,3.388) (4.1,3.448) (4.12,3.437) (4.14,3.481) (4.16,3.466) (4.18,3.464) (4.2,3.425) (4.22,3.447) (4.24,3.486) (4.26,3.489) (4.28,3.470) (4.3,3.535) (4.32,3.576) (4.34,3.570) (4.36,3.619) (4.38,3.615) (4.4,3.610) (4.42,3.577) (4.44,3.596) (4.46,3.557) (4.48,3.554) (4.5,3.556) (4.52,3.559) (4.54,3.422) (4.56,3.401) (4.58,3.386) (4.6,3.403) (4.62,3.444) (4.64,3.411) (4.66,3.393) (4.68,3.392) (4.7,3.283) (4.72,3.328) (4.74,3.374) (4.76,3.290) (4.78,3.133) (4.8,3.073) (4.82,3.047) (4.84,2.999) (4.86,2.921) (4.88,2.864) (4.9,2.893) (4.92,2.835) (4.94,2.798) (4.96,2.782) (4.98,2.638) (5,2.617) (5.02,2.604) (5.04,2.603) (5.06,2.617) (5.08,2.502) (5.1,2.516) (5.12,2.401) (5.14,2.328)
\psline[linecolor=red](1.42,1.534) (1.44,1.541) (1.46,1.543) (1.48,1.489) (1.5,1.390) (1.52,1.087) (1.54,0.763) (1.56,0.400) (1.58,0.318) (1.6,0.274) (1.62,0.618) (1.64,0.492) (1.66,0.340) (1.68,0.412) (1.7,0.417) (1.72,0.298) (1.74,0.389) (1.76,0.505) (1.78,0.494) (1.8,0.416) (1.82,0.528) (1.84,0.523) (1.86,0.267) (1.88,0.307) (1.9,0.405) (1.92,0.350) (1.94,0.335) (1.96,0.324) (1.98,0.330) (2,0.309) (2.02,0.350) (2.04,0.751) (2.06,0.748) (2.08,0.723) (2.1,0.963) (2.12,0.950) (2.14,0.934) (2.16,0.863) (2.18,0.868) (2.2,0.834) (2.22,0.798) (2.24,1.062) (2.26,1.097) (2.28,1.202) (2.3,1.410) (2.32,1.281) (2.34,1.414) (2.36,1.387) (2.38,1.378) (2.4,1.321) (2.42,1.373) (2.44,1.329) (2.46,1.285) (2.48,1.269) (2.5,1.302) (2.52,1.271) (2.54,1.249) (2.56,1.297) (2.58,1.309) (2.6,1.293) (2.62,1.311) (2.64,1.258) (2.66,1.241) (2.68,1.243) (2.7,1.218) (2.72,1.270) (2.74,1.308) (2.76,1.309) (2.78,1.381) (2.8,1.381) (2.82,1.408) (2.84,1.484) (2.86,1.487) (2.88,1.483) (2.9,1.550) (2.92,1.566) (2.94,1.534) (2.96,1.677) (2.98,1.670) (3,1.832) (3.02,1.822) (3.04,1.653) (3.06,1.723) (3.08,1.717) (3.1,1.809) (3.12,2.085) (3.14,2.089) (3.16,2.140) (3.18,2.118) (3.2,2.163) (3.22,2.183) (3.24,2.169) (3.26,2.181) (3.28,2.466) (3.3,2.482) (3.32,2.430) (3.34,2.572) (3.36,2.778) (3.38,2.913) (3.4,2.903) (3.42,2.948) (3.44,3.056) (3.46,2.828) (3.48,2.770) (3.5,2.950) (3.52,2.929) (3.54,3.009) (3.56,3.207) (3.58,3.165) (3.6,3.178) (3.62,3.138) (3.64,3.160) (3.66,3.163) (3.68,3.087) (3.7,3.060) (3.72,2.994) (3.74,3.234) (3.76,3.027) (3.78,3.036) (3.8,3.084) (3.82,3.080) (3.84,3.166) (3.86,3.315) (3.88,3.280) (3.9,3.311) (3.92,3.268) (3.94,3.216) (3.96,3.182) (3.98,3.153) (4,3.162) (4.02,3.127) (4.04,3.091) (4.06,3.140) (4.08,3.266) (4.1,3.255) (4.12,3.306) (4.14,3.265) (4.16,3.195) (4.18,3.137) (4.2,3.083) (4.22,3.094) (4.24,3.135) (4.26,3.134) (4.28,3.233) (4.3,3.323) (4.32,3.312) (4.34,3.367) (4.36,3.336) (4.38,3.241) (4.4,3.231) (4.42,3.171) (4.44,3.176) (4.46,3.184) (4.48,3.173) (4.5,3.247) (4.52,3.126) (4.54,2.924) (4.56,2.942) (4.58,2.976) (4.6,3.022) (4.62,2.943) (4.64,2.976) (4.66,2.960) (4.68,2.952) (4.7,2.747) (4.72,2.790) (4.74,2.786) (4.76,2.689) (4.78,2.479) (4.8,2.347) (4.82,2.315) (4.84,2.272) (4.86,2.212) (4.88,2.196) (4.9,2.247) (4.92,2.097) (4.94,2.050) (4.96,1.954) (4.98,1.751) (5,1.867) (5.02,1.850) (5.04,1.940) (5.06,1.942) (5.08,1.723) (5.1,1.710) (5.12,1.576) (5.14,1.543)
\end{pspicture}

{\noindent \small Figure 44: average volatility (blue) and average correlation (red) based on the log-returns for 1998, calculated in a moving window of 70 days and normalized so as to have mean two and standard deviation one, using Spearman's rank correlation.}

\vskip 0.2 cm

\begin{pspicture}(-2,0)(3.5,2.7)
\psset{xunit=2.5,yunit=0.6}
\psline{->}(0,0)(5.4,0) \psline{->}(0,0)(0,4) \rput(5.6,0){day} \rput(0.7,4){$<C>_n$,$<\text{vol.}>_n$}\scriptsize \psline(0.02,-0.2)(0.02,0.2) \rput(0.02,-0.6){01/02} \psline(0.46,-0.2)(0.46,0.2) \rput(0.46,-0.6){02/01} \psline(0.86,-0.2)(0.86,0.2) \rput(0.86,-0.6){03/01} \psline(1.3,-0.2)(1.3,0.2) \rput(1.3,-0.6){04/02} \psline(1.72,-0.2)(1.72,0.2) \rput(1.72,-0.6){05/01} \psline(2.18,-0.2)(2.18,0.2) \rput(2.18,-0.6){06/01} \psline(2.6,-0.2)(2.6,0.2) \rput(2.6,-0.6){07/02} \psline(3.04,-0.2)(3.04,0.2) \rput(3.04,-0.6){08/01} \psline(3.5,-0.2)(3.5,0.2) \rput(3.5,-0.6){09/03} \psline(3.9,-0.2)(3.9,0.2) \rput(3.9,-0.6){10/01} \psline(4.36,-0.2)(4.36,0.2) \rput(4.36,-0.6){11/01} \psline(4.8,-0.2)(4.8,0.2) \rput(4.8,-0.6){12/03} \psline(5.2,-0.2)(5.2,0.2) \rput(5.2,-0.6){12/31} \psline(-0.043,1)(0.043,1) \rput(-0.17,1){$1$} \psline(-0.043,2)(0.043,2) \rput(-0.17,2){$2$} \psline(-0.043,3)(0.043,3) \rput(-0.17,3){$3$}
\psline[linecolor=blue](1.62,2.128) (1.64,2.072) (1.66,2.038) (1.68,1.890) (1.7,1.962) (1.72,2.012) (1.74,2.011) (1.76,1.999) (1.78,2.004) (1.8,1.941) (1.82,1.943) (1.84,1.897) (1.86,1.800) (1.88,1.867) (1.9,1.852) (1.92,1.914) (1.94,1.935) (1.96,1.904) (1.98,2.013) (2,2.022) (2.02,2.077) (2.04,2.122) (2.06,2.102) (2.08,2.066) (2.1,2.047) (2.12,1.985) (2.14,1.984) (2.16,1.991) (2.18,1.998) (2.2,1.931) (2.22,1.949) (2.24,2.005) (2.26,1.932) (2.28,1.856) (2.3,1.744) (2.32,1.771) (2.34,1.871) (2.36,1.766) (2.38,1.807) (2.4,1.805) (2.42,1.817) (2.44,1.824) (2.46,1.839) (2.48,1.792) (2.5,1.798) (2.52,1.730) (2.54,1.708) (2.56,1.700) (2.58,1.705) (2.6,1.711) (2.62,1.525) (2.64,1.526) (2.66,1.422) (2.68,1.396) (2.7,1.430) (2.72,1.423) (2.74,1.426) (2.76,1.404) (2.78,1.238) (2.8,1.102) (2.82,0.985) (2.84,0.961) (2.86,0.933) (2.88,0.974) (2.9,0.973) (2.92,0.928) (2.94,0.798) (2.96,0.838) (2.98,0.672) (3,0.731) (3.02,0.733) (3.04,0.635) (3.06,0.606) (3.08,0.612) (3.1,0.633) (3.12,0.643) (3.14,0.629) (3.16,0.451) (3.18,0.508) (3.2,0.480) (3.22,0.417) (3.24,0.459) (3.26,0.490) (3.28,0.503) (3.3,0.510) (3.32,0.461) (3.34,0.483) (3.36,0.480) (3.38,0.417) (3.4,0.488) (3.42,0.490) (3.44,0.547) (3.46,0.533) (3.48,0.543) (3.5,0.541) (3.52,0.545) (3.54,0.552) (3.56,0.616) (3.58,0.619) (3.6,0.692) (3.62,0.725) (3.64,0.982) (3.66,1.037) (3.68,1.057) (3.7,1.307) (3.72,1.360) (3.74,1.348) (3.76,1.294) (3.78,1.543) (3.8,1.778) (3.82,2.024) (3.84,2.038) (3.86,2.089) (3.88,2.133) (3.9,2.308) (3.92,2.327) (3.94,2.248) (3.96,2.237) (3.98,2.328) (4,2.299) (4.02,2.260) (4.04,2.284) (4.06,2.393) (4.08,2.556) (4.1,2.580) (4.12,2.674) (4.14,2.690) (4.16,2.782) (4.18,2.833) (4.2,2.789) (4.22,2.852) (4.24,2.929) (4.26,2.938) (4.28,2.962) (4.3,2.933) (4.32,3.015) (4.34,3.126) (4.36,3.105) (4.38,3.059) (4.4,3.071) (4.42,3.156) (4.44,3.150) (4.46,3.117) (4.48,3.115) (4.5,3.100) (4.52,3.164) (4.54,3.271) (4.56,3.287) (4.58,3.296) (4.6,3.244) (4.62,3.302) (4.64,3.339) (4.66,3.325) (4.68,3.355) (4.7,3.348) (4.72,3.372) (4.74,3.387) (4.76,3.431) (4.78,3.340) (4.8,3.365) (4.82,3.383) (4.84,3.402) (4.86,3.567) (4.88,3.556) (4.9,3.459) (4.92,3.525) (4.94,3.501) (4.96,3.513) (4.98,3.618) (5,3.562) (5.02,3.612) (5.04,3.570) (5.06,3.556) (5.08,3.462) (5.1,3.504) (5.12,3.441) (5.14,3.416) (5.16,3.345) (5.18,3.339) (5.2,3.265)
\psline[linecolor=red](1.62,2.112) (1.64,2.119) (1.66,2.157) (1.68,2.002) (1.7,2.069) (1.72,2.074) (1.74,2.107) (1.76,2.144) (1.78,2.136) (1.8,2.032) (1.82,2.006) (1.84,2.021) (1.86,1.970) (1.88,1.970) (1.9,2.020) (1.92,2.057) (1.94,2.053) (1.96,2.258) (1.98,2.267) (2,2.260) (2.02,2.265) (2.04,2.266) (2.06,2.166) (2.08,2.143) (2.1,2.149) (2.12,2.083) (2.14,2.140) (2.16,2.100) (2.18,2.085) (2.2,2.145) (2.22,2.116) (2.24,2.135) (2.26,2.048) (2.28,2.006) (2.3,1.999) (2.32,2.083) (2.34,2.124) (2.36,2.039) (2.38,2.019) (2.4,2.068) (2.42,2.020) (2.44,2.020) (2.46,2.014) (2.48,2.037) (2.5,2.063) (2.52,2.087) (2.54,1.988) (2.56,1.955) (2.58,2.116) (2.6,2.084) (2.62,1.630) (2.64,1.473) (2.66,1.380) (2.68,1.582) (2.7,1.570) (2.72,1.555) (2.74,1.788) (2.76,1.691) (2.78,1.239) (2.8,1.245) (2.82,1.045) (2.84,1.074) (2.86,1.102) (2.88,1.021) (2.9,0.969) (2.92,1.048) (2.94,0.811) (2.96,0.788) (2.98,0.545) (3,0.546) (3.02,0.598) (3.04,0.429) (3.06,0.397) (3.08,0.404) (3.1,0.378) (3.12,0.397) (3.14,0.376) (3.16,0.150) (3.18,0.168) (3.2,0.205) (3.22,0.167) (3.24,0.169) (3.26,0.265) (3.28,0.419) (3.3,0.403) (3.32,0.411) (3.34,0.379) (3.36,0.362) (3.38,0.515) (3.4,0.594) (3.42,0.574) (3.44,0.568) (3.46,0.622) (3.48,0.522) (3.5,0.638) (3.52,0.508) (3.54,0.454) (3.56,0.540) (3.58,0.420) (3.6,0.492) (3.62,0.544) (3.64,0.628) (3.66,0.614) (3.68,1.108) (3.7,1.231) (3.72,1.188) (3.74,1.139) (3.76,1.457) (3.78,1.794) (3.8,1.866) (3.82,1.835) (3.84,1.885) (3.86,1.893) (3.88,2.224) (3.9,2.205) (3.92,2.166) (3.94,2.126) (3.96,2.231) (3.98,2.153) (4,2.213) (4.02,2.210) (4.04,2.264) (4.06,2.704) (4.08,2.673) (4.1,2.636) (4.12,2.831) (4.14,2.946) (4.16,2.985) (4.18,2.953) (4.2,2.786) (4.22,2.841) (4.24,2.886) (4.26,2.869) (4.28,2.905) (4.3,2.774) (4.32,3.022) (4.34,3.026) (4.36,2.907) (4.38,2.853) (4.4,2.984) (4.42,2.969) (4.44,2.958) (4.46,3.139) (4.48,3.098) (4.5,3.111) (4.52,3.169) (4.54,3.185) (4.56,3.197) (4.58,3.192) (4.6,3.203) (4.62,3.247) (4.64,3.229) (4.66,3.307) (4.68,3.278) (4.7,3.296) (4.72,3.278) (4.74,3.267) (4.76,3.257) (4.78,3.214) (4.8,3.209) (4.82,3.172) (4.84,3.457) (4.86,3.496) (4.88,3.438) (4.9,3.328) (4.92,3.335) (4.94,3.308) (4.96,3.435) (4.98,3.432) (5,3.336) (5.02,3.329) (5.04,3.299) (5.06,3.286) (5.08,3.229) (5.1,3.246) (5.12,3.193) (5.14,3.238) (5.16,3.361) (5.18,3.429) (5.2,3.276)
\end{pspicture}

\vskip 0.6 cm

{\noindent \small Figure 45: average volatility (blue) and average correlation (red) based on the log-returns for 2001, calculated in a moving window of 80 days and normalized so as to have mean two and standard deviation one, using Spearman's rank correlation.}

\begin{pspicture}(-1.5,-0.2)(3.5,3)
\psset{xunit=2.5,yunit=0.5}
\psline{->}(0,0)(5.4,0) \psline{->}(0,0)(0,5) \rput(5.6,0){day} \rput(0.5,5){$<C>_n$,$\text{vol.}_n$}\scriptsize \psline(0.02,-0.2)(0.02,0.2) \rput(0.02,-0.6){01/02} \psline(0.46,-0.2)(0.46,0.2) \rput(0.46,-0.6){02/01} \psline(0.88,-0.2)(0.88,0.2) \rput(0.88,-0.6){03/03} \psline(1.26,-0.2)(1.26,0.2) \rput(1.26,-0.6){04/01} \psline(1.7,-0.2)(1.7,0.2) \rput(1.7,-0.6){05/02} \psline(2.12,-0.2)(2.12,0.2) \rput(2.12,-0.6){06/02} \psline(2.54,-0.2)(2.54,0.2) \rput(2.54,-0.6){07/01} \psline(3,-0.2)(3,0.2) \rput(3,-0.6){08/01} \psline(3.42,-0.2)(3.42,0.2) \rput(3.42,-0.6){09/01} \psline(3.86,-0.2)(3.86,0.2) \rput(3.86,-0.6){10/02} \psline(4.3,-0.2)(4.3,0.2) \rput(4.3,-0.6){11/03} \psline(4.7,-0.2)(4.7,0.2) \rput(4.7,-0.6){12/01} \psline(5.06,-0.2)(5.06,0.2) \rput(5.06,-0.6){12/30} \psline(-0.043,1)(0.043,1) \rput(-0.17,1){$1$} \psline(-0.043,2)(0.043,2) \rput(-0.17,2){$2$} \psline(-0.043,3)(0.043,3) \rput(-0.17,3){$3$} \psline(-0.043,4)(0.043,4) \rput(-0.17,4){$4$}
\psline[linecolor=blue](2.02,1.531) (2.04,1.534) (2.06,1.527) (2.08,1.517) (2.1,1.505) (2.12,1.511) (2.14,1.505) (2.16,1.495) (2.18,1.500) (2.2,1.508) (2.22,1.480) (2.24,1.450) (2.26,1.466) (2.28,1.468) (2.3,1.361) (2.32,1.343) (2.34,1.329) (2.36,1.263) (2.38,1.254) (2.4,1.239) (2.42,1.234) (2.44,1.245) (2.46,1.256) (2.48,1.230) (2.5,1.241) (2.52,1.211) (2.54,1.197) (2.56,1.213) (2.58,1.226) (2.6,1.229) (2.62,1.200) (2.64,1.200) (2.66,1.210) (2.68,1.214) (2.7,1.210) (2.72,1.223) (2.74,1.210) (2.76,1.245) (2.78,1.231) (2.8,1.256) (2.82,1.250) (2.84,1.275) (2.86,1.273) (2.88,1.282) (2.9,1.259) (2.92,1.258) (2.94,1.245) (2.96,1.237) (2.98,1.242) (3,1.213) (3.02,1.210) (3.04,1.212) (3.06,1.192) (3.08,1.199) (3.1,1.137) (3.12,1.112) (3.14,1.112) (3.16,1.099) (3.18,1.065) (3.2,1.073) (3.22,1.062) (3.24,1.060) (3.26,1.088) (3.28,1.066) (3.3,1.050) (3.32,1.068) (3.34,1.069) (3.36,1.058) (3.38,1.050) (3.4,1.060) (3.42,1.058) (3.44,1.052) (3.46,1.030) (3.48,1.042) (3.5,1.056) (3.52,1.103) (3.54,1.126) (3.56,1.157) (3.58,1.173) (3.6,1.204) (3.62,1.219) (3.64,1.284) (3.66,1.348) (3.68,1.378) (3.7,1.388) (3.72,1.521) (3.74,1.521) (3.76,1.564) (3.78,1.559) (3.8,1.572) (3.82,1.590) (3.84,1.682) (3.86,1.683) (3.88,1.694) (3.9,1.693) (3.92,1.841) (3.94,1.885) (3.96,1.983) (3.98,1.984) (4,2.123) (4.02,2.241) (4.04,2.299) (4.06,2.410) (4.08,2.505) (4.1,2.503) (4.12,2.539) (4.14,2.530) (4.16,2.631) (4.18,2.669) (4.2,2.787) (4.22,2.855) (4.24,2.881) (4.26,2.947) (4.28,3.004) (4.3,3.045) (4.32,3.095) (4.34,3.177) (4.36,3.190) (4.38,3.273) (4.4,3.271) (4.42,3.275) (4.44,3.345) (4.46,3.392) (4.48,3.405) (4.5,3.380) (4.52,3.412) (4.54,3.448) (4.56,3.459) (4.58,3.537) (4.6,3.535) (4.62,3.606) (4.64,3.631) (4.66,3.606) (4.68,3.623) (4.7,3.623) (4.72,3.670) (4.74,3.682) (4.76,3.632) (4.78,3.638) (4.8,3.634) (4.82,3.715) (4.84,3.712) (4.86,3.707) (4.88,3.714) (4.9,3.714) (4.92,3.716) (4.94,3.720) (4.96,3.719) (4.98,3.707) (5,3.732) (5.02,3.713) (5.04,3.694) (5.06,3.694)
\psline[linecolor=red](2.02,1.586) (2.04,1.585) (2.06,1.581) (2.08,1.602) (2.1,1.608) (2.12,1.614) (2.14,1.617) (2.16,1.620) (2.18,1.636) (2.2,1.634) (2.22,1.633) (2.24,1.573) (2.26,1.582) (2.28,1.561) (2.3,1.397) (2.32,1.404) (2.34,1.422) (2.36,1.392) (2.38,1.392) (2.4,1.384) (2.42,1.373) (2.44,1.408) (2.46,1.419) (2.48,1.340) (2.5,1.311) (2.52,1.270) (2.54,1.305) (2.56,1.306) (2.58,1.311) (2.6,1.307) (2.62,1.243) (2.64,1.258) (2.66,1.229) (2.68,1.230) (2.7,1.207) (2.72,1.162) (2.74,1.240) (2.76,1.198) (2.78,1.233) (2.8,1.206) (2.82,1.198) (2.84,1.208) (2.86,1.254) (2.88,1.225) (2.9,1.182) (2.92,1.184) (2.94,1.143) (2.96,1.214) (2.98,1.179) (3,1.170) (3.02,1.208) (3.04,1.158) (3.06,1.146) (3.08,1.146) (3.1,0.978) (3.12,0.968) (3.14,0.983) (3.16,0.974) (3.18,0.937) (3.2,0.940) (3.22,0.946) (3.24,0.975) (3.26,0.977) (3.28,0.956) (3.3,0.915) (3.32,0.918) (3.34,0.909) (3.36,0.874) (3.38,0.901) (3.4,0.916) (3.42,0.914) (3.44,0.911) (3.46,0.897) (3.48,0.921) (3.5,0.958) (3.52,0.998) (3.54,0.992) (3.56,1.027) (3.58,1.072) (3.6,1.064) (3.62,1.251) (3.64,1.354) (3.66,1.340) (3.68,1.367) (3.7,1.563) (3.72,1.487) (3.74,1.571) (3.76,1.569) (3.78,1.566) (3.8,1.570) (3.82,1.653) (3.84,1.612) (3.86,1.591) (3.88,1.555) (3.9,1.784) (3.92,1.835) (3.94,2.030) (3.96,2.001) (3.98,2.204) (4,2.331) (4.02,2.470) (4.04,2.510) (4.06,2.664) (4.08,2.628) (4.1,2.636) (4.12,2.594) (4.14,2.760) (4.16,2.819) (4.18,2.997) (4.2,3.052) (4.22,3.042) (4.24,3.066) (4.26,3.101) (4.28,3.116) (4.3,3.145) (4.32,3.235) (4.34,3.195) (4.36,3.276) (4.38,3.259) (4.4,3.259) (4.42,3.363) (4.44,3.423) (4.46,3.416) (4.48,3.393) (4.5,3.425) (4.52,3.442) (4.54,3.455) (4.56,3.567) (4.58,3.545) (4.6,3.524) (4.62,3.556) (4.64,3.542) (4.66,3.588) (4.68,3.603) (4.7,3.627) (4.72,3.618) (4.74,3.605) (4.76,3.582) (4.78,3.579) (4.8,3.579) (4.82,3.582) (4.84,3.547) (4.86,3.555) (4.88,3.514) (4.9,3.520) (4.92,3.529) (4.94,3.523) (4.96,3.527) (4.98,3.485) (5,3.473) (5.02,3.475) (5.04,3.470) (5.06,3.493)
\end{pspicture}

\vskip 0.3 cm

{\noindent \small Figure 46: average volatility (blue) and average correlation (red) based on the log-returns for 2008, calculated in a moving window of 100 days and normalized so as to have mean two and standard deviation one, using Spearman's rank correlation.}

\vskip 0.2 cm

These four figures summarize what we have attempted here: to show that high correlation between world indices goes hand in hand with high volatility, possibly causing and definitely being caused by it.

\section{Conclusion and future research}

Using the correlation matrices of the log-returns of a diversity of market indices during times of crisis, we showed that markets tend to behave similarly during times of high volatility. In the process, we verified the results obtained in a diversity of articles, but now applied to world financial market indices, and not to equities. Some of those results are that the probability distributions of the eigenvalues of the correlation matrices show peaks that are far off the maximum values predicted by Random Matrix Theory. Another result was the presence of certain combinations of indices that emulate a joint movement of most indices in what is called a market mode. An analysis of the probability distributions of the correlation matrices obtained show that those distributions are not normal and tend to flatten (low kurtosis) in times of crisis.

We also showed that the relation of the average correlation and the average volatility (as calculated using the market mode) increases when one uses Spearman's rank correlation instead of Pearson's correlation, possibly highlighting nonlinear relations between them. The covariance between average correlation and the volatility of the market mode seems to be a good indicator of when periods of acute crises occur.

Some direction for future research is to analyze how the techniques used in this work are modified if we consider that the frequency distributions of the log-returns are not Gaussian. Another topic that is being pursued is to study the hierarchies between the many indices and its evolution in times of crisis. For that, we shall use a distance measure based on the correlation between indices and build Minimum Spanning Trees and also Asset Trees in order to study cluster formation between indices  \cite{Sand1}. Some of the results obtained here shall also be used in our studies of financial markets as coupled damped harmonic oscillators subject to stochastic perturbations \cite{SF}.

\vskip 0.6 cm

\noindent{\Large \bf Acknowledgements}

\vskip 0.4 cm

The authors thank for the support of this work by a grant from Insper, Instituto de Ensino e Pesquisa (L. Sandoval Jr.), and by a PIBIC grant from CNPq (I.P. Franca). This article was written using \LaTeX, all figures were made using PSTricks, and the calculations were made using Matlab and Excel. All data are freely available upon request on leonidassj@insper.edu.br.

\appendix

\section{Stock Market Indices}

The next table (table 2) shows the stock market indices we used, their original countries, the symbols we used for them in the main text, and their codes in Bloomberg. In the tables, we use ``SX'' as short for ``Stock Exchange''. Some of the indices changed names and/or method of calculation and are designated by the two names, prior to and after the changing date.

\[ \begin{array}{|l|l|c|c|} \hline \text{Index} & \text{Country} & \text{Symbol} & \text{Code in Bloomberg} \\ \hline \text{\bf North America} \\ \hline \text{S\&P 500} & \text{United States of America} & \text{S\&P} & \text{SPX} \\ \text{Nasdaq Composite} & \text{United States of America} & \text{Nasd} & \text{CCMP} \\ \text{S\&P/TSX Composite} & \text{Canada} & \text{Cana} & \text{SPTSX} \\ \text{IPC} & \text{Mexico} & \text{Mexi} & \text{MEXBOL} \\ \hline \text{\bf Central America} \\ \hline \text{Bolsa de Panama General} & \text{Panama} & \text{Pana} & \text{BVPSBVPS} \\ \text{BCT Corp Costa Rica} & \text{Costa Rica} & \text{CoRi} & \text{CRSMBCT} \\ \hline \text{\bf Caribbean} \\ \hline \text{Jamaica SX Market Index} & \text{Jamaica} & \text{Jama} & \text{JMSMX} \\ \hline \text{\bf British overseas territories} \\ \hline \text{Bermuda SX Index} & \text{Bermuda} & \text{Berm} & \text{BSX} \\ \hline \end{array} \]

\[ \begin{array}{|l|l|c|c|} \hline \text{Index} & \text{Country} & \text{Symbol} & \text{Code in Bloomberg} \\ \hline \text{\bf South America} \\ \hline \text{Ibovespa} & \text{Brazil} & \text{Braz} & \text{IBOV} \\ \text{Merval} & \text{Argentina} & \text{Arge} & \text{MERVAL} \\ \text{IPSA} & \text{Chile} & \text{Chil} & \text{IPSA} \\ \text{IGBC} & \text{Colombia} & \text{Colo} & \text{IGBC} \\ \text{IBC} & \text{Venezuela} & \text{Vene} & \text{IBVC} \\ \text{IGBVL} & \text{Peru} & \text{Peru} & \text{IGBVL} \\ \hline \text{\bf Western and Central Europe} \\ \hline \text{FTSE 100} & \text{United Kingdom} & \text{UK} & \text{UKX} \\ \text{ISEQ} & \text{Ireland} & \text{Irel} & \text{ISEQ} \\ \text{CAC 40} & \text{France} & \text{Fran} & \text{CAC} \\ \text{DAX} & \text{Germany} & \text{Germ} & \text{DAX} \\ \text{SMI} & \text{Switzerland} & \text{Swit} & \text{SMI} \\ \text{ATX} & \text{Austria} & \text{Autr} & \text{ATX} \\ \text{FTSE MIB or MIB-30} & \text{Italy} & \text{Ital} & \text{FTSEMIB} \\ \text{Malta SX Index} & \text{Malta} & \text{Malt} & \text{MALTEX} \\ \text{BEL 20} & \text{Belgium} & \text{Belg} & \text{BEL20} \\ \text{AEX} & \text{Netherlands} & \text{Neth} & \text{AEX} \\ \text{Luxembourg LuxX} & \text{Luxembourg} & \text{Luxe} & \text{LUXXX} \\ \text{OMX Stockholm 30} & \text{Sweden} & \text{Swed} & \text{OMX} \\ \text{OMX Copenhagen 20} & \text{Denmark} & \text{Denm} & \text{KFX} \\ \text{OMX Helsinki} & \text{Finland} & \text{Finl} & \text{HEX} \\ \text{OBX} & \text{Norway} & \text{Norw} & \text{OBX} \\ \text{OMX Iceland All-Share Index} & \text{Iceland} & \text{Icel} & \text{ICEXI} \\ \text{IBEX 35} & \text{Spain} & \text{Spai} & \text{IBEX} \\ \text{PSI 20} & \text{Portugal} & \text{Port} & \text{PSI20} \\ \text{Athens SX General Index} & \text{Greece} & \text{Gree} & \text{ASE} \\ \hline \text{\bf Eastern Europe} \\ \hline \text{PX or PX50} & \text{Czech Republic} & \text{CzRe} & \text{PX} \\ \text{SAX} & \text{Slovakia} & \text{Slok} & \text{SKSM} \\ \text{Budapest SX Index} & \text{Hungary} & \text{Hung} & \text{BUX} \\ \text{BELEX 15} & \text{Serbia} & \text{Serb} & \text{BELEX15} \\ \text{CROBEX} & \text{Croatia} & \text{Croa} & \text{CRO} \\ \text{SBI TOP} & \text{Slovenia} & \text{Slov} & \text{SBITOP} \\ \text{SASE 10} & \text{Bosnia and Herzegovina} & \text{BoHe} & \text{SASX10} \\ \text{MOSTE} & \text{Montenegro} & \text{Mont} & \text{MOSTE} \\ \text{MBI 10} & \text{Macedonia} & \text{Mace} & \text{MBI} \\ \text{WIG} & \text{Poland} & \text{Pola} & \text{WIG} \\ \text{BET} & \text{Romania} & \text{Roma} & \text{BET} \\ \text{SOFIX} & \text{Bulgaria} & \text{Bulg} & \text{SOFIX} \\ \text{OMXT} & \text{Estonia} & \text{Esto} & \text{TALSE} \\ \text{OMXR} & \text{Latvia} & \text{Latv} & \text{RIGSE} \\ \text{OMXV} & \text{Lithuania} & \text{Lith} & \text{VILSE} \\ \text{PFTS} & \text{Ukraine} & \text{Ukra} & \text{PFTS} \\ \hline \text{\bf Eurasia} \\ \hline \text{MICEX} & \text{Russia} & \text{Russ} & \text{INDEXCF} \\ \text{ISE National 100} & \text{Turkey} & \text{Turk} & \text{XU100} \\ \hline \end{array} \]

\[ \begin{array}{|l|l|c|c|} \hline \text{Index} & \text{Country} & \text{Symbol} & \text{Code in Bloomberg} \\ \hline \text{\bf Western and Central Asia} \\ \hline \text{KASE} & \text{Kazakhstan} & \text{Kaza} & \text{KZKAK} \\ \text{CSE} & \text{Cyprus} & \text{Cypr} & \text{CYSMMAPA} \\ \text{Tel Aviv 25} & \text{Israel} & \text{Isra} & \text{TA-25} \\ \text{Al Quds} & \text{Palestine} & \text{Pale} & \text{PASISI} \\ \text{BLOM} & \text{Lebanon} & \text{Leba} & \text{BLOM} \\ \text{ASE General Index} & \text{Jordan} & \text{Jord} & \text{JOSMGNFF} \\ \text{TASI} & \text{Saudi Arabia} & \text{SaAr} & \text{SASEIDX} \\ \text{Kwait SE Weighted Index} & \text{Kwait} & \text{Kwai} & \text{SECTMIND} \\ \text{Bahrain All Share Index} & \text{Bahrein} & \text{Bahr} & \text{BHSEASI} \\ \text{QE or DSM 20} & \text{Qatar} & \text{Qata} & \text{DSM} \\ \text{ADX General Index} & \text{United Arab Emirates} & \text{UAE} & \text{ADSMI} \\ \text{MSM 30} & \text{Ohman} & \text{Ohma} & \text{MSM30} \\ \hline \text{\bf South Asia} \\ \hline \text{Karachi 100} & \text{Pakistan} & \text{Paki} & \text{KSE100} \\ \text{SENSEX 30} & \text{India} & \text{Indi} & \text{SENSEX} \\ \text{Colombo All-Share Index} & \text{Sri Lanka} & \text{SrLa} & \text{CSEALL} \\ \text{DSE General Index} & \text{Bangladesh} & \text{Bang} & \text{DHAKA} \\ \hline \text{\bf Asia-Pacific} \\ \hline \text{Nikkei 25} & \text{Japan} & \text{Japa} & \text{NKY} \\ \text{Hang Seng} & \text{Hong Kong} & \text{HoKo} & \text{HSI} \\ \text{Shangai SE Composite} & \text{China} & \text{Chin} & \text{SHCOMP} \\ \text{MSE TOP 20} & \text{Mongolia} & \text{Mong} & \text{MSETOP} \\ \text{TAIEX} & \text{Taiwan} & \text{Taiw} & \text{TWSE} \\ \text{KOSPI} & \text{South Korea} & \text{SoKo} & \text{KOSPI} \\ \text{SET} & \text{Thailand} & \text{Thai} & \text{SET} \\ \text{VN-Index} & \text{Vietnam} & \text{Viet} & \text{VNINDEX} \\ \text{KLCI} & \text{Malaysia} & \text{Mala} & \text{FBMKLCI} \\ \text{Straits Times} & \text{Singapore} & \text{Sing} & \text{FSSTI} \\ \text{Jakarta Composite Index} & \text{Indonesia} & \text{Indo} & \text{JCI} \\ \text{PSEi} & \text{Philipines} & \text{Phil} & \text{PCOMP} \\ \hline \text{\bf Oceania} \\ \hline \text{S\&P/ASX 200} & \text{Australia} & \text{Aust} & \text{AS51} \\ \text{NZX 50} & \text{New Zealand} & \text{NeZe} & \text{NZSE50FG} \\ \hline \text{\bf Northern Africa} \\ \hline \text{CFG 25} & \text{Morocco} & \text{Moro} & \text{MCSINDEX} \\ \text{TUNINDEX} & \text{Tunisia} & \text{Tuni} & \text{TUSISE} \\ \text{EGX 30} & \text{Egypt} & \text{Egyp} & \text{CASE} \\ \hline \text{\bf Central and Southern Africa} \\ \hline \text{Ghana All Share Index} & \text{Ghana} & \text{Ghan} & \text{GGSEGSE} \\ \text{Nigeria SX All Share Index} & \text{Nigeria} & \text{Nige} & \text{NGSEINDX} \\ \text{NSE 20} & \text{Kenya} & \text{Keny} & \text{KNSMIDX} \\ \text{DSEI} & \text{Tanzania} & \text{Tanz} & \text{DARSDSEI} \\ \text{FTSE/Namibia Overall} & \text{Namibia} & \text{Nami} & \text{FTN098} \\ \text{Gaborone} & \text{Botswana} & \text{Bots} & \text{BGSMDC} \\ \text{FTSE/JSE Africa All Share} & \text{South Africa} & \text{SoAf} & \text{JALSH} \\ \text{SEMDEX} & \text{Mauritius} & \text{Maur} & \text{SEMDEX} \\ \hline \end{array} \]
\hskip 1.8 cm Table 2: names, codes, and abreviations of the stock market indices used in this article.


\begin{thebibliography}{99}


\bibitem{cont1} M. King and S. Wadhwani, {\sl Transmission of volatility between stock markets}, (1989) National Bureau of Economic Research working paper series, number 2910.

\bibitem{cont2} M. King, E. Sentana, and S. Wadhwani, {\sl Volatility and links between national stock markets}, (1990) National Bureau of Economic Research working paper series, number 3357.

\bibitem{cont3} J. Ammer and J. Mei, {\sl Measuring international economic linkages with stock market data}, (1993) Board of Governors of the Federal Reserve System, International finance discussion papers, number 449.

\bibitem{cont4} W-L Lin, R.F. Engle, and T. Ito, {\sl Do bulls and bears move across borders? International transmission of stock returns and volatility as the world turns}, Review of Financial Studies {\bf 7} (1994) 507-538.

\bibitem{cont5} C.B. Erb, C.R. Harvey, and T.E. Viskanta, {\sl Forecasting international equity correlations}, Financial Analyst Journal (November-December) (1994) 32-45.

\bibitem{cont6} T. Baig and I. Goldfajn, {\sl Financial market contagion in the Asian Crisis}, (1999) IMF Staff Papers {\bf 46}.

\bibitem{cont7} K. Forbes and R. Rigobon, {\sl No contagion, only interdependence: measuring stock market co-movements}, Journal of Finance {\bf 57} (2002) 2223-2261.

\bibitem{cont8} H. Jang and W. Sul, {\sl The Asian financial crisis and the co-movement of Asian stock markets}, Journal of Asian Economics {\bf 13} (2002) 94-104.

\bibitem{cont9} R. Rigobon, {\sl On the measurement of the international propagation of shocks: is the transmission stable?}, J. of Int. Economics {\bf 61} (2003), 261-283.

\bibitem{cont10} P. Hartmann, S. Straetmans, and C.G. De Vries, {\sl Asset market linkages in crisis periods}, The Review of Economics and Statitstics {\bf 86} (2004), 313-326.

\bibitem{cont11} G. Corsetti, M. Pericoli, and M. Sbracia, {\sl Some contagion, some interdependence. More pitfalls in tests of financial contagion}, J. of Int. Money and Finance {\bf 24} (2005), 1177-1199.

\bibitem{cont12} D. Baur and N. Schulze, {\sl Coexceedances in financial markets—a quantile regression analysis of contagion}, Emerging Markets Review {\bf 6} (2005), 21-43.

\bibitem{cont13} T.C. Chianga, B.N. Jeonb, and H. Lic, {\sl Dynamic correlation analysis of financial contagion: Evidence from Asian markets}, J. of Int. Money and Finance {\bf 26} (2007), 1206-1228.

\bibitem{cont14} D. Baur, and R.A. Fry, {\sl Multivariate contagion and interdependence}, J. of Asian Economics {\bf 20} (2009), 353-366.

\bibitem{cont15} A.G. Orlov, {\sl A cospectral analysis of exchange rate comovements during Asian financial crisis}, J. of Int. Financial Markets, Institutions and Money {\bf 19} (2009), 742-758.


\bibitem{time1} F. Longin and B. Solnik, {\sl Is the correlation in international equity returns constant: 1960-1990?}, J. of Int. Money and Finance {\bf 14} (1995) 3-26.

\bibitem{time2} G. Bekaert and C.R. Harvey, {\sl Time-varying world market integration}, The Journal of Finance {\bf 1} (1995) 403-444.

\bibitem{time3} G. de Santis and B. Gerard, {\sl International asset pricing and portfolio diversification with time-varying risk}, The Journal of Finance {\bf 52} (1997) 1881-1912.

\bibitem{time4} D.J. Fenn, M.A. Porter, S. Williams, M. McDonald, N.F. Johnsn, and N.S. Jones, {\sl Temporal evolution of financial market correlations}, preprint (2010), arXiv:1011.3225v1.


\bibitem{vol1} B. Solnik, C. Boucrelle, and Y. Le Fur, {\sl International market correlation and volatility}, Financial Analysts Journal {\bf 52} (1996) 17-34.

\bibitem{vol2} I. Meric and G. Meric, {\sl Co-movements of European equity markets before and after the 1987 crash}, Multinational Finance Journal {\bf 1} (1997) 137-152.

\bibitem{vol3} F. Longin and B. Solnik, {\sl Correlation structure of international equity markets during extremely volatile periods}, Le Cashiers de Reserche, HEC Paris {\bf 646} (1999).

\bibitem{vol4} P. Hartmann, S. Straetmans, and C.G. de Vries, {\sl Asset market linkages in crisis periods}, (2001) Tinbergen Institute Discussion Paper, TI 2001-71/2.

\bibitem{vol5} F. Lillo, G. Bonanno, and R.N. Mantegna, {\sl Variety of stock returns in normal and extreme market days: the August 1998 crisis}, Proceedings of Empirical Science of Financial Fluctuations, Econophysics on the Horizon, Edited by H. Takayasu (2001).

\bibitem{vol6} A. Ang and J. Chen, {\sl Asymmetric correlations of equity portfolios}, Journal of Financial Economics {\bf 63} (2002) 443-494.

\bibitem{vol7} F. Longin and B. Solnik, {\sl Extreme correlation of international equity markets}, The Journal of Finance {\bf 56} (2001) 649-675.

\bibitem{vol8} I. Meric, S. Kim, J.H. Kim, and G. Meric, {\sl Co-movements of U.S., U.K., and Asian stock markets before and after September 11, 2001}, Journal of Money, Investiment and Banking {\bf 3} (2008) 47-57.

\bibitem{vol9} P. Cizeau, M. Potters, and J-P Bouchaud, {\sl Correlation structure of extreme stock returns}, Quantitative Finance {\bf 1} (2001) 217-222.

\bibitem{vol10} Y. Malevergne and D. Sornette, {\sl Investigating extreme dependences: concepts and tools}, Extreme Financial Risks (From dependence to risk management) (2006) Springer, Heidelberg.

\bibitem{vol11} R. Marshal and A. Zeevi, {\sl Beyond correlation: extreme co-movements between financial assets}, Working Paper, Columbia Business School (2002).

\bibitem{vol12} S.M. Bartram and Y-H Wang, {\sl Another look at the relationship between cross-market correlation and volatility}, Finance Research Letters {\bf 2} (2005) 75-88.

\bibitem{vol13} J. Knif, J. Kolari, and S. Pynnönen, {\sl What drives correlation between stock market returns?}, IMF Working Paper WP/07/157 (2007).

\bibitem{vol14} J. Maskawa and W. Souma, {\sl Large correlations as a signal of instability in stock market}, peprint (2010).

\bibitem{vol15} P-A. Reigneron, R. Allez, and J-P. Bouchaud, {\sl Principal regression analysis and the index leverage effect}, (2011) arXiv:1011.5810.


\bibitem{port1} R. Campbell and K. Koedijk, {\sl Covariance and correlation in international equity returns: a value-at-risk approach}, (2000).

\bibitem{port2} S. Pafka and I. Kondor, {\sl Noisy covariance matrices and portfolio optimization}, Eur. Phys. J. B {\bf 27} (2002) 277-280.

\bibitem{port3} B. Rosenow, V. Plerou, P. Gopikrishnan, and H.E. Stanley, {\sl Portfolio optimization and the random magnet problem}, Europhys. Lett. {\bf 59} (2002) 500.

\bibitem{port4} A. Ang and G. Bekaert, {\sl International asset allocation with regime shifts}, The Review of Financial Studies {\bf 15} (2002) 1137-1187.

\bibitem{port5} S. Pafka and I. Kondor, {\sl Noisy covariance matrices and portfolio optimization II}, Physica A {\bf 319} (2003) 487-494.

\bibitem{port6} J-P. Onnela, A. Chakraborti, and K. Kaski, {\sl Dynamics of market correlations: taxonomy and portfolio analysis}, Phys. Rev. E {\bf 68} (2003) 056110.

\bibitem{port7} S. Sharifi, M. Crane, A. Shamaier, and H. Ruskin, {\sl Random matrix theory for portfolio optimization: a stability approach}, Physica A {\bf 335} (2004) 629-643.

\bibitem{port8} S. Pafka and I. Kondor, {\sl Estimated correlation matrices and portfolio optimization}, Physica A {\bf 343} (2004) 623-634.

\bibitem{port9} G. Papp, S. Pafka, N. Nowak, and I. Kondor, {\sl Random Matrix Filtering in Portfolio Optimization}, Acta Physica Polonica B {\bf 36} (2005) 2757-2766.

\bibitem{port10} T. Conlon, H.J. Ruskin, and M. Crane, {\sl Random Matrix Theory and fund of funds portfolio optimization}, Physica A {\bf 382} (2007) 565-576.

\bibitem{port11} V. Tola, F. Lillo, M. Gallegati, and R.N. Mantegna, {\sl Cluster analysis for portfolio optimization}, Journal of Economic Dynamics and Control {\bf 32} (2008) 235-258.

\bibitem{port12} E. Pantaleo, M. Tumminello, F. Lillo, and R.S. Mantegna, {\sl When do improved covariance matrix estimators enhance portfolio optimization? An empirical comparative study of nine estimators}, Working paper (2010) arXiv:1004.4272v1.

\bibitem{port13} F. Abergel and M. Politi, {\sl Optimizing a basket against the efficient market hypothesis}, preprint (2010) arXiv:1006.5230v1.


\bibitem{mod1} S.D. Bekiros and D.A. Goergoutsos, {\sl Estimating the correlation of international equity markets with multivariate extreme and GARCH}, CeNDEF Working paper 06-17, University of Amsterdam (2001).

\bibitem{mod2} S. Dro\.{z}d\.{z}, F. Grümer, R. Ruf, and J. Speth, {\sl Dynamics of competition between collectivity and noise in the stock market}, Physica A {\bf 287} (2000), 440-449.

\bibitem{mod3} C.A. Ball and W.N. Torous, {\sl Stochastic correlation across international stock markets}, Journal of Empirical Finance {\bf 7} (2000) 373-388.

\bibitem{mod4} B. Solnik and J. Roulet, {\sl Dispersion as cross-sectional correlation}, Financial Analyst Journal {\bf 56} (2000) 54-61.

\bibitem{mod5} P.C. Ivanov, B. Podobnik, Y. Lee, and H.E. Stanley, {\sl Truncaded Lévy process with scale-invariant behavior}, Physica A {\bf 299} (2001) 154-160.

\bibitem{mod6} T. Flavin, M.J. Hurley, and F. Rousseau, {\sl Explaining stock market correlation: a gravity model approach}, The Manchester School {\bf 70} (2002) 87-106.

\bibitem{mod7} F. Michael and M.D. Johnson, {\sl Financial market dynamics}, Physica A {\bf 320} (2003) 525-534.

\bibitem{mod8} P. Gopikrishnan, B. Rosenow, V. Plerou, and H.E. Stanley, {\sl Quantifying and interpreting collective behavior in financial markets}, Phys. Rev. E {\bf 64} (2001).

\bibitem{mod9} V.D. Skintzi, {\sl Dynamic correlation models}, (2003).

\bibitem{mod10} W-K Wong, J. Penm, R.D. Terrell, and K.Y.C. Lim, {\sl The relationship between stock markets of major developed countries and Asian emerging markets}, Journal of Applied Mathematics and Decision Sciences {\bf 8} (2004) 201-218.

\bibitem{mod11} J.D. Farmer, L. Gillemot, F. Lillo, S. Mike, and A. Sen, {\sl What really causes large price changes?}, Quantitative Finance {\bf 4} (2004) 383-397.

\bibitem{mod12} P. Repetowicz and P. Richmond, {\sl Removing noise from correlation in multivariate stock price data}, (2004) arXiv:cond-mat/0403177v1.

\bibitem{mod13} S.R.S. Durai and S.N. Bhaduri, {\sl Correlation dynamics in equity markets. Evidence from India}, (NSE) Working Paper No 51, (2009).

\bibitem{mod14} S.J. Hyde, D.P. Bredin, and N. Nguyen, {\sl Correlation dynamics between Asia-Pacific, EU and US stock returns}, Munich Personal RePEc Archive (2007), number 9681.

\bibitem{mod15} P. Ormerod, {\sl Random Matrix Theory and the evolution of business cycle synchronisation, 1886-2006}, Economics E-Journal 2 (2008).

\bibitem{mod16} M. Karanasos, {\sl The correlation structure of some financial time series models}, Quantitative and Qualitative Analysis in Social Sciences {\bf 1} (2007) 71-87.

\bibitem{mod17} A. Abdelwahab, O. Amor, and T. Abdelwahed, {\sl The analysis of the interdependence structure in international financial markets by graphical models}, International Research Journal of Finance and Economics {\bf 15} (2008) 291-306.

\bibitem{mod18} C. Genovese and R. Renò, {\sl Modeling international market correlations with high frequency data}, in Correlated Data Modelling 2004, Franco Angeli Editore, Milano, Italy (2008) 99-113.

\bibitem{mod19} T. Evans and D.G. McMillan, {\sl Financial co-movements and correlation: evidence from 33 international stock market indices}, International Journal of Banking, Accounting and Finance (2009) 215-241.


\bibitem{rmt1} E. P. Wigner, {\sl Characteristic vectors of bordered matrices with infinite dimensions}, Ann. Math. {\bf 62} (1955), 548-564.

\bibitem{rmt2} E. P. Wigner, {\sl On the distribution of the roots of certain symmetric matrices}, Ann. Math. {\bf 67} (1958), 325-327.

\bibitem{rmt3} V.A. Mar\v{e}nko and L.A. Pastur, USSR-Sb {\bf 1} (1967) 457-483.

\bibitem{rmt4} M. L. Mehta, {\sl Random Matrices}, (2004) Academic Press.

\bibitem{rmt5} L. Laloux, P. Cizeau, J-P Bouchaud, and M. Potters, {\sl Noise dressing of financial correlation matrices}, Phys. Rev. Letters {\bf 83} (1999) 1467-1470.

\bibitem{rmt6} L. Laloux, P. Cizeau, M. Potters, and J-P Bouchaud, {\sl Random Matrix Theory and financial correlations}, Mathematical Models and Methods in Applied Sciences (2000).

\bibitem{rmt7} V. Plerou, P. Gopikrishman, B. Rosenow, L. A.N. Amaral, and H. E. Stanley, {\sl Universal and non-universal properties of cross-correlations in financial time series}, Phys. Lett. {\bf 83} (1999) 1471-1474.

\bibitem{rmt8} V. Plerou, P. Gopikrishman, B. Rosenow, L. A.N. Amaral, and H. E. Stanley, {\sl Scaling of the distribution of fluctuations of financial market indices}, Phys. Rev. E {\bf 60} (1999) 5306-5316.

\bibitem{rmt9} V. Plerou, P. Gopikrishman, B. Rosenow, L. A.N. Amaral, and H. E. Stanley, {\sl A Random Matrix Theory approach to financial cross-correlations}, Physica A {\bf 287} (2000) 374-382.

\bibitem{rmt10} B. Rosenow, V. Plerou, P. Gopikrishnan, L.A.N. Amaral, and H.E. Sanley, {\sl Application of Random Matrix Theory to study cross-correlations of stock prices}, Int. J. of Theoret. and Appl. Finance {\bf 3} (2002) 399-403.

\bibitem{rmt11} V. Plerou, P. Gopikrishman, B. Rosenow, L. A.N. Amaral, T. Guhr, and H. E. Stanley, {\sl Random matrix approach to cross-correlations in financial data}, Phys. Rev. E {\bf 65} (2002) 066126.

\bibitem{rmt12} P. Ormerod, {\sl The Convergence of European Business Cycles 1980-2004}, Acta Physica Polonica B {\bf 36} (2005) 2747-2756.

\bibitem{rmt13} V. Kulkarni and N. Deo, {\sl Correlation and volatility of an Indian stock market: a random matrix approach}, Eur. Phys. J. B {\bf 60} (2007) 101-109.

\bibitem{rmt14} J. Kwapièn, S. Dro\.{z}d\.{z}, and P. O\'swi\c{e}cimka, {\sl The bulk of the stock market correlation matrix is not pure noise}, Physica A {\bf 359} (2006) 589-606.

\bibitem{rmt15} J. Kwapièn, S. Dro\.{z}d\.{z}, A.Z. Górski, and P. O\'swi\c{e}cimka, {\sl Asymmetric matrices in an analysis of financial correlations}, Acta Physica Polonica B {\bf 37} (2006) 3039-3048.

\bibitem{rmt16} M. Potters, J-P Bouchaud, and L. Laloux, {\sl Financial applications of Random Matrix Theory: old laces and new pieces}, Proceedings of the Cracow conference on ``Applications of Random Matrix Theory to economy and other complex systems'' (2005).

\bibitem{rmt17} J-P Bouchaud, L. Laloux, M.A. Miceli, and M. Potters, {\sl Large dimension forecasting models and random singular value spectra}, European Physical Journal B {\bf 2} (2007) 201-207.

\bibitem{rmt18} R. Rak, S. Dro\.{z}d\.{z}, J. Kwapien, and P. O\'swi\c{e}cimka, {\sl Correlation matrix decomposition of WIG20 intraday fluctuations}, Acta Physica Polonica B {\bf 37} (2006) 3123-3132.

\bibitem{rmt19} D. Wilcox and T. Gebbie, {\sl An analysis of cross-correlations in an emerging market}, Physica A {\bf 375} (2007) 584-598.

\bibitem{rmt20} G. Birolli, J-P Bouchaud, and M. Potters, {\sl On the top eigenvalue of heavy-tailed random matrices}, Europhysics Letters {\bf 78} (2007) 10001.1-10001.5.

\bibitem{rmt21} A. Chakraborti, {\sl An outlook on correlations in stock prices}, Econophysics of Stock and other Markets - Proceedings of the Econophys-Kolkata II (2007) Springer Milan.

\bibitem{rmt22} A.C.R. Martins, {\sl Non-stationary correlation matrices and noise}, Physica A {\bf 379} (2007) 552-558.

\bibitem{rmt23} A.C.R. Martins, {\sl Random but not so much. A parametrization for the returns and correlation matrix of financial time series}, Physica A {\bf 383} (2007) 527-532.

\bibitem{rmt24} R.K. Pan and S. Sinha, {\sl Collective behavior of stock price movements in an emerging market}, Phys. Rev. E {\bf 76} (2007) 1-9.

\bibitem{rmt25} A. Chakraborti, M. Patriarca, and M.S. Santhanam, {\sl Financial time-series analysis: a brief overview}, Proceedings of the International Workshop "Econophys-Kolkata III" (2007) Springer, Milan.

\bibitem{rmt26} I.I. Dimov, P.N. Kolm, L. Maclin, and D.Y.C. Shiber, {\sl Hidden noise structure and random matrix models of stock correlations}, (2009) arXiv:0909.1383v3.

\bibitem{rmt27} A. Chakraborti, I.M. Toke, M. Patriarca, and F. Abergel, {\sl Econophysics: empirical facts and agent-based models}, submitted to Quantitative Finance.

\bibitem{rmt28} J-P Bouchaud and M. Potters, {\sl Financial applications of Random Matrix Theory: a short review}, to appear in the ``Handbook on Random Matrix Theory'', Oxford University Press.

\bibitem{rmt29} J.D. Noh, {\sl A model of correlations in stock markets}, Phys. Rev. E {\bf 61} (2000) 5981.

\bibitem{rmt30} R. Rak, J. Kwapie\'n, S. Dro\.{z}d\.{z}, and P. O\'swi\c{e}cimka, {\sl Cross-correlations in Warsaw Stock Exchange}, Acta Physica Polonica A {\bf 114} (2008) 561-568.

\bibitem{rmt31} T. Conlon, H.J. Ruskin, and M. Crane, {\sl Cross-correlations dynamics in financial time series}, Physica A {\bf 388} (2009) 705-714.

\bibitem{rmt32} M.C. Münnix, R. Schäfer, and T. Guhr, {\sl Compensating asynchrony effects in the calculation of financial correlations}, Physica A {\bf 389} (2010) 767-779.

\bibitem{rmt33} A. Namaki, G.R. Jafari, and R. Raei, {\sl Comparing TEPIX as an emerging market with efficiet market by Random Matrix Theory}, preprint (2010).

\bibitem{rmt34} G. Oh, C. Eom, F. Wang, W-S Jung, H.E. Stanley, and S. Kim, {\sl Statistical properties of cross-correlation in the Korean stock market}, Eur. Phys. J. B {\bf 79} (2011), 55-60.


\bibitem{world1} S. Maslov, {\sl Measures of globalization based on cross-correlations of world financial indices}, Physica A {\bf 301} (2001) 397-406.

\bibitem{world2} S. Dro\.{z}d\.{z}, F. Grümmer, F. Ruf, and J. Speth, {\sl Towards identifying the world stock market cross-correlations: DAX versus Dow Jones}, Physica A {\bf 294} (2001), 226-234.


\bibitem{h1} R.N. Mantegna, {\sl Hierarchical structure in financial markets}, The European Phys. J. B {\bf 11} (1999) 193.

\bibitem{h2} G. Bonanno, N. Vandewalle, and R.N. Mantegna, {\sl Taxonomy of stock market indices}, Phys. Rev. E {\bf 62}, (2000) R7615–R7618.

\bibitem{h3} G. Bonanno, F. Lillo, and R.N. Mantegna, {\sl Levels of complexity in financial markets}, Physica A {\bf 299} (2001) 16-27.

\bibitem{h4} G. Bonanno, F. Lillo, and R.N. Mantegna, {\sl High-frequency cross-correlation in a set of stocks}, Quantitative Finance {\bf 1} (2001) 96-104.

\bibitem{h5} P. Gopikrishnan, B. Rosenow, V. Plerou, and H.E. Stanley, {\sl Identifying business sectors from stock price flutuations}, Phys. Rev E {\bf 64} (2001) 35106.

\bibitem{h6} S. Micchichè, G. Bonanno, F. Lillo, and R.N. Mantegna, {\sl Degree stability of a minimum spanning tree of price return and volatility}, Physica A {\bf 324} (2003) 66-73.

\bibitem{h7} J.-P. Onnela, A. Chakraborti, K. Kaski, and J. Kertész, {\sl Dynamic asset trees and Black Monday}, Physica A {\bf 324} (2003) 247-252.

\bibitem{h8} J.-P. Onnela, A. Chakraborti, and K. Kaski, {\sl Dynamics of market correlations: taxonomy and portfolio analysis}, Phys. Rev. E {\bf 68} (2003) 1-12.

\bibitem{h9} J.-P. Onnela, A. Chakraborti, K. Kaski, and A. Kanto, {\sl Asset trees and asset graphs in financial markets}, Physica Scripta {\bf T106} (2003) 48-54.

\bibitem{h10} J.-P. Onnela, A. Chakraborti, K. Kaski, J. Kertész, and A. Kanto, {\sl Asset trees and asset graphs in financial markets}, Phys. Scripta T {\bf 106} (2003) 48-54.

\bibitem{h11} J.-P. Onnela, A. Chakraborti, K. Kaski, and J. Kertész, {\sl Dynamic asset trees and Black Monday}, Physica A {\bf 324} (2003) 247-252.

\bibitem{h12} J.-P. Onnela, K. Kaski, and J. Kertész, {\sl Clustering and information in correlation based financial networks}, Eur. Phys. J. B {\bf 38} (2004) 353-362.

\bibitem{h13} G. Bonanno, G. Caldarelli, F. Lillo, S. Miccichè, N. Vandewalle, and R.N. Mantegna, {\sl Networks of equities in financial markets}, The European Phys. J. B {\bf 38}, (2004) 363–371.

\bibitem{h14} C. Coronnello, M. Tumminello, F. Lillo, S. Micchichè, and R.N. Mantegna, {\sl Sector identification in a set of stock return time series traded at the London Stock Exchange}, Acta Phys. Pol. B {\bf 36} (2005) 2653-2679.

\bibitem{h15} T. Aste and T. Di Matteo, {\sl Correlation filtering in financial time series}, {\sl Noise and Fluctuations in Econophysics and Finance}, Proceedings of the SPIE 5848 (2005) 100-109.

\bibitem{h16} S. Sinha and R.K. Pan, {\sl Uncovering the internal structure of the Indian financial market: cross-correlation behavior in the NSE}, in ``Econophysics of markets and business networks'', Springer (2007) 215-226.

\bibitem{h17} C. Coronnello, M. Tumminello, F. Lillo, S. Micchichè, and R.N. Mantegna, {\sl Economic sector identification in a set of stocks traded at the New York Stock Exchange: a comparative analysis}, Proceedings of the SPIE, vol. 6601, 66010T (2007).

\bibitem{h18} M. Tumminello, T. Di Matteo, T. Aste, and R.N. Mantegna, {\sl Correlation based networks of equity returns sampled at different time horizons}, Eur. Phys. J. B {\bf 55} (2007) 209–217.

\bibitem{h19} R. Coelho, C.G. Gilmore, B. Lucey, P. Richmond, and S. Hutzler, {\sl The evolution of interdependence in world equity markets - evidence from minimum spanning trees}, Physica A {\bf 376} (2007) 455-466.

\bibitem{h20} C. Borghesi, M. Marsili, and S. Miccichè, {\sl Emergence of time-horizon invariant correlation structure in financial returns by subtraction of the market mode}, Phys. Rev. E {\bf 76} (2007), 026104.

\bibitem{h21} M. Tumminello, F. Liloo, and R.N. Mantegna, {\sl Kullback-Leiber distance as a measure of the information filtered from multivariate data}, Phys. Rev. E {\bf 76} (2007) 031123.

\bibitem{h22} M. Tumminello, F. Liloo, and R.N. Mantegna, {\sl Shrinkage and spectral filtering of correlation matrices: a comparison via the Kullback-Leiber distance}, Acta Physica Polonica B {\bf 38} (2007) 4079-4088.

\bibitem{h23}D. Garlaschelli, T. Di Matteo, T. Aste, G. Caldarelli, and M.I. Loffredo, {\sl Interplay between topology and dynamics in the World Trade Web}, Europ. Phys. J. B {\bf 57} (2007) 159-164.

\bibitem{h24} M. Ausloos and R. Lambiotte, {\sl Clusters or networks of economies? A macroeconomy study through gross domestic product}, Physica A {\bf 382} (2007) 16-21.

\bibitem{h25} J.G. Brida and W.A. Risso, {\sl Multidimensional minimal spanning tree: the Dow Jones case}, Physica A {\bf 387} (2008) 5205-5210.

\bibitem{h26} C. Eom, G. Oh, W.-S. Jung, H. Jeong, and S. Kim, {\sl Topological properties of stock networks based on minimal spanning tree and random matrix theory in financial time series}, Physica A {\bf 388} (2009), 900-906.

\bibitem{h27} M. Eryi\v{g}it and R. Eryi\v{g}it, {\sl Topological properties of stock networks based on minimal spanning tree and random matrix theory in financial time series}, Physica A {\bf 388} (2009), 900-906.

\bibitem{h28} J.C. Wong, H. Lian, and S.A. Cheong, {\sl Detecting macroeconomic phases in the Dow Jones Industrial Average time series}, Physica A {\bf 388} (2009), 4635-4645.

\bibitem{h29} Y.W. Goo, T.W. Lian, W.G. Ong, W.T. Choi, and S.A. Cheong, {\sl Financial atoms and molecules}, preprint (2009) arXiv:0903.2009v1.

\bibitem{h30} P. Sieczka and J.A. Ho{\l}yst, {\sl Correlations in commodity markets}, Physica A {\bf 388} (2009), 1621-1630.

\bibitem{h31} J. Kwapie\'{n}, S. Gworek, S. Dro\.{z}d\.{z}, and A. Górski, {\sl Analysis of a network structure of the foreign currency exchange market}, J. of Economic Interaction and Coordination {\bf 4} (2009), 55-72.

\bibitem{h32} J. Kwapie\'{n}, S. Gworek, and S. Dro\.{z}d\.{z}, {\sl Structure and evolution of the foreign exchange networks}, Acta Physica Polonica B {\bf 40} (2009), 175-194.

\bibitem{h33} G. Fagiolo, {\sl The international-trade network: gravity equations and topological properties}, J. of Economic Interaction and Coordination {\bf 5} (2010), 1-25.

\bibitem{h34} M. Tumminello, F. Liloo, and R.N. Mantegna, {\sl Correlation, hierarchies, and networks in financial markets}, Journal of Economic Behavior \& Organizations (2010) Article in press.

\bibitem{h35} J. He and M.W. Deem, {\sl Structure response in the world trade network}, Phys. Rev. Lett. {\bf 105} (2010), 198701-1 to 198701-4.

\bibitem{h36} K.M. Lee, J.-S. Yang, J. Lee, K.-I. Goh, and I.- M. Kim, {\sl Impact of the topology of global macroeconomic network on the spreading of economic crises}, preprint (2010), arXiv:1011.4336v1.

\bibitem{h37} S. Dro\.{z}d\.{z}, J. Kwapie\'{n}, and J. Speth, {\sl Coherent patterns in nulcei and in financial markets}, preprint (2010), arXiv:1009.1105v1.

\bibitem{h38} D.J. Fenn, M.A. Porter, P.J. Mucha, M. McDonald, S. Williams, N.F. Johnson, and N.S. Jones, {\sl Dynamical clustering of exchange rates}, preprint (2010) arXiv:0905.4912v2.

\bibitem{h39} M. Keskin, B. Deviren, and Y. Kocalkaplan, {\sl Topology of the correlation networks among major currencies using hierarchical structure methods}, Physica A {\bf 390} (2011), 719-730.

\bibitem{h40} Y. Zhang, G.H.T. Lee, J.C. Wong, J.L. Kok, M. Prusty, and S.A. Cheong, {\sl Will the US economy recover in 2010? A minimal spanning tree study}, to be published in Physica A (2011).


\bibitem{ext1} B. Podobnik, P.C. Iavanov, Y. Lee, A. Chessa, and H.E. Stanley, {\sl Systems with correlations in the variance: generating power-law tails in probability distributions}, Europhysics Letters {\bf 50} (2000) 711-717.

\bibitem{ext2} P. Weber and B. Rosenow, {\sl Large stock price changes: volume or liquidity?}, Quantitative Finance {\bf 6} (2006) 7.

\bibitem{ext3} C. Biely and S. Thurner, {\sl Random matrix ensembles of time-lagged correlation matrices: derivation of eigenvalue spectra and analysis of financial time-series}, Quantitative Finance {\bf 8} (2008) 705-722.

\bibitem{ext4} G. Biroli, J-P Bouchaud, and M. Potters, {\sl The student ensemble of correlation matrices: eigenvalue spectrum and Kullback-Leibler entropy}, Acta Phys. Pol. B {\bf 38} (2007) 4009-4026.

\bibitem{ext5} Y. Malevergne and D. Sornette, {\sl Investigating extreme dependences: concepts and tools}, Extreme Financial Risks (From dependence to risk management) (2006) Springer, Heidelberg.

\bibitem{ext6} L. Borland, {\sl Statistical signatures in times of panic: markets as a self-organizing system}, (2010) arXiv:0908.0111v2.


\bibitem{w16} M. Rubinstein, {\sl Comments on the 1987 stock market crash: eleven years later}, Risks in Accumulation Products, Society of Actuaries (2000).


\bibitem{out1} B. Podobnik, P.C. Ivanov, Y. Lee, A. Chessa, and H.E. Stanley, {\sl Systems with correlations in the variance: generating power law tails in probability distributions}, Europhys. Lett. {\bf 50} (2000) 711-717.

\bibitem{out2} F. Lillo and R.N. Mantegna, {\sl Symmetry alteration on ensemble return distribution in crash and rally days of financial markets}, European Phys. J. B {\bf 15} (2000) 603-606.

\bibitem{out3} B. Podobnik, K. Matia, A. Chessa, P.C. Ivanov, Y. Lee, and H.E. Stanley, {\sl Time evolution of stochastic processes with correlations in the variance: stability in powerlaw tails of distributuions}, Physica A {\bf 300} (2001) 300-309.

\bibitem{out4} F. Lillo and J.D. Farmer, {\sl The long memory of the efficient market}, Studies in Nonlinear Dynamics
\& Econometrics {\bf 8} (2004) 1-33.

\bibitem{out6} J-P Bouchaud, J.D. Farmer, and F. Lillo, {\sl How markets slowly digest changes in supply and demand}, Handbook of financial markets: dynamics and evolution. Editors: H. Thorsten and K. Schenk-Hoppe (2008) Elsevier: Academic Press.

\bibitem{out7} F. Abergel, N. Huth, and I.M. Toje, {\sl Financial bubbles analysis with a cross-sectional estimator}, (2009) arXiv:0909.2885v1.

\bibitem{Elton} E.J. Elton, M.J. Gruber, S.J. Brown, and W. Goetzmann, {\sl Modern Portfolio Theory and Investment Analysis}, Eigth Edition , (2009) Wiley.

\bibitem{Bodie} Z. Bodie, A. Kane, and A.J. Marcus, {\sl Investments}, Eigth Edition , (2009) McGraww-Hill/Irwin.


\bibitem{JB} C.M. Jarque, and A.K. Bera, {\sl A test for normality of observations and regression residuals}, International Statistical Review, Vol. 55, No. 2, 1987, pp. 163-172.

\bibitem{Lilli} H.W. Lilliefors, {\sl On the Komogorov-Smirnov test for normality with mean and variance unknown}, Journal of the American Statistical Association, vol. 62, 1967, pp. 399-402.


\bibitem{Sand1} L. Sandoval Jr., {\sl Evolution of hierarchies and clustering of international financial market indices in times of crisis}, preprint (2011).

\bibitem{SF} L. Sandoval Jr. and I. De P. Franca, {\sl Shocks in financial markets, price expectation, and damped harmonic oscillators}, preprint (2011).

\end{thebibliography}
\end{document}